\documentclass[defaultstyle,11pt,english]{thesis}

\usepackage{color}

\usepackage[T1]{fontenc}

\usepackage{csquotes}
\usepackage{microtype}

\usepackage{ragged2e}

\usepackage{graphicx}
\usepackage{epstopdf}
\usepackage{caption}
\usepackage[export]{adjustbox}
\usepackage{longfbox}

\usepackage{enumitem}

\usepackage{amsmath}
\usepackage{amssymb}
\usepackage{bbold}
\usepackage{bm}
\usepackage{comment}
\usepackage{mathtools}
\usepackage{braket}

\usepackage{nicefrac}

\usepackage{units}

\usepackage{array}
\usepackage{booktabs}
\usepackage{multirow}
\usepackage{dcolumn}
\usepackage{tabularx}
\usepackage{makecell}
\newcolumntype{L}[1]{>{\raggedright\let\newline\\\arraybackslash\hspace{0pt}}m{#1}}
\newcolumntype{C}[1]{>{\centering\let\newline\\\arraybackslash\hspace{0pt}}m{#1}}
\newcolumntype{R}[1]{>{\raggedleft\let\newline\\\arraybackslash\hspace{0pt}}m{#1}}

\usepackage{float}

\let\Re\relax
\DeclareMathOperator{\Re}{Re}

\DeclareMathOperator{\Angstr}{\textrm{\AA}}

\usepackage{hyperref}
\usepackage{xcolor}
\hypersetup{
  colorlinks,
  linkcolor={red!50!black},
  citecolor={blue!50!black},
  urlcolor={blue!80!black}
}

\usepackage[capitalise,noabbrev]{cleveref}

\usepackage{rotating}

\usepackage{url}

\makeatletter
\g@addto@macro\@floatboxreset{\vspace{2pt}}
\makeatother

\usepackage{etoolbox}
\usepackage{xpatch}
\AtBeginEnvironment{quote}{\itshape}

\usepackage[ 
  backend=biber, 
  style=phys, 
  articletitle=true, 
  biblabel=brackets,
  pageranges=false, 
  backref=true, 
  sorting=none, 
  sortcites=true,
  sortcase=false,
  url=false, 
  eprint=false,
  doi=false, 
  date=year,
]{biblatex}
\usepackage{doi}
\DefineBibliographyStrings{english}{%
  backrefpage = {Cited on p.},
  backrefpages = {Cited on pp.},
}

\title{Lattice Dynamics of Energy Materials Investigated by Neutron Scattering}

\author{Tyler Chase} 
  {Sterling} 

\otherdegrees{
  B.S., Texas State University, 2016 \\
  M.S., University of Colorado Boulder, 2019 \\
  M.S., University of Colorado Boulder, 2023
}

\degree{Doctor of Philosophy} 
  {Ph.D., Physics}  

\dept{Department of} 
  {Physics} 

\advisor{Professor} 
  {Dmitry Reznik}  
\reader{Michael Hermele}
\readerThree{Minhyea Lee}
\readerFour{Sandeep Sharma}
\readerFive{Michael F. Toney}
\readerSix{Nicholas J. Weadock}

\abstract{ \OnePageChapter  

In this thesis, I discuss several basic science studies in the field of energy materials using neutron scattering as a probe for the lattice dynamics. To enable understanding of neutron scattering spectra, I also use computational and theoretical methods. These methods and neutron scattering in general are discussed in detail in \cref{chp:neutrons}. It is assumed that the reader is familiar with basic quantum mechanics as well as with solid state physics topics including the band theory of electrons, harmonic lattice dynamics, and molecular dynamics. For the unfamiliar reader, the details of electronic structure theory and lattice dynamics that are needed to understand the methods in \cref{chp:neutrons} are provided in \cref{chp:electrons,chp:phonons}. In the remaining chapters, these methods are applied to the study of several energy materials: cuprate La$_2$CuO$_4$, (hybrid) solar perovskite CH$_3$NH$_3$PbI$_3$, and thermoelectric clathrate Ba$_{8}$Ga$_{16}$Ge$_{30}$. 

In \cref{chp:lco_lda_u}, I calculate the lattice dynamics of La$_2$CuO$_4$, a parent compound of cuprate superconductors, with DFT including a correction applied to the electronic structure. Calculations usually predict the wrong electronic structure and, similarly, the wrong lattice dynamics. The correction improves the agreement between lattice dynamics calculations and inelastic neutron scattering data. This work was published in ref. \cite{sterling2021effect}.

In \cref{chp:mapi_diffuse}, I discuss how elastic neutron scattering can be used to understand the locally correlated lattice dynamics in hybrid perovskite CH$_3$NH$_3$PbI$_3$ (MAPI). MAPI has potential as a next-generation solar energy harvesting material and the local correlations have implications for the photoexcited carrier dynamics. Diffuse neutron and x-ray scattering spectra calculated with large supercell classical molecular dynamics are in excellent agreement with experiment. We use the details of the real space dynamics to understand the local correlations in real space at the atomic scale. This work and a related code were published in refs. \cite{weadock2023nature} and \cite{sterling_pynamic} respectively.

In \cref{chp:bgg}, I discuss the results of inelastic neutron scattering from thermoelectric clathrate Ba$_{8}$Ga$_{16}$Ge$_{30}$ (BGG). Clathrates are cubic crystals with many atoms in the unit cell. The lattice consists of a cage of connected "spheres" of atoms: in BGG the cage atoms are Ga and Ge. The structure is ordered \emph{on average}, but the occupations of Ga and Ge are randomly distributed. The Ga/Ge spheres contain well isolated and loosely bonded Ba atoms that "rattle" like the ball in a paint can, scattering phonons and limiting the phonon contribution to thermal conductivity. BGG is a semiconductor, so the low thermal conductivity makes it an efficient thermoelectric. Still, the microscopic mechanism of reduction in thermal conductivity isn't well understood. By comparing lattice dynamics calculations and inelastic neutron scattering we identify signatures of disorder in the spectra, and argue that the disorder of the Ga and Ge atoms, rather than anharmonicity (damping) due to the motion of the Ba atoms, is responsible for the reduction in thermal conductivity. This work was published in ref. \cite{roy2023occupational}.

In all materials studied here, anomalies in the neutron scattering data were explained in microscopic detail using computational methods, advancing the state of knowledge about energy relevant properties. Hopefully the reader agrees that neutron scattering combined with computation is a powerful method to study energy materials. 

}

\dedication[Dedication]{ 
  Thank you to all of the scientists that have come before me; cheers to all of the scientists that might come after me. 
}

\acknowledgements{ \OnePageChapter  
My parent, first and foremost, brought me into this world. Then they tried to raise me to be a decent, resilient person (it's not up to me to judge their success). In any case, I can't thank my mom and dad enough for helping me become me.

My parents also gave the world my brothers, Beau and Dillon. There is not enough room in this entire thesis to describe the positive ways my brothers have affected me, so I will just say this: they are my best friends and I love them. 

My uncle "Mackie" has shown me the small part of the world that I have seen so far. He is kind and fun. I would be a lot more closed minded (and broke) without him.  

Of course, there are countless other I have met along the way. Ben, Dylan, and Bob have been my friends since I was 6 years old. I have shared more laughs with them than any one person deserves. My first friends in grad school were Rob and Vitaly. I have ridden a lot of bikes with John, Jake, Pablo. I have had many influential (scientific and personal) discussions with John, Hope, and Seth. Claire looks out for my heart (and stomach). I am thankful for all the beautiful, creative culinary projects she has shared with me. I love all of you.

Many scientists have supported and encouraged me: Minhyea has treated me like a peer and introduced me to many other influential scientists in our field. Feng Ye took a turn as my advisor while I was at Oak Ridge. He is kind, supportive, and encouraging. 

Finally, I want to thank Dmitry. I was a stray engineer looking for a place in physics and he took me in. He has given me trust, patience, and the opportunity to succeed. His personal and professional support has been unwavering. Thanks!

The work presented in this dissertation was funced by Department of Energy Office of Basic Energy Sciences, Office of Science, under contracts No. DE-SC0006939 \& DE-SC0024117 and the U.S. Department of Energy, Office of Science, Office of Workforce Development for Teachers and Scientists, Office of Science Graduate Student Research (SCGSR) program. The SCGSR program is administered by the Oak Ridge Institute for Science and Education for the DOE under contract number DE‐SC0014664.  

}

\introductoryPage{
    All problems are materials science problems - \textit{unknown}
}

\LoFisShort  
\emptyLoT  
\begin{document}






\chapter{Introduction}
\label{chp:intro}

Just like the quote above says, \emph{all problems are materials science problems}. Our planet and its inhabitants continuously face many challenges: unreliable electricity, plastic pollution, dirty water, health crises etc. Let us focus on one of these examples: unreliable electricity. For some people, losing power might be life-threatening: e.g. going without life-saving medical equipment or not having heat during a blizzard. Advances in \emph{energy materials} could enable solar power systems, battery technologies, etc. that could power entire homes during blackouts. There are already "house batteries" that provide supplemental power to homes, reducing demand during peak-times and preventing rolling blackouts \cite{chen2020applications} and a huge amount of research effort is being invested into next generation solar materials that are more efficient and lower cost than ever before, improving accessibility \cite{huang2017understanding}.

While the applications above are unarguably important, much of the work that has advanced the state of the art today was not motivated by applications; rather, it was \emph{basic science research}. Consider, for example, superconducting magnets in MRI machines: MgB$_2$ is being investigated as a next generation material for this application \cite{parizh2016conductors}. However, discovery of superconducitivty in MgB$_2$ \cite{nagamatsu2001superconductivity} and the subsequent explanation of the physics, characterization of the material, etc. \cite{bohnen2001phonon,hinks2001complex,mazin2003electronic} were certainly not motivated by its potential application for MRI. Rather, many hard working and curious individuals were endeavouring to discover new physics and the applications emerged later. This is the position in which I find myself.

In this thesis, I will discuss several basic science studies in the field of energy materials. Most energy materials are ultimately of interest for their non-trivial electronic properties (e.g. Seebeck effect, superconductivity) and optoelectronic properties (photovoltaic effect, light emission by LED's); the electrons interact with the atoms, so the atoms usually behave non-trivially too. In many systems (e.g. cuprates), the dynamics of the crystalline lattice shows a variety of anomalies of interest. I study these anomalies using neutron scattering. Nowadays, neutron scattering produces enormous datasets and, since the materials of interest are usually complicated, the datasets are complicated. To make any progress requires computational and theoretical methods. In particular, the most straightforward way to understand the neutron scattering spectra\footnote{Or, in many cases, to straightforwardly \emph{not} understand the spectra... by this I mean that when the theory clearly doesn't agree with experimental data, we usually have discovered something interesting (or have done something wrong).} are to simulate them and try to understand the origin of spectral features in the simulated dynamics. In some cases, we are lucky and interesting features in the spectra agree with calculations. 
In others, the calculations don’t agree and we endeavor to understand why. In either case, we need to calculate the dynamics of atoms in materials. 

In \cref{chp:neutrons}, I introduce neutron scattering and explain its usefulness as a probe for condensed matter physics. I then explain how we can calculate the scattering intensity in two ways: (i) a harmonic phonon expansion and (ii) approximating the quantum position operators as classical variables. We use these results extensively in this thesis. In \cref{chp:neutrons}, it is assumed that the reader is familiar with basic quantum mechanics as well as with solid-state physics topics including the band theory of electrons, harmonic lattice dynamics, and molecular dynamics. If this describes you, then you can read only this chapter and skip \cref{chp:electrons,chp:phonons}. On the other hand, if you are unfamiliar with modern electronic structure theory methods or the theory of lattice dynamics, read \cref{chp:electrons,chp:phonons} \footnote{These chapters (\cref{chp:neutrons,chp:electrons,chp:phonons}) are rather long; I have assimilated a significant collection of personal notes during graduate school and have reproduced the most relevant content here. The idea is that this thesis could serve a "recipe" for a beginner to learn how to use the methods I have used. The uninterested reader should skim or skip these chapters.}. 

\cref{chp:neutrons,chp:electrons,chp:phonons} introduce the theory needed to understand the methods of neutron scattering and the corresponding calculations. In the remaining chapters, \cref{chp:lco_lda_u,chp:mapi_diffuse,chp:bgg}, these methods are applied to study several energy materials: in \cref{chp:lco_lda_u}, I calculate the lattice dynamics of La$_2$CuO$_4$, a parent compound of cuprate superconductors, with DFT including a correction applied to the electronic structure. Calculations usually predict the wrong electronic structure and, similarly, the wrong lattice dynamics. The correction improves the agreement between lattice dynamics calculations and inelastic neutron scattering data. This work was published in ref. \cite{sterling2021effect}.

In \cref{chp:mapi_diffuse}, I discuss how elastic neutron scattering can be used to understand the locally correlated lattice dynamics in hybrid-perovskite CH$_3$NH$_3$PbI$_3$ (MAPI). MAPI has potential as a next-generation solar material and the local correlations have implications for the photoexcited carrier dynamics. Diffuse neutron and x-ray scattering spectra calculated with large-supercell classical molecular dynamics are in excellent agreement with experiment. We use the details of the real space dynamics to understand the local correlations in real space at the atomic scale. This work and a related code were published in refs. \cite{weadock2023nature} and \cite{sterling_pynamic} respectively.

In \cref{chp:bgg}, I discuss the results of inelastic neutron scattering from thermoelectric clathrate Ba$_{8}$Ga$_{16}$Ge$_{30}$ (BGG). Clathrates are cubic crystals with many atoms in the unit cell. The lattice consists of a cage of connected "spheres" of atoms: in BGG the cage atoms are Ga and Ge. The structure is ordered \emph{on average}, but the occupations of Ga and Ge are randomly distributed. The Ga/Ge spheres contain well-isolated and loosely bonded Ba atoms that "rattle" like the ball in a paint can, scattering phonons and limiting the phonon contribution to thermal conductivity. The low thermal conductivity makes BGG an efficient thermoelectric, but the microscopic mechanism for low thermal conductivity aren't well understood. By comparing lattice dynamics calculations and inelastic neutron scattering we identify signatures of disorder in the spectra, and argue that the disorder of the Ga and Ge atoms, rather than anharmonicity (damping) due to the motion of the Ba atoms, is responsible for the reduction in thermal conductivity. This work was published in ref. \cite{roy2023occupational}.




\chapter{Neutron Scattering}
\label{chp:neutrons}

The importance of discovery of the neutron in 1932 was emphasized by the prompt awarding of the Nobel prize in 1935 \cite{chadwick1932existence}. Shorly afterwards, nuclear fission was realized and the first reactor was designed and patented in 1934. A reliable source of neutron became available and, being scientists, Clifford Shull and Bertram Brockhouse shot the neutrons at stuff to see what would happen \cite{mason2013early}. What they found was that neutrons were diffracted by matter, similar to x-rays. Moreover, it was possible to analyze the energy distribution of the scattered neutrons, enabling direct measurement of dynamics of atoms \cite{hauser1952inelastic}. Together, Shull and Brockhouse were awarded the Nobel prize in 1994. Today, neutron scattering has become one of the most powerful probes for studying condensed matter; applications include studying solids like crystals and glass, soft matter like polymers and liquids, and even biological materials like DNA. In this chapter, I explain (i) why neutrons are so useful, (ii) the basic theory of neutron scattering, (iii) two different methods for calculating the intensity, and (iv) how some of the neutron scattering instruments used in this thesis work in practice. The discussion is kept as general as possible so as to be applicable to solids, soft matter, liquids, and so on except for \cref{sec:harmonic_cross_section} which is only applicable to crystals. 

This chapter is intended as an introduction to the physics of neutron scattering and the ways to calculate the neutron scattering intensity that are used in this thesis. It is assumed that the reader is familiar with basic quantum mechanics as well as with solid state physics topics including the band theory of electrons and harmonic lattice dynamics. Mention will also be made of molecular dynamics. If this describes you, then you can read only this chapter and skip Chapters \ref{chp:electrons} and \ref{chp:phonons}. If this does \emph{not} apply to you, then it is recommended to read all chapters. Moreover, if you are interested in learning some of the details of modern density functional theory calculations, it is recommended to read \cref{chp:electrons}.

This chapter mostly follows the excellent book by Squires \cite{squires1996introduction}, but I have also looked at ref. \cite{boothroyd2020principles} from time to time.

\section{Neutrons}

Neutrons are neutral spin 1/2 particles with the roughly the same mass, $m$, as the proton ($m \approx 1.008$ AMU). This simple sentence encodes almost everything about neutrons that makes them a useful probe for condensed matter! Let's see why.

There are many "types" of neutrons such as fast, cold, thermal, etc. These types refer to the energy of the neutron beam. We only care about thermal neutrons in this work. Thermal neutrons are in thermal equilibrium with matter at or near room temperature. In this sense, we can talk about the temperature of "a" neutron. For a thermal neutron, i.e. at $T=300$ K, the thermal de-Broglie wavelength is $\lambda_{th} = \sqrt{2\pi \hbar^2 / m k_B T} \approx 1~\Angstr$ with $k_B$ Boltzmann's constant and $\hbar$ the reduced Planck's constant. The wavelength is roughly the inter-atomic spacing in condensed matter, so neutrons scatter off atoms like waves and interference effects tell us about the distribution of atoms. 

Thermal neutrons have a characteristic energy $E=k_B T \approx 25$ meV. The atomic fluctuations in condensed matter that we care about have energies on the order of meV too\footnote{We only care about phonons here, but magnons have a similar energy scale}. This means that when a neutron creates or absorbs an excitation, its energy charges by a large fraction, enabling very good energy resolution in inelastic scattering experiments. Compare to x-rays which have wavelength $\sim 1~\Angstr$ so also can probe the distribution of atoms, but have energies $> 1000$ eV.

Unlike x-rays, neutrons are neutral and don't scatter off the charge in matter. This means (i) that they can penetrate deep into bulk matter and thus aren't sensitive to the surface and (ii) they get very close to the nuclei. They scatter from nuclei via the strong nuclear force. The exact scattering problem can't be solved, but the strong force has length scale $\sim 10^{-15}$ m, i.e. 5 orders of magnitude smaller than the thermal wavelength. Thus, an extremely good approximation is the "Fermi pseudopotential" $V(\bm{r}) = b (2\pi\hbar^2/m) \delta(\bm{r})$. $b$ is called the scattering length and varies wildly from atom to atom and even between isotopes of the same atom. Compare to x-rays: x-rays scatter off the charge of the atom, so are more sensitive to heavy elements. On the other hand, neutrons scatter very strongly even from hydrogen, making neutrons an excellent probe for organic matter. 

Finally, neutrons have spin 1/2. This isn't relevant in this thesis, but we mention it here for fun. The spin of the neutron couples to the unpaired spins on atoms in e.g. ferromagnets, antiferromagnets, etc. so neutrons can be used to measure magnetic structure and excitations (so called "magnons"). In fact, Landau proposed antiferromagnetism in 1933 \cite{landau1933possible} to explain the susceptibility of certain materials, but this was widely disbelieved until Shull scattered neutrons from antiferromagnets in 1949 \cite{shull1949detection}.

\section{Neutron scattering}

\begin{figure}[t!]
    \centering
    \includegraphics[width=0.6\linewidth]{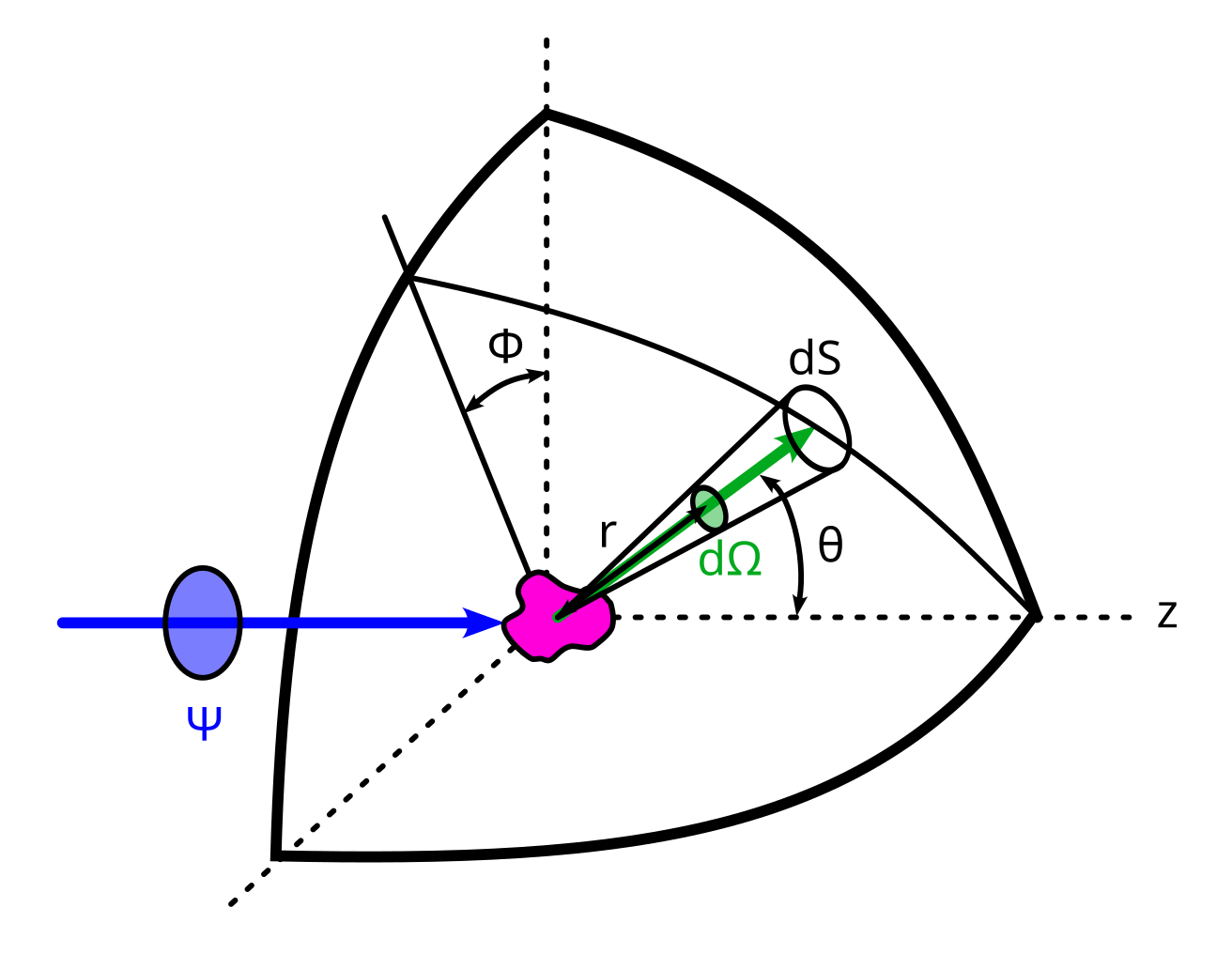}
    \caption{Diagram of a neutron scattering experiment as discussed in the text. The incident flux is $\psi$. The scattering system is the pink blob at the origin. The differential scattering angle, $d\Omega$, is labeled by the green circle.}
    \label{fig:scattering_diagram}
\end{figure}

Assume a scattering setup as shown in \cref{fig:scattering_diagram}. The incident beam of neutrons is along the $z$-axis. The incident momentum is $\bm{k}=k\hat{\bm{z}}$. The material we are probing is shown as a blob at the origin; it could be liquid, gas, etc. We call it the scattering system. The incident flux is $\Psi=$"number of incoming neutrons per second per unit area"; the area is perpendicular to the incident beam. The simplest quantity we can consider that characterizes the scattering system is the \emph{total cross section}, $\sigma$. It is defined as
\begin{equation}
    \sigma = \frac{\textrm{number of neutrons scattered per second}}{\Psi} .
\end{equation}
It has dimensions of area. It essentially describes how much of the incident beam is scattered by the scattering system. $\sigma$ really only tells us about the probability of the sample scattering neutrons; all of the atomic details are hidden. To make progress, we define another quantity called the \emph{differential cross section}. It is 
\begin{equation}
    \frac{d\sigma}{d\Omega} = \frac{\textrm{number of neutrons scattered into solid angle $d\Omega$ per second}}{\Psi d\Omega} .
\end{equation}
"Into solid angle $d\Omega$" means both in the direction of $d\Omega$ and into the infinitesimal area subtended by $d\Omega$. If we integrate over all angles, we recover the total cross section $\sigma = \int_\Omega (d\sigma/d\Omega) d\Omega$. The differential cross section is an experimentally accessible quantity; e.g. we can measure it to do neutron diffraction, since the change in direction of neutrons corresponds to change in momentum. Still, we don't know anything about the change in energy of the neutron, so we miss out on the dynamics. To that end, we finally define the \emph{doubly differential cross section}
\begin{equation}
    \frac{d^2\sigma}{d\Omega dE'} = \frac{\textrm{number of neutrons scattered into $d\Omega$ and into the interval $E'+dE'$ per second}}{\Psi d\Omega dE'} ,
\end{equation}
which integrates to the differential cross section: $d\sigma /d\Omega = \int (d^2 \sigma / d\Omega dE') dE'$ . The $E'$ is not the incident energy; it is the final energy. Usually the incident beam is \emph{monochromated} so that there is only a single incident energy. This quantity is also experimentally accessible and tells us about the excitations in the scattering system. It is the goal of inelastic scattering to measure this quantity. Let us now see how this tells about the excitations. 

\subsection{Expressions for the doubly differential cross section}

This section follows Chapter 2 in Squires \cite{squires1996introduction}. 

We place our system and neutrons in a really big box with volume $Y$ so that things are normalizable. The volume $Y$ drops out later, so everything is well and good and we can assume an infinite box. Anyway, we want to calculate $d^2\sigma/d\Omega dE'$. Let us now do that concretely. Let the scattering system be an arbitrary assembly of atoms in gas, liquid, solid phase, whatever. There are $N$ atoms. The coordinates of the atoms are represented by position operators $\hat{\bm{R}}$. The initial state of the scattering system is $|\lambda\rangle$. $\lambda$ labels the state: e.g. for phonons, it could represent the set of occupation numbers. The incident neutron state is $|\bm{k}\rangle$, which is just $\langle \bm{r} | \bm{k}\rangle 
 = \psi_{\bm{k}}(\bm{r})/\sqrt{Y}$. Let $|\bm{k}'\rangle$ and $|\lambda'\rangle$ represent the final states. 
 The differential cross section for a monochromatic beam with energy $E=\hbar^2k^2/2m$ is
 \begin{equation}
     \frac{d\sigma}{d\Omega} = \frac{1}{\Psi d\Omega} \sum_{\bm{k}'\parallel d\Omega} W_{\bm{k}\lambda\rightarrow\bm{k}'\lambda'}
 \end{equation}
 where the sum is over all $\bm{k}'$ that point in the direction of $d\Omega$ and $W_{\bm{k}\lambda\rightarrow\bm{k}'\lambda'}$ is the transition rate, i.e. number of neutrons scattered per second into $d\Omega$. We want to calculate this for a particle fixed set of $\bm{k}$, $\lambda$, and $\lambda'$. I.e. we want to know the number of neutrons scattered for a particular transition. Later, we can sum over all possible transitions $\lambda\rightarrow\lambda'$. For a particular transition, the sum is readily calculated with Fermi's golden rule \cite{sakurai2020modern}:
 \begin{equation}
    \sum_{\bm{k}'\parallel d\Omega} W_{\bm{k}\lambda\rightarrow\bm{k}'\lambda'} = \frac{2\pi }{\hbar} \rho_{\bm{k}'} |\langle \bm{k}' \lambda'| V | \bm{k} \lambda \rangle |^2 .
 \end{equation}
$\rho_{\bm{k}'}$ is the density of final neutron states with wave vector in the vicinity of $\bm{k}'$. Since the final state is a free particle, this is easy to calculate. The volume occupied by $\bm{k}'$ is (in spherical coordinates) $k'^2 dk' d\Omega$, while the number of states per volume in $k$-space is given $(2\pi)^3/Y$. This follows from our choice of normalization to a big box. Then 
\begin{equation}
    \rho_{\bm{k}'}=\frac{Y}{(2\pi)^3}k'^2dk'd\Omega \delta(E_{\bm{k}\lambda}-E_{\bm{k}'\lambda'}) 
\end{equation} 
where the $\delta$-function ensures energy conservation\footnote{In the derivation of Fermi's golden rule, it is shown that in the limit of long times compared to the time-scale of the scattering potential (in our case, the scattering system), the density of states limits to a $\delta$-function. See e.g. \S5.6 in Sakurai, Modern Quantum Mechanics, revised 2nd ed. In our case, there are prefactors associated with converting from momentum to energy.}. $E_{\bm{k}\lambda} = E_{\lambda} + E_{\bm{k}}$ etc, with $E_{\bm{k}}\equiv E$ and $E_{\bm{k}'}\equiv E'$.

The incident neutron flux, $\Psi$, is just the density of neutrons times their velocity. We assume all incident neutrons are along $z$ with momentum $\hbar k$ and there is only one at a time. Then there is one incident neutron per volume $Y$ and the speed is $\hbar k / m$ so that $\Psi = \hbar k / m Y$ in the direction of $z$. 

 Using the neutron free particle wavefunctions, the matrix elements are 
 \begin{equation}
     \langle \bm{k}' \lambda'| V | \bm{k} \lambda \rangle = \frac{1}{Y}\int_Y \langle \lambda' | V | \lambda \rangle \exp(-i(\bm{k}'-\bm{k})\cdot\bm{r}) d\bm{r} .
     \label{eq:neutron_matrix_element}
 \end{equation}
 For convenience, let us redefine $(1/Y) \int_Y \langle \lambda' | V | \lambda \rangle \exp(-i(\bm{k}'-\bm{k})\cdot\bm{r}) d\bm{r} \equiv \langle \bm{k}' \lambda'| V | \bm{k} \lambda \rangle / Y $. I.e. we place the normalization to the box outside of the matrix element. Then we arrive at the intermediate result
 \begin{equation}
     \frac{d\sigma}{d\Omega} = \frac{m}{(2\pi)^2 \hbar^2 k } k'^2dk' |\langle \bm{k}' \lambda'| V | \bm{k} \lambda \rangle |^2 \delta(E_{\bm{k}\lambda}-E_{\bm{k}'\lambda'})
     \label{eq:diff_cross_sec_1}
 \end{equation}
which is independent of $Y$ as it should be. This is almost the result that we want! All that is left is to convert to an expression involving $E$. To that end, note that the neutron's final energy is $E'=\hbar^2k'^2/2m$. Then the momentum interval $dk'$ is equivalent to the energy interval $dE'=\hbar^2 k' dk' /m$ energy. Substituting this into \cref{eq:diff_cross_sec_1} and dividing by $dE'$, we find
\begin{equation}
    \frac{d^2\sigma}{d\Omega dE'} = \left(\frac{m}{2\pi \hbar^2} \right)^2\frac{k'}{k} |\langle \bm{k}' \lambda'| V | \bm{k} \lambda \rangle |^2 \delta(E_{\bm{k}\lambda}-E_{\bm{k}'\lambda'}) .
    \label{eq:doubly_diff_cross_sec_intermediate} 
\end{equation}

This is great! However, we still need to evaluate the matrix elements. Earlier, we introduced the so-called Fermi pseudopotential for neutron-nuclear scattering. We extend that to an arbitrary assembly of atoms. $b_i$ is the scattering length of atom $i$. Recall that the real interaction is the strong nuclear force and the exact scattering problem is insoluble. We parameterize the scattering with the Fermi pseudopotential. $b_i$ is different for different elements and even different isotopes of the same elements. We can't calculate the scattering lengths, but they have been very accurately measured and tabulated \cite{sears1992neutron}. We note that it is possible for $b_i$ to be complex with the imaginary part corresponding to absorption of neutrons. This is bad for two reasons (i): in a experiment, the incident flux is absorbed, reducing the scattered signal and screwing up experiments and (ii) absorbing elements become "activated" and are radioactive. Fortunately, not many elements are strong absorbers of neutrons. For most elements and isotopes, $b_i$ are real but may be positive or negative.

There is also a distinction between "coherent" and "incoherent" scattering that is usually made that we ignore here. See Section 2.4 in Squires \cite{squires1996introduction}. The nuclei are quantum particles and the nuclear+neutron system has a spin. The usual assumption is that the total nuclear+neutron spins are spatially uncorrelated and that every configuration of spins is equally likely. The cross section of a single species is separated into a "coherent" part, $\sim b_{coh}^2$ with $i\neq j$, and an "incoherent" part, $\sim b_{inc}^2$ with $i\equiv i$. The incoherent part is independent of momentum, so is a nuisance for diffraction experiments where momentum dependence is essential information. Some experiments use the incoherent scattering for physical interpretation, but in this thesis, we just lump the incoherent part into background contributions. From here on out, we only explicitly consider the coherent contribution.

We return to the problem at hand. $b_i$ is the (coherent) scattering length for atom $i$ and $\hat{\bm{R}}_i$ is it's coordinate. Then 
\begin{equation}
    V = \frac{2\pi \hbar^2}{m} \sum_i  b_i \delta(\bm{r}-\hat{\bm{R}}_i)
    \label{eq:fermi_pseudopotential}
\end{equation}
and 
\begin{equation}
    \langle \bm{k}' \lambda'| V | \bm{k} \lambda \rangle = \frac{2\pi\hbar^2}{m} \sum_i b_i \int_Y \langle \lambda' | \delta(\bm{r}-\hat{\bm{R}}_i) | \lambda \rangle \exp(-i(\bm{k}'-\bm{k})\cdot\bm{r}) d\bm{r} .
\end{equation}
Remember that $|\lambda\rangle$ is the state of the scattering system; $\hat{\bm{R}}_i$ is an operator acting on $|\lambda\rangle$. Define $\bm{Q} = \bm{k}-\bm{k}'$ and use the Fourier transform representation of the $\delta$-function to find
\begin{equation}
\begin{gathered}
    \langle \bm{k}' \lambda'| V | \bm{k} \lambda \rangle =  \frac{2\pi\hbar^2}{m} \sum_i b_i \langle \lambda' | \exp(i\bm{Q}\cdot\hat{\bm{R}}_i) | \lambda \rangle   .
\end{gathered}
\end{equation}
Plugging this into \cref{eq:doubly_diff_cross_sec_intermediate}, we find
\begin{equation}
    \frac{d^2\sigma}{d\Omega dE'} = \frac{k'}{k} \left| \sum_i b_i \langle \lambda' | \exp(i\bm{Q}\cdot\hat{\bm{R}}_i) | \lambda \rangle \right|^2 \delta(E_{\bm{k}\lambda}-E_{\bm{k}'\lambda'}) .
\end{equation}
Next, introduce the Fourier transform representation of the energy $\delta$-function: 
\begin{equation}
    \delta(E_{\bm{k}\lambda}-E_{\bm{k}'\lambda'}) = \delta(E-E' + E_\lambda - E_{\lambda'}) = \frac{1}{2\pi\hbar} \int \exp(-i(E_\lambda - E_{\lambda'})t) \exp(-i\omega t) dt
\end{equation}
where $\omega = (E-E')/\hbar$ and $E$ and $E'$ are short-hand for $E_{\bm{k}}$ and $E_{\bm{k}'}$ respectively. Then 
\begin{equation}
    \frac{d^2\sigma}{d\Omega dE'} = \frac{k'}{k} \frac{1}{2\pi\hbar} \sum_{ij} b^*_i b_j \int  \langle \lambda | \exp(-i\bm{Q}\cdot\hat{\bm{R}}_i(0)) | \lambda' \rangle \langle \lambda' |  \exp(i\bm{Q}\cdot\hat{\bm{R}}_j(t)) | \lambda \rangle  \exp(-i\omega t) dt 
\end{equation}
where 
\begin{equation}
\begin{gathered}
    \langle \lambda' |  \exp(i\bm{Q}\cdot\hat{\bm{R}}_j(t)) | \lambda \rangle  =  \langle \lambda' | \exp(i\hat{H} t) \exp(i\bm{Q}\cdot\hat{\bm{R}}_j) \exp(-i \hat{H} t) | \lambda  \rangle \\
    =  \langle \lambda' | \exp(iE_\lambda't) \exp(i\bm{Q}\cdot\hat{\bm{R}}_j) \exp(-iE_\lambda t) | \lambda \rangle
\end{gathered}
\end{equation}
is the time evolved expectation value of the Fourier representation of the nuclear density of the scattering system. 

\subsection{Scattering from real systems}

So far, we have been considering the case that the only transition is $\lambda \rightarrow \lambda'$ and the associated constraints this imposed on the final state of the neutron. In real life, many different transitions can occur. Here, we relax the assumption and sum over all possible transitions. I am going to use the same symbol $d^2\sigma/d\Omega dE'$ to represent both the cross section for a single transition and the cross section summed over all possible transitions. From here on, unless otherwise specified, $d^2\sigma/d\Omega dE'$ represents the cross section summed over all possible transitions.  

The real cross section is a sum over all possible transitions to final states from a given initial state, averaged over all possible initial states. The first step, summing over all final states, is simply a sum over $\lambda'$. The later step, averaging over all initial states, is easily done if our system is in thermal equilibrium. We assume a canonical ensemble here. Then the average is a sum over all initial states weighted by the Bolztmann factor, $w_\lambda = \exp(-\beta E_\lambda)$ where $\beta = 1/ k_B T$, divided by the partition function $Z=\sum_\lambda w_\lambda$ for normalization. Then
\begin{equation}
\begin{gathered}
    \frac{1}{Z} \sum_\lambda w_\lambda \sum_{\lambda'} \langle \lambda | \exp(-i\bm{Q}\cdot\hat{\bm{R}}_i(0)) | \lambda' \rangle \langle \lambda' |  \exp(i\bm{Q}\cdot\hat{\bm{R}}_j(t)) | \lambda \rangle) \\
    = \frac{1}{Z} \sum_\lambda  w_\lambda  \langle \lambda | \exp(-i\bm{Q}\cdot\hat{\bm{R}}_i(0)) \exp(i\bm{Q}\cdot\hat{\bm{R}}_j(t)) | \lambda \rangle \\
    \equiv \langle  \exp(-i\bm{Q}\cdot\hat{\bm{R}}_i(0)) \exp(i\bm{Q}\cdot\hat{\bm{R}}_j(t))  \rangle 
\end{gathered}
\end{equation}
where $\langle \hat{A} \rangle = Z^{-1} \sum_\lambda \exp(-\beta E_\lambda)  \langle \lambda | \hat{A} | \lambda \rangle $ is the thermal average of some operator $\hat{A}$. Then the final expression for the doubly differential cross section is 
\begin{equation}
    \frac{d^2\sigma}{d\Omega dE'} = \frac{k'}{k}  \sum_{ij} b^*_i b_j \int \frac{dt}{2\pi\hbar}\langle \exp(-i\bm{Q}\cdot\hat{\bm{R}}_i(0))  \exp(i\bm{Q}\cdot\hat{\bm{R}}_j(t))  \rangle  \exp(-i\omega t) 
    \label{eq:doubly_differential_cross_section}
\end{equation}
This is the most important equation in neutron scattering. At face value, it tells us that the probability of a neutron scattering into a particular direction with a particular change in energy is proportional to the time and space Fourier transform of the atomic density. \cref{eq:doubly_differential_cross_section} is applicable to liquids and solids. 

We mention that very often what is discussed in the literature is not the "cross section" in \cref{eq:doubly_differential_cross_section} but the "dynamical structure factor" $S(\bm{Q},\omega)$:
\begin{equation}
    S(\bm{Q},\omega) = \sum_{ij} b^*_i b_j \int \frac{dt}{2\pi\hbar}\langle \exp(-i\bm{Q}\cdot\hat{\bm{R}}_i(0))  \exp(i\bm{Q}\cdot\hat{\bm{R}}_j(t))  \rangle  \exp(-i\omega t) 
\end{equation}
with $d^2\sigma/(d\Omega dE') = (k'/k) S(\bm{Q},\omega)$. The reason to define this quantity is that it only depends on the microscopic details of that scattering system where as the cross section depends on the incident momentum. These two quantities are usually talked about interchangeably.

\cref{eq:doubly_differential_cross_section} is tidy looking but is deceptively complicated. The complication comes from the fact that $\hat{\bm{R}}_i(0)$ and $\hat{\bm{R}}_j(t)$ are operators at different times; in general, they don't commute and calculating the thermal average is a complicated many-body problem. \cref{eq:doubly_differential_cross_section} can't be calculated exactly for realistic physical systems. Instead, we have to make approximations. How good the approximation is varies from material to material. The two approximations we use in this work are the harmonic approximation (i.e. applicable to crystals) and the classical approximation (i.e. $\hat{\bm{R}}\rightarrow \bm{R}$ are demoted from operators to classical coordinates) which is true for liquids, crystals, etc. but only at sufficiently high temperature. 

In the remainder of this chapter, we see how to solve \cref{eq:doubly_differential_cross_section} in detail. But for now, let us just make some remarks to conclude this section. Neutrons measure (in an average sense), the dynamics of the atoms in matter. The dynamics of the atoms are driven by the nuclear-nuclear and nuclear-electron interactions via the Hamiltonian, $\hat{H}$. Even for systems with strong electron-electron and electron-nuclear interactions, \cref{eq:doubly_differential_cross_section} is valid. The signatures of the interactions are readily measured with neutron scattering. All the neutrons see is the dynamics of the atoms. Thus, it should be clearer than ever that we need to know how to calculate the dynamics of atoms!

\section{The harmonic approximation for the cross section}\label{sec:harmonic_cross_section}

We first calculate the cross section assuming a crystal, i.e. that the atoms form an ordered, equilibrium, arrangement with average position $\bm{R}_i+\bm{\tau}_a$ and displacement from equilibrium $\hat{\bm{u}}_{ia}(t)$. The displacement is assumed to be small enough that the dynamics consists of oscillations around the equilibrium structure. The index $i$ labels the unit cell and $a$ the atom within the unit cell; $\bm{R}_i$ is the coordinate of origin of the $i^{th}$ unit cell and $\bm{\tau}_a$ the atom's position relative to $\bm{R}_i$. The instantaneous position is $\hat{\bm{R}}_{ia}(t) = \bm{R}_i + \bm{\tau}_a + \hat{\bm{u}}_{ia}(t)$.
Substitute into \cref{eq:doubly_differential_cross_section}:
\begin{equation}\begin{split}
    \frac{d^2\sigma}{d\Omega dE'} & = \frac{k'}{k} N \sum_{\bm{R}ab} b_a b_b \exp(i\bm{Q}\cdot \bm{R}) \exp(i\bm{Q}\cdot(\bm{\tau}_b-\bm{\tau}_a)) \\
    &\qquad \int \frac{dt}{2\pi\hbar} \exp(-i\omega t) \langle \exp( -i\bm{Q}\cdot\hat{\bm{u}}_{\bm{0}a}(0)) \exp( i\bm{Q}\cdot\hat{\bm{u}}_{\bm{R}b}(t))\rangle 
    \label{eq:cross_section_small_disp}
\end{split}\end{equation}
where we used translational invariance to reduce the sum over $i,j$ to a single sum over the whole crystal. From here on, we assume that the scattering lengths $b_a,~b_b$ are real. Besides assuming that the crystal oscillates around the equilibrium structure, we have made no approximations up to this point. The thermal expectation value in \cref{eq:cross_section_small_disp} is a complicated many-body expectation value whose evaluation is an arduous task. However, if we assume a harmonic lattice, we can readily evaluate the expectation value with a phonon expansion.

In Squires \cite{squires1996introduction}, it is shown that, for harmonic phonons, the expectation value can be simplified. Define $\hat{U}_{\bm{0}a} = -i\bm{Q}\cdot\hat{\bm{u}}_{\bm{0}a}(0)$ and $\hat{V}_{\bm{R}b} = i\bm{Q}\cdot\hat{\bm{u}}_{\bm{R}b}(t)$. Using some fancy operator identities\footnote{They aren't really relevant to understand the physics. The interested reader should refer to Sections 3.3 and 3.4 in Squires \cite{squires1996introduction}.}, the expectation value can be rewritten exactly (in the harmonic approximation) as 
\begin{equation}\begin{split}
    \langle \exp( \hat{U}_{\bm{0}a}) \exp( \hat{V}_{\bm{R}b})\rangle = \exp(\langle\hat{U}_{\bm{0}a}\hat{V}_{\bm{R}b}\rangle) \exp\left(\frac{1}{2}\left[\langle \hat{U}^2_{\bm{0}a} \rangle + \langle \hat{V}^2_{\bm{R}b}  \rangle \right]\right) .
    \label{eq:exp_UV_exact}
\end{split}\end{equation}
The second quantity, $\exp([\langle \hat{U}^2_{\bm{0}a} \rangle + \langle \hat{V}^2_{\bm{R}b}  \rangle ]/2)$, can be simplified further by noting that each expectation value contains only a single space-time coordinate. By assuming a crystal, we are assuming that each unit cell is identical on average so that we can shift $\bm{R}\rightarrow 0$. Moreover, we are assuming time-translational invariance so that we can also shift time $t\rightarrow 0$. Then $\hat{V}_{\bm{R}b} \rightarrow \hat{U}_{\bm{0}b}$ and $ \exp([\langle \hat{U}^2_{\bm{0}a} \rangle + \langle \hat{U}^2_{\bm{0}b}  \rangle ]/2 \equiv \exp(W_{ab})$. This expression is closely related to what will be called the \emph{Debye-Waller factor} (DWF), which corresponds to the attenuation of intensity from thermal motion of the atoms. I.e. as the atoms wiggle around, they stray away from the average positions and the intensity at e.g. Bragg peaks, which is scattering from the average positions, decreases. This expression is exact (in the harmonic approximation), but is intractable as it stands. To make progress, we further expand 
\begin{equation}
\begin{gathered}
    \exp(\langle \hat{U}_{\bm{0}a}\hat{V}_{\bm{R}b}\rangle) = \sum_n \frac{\langle \hat{U}_{\bm{0}a}\hat{V}_{\bm{R}b}\rangle^n}{n!} = 1 + \langle \hat{U}_{\bm{0}a}\hat{V}_{\bm{R}b}\rangle + \frac{1}{2}\langle \hat{U}_{\bm{0}a}\hat{V}_{\bm{R}b}\rangle^2 + \cdots
    \label{eq:exp_UV}
\end{gathered}
\end{equation}
Each term of order $n$ corresponds to a scattering process where $n$ phonons are created or annihilated. E.g. the first term, $n=0$, corresponds to creating $0$ phonons; this is Bragg scattering. The next term, $n=1$, is one-phonon scattering. This corresponds to creating/annihilating a phonon with momentum\footnote{We will often call the wave vector $\bm{Q}$ momentum which isn't really correct, but it should be easily understood that we actually mean $\bm{p}=\hbar\bm{Q}$.} $\pm\bm{Q}$ and energy $\pm
\omega$, changing the momentum of the neutron by $\mp\bm{Q}$ and its energy by $\mp \omega$. We will study the $0^{th}$ and $1^{st}$ order terms in detail. Higher order terms correspond to creating/annihilating multiple phonons at once which is a very low probability phenomenon. We won't consider terms beyond one-phonon.

\subsection{Bragg scattering}\label{sec:bragg_scattering}

According to the phonon expansion, the elastic cross section, corresponding to 0-phonon processes, is
\begin{equation}\begin{split}
    \left(\frac{d^2\sigma}{d\Omega dE'}\right)_0 & = N^2 \delta(\bm{Q}-\bm{G}) \sum_{ab} b_a b_b \exp(W_{ab}) \exp(i\bm{Q}\cdot(\bm{\tau}_b-\bm{\tau}_a))\delta(E-E')
\end{split}\end{equation}
where we used $\delta$-function identities to show that $\int \exp(-i\omega t ) dt/(2\pi\hbar) = \delta(\hbar\omega)$, with $\hbar\omega = E-E'$, i.e. the incident and scattered neutrons have the same energy, and $\sum_{\bm{R}}\exp(i\bm{Q}\cdot\bm{R}) = N \delta(\bm{Q}-\bm{G})$, which expresses conservation of crystal momentum. Note that since energy is conserved, $k=k'$ and $k'/k=1$. This expression corresponds to a neutron bouncing elastically off the average positions of the lattice without creating or absorbing any vibrations. This type of scattering is called "Bragg scattering" \cite{bragg1913reflection}.

We mention that capital $\bm{Q}$ is an "extended" wave vector, i.e. not confined to the first Brillouin zone of the scattering system. This is because the momentum transfer is not periodic with respect to the Brillouin zone (see \cref{sec:blochs_thm}). However, we can parameterize $\bm{Q}$ with respect to the crystal coordinates by writing $\bm{Q}=\bm{q}+\bm{G}$, with $\bm{q}$ a "reduced" wave vector confined to the first Brillouin zone and $\bm{G}$ the nearest reciprocal lattice vector to $\bm{Q}$. 

Since the scattering is elastic, it is convenient to integrate over all energies to calculate the (singly) differential cross section. The result is 
\begin{equation}\begin{split}
    \left(\frac{d\sigma}{d\Omega} \right)_0 & = N^2 \delta(\bm{Q}-\bm{G}) \left| \sum_{a} b_a \exp(W_{a}) \exp(i\bm{Q}\cdot\bm{\tau}_a) \right|^2
    \label{eq:elastic_cross_section}
\end{split}\end{equation}
with $W_a = -\langle ( \bm{Q}\cdot\hat{\bm{u}}_{\bm{0}a}(0) )^2 \rangle /2$ the \emph{Debye-Waller factor} (DWF). $\bm{G}$ is a reciprocal lattice vector of the crystal, i.e. an integral combination of the primitive lattice vectors (see \cref{sec:crystals}). We call the 0-phonon process "elastic" because the energy of the neutron is unchanged. The $\delta$-function enforces (crystal) momentum conservation.

\begin{figure}[t!]
    \centering
    \includegraphics[width=0.65\linewidth]{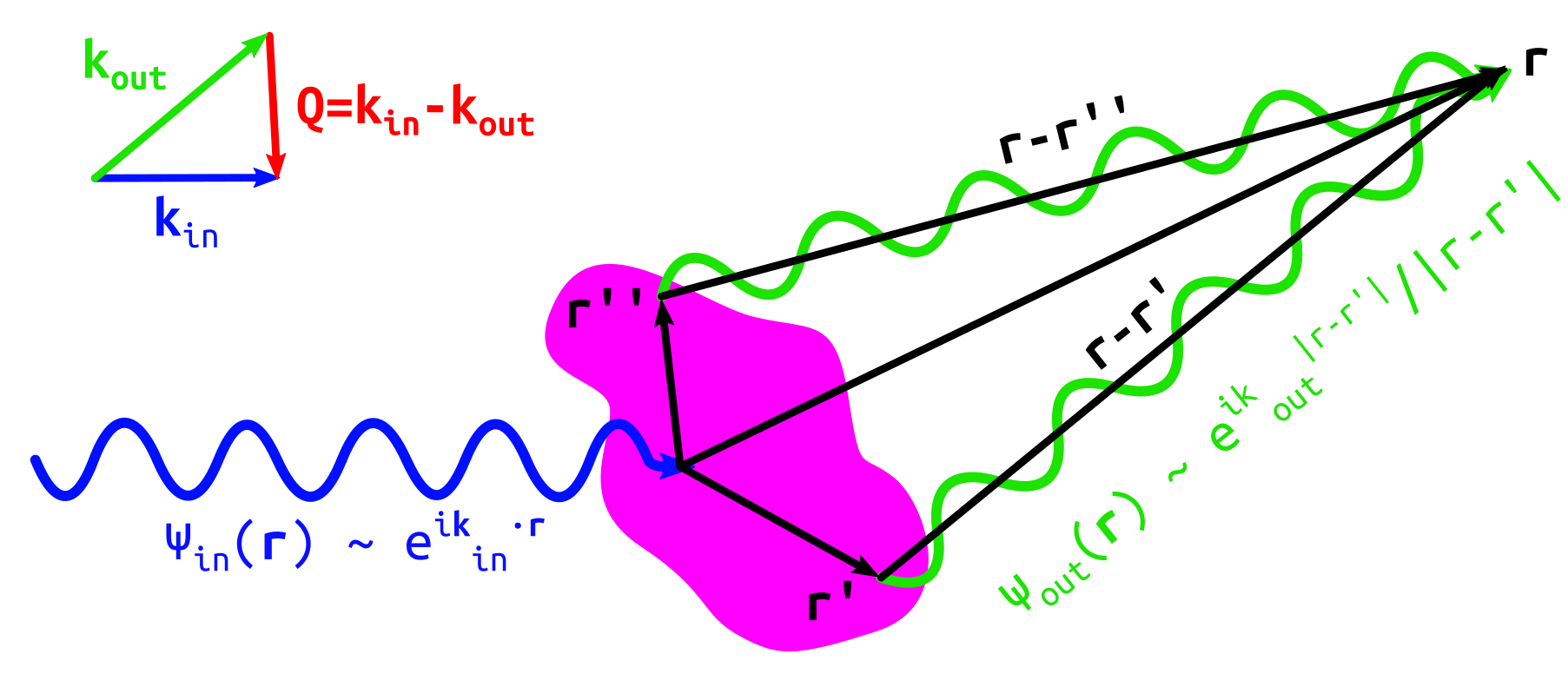}
    \caption{Diagram of plane waves scattering from matter. The magenta blob is the matter. The incident wave, $\psi_{in}(\bm{r})$, is a plane wave. The scattered waves, $\psi_{out}(\bm{r})$, are spherical waves. The detector is at $\bm{r}$.}
    \label{fig:classical_scattering_diagram}
\end{figure}

\cref{eq:elastic_cross_section} describes the net intensity measured at a detector for scattering from all atoms in the crystal; the net amplitude at the detector is, in general, the sum of all spherical waves scattered from every point in the sample. $\bm{Q} = \bm{k}_{in} - \bm{k}_{out}$ is the momentum transferred from the neutron into the sample. The detector is in the direction $\hat{\bm{k}}_{out}$ from the crystal
(see \cref{fig:classical_scattering_diagram}). 

The Bragg intensity is due to the interference of waves scattered from all of the scattering centers in the crystal. Since the crystal is periodic, we only have to sum over a single unit cell. The phase is the same from unit cell to unit cell since $\exp( i\bm{G}\cdot\bm{R}) = \exp(i2\pi n) = 1$. The DWF factor lowers the amplitude for scattering from a given atom at finite temperature. The DWF essentially describes the thermal fluctuations around the mean and, for larger fluctuations, the atom spends less time near the equilibrium positions and the amplitude for scattering from the mean positions is reduced. 

\begin{figure}[t!]
    \centering
    \includegraphics[width=0.9\linewidth]{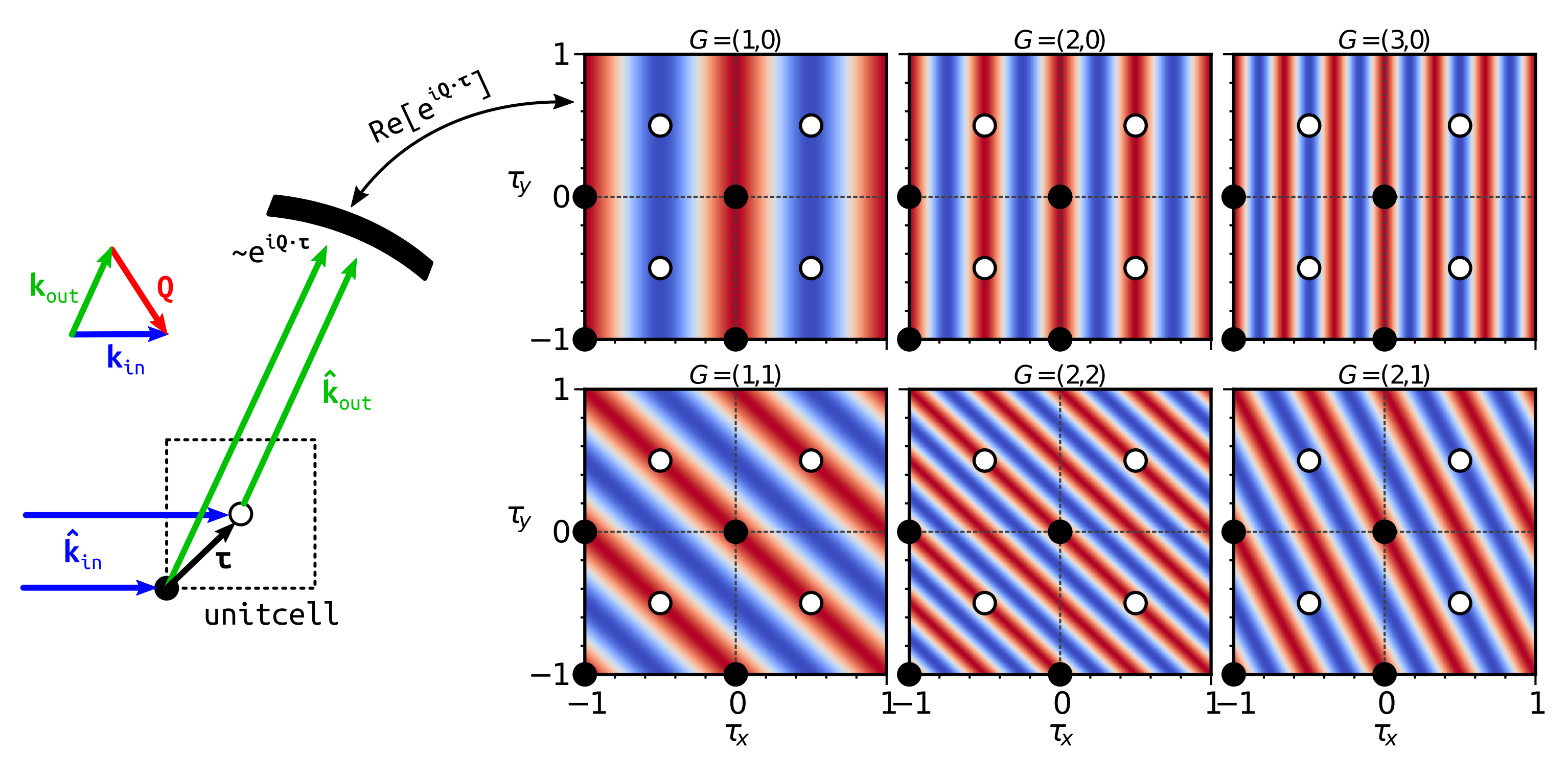}
    \caption{Schematic showing how the scattered neutron amplitude at the detector varies with position in the unit cell for fixed momentum transfer, $\bm{Q} = \bm{G}$ (Bragg peaks). The colormaps plot the real part of $\exp(i\bm{G}\cdot\bm{\tau})$, where $\bm{\tau}$ is a continuous coordinate spanning the unit cell, for several unit cells. The black and white circles label different atoms in the unit cell.}
    \label{fig:interference_diagram}
\end{figure}

The factor $|F(\bm{Q})|^2 = |\sum_{a} b_a \exp(i\bm{Q}\cdot\bm{\tau}_a) \exp(W_a)|^2$ is the \emph{unit cell structure factor}; it is the Fourier transform of the scattering centers in a single unit cell. For a Bravais crystal, $F(\bm{Q}) = b$ and the only scattering is from the unit cell origins and $(d\sigma/d\Omega)_0 \sim N^2 b^2$. With a basis, there is relative separation between the atoms in each unit cell that can further lead to inference effects at the detector. For particular $\bm{Q}$ and particular unit cells, it is possible to have purely constructive interference (very intense scattering peaks), purely destructive interference (no scattering), or something in between. The variation of the structure factor with $\bm{Q}$ is what allows scientists to determine the space group of crystals. This field is called "crystallography" and, while useful and intersting, has surprisingly little overlap with the contents of this thesis. We only provide a simple discussion here in the context of an example.

Consider the unit cell in \cref{fig:interference_diagram}. We consider a $2D$ model for simplicity. Suppose that the two atoms in the basis, the white and black cirlces, have scattering lengths $b_0$ and $b_1$ respectively. The colormaps show the phase (at the detector), $\exp(i\bm{Q}\cdot\bm{\tau})$, as a continuous function within each unit cell: i.e. the colormap plots the phase of the wave at the detector scattered from each point in the unit cell. The amplitude at the detector for scattering from the discrete set of atoms is a sum over the phases from the atoms' coordinates weighted by the scattering lengths: $F(\bm{Q}) = b_0 + b_1 \exp(i\bm{Q}\cdot\bm{\tau})$. 

We work in reduced units such that $\bm{\tau} = \tau_x \bm{a}_x + \tau_y \bm{a}_y$ with $\bm{a}_i$ the $i^{th}$ primitive lattice vector of the unit cell and $\bm{Q} = Q_x \bm{b}_x + Q_y \bm{b}_y$ with $\bm{b}_i \cdot \bm{a}_j = 2\pi\delta_{ij}$. $\bm{b}_i$ are the primitive vectors of the reciprocal lattice (see \cref{sec:crystals}). We only see intensity when momentum transfer is equal to a Bragg peak, so we only consider $\bm{Q} = \bm{G}$ in \cref{fig:interference_diagram}. The Bragg peaks are indexed by $\bm{G} = h \bm{b}_x + k \bm{b}_y $ with $h,k \in \mathbb{Z}$ integers. The coordinate of the basis atoms are $\bm{\tau}_0 = (0,0)$ and $\bm{\tau}_1 = (1/2,1/2)$. Then 
\begin{equation}\begin{split}
    |F(\bm{G})|^2 &= (b_0 + b_1 \exp(-i\bm{G}\cdot\bm{\tau})) (b_0 + b_1 \exp(i\bm{G}\cdot\bm{\tau})) \\
    &= b^2_0 + b^2_1 + 2 b_0 b_1 \cos( \pi (h + k)) .
\end{split}\end{equation}
In general, $b_0 \neq b_1$ and $|F(\bm{G})|^2 \neq 0$. As an example, consider $\bm{G}_{odd} = (1,0)$ and $\bm{G}_{even} = (2,0)$. Explicitly,
$|F(\bm{G}_{odd})|^2 = b^2_0 + b^2_1 - b_0 b_1 $ while $|F(\bm{G}_{even})|^2 = b^2_0 + b^2_1 + b_0 b_1 $. If $\sum_i G_i$ is even, $|F(\bm{G})|^2$ is maximized, while if it is odd, $|F(\bm{G})|^2$ is minimized. The point is that the the variation of $(d\sigma/d\Omega)_0$ with $\bm{Q}$ depends microscopically on the of the underlying crystal structure. Bragg scattering (and lots of hard work) can be used to identify crystal structures.

What if the white and black atoms in \cref{fig:interference_diagram} are identical? In this case, a primitive unit cell containing only one atom could be used, but consider using the conventional unit cell with two atoms and setting $b_0 = b_1 = b$. Then 
\begin{equation}\begin{split}
    |F(\bm{G})|^2 &= 2b^2[1+\cos( \pi (h + k))] .
\end{split}\end{equation}
If $\sum_i G_i $ is even, $|F(\bm{G}_{even})|^2 = (2b)^2$. If $\sum_i G_i$ is odd, $|F(\bm{G}_{odd})|^2 = 0$ vanishes! In the context of "Brillouin zone folding" (see \cref{sec:unfolding} below), one can show that $\bm{G}_{even}$ coincide with the reciprocal lattice vectors of the underlying primitive unit cell, while $\bm{G}_{odd}$ do not and thus should show no intensity. For $\bm{G}_{even}$, the cross section is proportional to $(2b)^2$, i.e. twice the number of scatterers in the unit cell results in the square of twice the scattering lengths. We could even pick a larger unit cell containing any number of atoms and only the Bragg peaks that coincide with the Bragg peaks of the underlying primitive cell are non vanishing.

\subsection{Diffuse scattering}\label{sec:diffuse_scattering}

The elastic intensity can be used for (elastic) diffuse scattering measurements. In writing \cref{eq:elastic_cross_section}, we are assuming a periodically ordered material; all scattering should reflect the periodic nature of the material. In a perfect material, we only see intensity at a discrete set of points, $\bm{Q}=\bm{G}$, called "Bragg peaks". If we include disorder or defects in the material, we are essentially adding volumes of a different (not necessarily periodic) structure than the underlying host material, so we see intensity at new wave vectors that aren't Bragg peaks of the underlying material in general. Since the defect volumes are often smaller than the sample volume, the new intensity is usually weak compared to Bragg peaks. Moreover, scattering from disorder or defects often results in a broad distribution of intensity: the scattering from defects/disorder is called "diffuse scattering". Diffuse intensity can be at points in momentum space, or can cover lines, planes, or other complicated shapes. We show examples of other shapes in this thesis (see \cref{chp:mapi_diffuse}).

\subsection{One-phonon excitations}

The one-phonon term in \cref{eq:exp_UV} is
\begin{equation}\begin{gathered}
    \left(\frac{d^2\sigma}{d\Omega dE'}\right)_{\pm 1} = \frac{k'}{k} N \sum_{\bm{R}ab} b_a b_b \exp(W_{ab}) \exp(i\bm{Q}\cdot \bm{R}) \exp(i\bm{Q}\cdot(\bm{\tau}_b-\bm{\tau}_a)) \int \frac{dt}{2\pi\hbar} \exp(-i\omega t) \langle \hat{U}_{\bm{0}a} \hat{V}_{\bm{R}b} \rangle .
\end{gathered}\end{equation}
We can use the usual expressions for displacements in terms of creation and annihilation operators (See \cref{eq:quantum_normal_mode_displacements} or e.g. refs. \cite{squires1996introduction,dove1993introduction,mahan2013many}):
\begin{equation}
\begin{gathered}
    \hat{\bm{u}}_{ia}(t) = \sum_{\bm{q}\nu} \sqrt{\frac{\hbar}{2 m_a \omega_{\bm{q}\nu} N}} \bm{\epsilon}^a_{\bm{q}\nu} \left[\hat{a}^\dagger_{-\bm{q}\nu}\exp(i\omega_{\bm{q}\nu}t)+\hat{a}_{\bm{q}\nu}\exp(-i\omega_{\bm{q}\nu}t)\right]
    \exp(i\bm{q}\cdot\bm{r}^{(0)}_{ia}) .
\end{gathered}
\end{equation} 
$\bm{q}$ is the a wave vector in the first Brillouin zone, $\nu$ labels the modes, $\omega_{\bm{q}\nu}$ is the frequency, $\bm{\epsilon}^a_{\bm{q}\nu}$ is the component of eigenvector for atom $a$, and $\bm{r}^{(0)}_{ia}=\bm{R}_i+\bm{\tau}_a$ is the equilibrium position of the atom\footnote{Note that this expression is a Fourier expansion; there is an ambiguity in the phase. It is perfectly valid to expand with respect to either $\bm{R}_i$ or $\bm{r}^{(0)}_{ia}$ with the latter preferred by me since it is easier to work with computationally. I.e. it is easier to directly calculate relative positions than to have to tabulate the unit cell coordinate for all atoms to calculate the relative unit cell vectors.}, and $m_a$ is the atom's mass. $N$ is the number of unit cells. For convenience, define $\hat{A}_{\bm{q}\nu}(t) = \hat{a}^\dagger_{-\bm{q}\nu} \exp(i\omega_{\bm{q}\nu}t) + \hat{a}_{\bm{q}\nu} \exp(-i\omega_{\bm{q}\nu}t)$. Plugging in,
\begin{equation}\begin{split}
    \left(\frac{d^2\sigma}{d\Omega dE'}\right)_{\pm 1} & = \frac{k'}{k} \frac{\hbar}{2} \sum_{\bm{q}\nu}\sum_{\bm{p}\mu}\sum_{ab} b_a b_b \frac{ (\bm{Q}\cdot\bm{\epsilon}^a_{\bm{q}\nu}) (\bm{Q}\cdot\bm{\epsilon}^b_{\bm{p}\mu}) }{\sqrt{\omega_{\bm{q}\nu}\omega_{\bm{p}\mu}m_a m_b}} \exp(W_{ab}) \exp(i\bm{Q}\cdot(\bm{\tau}_b-\bm{\tau}_a)) \\
    & \exp(i\bm{q}\cdot\bm{\tau}_a)\exp(i\bm{p}\cdot\bm{\tau}_b) \left(\sum_{\bm{R}} \exp(i(\bm{Q}+\bm{p})\cdot\bm{R}) \right) \int \frac{dt}{2\pi\hbar} \exp(-i\omega t) \langle \hat{A}_{\bm{q}\nu}(0)\hat{A}_{\bm{p}\mu}(t) \rangle   .
\end{split}\end{equation}
and we can calculate the expectation value:
\begin{equation}\begin{gathered}
    \int \frac{dt}{2\pi\hbar} \exp(-i\omega t) \langle \hat{A}_{\bm{q}\nu}(0)\hat{A}_{\bm{p}\mu}(t) \rangle = \\
    \delta_{-\bm{q},\bm{p}}\delta_{\nu\mu} n_{BE}(\omega_{\bm{q}\nu}) \delta(\hbar(\omega+\omega_{\bm{q}\nu})) + \delta_{\bm{q},-\bm{p}}\delta_{\nu\mu} [n_{BE}(\omega_{\bm{q}\nu})+1] \delta(\hbar(\omega-\omega_{\bm{q}\nu}))  .
\end{gathered}\end{equation}
$n_{BE}(\omega_{\bm{q}\nu}) = [\exp(\hbar\omega_{\bm{q}\nu}/k_B T)-1]^{-1}$ the Bose-Einstein function. The first term corresponds to phonon annihilation, i.e. the neutrons gain energy from phonons. The second term corresponds to phonon creation, i.e. the neutrons loses energy by creating phonons. Plugging back in for phonon creation,
\begin{equation}\begin{split}
    \left(\frac{d^2\sigma}{d\Omega dE'}\right)_{+1} & =  N \frac{k'}{k} \sum_{\bm{q}\nu} \left| \sum_{a}  \sqrt{\frac{\hbar}{2m_a \omega_{\bm{q}\nu}}} b_a \exp(W_{a}) \exp(i(\bm{Q}-\bm{q})\cdot \bm{\tau}_a) (\bm{Q}\cdot \bm{\epsilon}^{a}_{-\bm{q}\nu}) \right|^2 \\
    & \qquad [n_{BE}(\omega_{\bm{q}\nu})+1] \delta(\hbar(\omega-\omega_{\bm{q}\nu}) \delta(\bm{Q}-\bm{q}-\bm{G}) ,
\end{split}\end{equation}
where we used $\omega_{-\bm{q}\nu}=\omega_{\bm{q}\nu}$. Note that the phase convention we use is different than Squire's convention, so our expression differs from Squire's; rather, it agrees with the convention used in the \texttt{phonopy} package \cite{phonopy}. $\exp(W_{a})$ is the DWF. \cref{eq:one_phonon_creation} describes the intensity of neutrons that are scattered by creating a phonon in the scattering system. The $\delta$-functions enforce energy and momentum conservation. Energy conservation requires that $E'=E-\hbar\omega_{\bm{q}\nu}$, i.e. the scattered neutron loses an amount of energy, $\hbar \omega_{\bm{q}\nu}$, into the sample. Crystal momentum conservation requires that $\bm{k}-\bm{k}'-\bm{q}=\bm{G}$, i.e. that the momentum lost into the sample must be equal the phonon momentum (up to a reciprocal lattice vector). We define the one-phonon structure factor 
\begin{equation}\begin{split}
    F_{\bm{q}\nu}(\bm{Q}) = \sum_{a}  \sqrt{\frac{\hbar}{2m_a \omega_{\bm{q}\nu}} } b_a \exp(W_{a}) \exp(i(\bm{Q}+\bm{q})\cdot\bm{\tau}_a) (\bm{Q}\cdot \bm{\epsilon}^{a}_{\bm{q}\nu}) .
    \label{eq:form_factor}
\end{split}\end{equation}
Note that $\bm{\epsilon}^b_{-\bm{q}\nu}=\bar{\bm{\epsilon}}^{b}_{\bm{q}\nu}$ (the over-bar denotes complex conjugation). Then the phonon creation cross section is
\begin{equation}\begin{split}
    \left(\frac{d^2\sigma}{d\Omega dE'}\right)_{+1} & =  N \frac{k'}{k} \sum_{\bm{q}\nu} \left| F_{\bm{q}\nu}(\bm{Q}) \right|^2 [n_{BE}(\omega_{\bm{q}\nu})+1] \delta(\hbar(\omega-\omega_{\bm{q}\nu})) \delta(\bm{Q}-\bm{q}-\bm{G}) 
    \label{eq:one_phonon_creation}
\end{split}\end{equation}
The same analysis above applied to phonon absorption gives
\begin{equation}\begin{split}
    \left(\frac{d^2\sigma}{d\Omega dE'}\right)_{-1} & =  N \frac{k'}{k} \sum_{\bm{q}\nu} \left| F_{\bm{q}\nu}(\bm{Q}) \right|^2 n_{BE}(\omega_{\bm{q}\nu}) \delta(\hbar(\omega+\omega_{\bm{q}\nu})) \delta(\bm{Q}+\bm{q}-\bm{G}) .
    \label{eq:one_phonon_absorption}
\end{split}\end{equation}
Energy conservation requires that $E'=E+\hbar\omega_{\bm{q}\nu}$, i.e. the scattered neutron gains energy from the sample. Crystal momentum conservation requires that $\bm{k}-\bm{k}'+\bm{q}=\bm{G}$, i.e. that the momentum gained from the sample must be equal to the phonon momentum (up to a reciprocal lattice vector).

The \cref{eq:one_phonon_creation,eq:one_phonon_absorption} are nice because they make the momentum conservation obvious, but they can be simplified a little. The result is  
\begin{equation}\begin{gathered}
    \left(\frac{d^2\sigma}{d\Omega dE'}\right)_{\pm 1} =  N \frac{k'}{k} \sum_{\nu} \left| F_{\bm{q}\nu}(\bm{Q}) \right|^2 \left[ n_{BE}(\omega_{\bm{q}\nu}) + \frac{1}{2} \pm \frac{1}{2} \right] \delta(\hbar(\omega \mp \omega_{\bm{q}\nu}))  \\
    F_{\bm{q}\nu}(\bm{Q}) = \sum_a \sqrt{\frac{\hbar}{2 m_a \omega_{\bm{q}\nu}}} b_a \exp(W_a) \exp(i\bm{G}\cdot\bm{\tau}_a) (\bm{Q}\cdot\bm{\epsilon}^a_{\bm{q}\nu}) \\
    \bm{Q}=\bm{q}+\bm{G}
    \label{eq:one_phonon_cross_section}
\end{gathered}\end{equation}
where the upper/lower sign corresponds to one-phonon creation/annihilation and we used $\omega_{-\bm{q}\nu} = \omega_{\bm{q}\nu}$ and $\bm{\epsilon}^a_{-\bm{q}\nu} = \bar{\bm{\epsilon}}^a_{\bm{q}\nu}$. Since $\bm{q}$ is restricted to the first Brillouin zone (see \cref{sec:blochs_thm}) $\bm{G}$ is uniquely defined by $\bm{Q}$. Note that in the harmonic approximation, the one-phonon dynamic structure factor, $S_{\pm 1}(\bm{Q},\omega)$, is defined by 
\begin{equation}\begin{gathered}
    \left( \frac{d^2\sigma}{d \Omega dE'} \right)_{\pm 1} = N \frac{k'}{k} S_{\pm 1}(\bm{Q},\omega) .
\end{gathered}\end{equation}

Consider the occupation factors in \cref{eq:one_phonon_cross_section}. For phonon absorption ($-1$), the scattering probability is proportional to $n_{BE}(\omega_{\bm{q}\nu})$, i.e. the probability of absorbing an excitation is proportional to the number of excitations already in the sample. At low temperature, there aren't many excitations and the phonon absorption cross section is much smaller than the phonon creation cross section. Usually in an experiment, we only measure the creation cross section ($+1$) which, at low temperature, limits to unity for all modes. At high temperature, $n_{BE}(\omega)+1 \approx n_{BE}(\omega)$ and both cross sections are equally large.

The harmonic, inelastic cross section is proportional to a $\delta$ function of energy. Thus, phonons should show up as sharp peaks in the measured intensity in principle. In practice, interactions (electron-phonon, phonon-phonon, etc.) and experimental resolution broaden the peaks substantially. Moreover, the variation of the form factor (see below) modulates the "heights" of peaks with $\bm{Q}$, complicating matters. Still, the \emph{energy} dependence of the cross section can be used to determine the phonon energies in a relatively straightforward way. On the other hand, the \emph{momentum} dependence can be used to analyze the polarization of the phonons, as we now see.

\subsection{The one-phonon structure factor}\label{sec:one_phonon_form_factor}

Just like in the elastic cross section, the one-phonon structure factor depends on the symmetry of the crystal; however, the inelastic one-phonon cross section additionally depends on the phonon polarization. We can apply the same analysis from elastic scattering to one-phonon scattering if we define $b^a_{\bm{q}\nu}(\bm{Q}) =  b_a \sqrt{\hbar /(2m_a \omega_{\bm{q}\nu})} (\bm{Q} \cdot \bm{\epsilon}^a_{\bm{q}\nu})  $. Then $|F_{\bm{q}\nu}(\bm{Q})|^2 = |\sum_a b^a_{\bm{q}\nu}(\bm{Q}) \exp(W_a) \exp( i\bm{G} \cdot \bm{\tau}_a )|^2$ looks just like the structure factor in \cref{eq:elastic_cross_section} with a scattering length that depends on $\bm{Q}$, $\bm{q}$, $\nu$, and $a$ rather than simply on the atoms, $a$. The $\bm{Q}$ dependence of structure factor depends on the symmetry of the underlying crystal structure, analogous to Bragg scattering, and on the symmetry of the atomic displacements associated with the phonon modes. Similar arguments to \cref{sec:bragg_scattering} apply, but let's not dwell on this any further and instead focus on just the phonon polarization parts.

\begin{figure}[t!]
    \centering
    \includegraphics[width=0.9\linewidth]{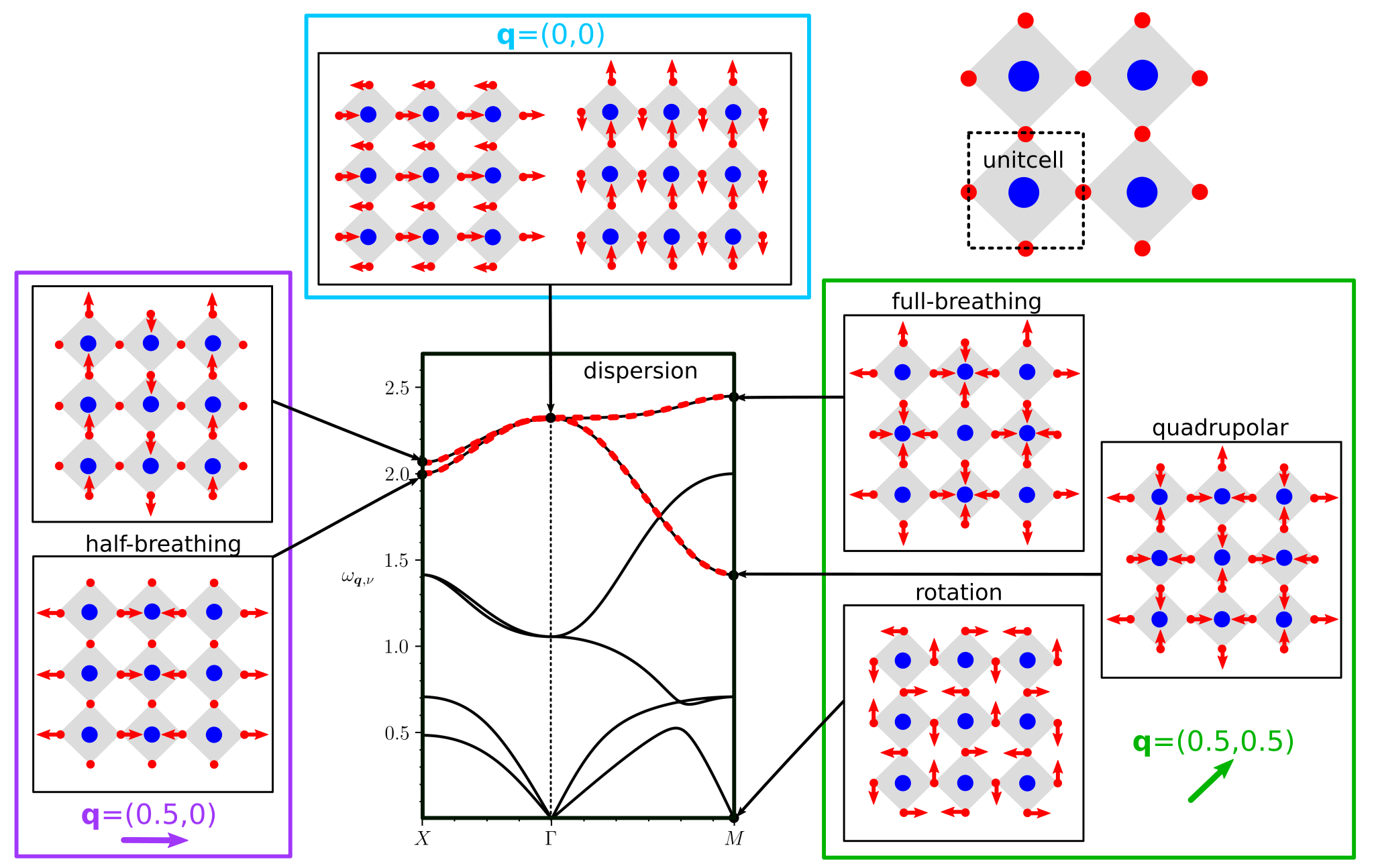}
    \caption{Dispersion and diagram of bond-stretching phonon eigenvectors of the model in \cref{sec:2d_perovskite}, \cref{fig:2d_square_lattice} at several wave vectors: $\bm{q}=\Gamma=(0,0)$, $\bm{q}=X=(1/2,0)$, and $\bm{q}=M=(1/2,1/2)$. The arrows next to the displacement diagrams show the wave vector of the plotted modes.}
    \label{fig:2d_model}
\end{figure}

The polarization term is $\sim | \bm{Q}\cdot \bm{\epsilon}^a_{\bm{q}\nu}|^2$; it depends on the direction of phonon propagation (i.e. its momentum, $\bm{q}$), phonon polarization ($\bm{\epsilon}^a_{\bm{q}\nu}$), and on the neutron's momentum transfer ($\bm{Q}$). Note that the structure grows as $\sim|\bm{Q}|^2$; intensity from phonons is larger at large $|\bm{Q}|$. Next, recall that the phonon polarization (or eigenvector) describes the direction in which the atoms are oscillating (see \cref{chp:phonons}). $\bm{\epsilon}^a_{\bm{q}\nu}$ is only defined in the first Brillouin zone, while $\bm{Q}$ can be anything. The variation of the cross section, \cref{eq:one_phonon_cross_section}, with $\bm{Q}$ allows us to determine the direction of polarization from an experiment. 

\begin{figure}[t!]
    \centering
    \includegraphics[width=0.9\linewidth]{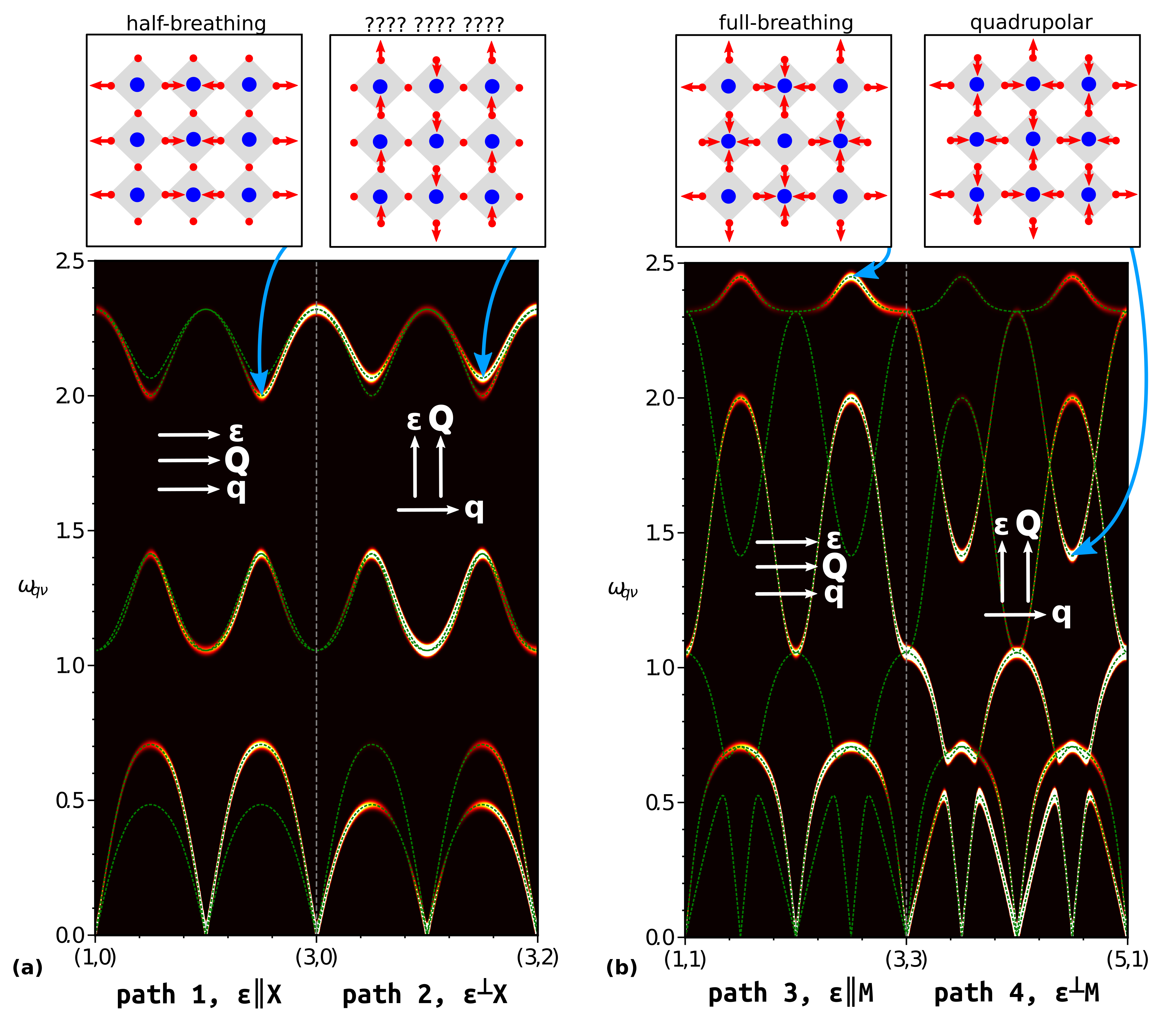}
    \caption{Inelastic neutron scattering cross section calculated from \cref{eq:one_phonon_cross_section} for the model in \cref{fig:2d_model}. The paths through (neutron) momentum space are plotted in \cref{fig:bz_paths} on top of the extended Brillouin zones of the model in \cref{fig:2d_model}. The polarizations are shown at the top; the arrows from the polarizations label where these modes are strong in $(\bm{Q},\omega)$ space. (a) Phonons from $\Gamma \rightarrow X$ along two paths: path 1, which measures longitudinal modes and path 2, which measures transverse modes. (b) Phonons from $\Gamma \rightarrow M$ along two paths: path 3, which measures longitudinal modes and path 4, which measures transverse modes.}
    \label{fig:htt_sqw}
\end{figure}

We will specialize to the simple, $2D$ model in \cref{fig:2d_model}\footnote{This is supposed to model the CuO plane in a cuprate. This Figure is replicated from \cref{sec:2d_perovskite}}. We work in $2D$ since it is easier to plot displacements; all of the nuances between $2D$ and $3D$ are the same. We can't actually measure this model system, but we can calculate the cross section (\cref{eq:one_phonon_cross_section}) as I have done in \cref{fig:htt_sqw} and results of calculated cross sections are usually in good agreement with experiment (\cref{chp:lco_lda_u,chp:mapi_diffuse,chp:bgg} verify this!). In this section, we can just pretend that \cref{fig:htt_sqw} represents a real experiment. The bonus of computation is that we can directly analyze the dispersions and displacements by simply plotting them as shown in \cref{fig:2d_model} and this combination (theory+experimental neutron scattering) is a powerful tool.

\begin{figure}[t!]
    \centering
    \includegraphics[width=0.9\linewidth]{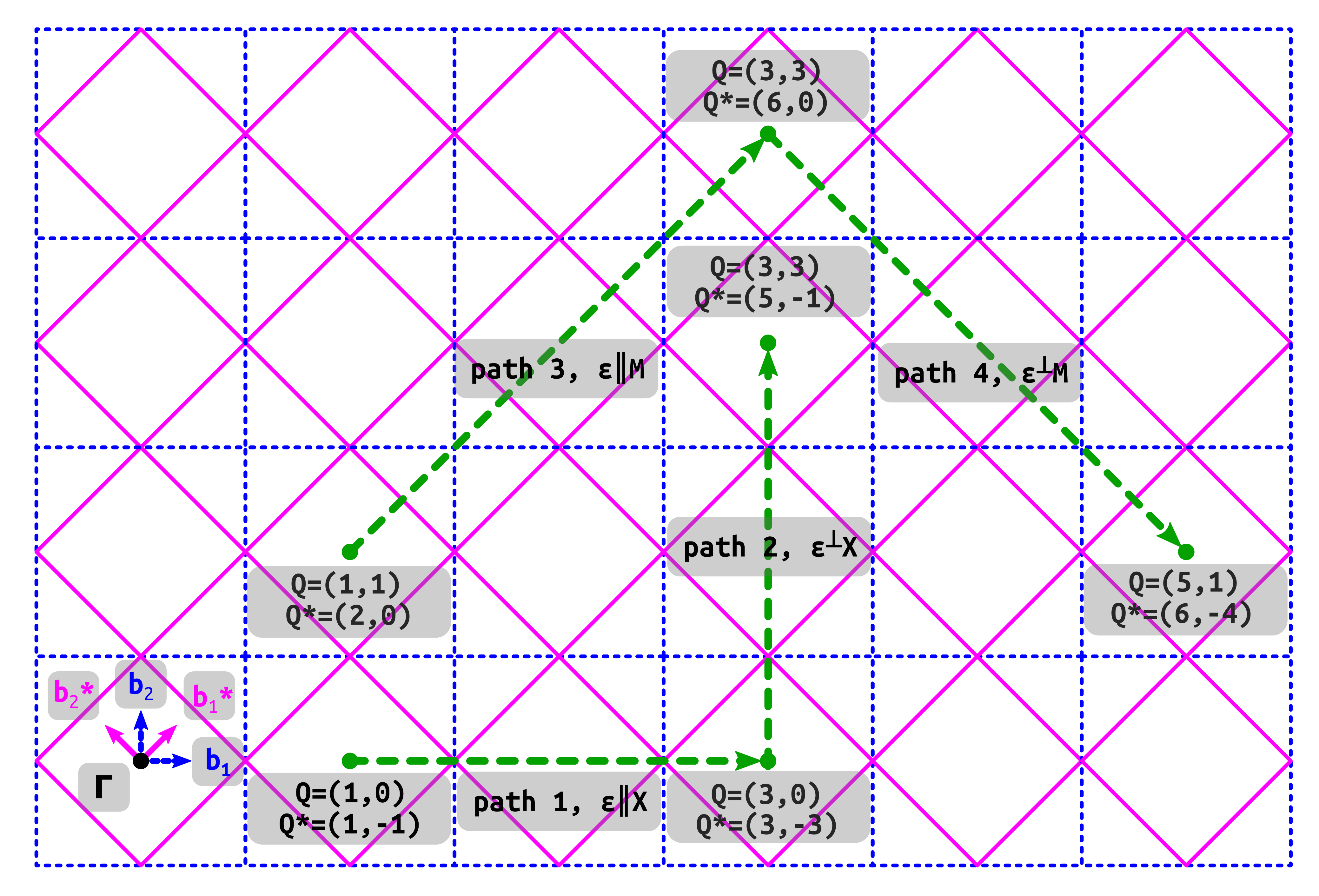}
    \caption{The paths (dashed green arrows) through neutron momentum space measured in \cref{fig:htt_sqw}. The origin, $\bm{Q} = \bm{Q}^* = \Gamma = (0,0)$, is at the lower left. $\bm{Q}$ and $\bm{Q}^*$ label the neutron momentum transfer in the primitive cell basis in \cref{fig:2d_model} and the AFM supercell respectively (see \cref{sec:unfolding}). The dashed blue lines label the extended Brillouin zones for the primitive cell. The solid magenta lines label the Brillouin zones of the AFM supercell. Note that an extended zone scheme is needed since the neutron scattering intensity is \emph{not} periodic with respect to momentum.}
    \label{fig:bz_paths}
\end{figure}

Consider the neutron scattering cross sections for momentum transfers that follow the paths 1-4 in \cref{fig:bz_paths}. We label the underlying Brillouin zones of the model in \cref{fig:2d_model} as dashed blue lines, but since $\bm{Q}$ is not periodic, we must use an extended zone scheme rather than measuring the cross section in only the first Brillouin zone. The colormaps in \cref{fig:htt_sqw} portray the cross section as a function of $\bm{Q}$ (momentum transferred into the sample) and $\omega$ (energy/$\hbar$ transfered into the sample) for paths 1-4. Our goal is to identify the half-breathing (HB), full-breathing (FB), and quadrupolar (QP) bond-stretching modes labeled in \cref{fig:2d_model}. 

Paths 1-2 show the cross section for phonons along the $\Gamma \rightarrow X$ path. Focus on the $\bm{q} = X = (1/2,0)$ point. Along path 1, $\bm{Q}$ is parallel to $\bm{q}$, so only longitudinal modes have a strong cross section. Recall, longitudinal modes have $\bm{\epsilon}_L \parallel \bm{q}$ and transverse modes have $\bm{\epsilon}_T \perp \bm{q}$. The HB mode is longitudinal, so the intensity from the HB mode is strongest. The arrow from the HB diagram in \cref{fig:htt_sqw}(a) labels where the HB mode intensity is strongest along path 1. Along path 2, $\bm{Q}$ is (roughly) perpendicular to $\bm{q}$, so the intensity from the HB mode is small. Rather, the other transverse mode\footnote{I can't remember it's colloquial name.} that is degenerate with the HB mode at the zone center is strong along path 2. This logic can be extended to arbitrary $\bm{Q}$ and is an extremely useful concept when interpreting inelastic scattering data. 

Of course, the separation into "purely" longitudinal and transverse modes is an idealization and in a non-Bravais crystal, many modes are mixed character. \cref{fig:htt_sqw}(b) shows the phonons along $\Gamma \rightarrow M$. Using the same arguments as above, along path 3, $\bm{Q}\parallel \bm{q}$ and longitudinal modes are strongest. On the other hand, along path 4, $\bm{Q} \perp \bm{q}$ and transverse modes are strongest. The FB and QP modes are mixed character, but the QP mode is very weak along path 3 while the FB mode is strong. Conversely, the FB mode is weak along path 4 while the QP mode is strong. The FB mode is usually labeled as longitudinal, while the QP mode is labeled as transverse.

\subsection{Calculating the one-phonon cross section}\label{sec:finite_diffs}

If one can calculate the phonon eigenvectors and phonon frequencies, it is fairly straightforward to write a program to calculate \cref{eq:one_phonon_creation} to model an experiment\footnote{I actually wrote a code to solve the lattice dynamics equations for the model in \cref{fig:2d_model} (see \cref{sec:2d_perovskite}) and to calculate the inelastic cross section, \cref{eq:one_phonon_cross_section}. The results from my code are what is shown in \cref{fig:2d_model,fig:htt_sqw,fig:ltt_sqw}.}. In practice, I usually use existing codes to study real materials. The phonons are calculated with density functional theory (DFT) and the cross section is calculated with a post processing code: I use \textsc{euphonic} \cite{fair2022euphonic}. 

There are several ways to calculate phonons with DFT. The basics of DFT are discussed in \cref{chp:electrons} and the basics of harmonic lattice dynamics in \cref{chp:phonons}. It is assumed the reader is familiar with these topics. Ultimately, DFT is used to calculate the force constant matrix, $\hat{\Phi}^{ab}_{ij} = \partial^2 E / (\partial \bm{u}_{ia} \partial \bm{u}_{jb})$ and then all of the lattice dynamics bits are done as a post processing step. In DFT, the total energy, $E$, is the Kohn-Sham energy plus the ion-ion Coulomb repulsion. 

The most common method to calculate force constants is with "finite displacements". Note that $\bm{F}_{ia}=-\partial E/\partial \bm{u}_{ia}$. Thus, the force constant is negative the derivative of the force: $\hat{\Phi}^{ab}_{ij} = -\partial \bm{F}_{ia}/\partial \bm{u}_{jb}$. A very simple way to calculate the force constant $\hat{\Phi}_{ij}^{ab}$ is to move atom $(jb)$ by a small amount in each direction and calculate the force on atom $(ia)$. Let $\bm{\delta}_{jb}$ represent the small, finite displacement of atom $(jb)$. The derivative can be approximated by finite differences as
\begin{equation}\begin{split}
    \hat{\Phi}_{ij}^{ab} = -\frac{\partial \bm{F}_{ia}}{\partial \bm{u}_{jb} } \approx -\frac{\bm{F}_{ia}(\bm{\delta}_{jb}) - \bm{F}_{ia}(-\bm{\delta}_{jb})}{2\bm{\delta}_{jb}}.
    \label{eq:finite_diff_force_constant}
\end{split}\end{equation}
When atoms are sufficiently far apart (i.e. $\bm{R}_j-\bm{R}_i$ is big), the force constants become vanishingly small, so we only need to calculate force constants between atoms in a relatively small "sphere" centered on the origin (we use a supercell, see below). Still, for a crystal with many atoms in the unit cell, calculating all of the force constant matrices from \cref{eq:finite_diff_force_constant} is expensive. Fortunately, symmetry can be used to reduce the number of calculations to an irreducible set. In high symmetry structures, many atoms are equivalent by translation and rotations. The corresponding force constants are also equivalent under such operations. Thus only a subset of displacements is needed to calculate the entire set of a force constants. 

The finite displacement method is implemented as a "wrapper" around a typical DFT code. Many packages implement this with the most famous being \texttt{phonopy} \cite{phonopy-phono3py-JPCM,phonopy-phono3py-JPSJ}, which I use in this thesis. \texttt{phonopy} handles all of the symmetry stuff and automatically reduces the number of independent calculations that have to be done. The work flow is as follows: first, a DFT calculation is performed to "relax" a structure so that all forces on all atoms in the calculation are smaller than some numerical tolerance. This ensures that the harmonic approximation is valid. Then
\begin{enumerate}
    \item a minimal set of "distorted" structures is created. In each structure, a single atom is moved and the force on all other atoms is calculated. 
    \item The changes in forces are read from the outputs of the calculations and used to approximate the force constants.
    \item The process is repeated for every independent atomic displacement in the unit cell.
\end{enumerate}
Once all of the independent calculations are done, \texttt{phonopy} uses symmetry to create the full set of force constants between all atoms. The force constants are conveniently written to a file that is read by \textsc{euphonic}, which calculates the one-phonon cross sections in \cref{eq:one_phonon_creation}.

The finite displacement method is the most straightforward way to calculate phonons and all DFT codes are applicable. However, there are some drawbacks: (i) the displacement amplitudes must be converged to guarantee the dispersions are correct and (ii) supercells must be used. It turns out that (i) isn't an issue in practice and the default used by \textsc{phonopy} is usually sufficient. (ii) is the biggest challenge in the finite displacement method. 

A "supercell" is an integer multiple of the primitive unit cell, containing the primitive cell within it. In other words, a supercell is an expanded model with the primitive cell embedded in the crystal and we explicitly treat the surrounding unit cells in the calculation. We must use a supercell in the finite displacement method because the force constants describe isolated displacements in an infinite crystal, but DFT models the crystal using Bloch functions so that only a single unit cell with periodic boundary conditions is explicitly needed. In other words, the displacements used for finite displacement calculations are a periodic modulation of the crystal in DFT calculations. The way around this discrepancy is to use a supercell: we embed the unit cell in a big copy of the crystal and move one atom at the middle. The idea is that the effects of the displacement are negligibly small at the boundary of the supercell so that the displaced atom is effectively isolated. 

The correct way to choose the supercell size is with a convergence study: larger and larger supercells are used to calculate phonon dispersions and the smallest supercell for which the results no longer change with increasing size is used for further analysis. However, DFT scales very poorly with the number of atoms in the system (nominally as $\mathcal{O}[N^3_{at}]$), so such a convergence study is extremely expensive. In practice, usually as large a supercell as is practical to calculate is chosen and the phonons are calculated only once (cf. \cref{chp:lco_lda_u}). Moreover, if the primitive unit cell is already large, as is the case with e.g. clathrates, a single unit cell is sufficient rather than a supercell (cf. \cref{chp:bgg}).

As the simplest option, the finite displacement method is also the oldest method for phonon calculations. A more recent and very popular way to calculate the phonons is with the so called "density functional perturbation theory" (DFPT) \cite{baroni2001phonons}. Basically, the force constants depend on the linear response of the electronic density to displacements of the atoms. Perturbation theory is used to write a self consistent equation for the response of the density, which can be solved with various methods. Many existing codes have implemented DFPT. One of the biggest advantages is that we can directly calculate dynamical matrices, rather than force constants, without resorting to supercells. Just like with finite differences, symmetry can be used to reduce the number of independent calculations needed and DFPT calculations are competitive with respect to computational cost in many cases. The main drawbacks with DFPT are (i) the implementation is very complicated and (ii) for large unit cells, DFPT actually scales worse than finite displacements. 

The first issue limits DFPT to simple calculations: e.g. most codes don't implement DFPT for DFT+U or spin-orbit coupling. Even spin polarized DFPT calculations are unavailable in some codes. The second issue limits DFPT to materials with small primitive unit cells, since finite differences is preferred for large primitive unit cells. A significant advantage of DFPT over finite differences is that electron-phonon interaction matrix elements are calculated as a by product of solving the self consistent equations. In this thesis, we discuss calculations of phonons with DFT+U and spin polarization in a large unit cell and we don't need electron-phonon matrix elements. Finite differences is the natural choice and we use it through out.

\subsection{Unfolding}\label{sec:unfolding}

\begin{figure}[t!]
    \centering
    \includegraphics[width=1\linewidth]{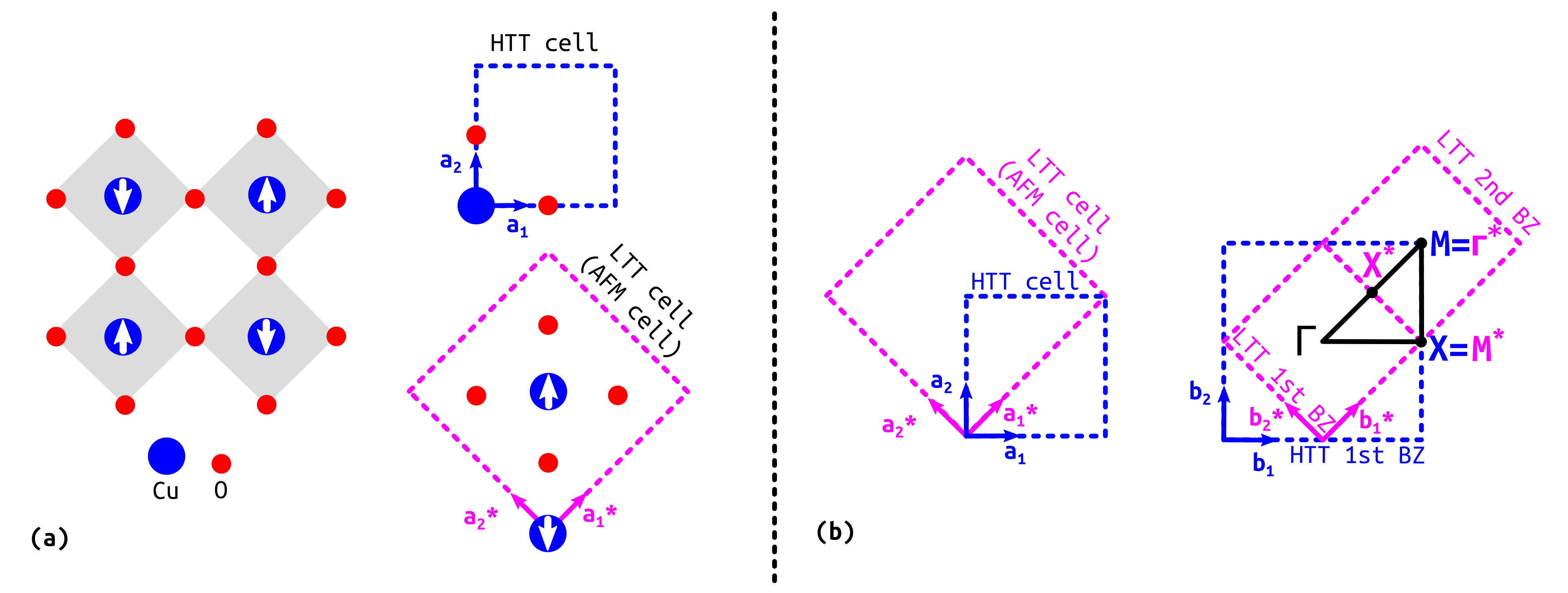}
    \caption{(a) Schematic demonstrating the antiferromagnetic (AFM) order in an CuO plane of La$_2$CuO$_4$. The arrows label the spin order. At high temperature, the AFM order vanishes and the primitive cell is tetragonal with a single Cu atom. We call it the HTT cell. At low temperature, the AFM order requires using a supercell with (atleast) two Cu atoms of opposite spin. The primitive cell of the low temperature structure is a $\sqrt{2}$ expansion of the HTT cell and a rotation by $45^\circ$. The corresponding Brillouin zones are shown in (b).}
    \label{fig:htt_vs_ltt}
\end{figure}

In both calculations and experiments, it is often convenient to pick a "conventional" cell rather than the primitive cell of a crystal (see \cref{sec:crystals}). E.g. the primitive cell of silicon has a two atom basis but has non-orthogonal lattice vectors, while the conventional cell has an eight atom basis but has orthogonal lattice vectors. More atoms is harder to handle, but orthogonal lattice vectors are easier, so the conventional cell is usually what is discussed in neutron scattering experiments. 

We also need enlarged unit cells (supercells) in other cases. Consider a disordered solid: e.g. in Ba$_8$Ga$_{16}$Ge$_{30}$, the Ga and Ge atoms are scrambled so that there is occupational disorder\footnote{We study Ba$_8$Ga$_{16}$Ge$_{30}$ in \cref{chp:bgg}, but the primitive unit cell even for the ordered structure is huge (54 atoms in the basis) so we use a disordered primitive cell since a supercell would have been too expensive to model. Still, the concepts discussed here are valid.}. In this case, periodicity is lost, so a unit cell can't be rigorously defined. Still, neutron scattering shows that there is \emph{average} long ranged order, consistent with a simple-cubic clathrate structure. We invariably use periodic boundary conditions for modelling solids, so a computational model is periodic whether we like or not. The way to handle disorder computationally is to use a supercell that is large enough that the defect-defect correlations decay within the simulation box \cite{zunger1990special}.

In other situations, some sort of spontaneous symmetry breaking occurs, requiring a supercell. A salient example is antiferromagnetism (AFM), which is shown schematically in \cref{fig:htt_vs_ltt} for a simple model applicable to e.g. cuprate La$_2$CuO$_4$. The high temperature, non-magnetic cell (HTT) is primitive tetragonal, while the corresponding low temperature AFM cell (LTT) is body-centered tetragonal\footnote{There are structural distortions that make the real structure orthorhombic, but it is sufficient to discuss our simple model as tetragonal.}. We focus on this example in this section. The primitive (HTT) cell was discussed in detail in \cref{sec:one_phonon_form_factor}. 

\begin{figure}[t!]
    \centering
    \includegraphics[width=1\linewidth]{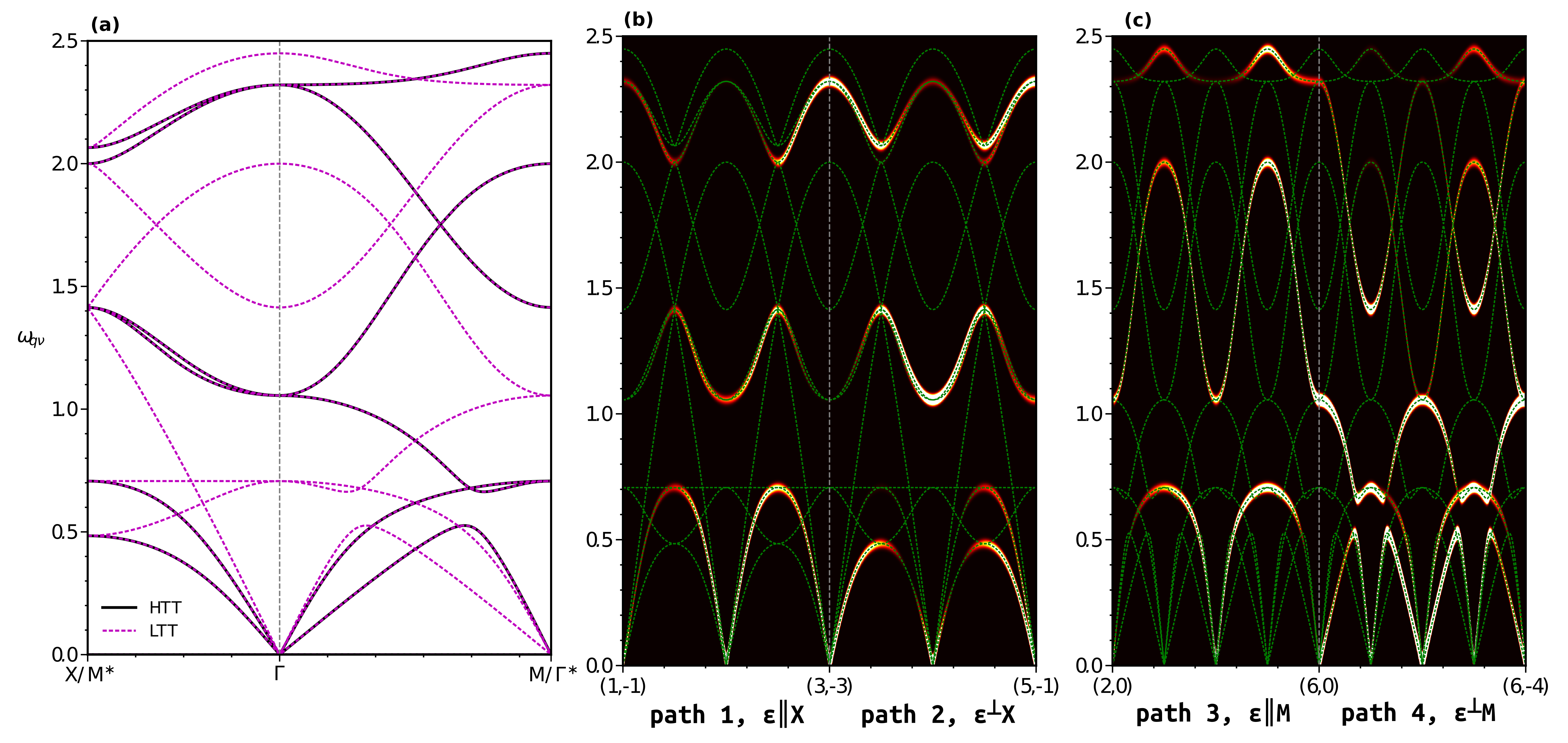}
    \caption{(a) Primitive and folded phonon dispersions for the HTT and LTT cells in \cref{fig:htt_vs_ltt}. $X$ and $M$ refer to the HTT basis, while $M^*$ and $X^*$ refer to the LTT basis. The HTT dispersions are the same as in \cref{fig:2d_model}. (b)-(c) The calculated cross section for the LTT cell along the same paths as in \cref{fig:htt_sqw}. The coordinates in (b)-(c) are in the LTT basis.}
    \label{fig:ltt_sqw}
\end{figure}

Computationally, the AFM order is handled by using the LTT cell in a calculation. Since we are increasing the number of atoms in the cell, the number of modes in the dispersion also increases. This is called "Brillouin zone folding" (cf. \cref{sec:diatomic_chain}). In short, if we double the simulation cell size, we are halving the first Brillouin zone size and doubling the number of modes. 

Consider \cref{fig:htt_vs_ltt}(b), where we can see see how the paths through the HTT and LTT first Brillouin zones correspond to one another: the LTT cell is bigger than the HTT cell and, similarly, the LTT Brillouin zone is smaller than the HTT Brillouin zone. \cref{fig:ltt_sqw}(a) plots the dispersions for the corresponding paths in the HTT and LTT cells on top of one another. Note that there are more modes in the LTT calculation and they are more closely spaced in energy. We don't explicitly consider AFM order so the phonons are completely unchanged and the LTT dispersions don't contain any new information; to first order, this is actually a good approximation to reality. In the limit of an infinitely large supercell, the first Brillouin zone becomes a point and the number of modes is infinite; i.e. all we have access to is the density of states, and we have lost a vast amount of information. The goal of "unfolding" is to project the densely packing information calculated in a small Brillouin zone with a supercell back onto the large, sparse Brillouin zone of the primitive cell. This depends on the assumption that the symmetry breaking is "small" enough that the disordered supercell is still similar to the primitive cell.

Well developed techniques exist to unfold supercell phonon dispersions \cite{allen2013recovering,ikeda2018temperature,ikeda2017mode,samolyuk2021role,mu2020unfolding,kormann2017phonon} and electronic band structures \cite{ku2010unfolding,popescu2010effective,popescu2012extracting} back to the primitive cell. In short, the eigenvectors of the modes in the supercell are projected onto the primitive cell eigenvectors, resulting in a "spectral function". Disorder and other effects show up as broadening of the peaks in the unfolded spectral function. 

We propose an alternative, intuitive method that achieves the same results as unfolding from a neutron scattering perspective (cf. \cref{chp:lco_lda_u}). To that end consider that what we actually measure in an experiment is the cross section as a function of neutron momentum transfer. It is \emph{convenient} to work in a basis of "reduced coordinates", i.e. in reciprocal lattice units such that a wave vector is $\bm{Q}=(Q_1,Q_2,Q_3)=Q_1 \bm{b}_1 + Q_2 \bm{b}_2 + Q_3 \bm{b}_3$ with $\bm{b}_i$ the primitive reciprocal lattice vectors. (see \cref{sec:PBC}). But, since the choice of unit cell is arbitrary, this basis for wave vectors is also arbitrary. The neutron doesn't know about the unit cell. It simply scatters from the sample, changing momentum by an amount that is most conveniently measured in the lab frame with units 1/length or, equivalently, $\hbar$/length. We are free to convert from one unit cell choice to the other as follows. 

Define $A$ and $A'$ as column-matrices of the lattice vectors for two different unit cell choices for the same crystal\footnote{By column-matrix, I mean matrix of column vectors, where $\bm{a}_i$ are the the lattice vectors, $a_{ij}$ is the $j^{th}$ component of the $i^{th}$ lattice vector, and
\begin{equation}\begin{gathered}
    A = \begin{pmatrix}
         a_{1x}  &  a_{2x}  &  a_{3x}  \\
         a_{1y}  &  a_{2y}  &  a_{3y}  \\
         a_{1z}  &  a_{2z}  &  a_{3z}          
         \end{pmatrix} 
\end{gathered}\end{equation}
and so on.}. Suppose that $A$ are the lattice vectors for the primitive cell and $A'$ the lattice vectors for the supercell. The corresponding matrices of reciprocal lattice vectors are $B$ and $B'$, defined by $A^T B = 2\pi I$ with $I$ the identity matrix and $A^T$ the transpose of $A$. 

Let $\tilde{\bm{Q}}$ be the actual neutron momentum transfer in the lab frame and $\bm{Q}$ and $\bm{Q}'$ be the momentum transfer in the basis of $A$ and $A'$ respectively. $\tilde{\bm{Q}}$ is related to $\bm{Q}$ and $\bm{Q}'$ by $\tilde{\bm{Q}} = B \bm{Q} = B' \bm{Q}'$, i.e. $B \bm{Q} = B' \bm{Q}'$. 

To achieve the same results as unfolding, we want to calculate a quantity that is directly comparable to some spectrum of the primitive cell. For the primitive cell spectrum, we assume that we have calculated or measured the inelastic cross section, \cref{eq:one_phonon_cross_section}, in the primitive basis (\cref{fig:htt_sqw}). We want the analogous spectrum from the supercell. We can equate momentum in the different bases from $\bm{Q}' = [B']^{-1} B \bm{Q}$. E.g. for the basis in \cref{fig:htt_vs_ltt}(a), 
\begin{equation}\begin{gathered}
    [B']^T B = \begin{pmatrix}
         \quad  1  & \quad 1  & \quad 0 ~~~~ \\
         \quad -1  & \quad 1  & \quad 0 ~~~~ \\
         \quad  0  & \quad 0  & \quad 1 ~~~~
         \end{pmatrix} 
\end{gathered}\end{equation}
We used this matrix to convert paths 1-4 in \cref{fig:bz_paths} from the HTT basis to the LTT basis. The inelastic cross section calculated from the LTT cell is shown in \cref{fig:ltt_sqw}, in perfect agreement with the equivalent path in the HTT calculation (see \cref{fig:htt_sqw}). Note that, even though there are more modes in the LTT cell, the neutron scattering intensity is only peaked near the same modes as in the HTT cell. There are frequency "sum rules" for the dynamical structure factor and it can be shown that the integral over frequency at equivalent $\bm{Q}=\bm{Q}'$ points are the same whether using supercells or primitive cells.

\section{The classical approximation for the cross section}
\label{sec:classical_cross_section}

In the last section, we saw how to model the neutron scattering intensity for a crystal in the harmonic approximation. This was straightforward enough and we derived a nice result in terms of the phonon eigenvectors and dispersions, \cref{eq:one_phonon_creation}. However, in disordered systems or materials with large fluctuations, the assumption of a harmonic crystal breaks down. In this section, we show how to calculate the cross section by demoting the position operators to classical coordinates, but without assuming order or small displacements. Thus, this method is applicable to crystals, soft matter, liquids, etc. This method and the resulting code \cite{sterling_pynamic} were developed for the research presented in \cref{chp:mapi_diffuse} and in ref. \cite{weadock2023nature}. The code, working examples, and a very detailed manual for the code and data handling procedures are provided on the Github \cite{sterling_pynamic}.

The cross section we want to calculate is given in \cref{eq:doubly_differential_cross_section}. In general, positions do not commute at different times, so evaluating \cref{eq:doubly_differential_cross_section} is difficult. To make progress, we approximate the positions as classical coordinates so that $\hat{\bm{R}} \rightarrow \bm{R}$ is classical. The classical trajectories $\bm{R}(t)$ can be straightforwardly calculated using the molecular dynamics, so this approach is computationally useful. Importantly we have made no assumptions about the configuration of the material, so this method is valid for liquids, disordered compounds, molecular crystals, etc. The main error in the classical approximation is that the scattering function $S(\bm{Q},\omega)$ does not satisfy the principle of detailed balance, i.e. it is not consistent with Bose-Einstein statistics \cite{squires1996introduction,dove1993introduction,harrelson2021computing}. It is possible to systematically add corrections that include quantum mechanical effects by expanding the many-body path integral in \cref{eq:doubly_differential_cross_section} around the classical trajectory \cite{tuckerman2023statistical}. I do not pursue this here\footnote{The classical approximation used here can be shown to be (i) a Bolztmann approximation where we neglect exchange affects by approximating the density matrix as diagonal, and (ii) a stationary phase approximation to the many-body path integral in \cref{eq:doubly_differential_cross_section}. To neglect exchange, we require that particles are "distinguishable" which is valid at high temperature, where "high" is above $\sim30$ K for atoms heavier than carbon. For helium, it is true above $\sim75$ K. For deuterium and hydrogen (protium), it is true above $\sim175$ K and $\sim300$ K respectively. Thus, for heavy elements, this is a very good approximation even at what is typically considered "low" temperature. This is the subject of work I plan to publish elsewhere, but is not included in this thesis. }. In any case, the classical approximation becomes valid at high enough temperature where quantum effects on nuclear motion are negligible and the particles follow Maxwell-Boltzmann statistics (i.e $n_{BE}(\omega,T)+1 \approx n_{BE}(\omega,T)$ at high temperature). Moreover, since we are sampling finite temperature trajectories, this method naturally includes anharmonic effects to all orders. This is in contrast to harmonic lattice dynamics where phonon life times are infinite.

In the classical approximation, we assume that we know the trajectories for all times since they can be calculated in principle from e.g. molecular dynamics. In this case, we can use ergodicity to replace the thermal average by a time average
\begin{equation}\begin{gathered}
    \int \frac{dt}{2\pi\hbar}\langle \exp(-i\bm{Q}\cdot \bm{R}_i(0))  \exp(i\bm{Q}\cdot \bm{R}_j(t))  \rangle  \exp(-i\omega t) = \\
    \int \frac{dt'}{2\pi\hbar} \exp(-i\bm{Q}\cdot \bm{R}_i(t')) \int \frac{dt}{2\pi\hbar}  \exp(i\bm{Q}\cdot \bm{R}_j(t+t')) \exp(-i\omega t)  = \\
    \left( \int \frac{dt}{2\pi\hbar} \exp(-i\bm{Q}\cdot \bm{R}_i(t)+i\omega t) \right) \left( \int \frac{dt'}{2\pi\hbar} \exp(i\bm{Q}\cdot \bm{R}_j(t')-i\omega t') \right)
\end{gathered}\end{equation}
where we used time-translational invariance to shift $t'\rightarrow t-t'$. Finally, we arrive at
\begin{equation}\begin{split}
    \frac{d^2\sigma}{d\Omega dE'} = \frac{k'}{k} \left| \sum_{i} b_i \int \frac{dt}{2\pi\hbar} \exp(i\bm{Q}\cdot \bm{R}_i(t)-i\omega t) \right|^2 .
    \label{eq:classical_cross_section}
\end{split}\end{equation}
\cref{eq:classical_cross_section} can be straight forwardly evaluated from molecular dynamics trajectories, $\bm{R}_i(t)$. This is the expression that my code, \texttt{pynamic-structure-factor} \cite{sterling_pynamic}, calculates. The physical interpretation of \cref{eq:classical_cross_section} is as the probability of a neutron scattering from the scattering system by changing momentum by $\hbar\bm{Q}$ and energy by $\hbar \omega$. A similar expression to \cref{eq:classical_cross_section} has been used in the literature in the past \cite{zushi2015effect,xiong2017native}. However, the expressions in these references are not "squared" as they should be: for the scattering cross section to have the right dimensions, it must be proportional to $b^2$. This is probably just a typo in \cite{zushi2015effect} which was propagated into ref. \cite{xiong2017native} when quoting ref. \cite{zushi2015effect}. 

If the experiment uses x-rays instead of neutrons, the only change we need to make is to replace the scattering lengths $b_i$ in \cref{eq:classical_cross_section} by the atomic form factors $f_i(Q)$. This can be shown explicitly by using the correct interaction potential for x-ray/atom scattering in \cref{eq:fermi_pseudopotential}. The form factors can be approximated by a sum of Gaussians:
\begin{equation}
    f_i(Q)=\sum_j^4 p_{i,j} \exp \left(-q_{i,j} \left( \frac{Q}{4\pi}\right)^2\right)+s_i.
    \label{eq:fQ}
\end{equation}
The parameters $p_{i,j}$, $q_{i,j}$, and $s_i$ for x-rays and the scattering lengths $b_i$ for neutrons are tabulated and can be looked up \cite{sears1992neutron,brown2006intensity}. The data in these references are what are used by \texttt{pynamic-structure-factor}. The index $i$ runs over all atoms in the simulation cell and $f_i(Q)$ is different in general for different elements (and even different charge states of the same elements). $Q$ is the (magnitude of) momentum transferred from the incident (monochromatic) beam into the scattering system.

\subsection{Calculating the classical cross section}
\label{sec:finite_size_effects}

The trajectories in \cref{eq:classical_cross_section} can be generated with any code that can do molecular dynamics (MD), either classical or ab initio. See \cref{sec:md} for details of MD and \cref{chp:electrons} for details of ab initio calculations. \cref{eq:classical_cross_section} is calculated as a post processing step, so works with any MD codes in principle. The codes write trajectories to a file with metadata describing what elements are in the simulation. In this section, we describe some important considerations when calculating the classical cross section as a post processing step. Specifically, we talk about finite size effects. "Finite size effects" is a blanket term describing errors introduced into a calculation by using a finite supercell and finite length trajectory.

Computationally, the the classical cross section is calculated as a timez and space-Fourier transform (FT) of the trajectories. The trajectory is spaced evenly on a grid in time, so the time-FT is done using numerical fast-Fourier transforms (FFT). The temporal frequencies are generated automatically by the FFT routines on a regular grid. The positions aren't on a "grid" in space so it is not practical to use FFTs for the space-Fourier transform. We \emph{could} put the scattering lengths onto an extremely fine grid: e.g. discretize the simulation cell and put 0's on all empty voxels and the scattering lengths on all voxels containing atoms. However, since the coordinates are $\delta$-functions, we need an \emph{extremely} fine grid to do this accurately. We could then use FFT's; however, this scheme would be prohibitively costly, both with computer memory and computational speed. This would be absurd. Instead, the space-FT is calculated directly by summing over atoms and the user has to specify the $\bm{Q}$-points manually.

This is an important "operational" distinction between FFT's and direct FT's: in the case of the time FFT, the frequency "grid" is fixed by the spacing of the time steps. In the case of the direct space FT, the user gives the $\bm{Q}$-points as arguments. This has advantages, e.g. one can only calculate only the $\bm{Q}$-points of interest rather than a huge grid. But there is a pitfall: we must pick $\bm{Q}$-points that are commensurate with the simulation cell or we get nonsense data. 

What does "commensurate" mean? Suppose that the density in the simulation cell is $\rho(x)$. $1d$ notation is used for convenience. Let the size of the simulation box in $1d$ be $R$. We aren't making any assumptions about the system; it could be a crystal, liquid, etc. All we assume is the simulation uses periodic boundary conditions (PBC). Due to PBC, the density satisfies $\rho(x)=\rho(r+R)$. The FT is given by $\rho(G) \propto  \sum_x \rho(x) \exp(-i G x)$ and inverse FT by $\rho(x) \propto \sum_G \rho(G) \exp(i G x)$. Since $\rho(x)=\rho(x+R)$ is periodic, we require that $\exp(iGx)=\exp(iG(x+R))=\exp(iGx)\exp(iGR)$. In other words, we require $G = 2 \pi  n/ R$ with $n$ an integer. In the limit of an infinite system, $\rho(G) = 0$ if $n$ is \emph{not} an integer; for a \emph{finite} system, $\delta$-functions have finite width and there is "spectral leakage", i.e. peaks acquire finite width. Commensurate $\bm{Q}$-points are still correct, but incommensurate $\bm{Q}$-points acquire an oscillatory intensity that contains no useful information. See Section 4.1 in ref \cite{neder2008diffuse} for more details: they explain the same concept in terms of "windowing functions" and it is more clear in that context what happens at incommensurate $\bm{Q}$-points. In any case, what happens at incommensurate $\bm{Q}$-points depends on the system size, so is called a "finite size effect".

\section{Experimental neutron scattering}

Neutron scattering "beam time" is a shared resource at user facilities, so an experiment is usually as simple, in principle, as taking a crystal to the facility, loading the sample into a holder, and switching the beam on. E.g. at the SNS, each instrument has a GUI application that the user can use to rotate the sample, change incident energy, etc. The GUI can even be accessed in a web browser and experiments can be controlled remotely. Of course, physical intuition is needed to pick certain quantities: e.g. incident energy, sample orientation, and there is no general "formula" for choosing these quantities.  The correct choice is usually made "on the fly" to maximize a particular quantity: e.g. incident intensity or energy resolution. The choice even depends on the particular instrument and material being studied. With that in mind, it isn't especially important to know the details of any particular neutron scattering experiment. Rather, in this section, we describe the general operating principles of the instruments used in this thesis.

\subsection{Neutron scattering instruments}

In the context of this thesis, a neutron scattering "instrument" is the combination of incident neutron energy selector, sample holder/manipulator, and detectors. Nowadays, the neutron source, i.e. where the neutrons actually come from, is more or less irrelevant at the user level. Still, we briefly discuss neutron sources before describing instruments in more detail. 

\begin{figure}[t!]
    \centering
    \includegraphics[width=0.8\linewidth]{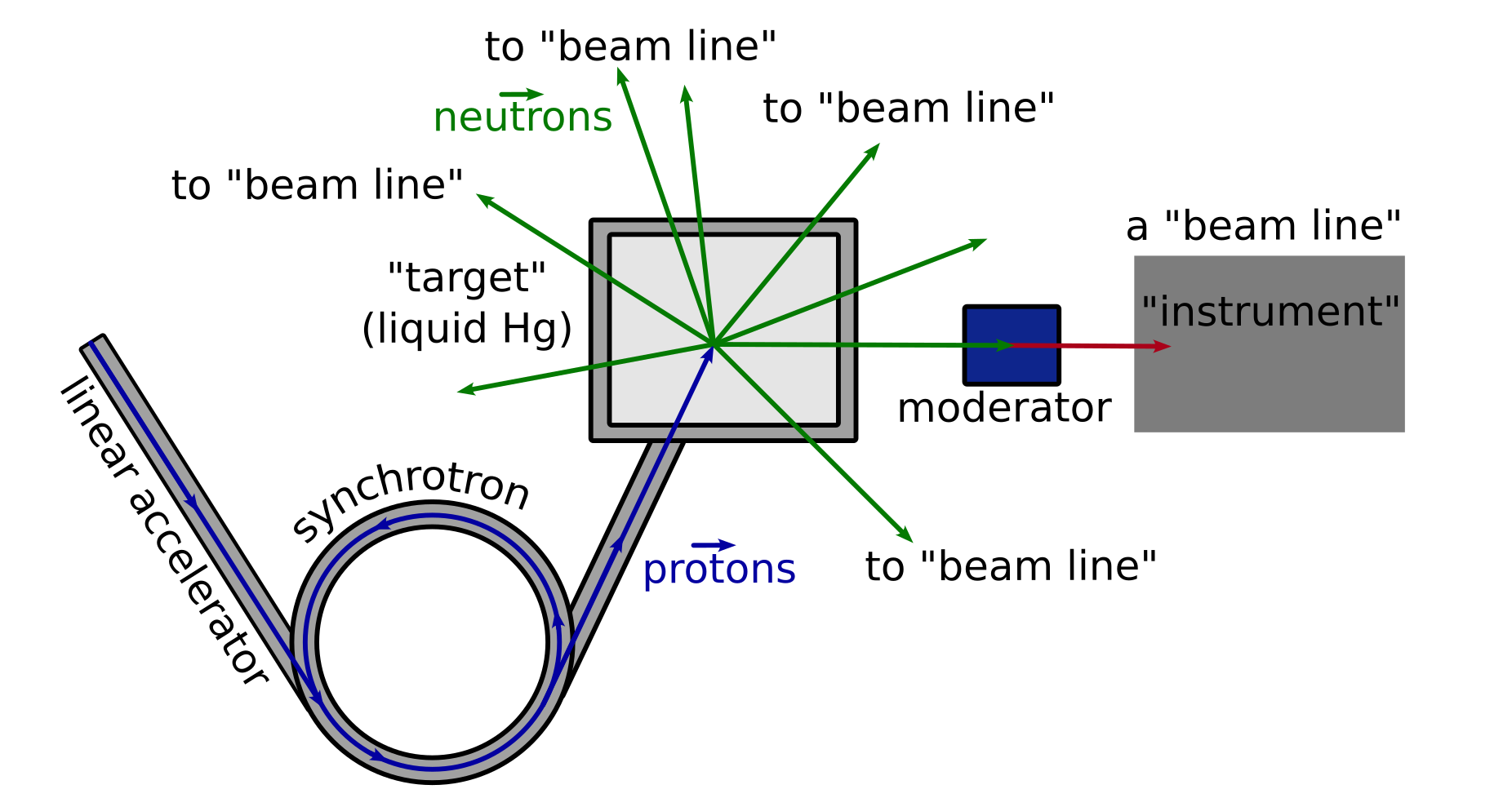}
    \caption{Schematic of a spallation neutron source such as the Spallation Neutron Source (SNS) at Oak Ridge National Lab. Charged particles are accelerated to near the speed of light in a linear accelerator and then a pulse of particles is accumulated in a synchrotron. At a fixed frequency (60 Hz at the SNS), pulses are delivered to the "target". The extremely high energy particles collide with target, knocking neutrons off the atoms in a process called "spallation". The neutrons are directed towards "beam lines", where they are slowed in moderators and then in e.g. a spectrometer, the beam is monochromated and scattered off a sample.}
    \label{fig:spallation_source}
\end{figure}

\subsection{Neutron sources}

The neutron sources in the earliest experiments were reactors \cite{rumsey2018history}. I.e. radioactive materials are combined in a reactor until a chain reaction starts, producing a steady stream of neutrons. The benefits of reactor sources are high neutron flux, since there is a continuous stream of neutrons. The neutrons produced in a reactor are shot into a moderator that thermalizes them, and then they are directed to instruments. The most serious drawbacks of reactors are the safety hazards: the most tangible hazard is the production of nuclear waste. The most catastrophic (and unlikely) hazard is melt down. Reactor sources are cheaper and simpler to operate than the alternatives, but don't generally have the support of public opinion because of their hazards. For this reason, neutron spallation sources are becoming more common. 

Spallation sources operate on a very different principle from reactors: a beam of charged particles is accelerated to near the speed of light and then crashed into a target \cite{bauer2001physics}. A schematic of a spallation source is shown in \cref{fig:spallation_source}. At the Spallation Neutron Source (SNS) at Oak Ridge National Lab (ORNL), the incident beam is H$^-$ ions and the target is liquid mercury \cite{ornlWorksNeutron}. The ions knock neutrons off the Hg atoms and those neutrons are directed at moderators to thermalize them and then are directed to samples/detectors. Spallation sources are usually pulsed: e.g. at the SNS, ion pulses are shot into the target at 60 Hz. The intensity of the resulting neutron pulses is high, but the time averaged flux is lower than reactor sources. But since there is no risk of melt down, spallation sources are considered safer and are becoming more prevalent. Still, the target, moderators, and many other materials used to direct and control neutrons become radioactive; spallation sources produce nuclear waste that must be handled with care. 

\subsection{Neutron spectrometers}\label{sec:neutron_instruments}

Spectrometers do spectroscopy, i.e. measure the excitations in condensed matter. Neutrons transfer momentum into a crystal by changing speed and direction; the change in speed results in a change in energy. Spectrometers measure these changes in the scattered beam of neutrons, telling us about the created or absorbed excitations. 

\begin{figure}[t!]
    \centering
    \includegraphics[width=0.9\linewidth]{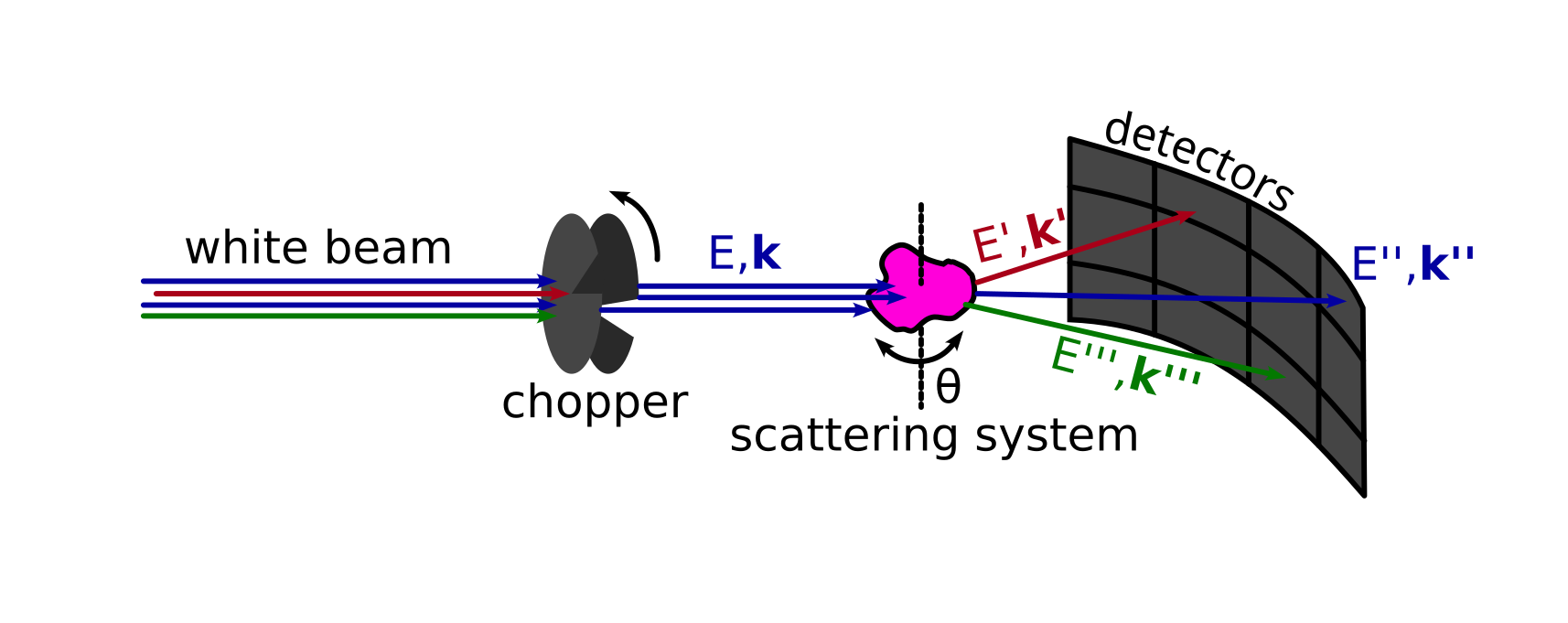}
    \caption{A schematic of a time-of-flight spectrometer. The chopper monochromates the beam; the phase and speed of the disk determines the energy allowed through. The monochromatic beam hits the sample and is scattered to a detector bank; each detector is like a pixel in a CCD. The speed of the input neutrons and the distances to the sample from the chopper and to the detectors from the sample are known and the time-of-flight of each pulse of neutrons is measured; this enables us to calculate the speed of the scattered neutrons, which tells us their energy. Combined with the speed, the positions of the detectors tell us the change in momentum.}
    \label{fig:time_of_flight}
\end{figure}

There are two main types of spectrometers used to measure bulk excitations in crystals: (i) triple axis spectrometers (TAS) and (ii) time-of-flight (TOF) spectrometers. We don't use TAS instruments in this thesis, so we only describe TOF instruments here. A schematic of a TOF spectrometer is shown in \cref{fig:time_of_flight}. Examples of TOF spectrometers used in this thesis are MERLIN \cite{bewley2006merlin,bewley2009merlin} at the ISIS Neutron and Muon Source in the UK and CORELLI \cite{ye2018implementation} at the SNS in Oak Ridge, Tennessee, USA.

The principles of time-of-flight spectrometers are simple. Assume that we know the energy and momentum of incident neutrons. In short TOF spectroscopy uses the time it takes for a neutron to traverse the sample chamber to measure change in speed (i.e the change in energy) of the scattered neutrons. Combined with the change in speed, the coordinate of the detectors are used to determine the change in momentum. 

In more detail: a pulsed beam of neutrons is introduced into the sample chamber, a large ($\sim$m) vacuum vessel holding the sample. The pulsed beam is created by a "chopper" (see \cref{fig:time_of_flight}). The chopper also selects the speed of the incoming neutrons (see below). Since we know the incoming speed, we know the incident energy according to $E_{in}= m v_{in}^2/2$. The distance from the chopper to the sample, $l_{in}$, and from the sample to each detector, $l_{out}$, is known. The time-of-flight from the chopper to the sample, $\tau_{in}=l_{in}/v_{in}$, is known since we know the incident speed. Fancy electronics record the total flight time, $\tau_{tot}$, for every neutron that enters the sample chamber from the duration it takes a neutron to traverse from the chopper to any given detector. The flight time from the sample to the detector is $v_{out} = l_{out}/\tau_{out} = l_{out}/(\tau_{tot}-l_{in}/v_{in})$ and the energy of the scattered neutron is $E_{out} = m v^2_{out}/2$. The change in energy is readily available, corresponding to the phonon energy. The change in speed and the coordinate of the detector are used to determine the change momentum, corresponding to the phonon momentum in the sample. The incident flux and the number of neutrons counted at each detector determine the cross section, i.e \cref{eq:doubly_differential_cross_section}.

If the incident beam is monochromatic, the chopper consists of spinning disks with slits in them; the relative phase between the slits selects the incident neutron energy, $E_{in}= m v_{in}^2/2$. This is the case with e.g. the MERLIN and ARCS \cite{abernathy2012design} instruments, at the ISIS and SNS facilities respectively, which are used to measure phonon dispersions. Since all neutrons have the same incident energy, it is easy to determine the energy change. We used the MERLIN TOF spectrometer for the measurements in \cref{chp:mapi_diffuse,chp:bgg}.

In a "correlation spectrometer" like CORELLI \cite{ye2018implementation} at the SNS, a pseudorandom distribution of slits in the chopper sends in a pulse of neutrons with a time varying distribution of energy. This is advantageous because the intensity of the incident beam is peaked at a certain energy and admitting the whole spectrum vastly increases the intensity relative to a monochromatic chopper. The arrival time of the neutrons at the detector can be correlated with the phase of the chopper to discriminate between elastically and inelastically scattered neutrons. The detector coordinate determines the change in momentum, so correlation spectroscopy can be used to measure elastic diffuse scattering in crystals (see \cref{sec:diffuse_scattering}). We used CORELLI for the research in \cref{chp:mapi_diffuse}.

\section{Summary}

We mentioned in the introduction that a useful way to study the properties of energy materials is through their atomic dynamics. In this section, we saw that neutron scattering can be used to measure the atomic dynamics. The exact expression for the doubly differential neutron scattering cross section in  \cref{eq:doubly_differential_cross_section} (roughly speaking, it is the fraction of incident neutrons scattered into a particular direction with a particular change in energy) contains a complicated many-body expectation value that can't be calculated exactly. We introduced two approximations: (i) the harmonic approximation which is valid for crystals at low enough temperature and (ii) the classical approximation which is valid for any structure at high enough temperature. 

Detailed calculations show that the cross section in the harmonic approximation depends on the phonon dispersions and eigenvectors in a non-trivial but tractable way. Thus, neutron scattering measurements can in principle be used to measure phonon dispersions and eigenvectors. In interesting materials, e.g. cuprates, the measured dispersions show anomalies that don't agree with calculations (see \cref{chp:lco_lda_u} and ref. \cite{sterling2021effect}). These anomalies are related to the electronic structure, enabling us to study the energy relevant physics. In Ba$_8$Ga$_{16}$Ge$_{30}$ (BGG), there are anomalies that are related to occupational disorder, revealing the mechanism of low thermal conductivity (see \cref{chp:bgg} and ref. \cite{roy2023occupational}).

Calculations in the classical approximation show that the cross section is a essentially the space- and time-Fourier transform of the atomic density. We can directly calculate atomic trajectories with molecular dynamics calculations. When the calculated cross section agrees well with measurements, or doesn't agree in an interesting way, we can directly look at the trajectories to understand the atomic dynamics. E.g. in hybrid perovskite CH$_3$NH$_3$PbI$_3$, molecular dynamics predicts the same diffuse scattering measured in experiments and the corresponding atomic dynamics can be directly analyzed (see \cref{chp:mapi_diffuse} and ref. \cite{weadock2023nature}).

\chapter{The Electronic Structure of Matter}
\label{chp:electrons}

By now, it is apparent that we need to model the dynamics of atoms to simulate neutron scattering spectra. Whether or not we do this quantum mechanically or classically, we need to know the potential energy seen by the atoms. In the quantum case, the potential energy shows up in the Hamiltonian; in the classical case, derivatives of the potential energy show up in the classical equations of motion equation. It turns out that directly calculating the potential energy of the ions is difficult. To the same extent, so is calculating the forces. In this chapter, we see how to calculate the energy and forces.

\section{The Condensed matter theory of everything}
\label{sec:condensed_matter}

In the absence of an external potential\footnote{e.g. an external electric field; later, the potential of the ions themselves is discussed as an "external" field acting on the electrons.}, the most general Hamiltonian to describe condensed matter is \cite{vanderbilt2018berry,giustino2017electron}
\begin{equation}
    \hat{H} = \hat{H}_{e}+\hat{H}_{n}+\hat{H}_{en}.
    \label{eq:cm_hamiltonian}
\end{equation}
It is the sum of the electronic Hamiltonian, $\hat{H}_e$, the nuclear Hamiltonian, $\hat{H}_n$, and the Hamiltonian for their interaction, $\hat{H}_{en}$. This is basically the \emph{the theory of everything} in condensed matter physics. Anything interesting that we can observe in materials science is buried in this equation: magnetism, superconductivity, metal-insulator transitions, etc.


In the position representation, the electronic Hamiltonian is 
\begin{equation}
\begin{gathered}
    \hat{H}_e = \hat{T}_{e}+\hat{V}_{ee} \qquad
    \hat{T}_e = \sum^{N_e}_i\frac{\hat{\bm{p}}^2_i}{2 m_e} \qquad
    \hat{V}_{ee} = \frac{k}{2} \sum^{N_e}_{i \neq j} \frac{e^2}{\vert \bm{r}_i-\bm{r}_j \vert} .
\end{gathered}
\end{equation}
$\hat{T}_e$ is the electronic kinetic energy operator with $\hat{\bm{p}}_i\equiv -i \hbar \bm{\nabla}_{i}$ the momentum operator acting on the $i^{th}$ electron. The kinetic energy sum runs over all electrons in the system. $\hat{V}_{ee}$ is the mutual Coulomb interaction between all the electrons, with $k$ Coulomb's constant. $m_e$ is the electronic mass and $e$ is the elementary charge. The denominator is the distance between the electrons, with lower case $\bm{r}$ denoting electronic coordinates. $i$ is a composite index labelling the electron and its spin, $\sigma$. When spin is relevant later, we label it explicitly.


Similarly, the nuclear Hamiltonian is
\begin{equation}
\begin{gathered}
    \hat{H}_n = \hat{T}_{n}+\hat{V}_{nn} \qquad
    \hat{T}_n = \sum^{N_n}_I\frac{\hat{\bm{P}}^2_I}{2 M_I} \qquad
    \hat{V}_{nn} = \frac{k}{2} \sum^{N_n}_{I \neq J} \frac{e^2 Z_I Z_J}{\vert \bm{R}_I-\bm{R}_J \vert} .
\end{gathered}
\end{equation}
Here, $\hat{\bm{P}}_i\equiv -i \hbar \bm{\nabla}_{I}$ is the nuclear momentum operator with the kinetic energy sum running over all atoms. $\hat{V}_{nn}$ is the nuclear Coulomb interaction. $M_I$ is the mass of the $I^{th}$ atom and $eZ_I$ is its charge. The denominator is the distance between the atoms with upper case $\bm{R}$ denoting nuclear coordinates. 


Finally, we have the interaction between electrons and nuclei:
\begin{equation}
\begin{gathered}
    \hat{H}_{en} = \hat{V}_{en} = - k \sum^{N_e}_i \sum^{N_n}_I \frac{e^2 Z_I}{\vert \bm{r}_i-\bm{R}_I \vert} .
\end{gathered}
\end{equation}

Solutions of the Schr\"odinger's equation (SE) for Hamiltonian \cref{eq:cm_hamiltonian} in coordinate space are the wave functions $\Psi(\{ \bm{r} , \bm{R} \})$ which depend explicitly on both the coordinates of the electrons and the nuclei.
\begin{equation}
    \hat{H} \Psi(\{ \bm{r} , \bm{R} \}) = E \Psi(\{ \bm{r} , \bm{R} \}) .
    \label{eq:mb_schrodingers}
\end{equation}
The notation $(\{ \bm{r} , \bm{R}  \})$ is supposed to imply "... depends explicitly on the combined set of nuclear and electronic coordinates." This is all well-and-good, but it turns out that if the number of the particles is more than a few, \cref{eq:mb_schrodingers} is unsolvable on even the best supercomputer. So we will need to make some simplifying assumptions.

\subsection{The Born-Oppenheimer approximation}\label{sec:born_opp}

The first simplifying assumption is the \emph{Born-Oppenheimer} or adiabatic approximation \cite{sham1963electron,chester1959electron,martin2020electronic,thijssen2007computational}. We make the separable ansatz that $\Psi(\{ \bm{r} , \bm{R} \}) \equiv \Psi(\{ \bm{r} \} ,\{ \bm{R} \}) =\chi(\{ \bm{R} \}) \Phi(\{ \bm{r} \}, \{ \bm{R} \}) $. The new notation $(\{\bm{r} \} , \{ \bm{R} \})$ means "... depends explicitly on the set of electronic coordinates but only parametrically on the set of nuclear coordinates." While this type of notation is very explicit, it's more cumbersome to typeset. Let's be lazy and instead use $\Psi(\{ \bm{r}, \bm{R} \}) \equiv \Psi(\bm{r},\bm{R})$ and $\Psi(\{ \bm{r} \}, \{ \bm{R} \}) \equiv \Psi(\bm{r};\bm{R})$; the semi-colon separates the explicit from parameteric dependence on the coordinates. 

Define the Born-Oppenheimer Hamiltonian for the electrons:
\begin{equation}
    \hat{\mathcal{H}}_e = \sum^{N_e}_i\frac{\hat{\bm{p}}^2_i}{2 m_e}  + \frac{k}{2} \sum^{N_e}_{i \neq j} \frac{e^2}{\vert \bm{r}_i-\bm{r}_j \vert} - k \sum^{N_e}_i \sum^{N_n}_I \frac{e^2 Z_I}{\vert \bm{r}_i-\bm{R}_I \vert} .
    \label{eq:bo_hamiltonian}
\end{equation}
For now, the curly font on $\hat{\mathcal{H}}$ and $\mathcal{E}$ means "Born-Oppenheimer". Suppose $\Phi(\bm{r};\bm{R})$ is an eigenstate of $\hat{\mathcal{H}}_{e}$ with eigenvalue $\mathcal{E}_e(\bm{R})$. $\Phi(\bm{r};\bm{R})$ is an approximation to the true many-body wave function of the electrons; it is still "interacting" in the electronic subspace. It's clear that $\Phi(\bm{r};\bm{R})$ and $\mathcal{E}_e(\bm{R})$ depend only parametrically on the positions of the nuclei. It's common to call the electron-nuclear Coulomb interaction in \cref{eq:bo_hamiltonian} $V_n(\bm{r})$ and simply treat it as an external potential in the electronic problem. This is the electronic Hamiltonian that is usually solved\footnote{Actually, this is still an interacting problem. It is what we \emph{try} to solve in practice.} in practice and its solution provides us with the potential energy on the nuclei. We will look at this in more detail soon. 

We want to insert $\Psi(\bm{r}; \bm{R} )$ into the full Hamiltonian in \cref{eq:cm_hamiltonian}, but we have to be careful and remember that $\hat{\bm{P}}^2= -\hbar^2 \bm{\nabla}^2_{\bm{R}} $ acts on both $\chi(\bm{R})$ and $\Phi(\bm{r};\bm{R})$. The result is
\begin{equation}
\begin{gathered}
    \hat{H} \Psi(\bm{r};\bm{R}) =  \Phi(\bm{r};\bm{R}) \left[ \sum^{N_n}_I\frac{\hat{\bm{P}}^2_I}{2 M_I}  + \frac{k}{2} \sum^{N_n}_{I \neq J} \frac{e^2}{\vert \bm{R}_I-\bm{R}_J \vert} + \mathcal{E}_e(\bm{R})\right] \chi(\bm{R}) \\
    -\chi(\bm{R}) \sum^{N_n}_I \frac{\hbar^2}{2 M_I} \nabla^2_{R_I} \Phi(\bm{r};\bm{R}) -\sum^{N_n}_I \frac{\hbar^2}{M_I} \bm{\nabla}_{R_I} \Phi(\bm{r};\bm{R}) \cdot \bm{\nabla}_{R_I} \chi(\bm{R})  \\ = \mathcal{E} \Psi(\bm{r};\bm{R}) 
    \label{eq:bo_2}
\end{gathered}
\end{equation}
The curly font, $\mathcal{E}$, is to distinguish it from the exact many-body eigenvalue  $E$ in \cref{eq:mb_schrodingers}. The electronic energy, $\mathcal{E}_{e}(\bm{R})$, is the eigenvalue of the Born-Oppenheimer Hamiltonian \cref{eq:bo_hamiltonian}, but its effect in \cref{eq:bo_2} is to act as a "potential energy" for the nuclei. The essence of the Born-Oppenheimer approximation is to ignore the second line in \cref{eq:bo_2}. For the first term, this is justified since the variation of electronic kinetic-energy with respect to atomic coordinates is small. This might be violated at large temperature, but in most condensed matter settings, it's safe to ignore this term. Ignoring the second term is the same as neglecting electron-phonon coupling. For more detailed justification, see ref. \cite{thijssen2007computational}, exercise 4.1, or \cite{giustino2017electron,ziman1979principles}). Electron-phonon coupling is very important in many cases and should not be neglected, but we won't look these in detail in this thesis.

\subsection{Summary}

To summarize, we have split the problem of the full interacting system of electrons and nuclei into the following set of (approximate) equations that describe the dynamics of the set of nuclei and electrons separately. For the electrons
\begin{equation}
\begin{gathered}
    \hat{\mathcal{H}}_{e} \Phi(\bm{r};\bm{R}) = \left[\sum^{N_e}_i\frac{\hat{\bm{p}}^2_i}{2 m_e}  + \frac{k}{2} \sum^{N_e}_{i \neq j} \frac{e^2}{\vert \bm{r}_i-\bm{r}_j \vert} + V_n(\bm{r}) \right] \Phi(\bm{r};\bm{R}) = \mathcal{E}_e(\bm{R}) \Phi(\bm{r};\bm{R})
    \label{eq:e_bo_ham}
\end{gathered}
\end{equation}
The external potential acting on the electrons is the Coulomb field of the nuclei
\begin{equation}
    V_{n}(\bm{r}) = - k \sum^{N_e}_i \sum^{N_n}_I \frac{e^2 Z_I}{\vert \bm{r}_i-\bm{R}_I \vert}
\end{equation}
and for the nuclei,
\begin{equation}
\begin{gathered}
    \hat{\mathcal{H}}_{n} \chi(\bm{R}) = \left[ \sum^{N_n}_I\frac{\hat{\bm{P}}^2_I}{2 M_I}  + \frac{k}{2} \sum^{N_n}_{I \neq J} \frac{e^2}{\vert \bm{R}_I-\bm{R}_J \vert} + \mathcal{E}_e(\bm{R}) \right] \chi(\bm{R}) = \mathcal{E}_n(\bm{R}) \chi(\bm{R})
    \label{eq:e_n_ham}
\end{gathered}
\end{equation}
Note the dependence of the nuclear energy $\mathcal{E}_n(\bm{R})$ on the electronic energy $\mathcal{E}_e(\bm{R})$. The electronic energy in the nuclear Hamiltonian is often called "the Born-Oppenheimer energy surface" or just the "Born-Oppenheimer" potential. It is a crucial part of the nuclear potential energy mentioned multiple time so far. It acts like an "external" potential on the ions. It contains all of the effects of \emph{instantaneous} electronic screening of the nuclear-nuclear Coulomb interaction. When making the adiabatic approximation, we are assuming the electrons are so light that they can instantly adjust to any change in the nuclear configuration without being excited. We won't need to know any more about this in this thesis: see refs. \cite{giustino2017electron,chester1959electron,sham1963electron} for more details.

\subsection{Remarks}

We've simplified the problem quite a bit, but solving either one of these equations is still pretty much impossible because they are interacting systems (electron-electron and nuclear-nuclear interactions). For now, suppose that we can, in principle, solve this system of equations. What does it tell us? A few examples come to mind in materials science. For instance, the derivative of the potential energies in the nuclear Hamiltonian are the \emph{forces} acting on the nuclei as mentioned before. If we minimize these forces, we can determine the equilibrium structure of the material we are modelling. If the equilibrium structure is ordered, second derivatives are force-constants that can be used to model "phonons", the quanta of lattice vibrations. These results can be used to calculate thermodynamics and other properties of crystals as well as to simulate the neutron scattering spectra. 

Moreover, solving the electronic Hamiltonian at a given set of atomic coordinates gives us the electronic energy; i.e. the "band structure." If we solve it at the equilibrium nuclear configuration, we can calculate thermodynamic and, to some extent, electronic transport quantities. Regardless of whether or not the material is ordered, we can use the forces to study lattice dynamics and other properties with molecular dynamics: this approach, where we calculate forces from the quantum mechanical Hamiltonian, is called \emph{ab-initio} molecular dynamics (AIMD).

The rest of this chapter is dedicated to discussing how to solve the electronic equations in practice. 


\section{Single-particle methods: the variational method and self consistency}

This section follows refs. \cite{girvin2019modern,thijssen2007computational,martin2020electronic,giannozzi2020numerical}. The discussion is left general so that (i) the notation is simpler and (ii) is applicable to gases, liquids, ordered and disordered solids. We are mainly concerned with crystals in this thesis, so we specialize to crystals later. Conveniently, all of the theory developed here is concisely summarized for periodic systems. 

In \cref{sec:condensed_matter}, we derived the the most general condensed matter electronic Hamiltonian in the Born-Oppenheimer approximation. We restate it here:
\begin{equation}
    \hat{H} = \sum_i\frac{\hat{\bm{p}}^2_i}{2 m}  + \frac{k}{2} \sum_{i \neq j} \frac{e^2}{\vert \bm{r}_i-\bm{r}_j \vert} - k \sum_i \sum_I \frac{e^2 Z_I}{\vert \bm{r}_i-\bm{R}_I \vert} .
    \label{eq:e_bo_hamiltonian}
\end{equation}
$\hat{\bm{p}}_i$ and $\bm{r}_i$ are the momentum operator and coordinates of the $i^{th}$ electron, $k$ is Coulombs constant, $e$ is the elementary charge, $eZ_I$ is the charge of the $I^{th}$ ion, and $\bm{R}_I$ is its coordinate. $i$ is a composite index labelling the electron and its spin, $\sigma$. When spin is relevant later, we label it explicitly. Note, we dropped the "curly" notation and assume from now on that everything is in the Born-Oppenheimer approximation.

\subsection{The single-particle Schr\"{o}dinger's equation}

The electronic Hamiltonian in \cref{eq:e_bo_ham} is still insoluble. We face the same problem as in \cref{eq:cm_hamiltonian}. The electronic Coulomb potential is a two-particle operator and in the operator many-body formalism, these are non-linear operator equations; at present, they \emph{can not} be solved. The field of many-body theory is focused around the approximate solution of these equations within the framework of quantum field theory; see e.g. \cite{altland2010condensed,ziman1969elements,giustino2017electron}. The formalism of many-body physics is gnarly and the concepts can be difficult to grasp. So for now, let us take a step back.

Our plan of attack will be to try to reduce the many-problem with the two-particle operators into an effective single-particle theory. To be specific, we want to map the Hamiltonian in \cref{eq:e_bo_ham} onto a set of single-particle Hamiltonians with an arbitrary external single-particle potential. This will be directly accessible to the tools of \emph{single-particle} quantum-mechanics\footnote{From now on, "quantum mechanics" will always mean single-particle.} and we can more easily solve the effective theory.

There are varying degrees of sophistication to this and a few different types of theories can be derived. We want to find single-particle equations with all of the effects of the interactions placed into an effective potential. This might sound like "free lunch." However, while the effective potentials will be single-particle type, they will depend explicitly on the wave functions (the Hartree-Fock theory) or the density (density functional theories); such potentials are called "self consistent". self consistency is the price to be paid if we want to accurately include the effects of interactions in a single-particle theory. Before we dive into these theories, let's warm up with a simpler example: the Hartree approximation.

\subsection{The Hartree approximation}

The issue in solving \cref{eq:e_bo_ham} lies in the Coulomb interaction, which is a two-particle operator in a second-quantized description. We want to replace it with an approximate single-particle operator. Re-write the electronic many-body Hamiltonian as 
\begin{equation}\begin{split}
    \hat{H} = \sum^{N}_i \left( \frac{\hat{p}^2_i}{2 m_e} + v_{ext}(\bm{r}_i) \right)+ \frac{1}{2} \sum^N_{i \neq j} u(\bm{r}_i,\bm{r}_j)
    \label{eq:e_ham}
\end{split}\end{equation}
For now, we will only talk about electrons so we drop the subscripts and don't write nuclear coordinates. Moreover, $v_{ext}(\bm{r}_i)$ can be any sensible \emph{external} potential. The only restriction is that it can only depend on one electronic coordinate at a time. On the other hand, $u(\bm{r}_i,\bm{r}_j)$ is a two-particle operator. It depends on two electronic coordinates, coupling particles. It doesn't have to be the Coulomb interaction for now; it can be anything sensible in what follows.

Let's re-write the interaction $u(\bm{r}_i,\bm{r}_j)$ in terms of the electronic density operator $\hat{n}(\bm{r})=\sum_i^N \delta(\bm{r}-\bm{r}_i)$ \cite{girvin2019modern}\footnote{Redo this derivation using the density \emph{matrix} operator.} First of all
\begin{equation}
    u(\bm{r}_i,\bm{r}_j)=\int \int d\bm{r} d\bm{r}' \delta(\bm{r}-\bm{r}_i) \delta(\bm{r}'-\bm{r}_j)u(\bm{r},\bm{r}')
\end{equation}
If we sum this over all possible pairs of atoms, we have
\begin{equation}
    \frac{1}{2} \sum^N_{i,j} u(\bm{r}_i,\bm{r}_j)=\frac{1}{2}\int \int d\bm{r} d\bm{r}' \hat{n}(\bm{r}) \hat{n}(\bm{r}') u(\bm{r},\bm{r}')
\end{equation}
This looks almost just like the two-particle potential in \cref{eq:e_ham}, except it includes interactions between an electron and itself. Just subtract out the single-electron term: 
\begin{equation}
    \hat{U} \equiv \frac{1}{2} \sum^N_{i \neq j} u(\bm{r}_i,\bm{r}_j)=\frac{1}{2}\int \int d\bm{r} d\bm{r}' \hat{n}(\bm{r}) \hat{n}(\bm{r}') u(\bm{r},\bm{r}')-\sum^N_i u(\bm{r}_i,\bm{r}_i)
\end{equation}
The last term is a constant and we can just cram it into the definition of the external potential.

Lets now rewrite the density as its mean plus fluctuations. $\hat{n}(\bm{r})=n(\bm{r})+\delta \hat{n}(\bm{r})$. The mean term is just the expectation value of the density $n(\bm{r})\equiv \langle \hat{n}(\bm{r}) \rangle$ and the fluctuation part is $\delta \hat{n}(\bm{r})= \hat{n}(\bm{r})- n(\bm{r})$. Stick this into the two-particle operator (ignoring the constant term). 
\begin{equation}
\begin{gathered}
    \hat{U} = \frac{1}{2}\int \int d\bm{r} d\bm{r}' [ n(\bm{r})+\delta\hat{n}(\bm{r}) ] [n(\bm{r}')+\delta\hat{n}(\bm{r}')] u(\bm{r},\bm{r}') \\
    = \int d\bm{r} \hat{n}(\bm{r}) \left( \int d\bm{r}' n(\bm{r}') u(\bm{r},\bm{r}') \right) \\
    + \frac{1}{2}\int \int d\bm{r} d\bm{r}' \left[ \delta \hat{n}(\bm{r}) \delta \hat{n}(\bm{r}') -  n(\bm{r}) n(\bm{r}') \right] u(\bm{r},\bm{r}')
\end{gathered}
\end{equation}
The last term in the last line is constant, so just like with the bonus part of \cref{eq:e_ham}, we suppose we can just stick it into our external potential. For the remaining term (which is 2$^{nd}$ order in the fluctuations), let's suppose that the fluctuations are small and just throw it away. Define the \emph{Hartree potential}
\begin{equation}\begin{split}
    v_H(\bm{r}_i) = \int d\bm{r} n(\bm{r}) u(\bm{r}_i,\bm{r}) 
    \label{eq:hartree_potential}
\end{split}\end{equation}
The name "Hartree potential" is usually used when $u(\bm{r}_i,\bm{r})$ is the Coulomb interaction, but let's leave $u(\bm{r}_i,\bm{r})$ unspecified for now. An important aspect of the Hartee potential is that it includes \emph{self-interaction}: i.e. the electron feels the potential of its own density. This self-interaction leads to important errors in e.g. density functional theory: the most famous example is the \emph{band gap problem}\cite {perdew2017understanding}. We defer the discussion until later. For now, lets rewrite \cref{eq:e_ham} as
\begin{equation}\begin{split}
    \hat{H}_H = \sum^{N}_i \left( \frac{\hat{p}^2_i}{2 m_e} + v_{ext}(\bm{r}_i) +v_H(\bm{r}_i) \right).
    \label{eq:hartree_hamiltonian}
\end{split}\end{equation}

We have removed the two-particle nature of the Coulomb interaction by coupling each electron to the \emph{mean-field} of the electronic density. This kind of approximation is called a mean-field approximation. We replaced a non-linear many-body problem with a self consistent single-particle one. We threw away all effects of both exchange and dynamical correlation. This is about as simple an approximation that we can make to the true interacting many-body problem. It's a bad approximation because it neglects dynamical screening effects and other things, but it's a good starting point. Later on when discussing density functional theory, we try to put exchange and correlation back in while still keeping a single-particle description. But for now, we have made enough progress to discuss important qualitative features shared by the Hartree approximation, the Hartree-Fock method, and density functional theory.

Specifically, the methods for attacking any of these single-particle Hamiltonian are the same. Solving a single-particle equation like \cref{eq:hartree_hamiltonian} is mostly straightforward. We say \emph{mostly} because the Hartree potential depends on the electronic density which is determined only after solving the Schr\"{o}dinger's equation of the electrons. So this is a so-called "self consistent field" (SCF) problem. The same situation will arise again when we study density functional theory, so it's worth discussing how we approach these problems in practice. But first, let's ignore the density dependence and look at how the single-particle problems are solved in the first place.

\subsection{The variational method}\label{sec:variational_method}

It's useful to phrase the single-particle problem in another way. We need to solve a single-particle Schr\"{o}dinger's equation under an arbitrary (but sensible) external potential. We will only focus on time-independent potentials here, so we will specialize to the time-independent Schr\"{o}dinger's equation (TISE). The TISE in Dirac notation is
\begin{equation}\begin{split}
    \hat{H}| \psi_\alpha \rangle = \epsilon_\alpha |\psi_\alpha \rangle
    \label{eq:tise}
\end{split}\end{equation}
i.e. an eigenvalue problem. We are using lower-case Greek letters for single-particle wave functions.  $\hat{H}$ is a single-particle operator and the eigenvectors are single-particle states. Our goal is to find the eigenvalues and eigenvectors. 

It turns out that direct numerical integration isn't really practical in most cases\footnote{These methods are currently under active research.}. So suppose instead that we do the following. Let $| \psi \rangle$ be some arbitrary trial state that we can somehow vary continuously. We only need to consider a single state for now, so we drop the subscript. The expectation value of $\hat{H}$ with respect to this function is
\begin{equation}
    \langle \hat{H} \rangle = \frac{\langle \psi | \hat{H} | \psi \rangle }{\langle \psi | \psi \rangle} .
\end{equation}
i.e. $ \langle \hat{H} \rangle$ is the energy of the trial state. For the trial function to be sensible, it has to be normalized. So we impose the constraint that this expectation value is normalized by dividing by the overlap. On physical grounds, we want to find the state that minimizes the energy. So we need to find the state that minimizes this expectation value. Let the trial state $|\psi + \delta \psi \rangle = | \psi \rangle+|\delta\psi\rangle$ include variations around the true solution, $|\psi \rangle$. We expect the variation $|\delta \psi \rangle$ to be small, but it is allowed to be arbitrary. The new expectation value with the trial state is
\begin{equation}
    \langle \hat{H} \rangle + \delta\langle \hat{H} \rangle  = \frac{\langle \psi + \delta\psi | \hat{H} | \psi + \delta \psi \rangle }{\langle \psi + \delta \psi | \psi + \delta \psi \rangle} \approx \frac{\langle \psi | \hat{H} | \psi \rangle + \langle \delta \psi | \hat{H} | \psi \rangle + \langle \psi | \hat{H} | \delta \psi \rangle}{ \langle \psi | \psi \rangle + \langle \delta \psi | \psi \rangle + \langle \psi | \delta \psi \rangle } .
\end{equation}
We ignore second order terms, since they are small. Use the expansion $(1+x)^{-1}\approx 1-x$ for small $x$ to make the replacement
\begin{equation}
    \langle \psi | \psi \rangle^{-1} \left[ 1 + \frac{\langle \delta \psi | \psi \rangle + \langle \psi | \delta \psi \rangle} {\langle \psi | \psi \rangle} \right]^{-1} \approx  \langle \psi | \psi \rangle^{-1} \left[ 1 - \frac{\langle \delta \psi | \psi \rangle}{\langle \psi | \psi \rangle} - \frac{\langle \psi | \delta \psi \rangle} {\langle \psi | \psi \rangle} \right] .
\end{equation}
Plug this in and drop second order terms again
\begin{equation}
    \delta\langle \hat{H} \rangle = \frac{\langle \delta \psi | \hat{H} | \psi \rangle}{ \langle \psi | \psi \rangle} + \frac{\langle \psi | \hat{H} | \delta \psi \rangle}{ \langle \psi | \psi \rangle} - \frac{\langle \psi | \hat{H} | \psi \rangle}{ \langle \psi | \psi \rangle} \left[\frac{\langle \delta \psi | \psi \rangle}{\langle \psi | \psi \rangle} + \frac{\langle \psi | \delta \psi \rangle} {\langle \psi | \psi \rangle} \right] .
\end{equation}
We are looking for the minimum of $\langle\hat{H}\rangle$ with respect to variation of $| \psi \rangle$. Just like in ordinary calculus, the minimum is an extrema and $\delta\langle \hat{H} \rangle \equiv 0$. Then
\begin{equation}
    \langle \delta \psi | \hat{H} - \langle \hat{H} \rangle | \psi \rangle + \langle \psi | \hat{H} - \langle \hat{H} \rangle | \delta \psi \rangle \equiv 0
\end{equation}
Now in general $| \delta \psi \rangle$ is complex. It has two components that can be varied independently; the real and imaginary parts. Equivalently, $| \delta \psi \rangle$ and  $| \delta \psi \rangle^\dagger \equiv \langle \delta \psi |$ can be varied independently. So, considering the arbitrariness of the variation $|\delta \psi \rangle$, we find that
\begin{equation}
    \hat{H} |\psi \rangle = \langle \hat{H} \rangle | \psi \rangle .
\end{equation}
I.e. if the change in the expectation value $\delta\langle\hat{H}\rangle$ vanishes with respect to changes in the trial function, the trial function is an eigenstate of $\hat{H}$. Minimizing $\langle\hat{H}\rangle$ is the same as solving the eigenvalue equation.

In other words, the variational method can be used to solve the TISE in the following way: we make a guess for the wave function that depends on parameters. We calculate $\langle \hat{H} \rangle$ and vary the parameters defining the trial funciton until the energy is extremized. In math, we calculate the following quantity
\begin{equation}\begin{split}
    E(\alpha_1,\cdots,\alpha_n) \equiv \langle \hat{H} \rangle = \int d\bm{r} \psi^*(\bm{r};\alpha_1,\cdots,\alpha_n) \hat{H}  \psi(\bm{r};\alpha_1,\cdots,\alpha_n)
    \label{eq:energy_expectation}
\end{split}\end{equation}
and look for the set $\{ \alpha_1,\cdots,\alpha_n \}$ that satisfies 
\begin{equation}
    \frac{\partial E}{\partial \alpha_1}=\cdots=\frac{\partial E}{\partial \alpha_n}=0 .
\end{equation}

Now let's see what this all tells us. Call $\epsilon_0\equiv E_0$ the ground state energy. The eigenvectors of the TISE \cref{eq:tise} are a complete set, so we can expand our \emph{trial state} as $| \psi \rangle = \sum_\alpha c_\alpha | \psi_\alpha \rangle$. The expectation value is
\begin{equation}
    E = \langle \psi | \hat{H} | \psi \rangle = \frac{\sum_\alpha |c_\alpha|^2 \epsilon_\alpha}{\sum_\alpha |c_\alpha|^2} = E_0+\frac{\sum_\alpha |c_\alpha|^2 (\epsilon_\alpha-E_0)}{\sum_\alpha |c_\alpha|^2} 
\end{equation}
Clearly $\epsilon_\alpha \geq E_0$, where it is a strict inequality if the ground state is non-degenerate. So the whole second term is a positive number (or zero). Lets call it $\Delta E$. Then it turns out that 
\begin{equation}\begin{split}
    E\geq E_0+\Delta E
    \label{eq:gstate_inequality}
\end{split}\end{equation}
The variational method puts an upper bound on the true ground state energy, a neat result. If we can't solve a particular problem, we can approximate the solution and try to get as close as possible. In practice, we usually pick a set of functions, called the \emph{basis}, and use linear combinations of them as trial solutions.

\subsection{The secular equation}\label{sec:secular_eq}

We have convinced ourselves that we can approximate the solution to the TISE with a variational calculation, but we really still haven't said \emph{how} to solve the problem.  

We can solve \cref{eq:tise} by first choosing a variational basis and expanding the eigenvectors in that basis. Let the new basis vectors be called $|\phi_\alpha \rangle$. In the new basis, the kets that extremize $\langle\hat{H}\rangle$ are 
\begin{equation}
    |\psi\rangle = \sum_\alpha c_{\alpha} |\phi_\alpha\rangle
\end{equation}
where $c_{\alpha}$ are complex expansion coefficients and the sum runs over all basis vectors: possibly infinitely many. The expectation value in \cref{eq:energy_expectation} is
\begin{equation}
     \sum_{\alpha,\beta}  c_{\alpha}c^*_\beta \langle \phi_\beta | \hat{H} | \phi_\alpha \rangle =  \sum_{\alpha,\beta} c_{\alpha} c^*_\beta E \langle \phi_\beta |\phi_\beta \rangle .
\end{equation}
It is common to write the matrix elements as $ \langle \phi_\alpha | \hat{H} | \phi_\beta \rangle \equiv H_{\alpha,\beta}$ and $\langle \phi_\alpha | \phi_\beta \rangle\equiv S_{\alpha,\beta}$. The elements $S_{\alpha,\beta}$ are called "overlaps" and the matrix of overlaps, $\hat{S}$, is the "overlap matrix". Note that if the basis functions are orthonormal (e.g. suitably chosen plane waves), the overlap operator is just the identity matrix.

The coefficients $c_\alpha$ are the variational parameters. If we want to calculate the wave function, we minimize with respect to the complex-conjugate coefficients (i.e. the variation is $\langle \delta \psi|$). The derivative of \cref{eq:energy_expectation} is
\begin{equation}\begin{split}
    \sum_\alpha \left( H_{\alpha,\beta} -ES_{\alpha,\beta} \right) c_\alpha
    \label{eq:secular_elements}
\end{split}\end{equation}
After staring carefully at \cref{eq:secular_elements} for a minute, we recognize this is the Schr\"{o}dinger's equation in matrix notation:
\begin{equation}\begin{split}
    \hat{H} \bm{c} = E \hat{S} \bm{c}
    \label{eq:secular_matrix}
\end{split}\end{equation}
where we introduced $\bm{c}$ as the column matrix of expansion coefficients. So we've shown that the TISE can be represented as an auxiliary TISE for the variational basis orbitals. But we have written it in the secular form of \cref{eq:secular_matrix} which we know how to solve: by diagonalization. The eigenvectors are stationary-states of the original TISE. The lowest eigenvalue is always greater than or equal to the true (single-particle) ground state energy. If we use a complete set of basis functions, e.g. plane waves, we can in principle calculate the exact solution. However, we would have to diagonalize an infinite dimensional matrix... which we can't do. 

What if we just truncate the basis set? \cref{eq:secular_elements} still holds and can be diagonalized on a computer. It's a generalized eigenvalue problem, but know how to solve those too (at worst, it is just diagonalizing two matrices instead of one: see \S 3.3 in Thijssen \cite{thijssen2007computational}). The method laid out in this section is how the single-particle problem is usually solved in modern computational condensed matter physics. The discussion was pretty general; in practice, we still have to choose a specific basis and figure out how to calculate the matrix elements. The basis functions can be plane waves (\textsc{abinit,qe,vasp}), localized atomic-like orbitals (\textsc{gaussian}), or their generalization in solids, called linear combinations of atomic orbitals (\textsc{cp2k}), or even tabulated numerical functions (\textsc{siesta}). There are plenty of reasons to choose one or the other, but that is out of the scope of the present discussion.

\subsection{Pseudopotentials}\label{sec:pseudo}

Let $N_B$ be the number of basis functions. The computational cost of solving e.g. \cref{eq:secular_matrix} goes like $\mathcal{O}(N_B^3)$ for direct diagonalization based algorithms and like $\mathcal{O}(N_B^2 \ln N_B)$ for iterative diagonalization algorithms \cite{thijssen2007computational,5517}. As shown above, the variational method puts an upper bound on the true ground state energy. Since the computational cost scales poorly with $N_B$, we want to pick $N_B$ as small as possible while still being accurate (i.e. close to the true ground state). This is hard for several reasons: (i) the core electrons are tightly bound, so have very short-ranged length scales. (ii) The valance electrons are loosely bound (or even delocalized), so have long-ranged length scales. (iii) The valence states have to be orthogonal to all lower lying states, so valence states must oscillate very rapidly near the core region where electrons are highly localized; i.e. again, there are very short-ranged components. The result is that, to cover both length scales, we need to include both long- and short-ranged basis functions into our basis set to accurately represent solutions of the Hamiltonian: $N_B$ is large. 

It turns out we can overcome all of the problems above and reduce $N_B$ to a very small number if we replace the nuclear potential in \ref{eq:e_ham} with a \emph{pseudopotential} \cite{singh2006planewaves,giustino2017electron,thijssen2007computational,martin2020electronic}. Pseudopotentials combine the core electronic states and nuclear potential into a singe effective ("pseudo") potential, removing the core states from the Hilbert space of the valence electrons. Why this works can be understood as follows: (i) the core electrons around atoms are so tightly bound that they are almost entirely unaffected by the dynamics of valence electrons. Even in a crystal potential, they are nearly identical to the isolated atom wave functions. We don't need to include them explicitly in the Hamiltonian. (ii) The core electrons screen the Coulomb potential of the nuclei. If we combine the potential of the core electrons and nuclei together, we get "shallow" potentials acting on the valence electrons. (iii) Combining the core electrons and nuclei into an effective potential removes the requirement that valence electrons be orthogonal to core states. These facts combine so that valence states are smooth in the vicinity of the nuclear cores and we only need long-ranged, smooth basis functions. This is the essence of the "pseudopotential" approximation and its success in electronic structure theory is indisputable. 

It is fairly easy to understand the pseudopotential method at a conceptual level. The origin of the pseudopotential method is in the \emph{orthogonalized planewave method} (OPW) \cite{herring1940new}. This discussion follows Giustino \cite{giustino2017electron} (look there for more references). We separate valence and core electrons into subsets and project out the space spanned by core electrons. We won't specialize to planewaves, but the original OPW method was derived for planewaves, hence the name.

Let $|\phi_n\rangle$ be a set of variational basis functions for the valence electrons:
\begin{equation}\begin{gathered}
    |\tilde{\phi}_n\rangle = | \phi_n \rangle - \sum_c |\psi_c\rangle \langle \psi_c | \phi_n \rangle \equiv \hat{\mathcal{T}} |\phi_n \rangle,
\end{gathered}\end{equation}
with $ \hat{\mathcal{T}} = 1 -  \sum_c |\psi_c\rangle \langle \psi_c |$, be the OPW basis functions. $|\psi_c\rangle$ are the atomic-like eigenstates of the core electrons. The OPW basis functions are, by definition, orthogonal to the core electrons, so we only need a small number of long-ranged basis functions to accurately represent the valence states. 

The "all-electron" valence-state variational solutions to the secular equation are $|\psi_v\rangle = \sum_n c_n |\tilde{\phi}_n \rangle$. They are "all-electron" because the Hilbert space contains all of the core and valence electrons. Define the "pseudo-wave functions" $|\tilde{\psi}_v\rangle = \sum_n c_n |\phi_n \rangle$ which use the original basis functions. These are related to the all-electron wave functions by $|\psi_v\rangle = \hat{\mathcal{T}}|\tilde{\psi}_v\rangle$. The TISE for the pseudo-wave functions is
\begin{equation}\begin{gathered}
    \hat{\mathcal{T}}^\dagger \hat{H} \hat{\mathcal{T}} | \tilde{\psi}_v \rangle =  \epsilon_v \hat{\mathcal{T}}^\dagger \hat{\mathcal{T}} | \tilde{\psi}_v \rangle
\end{gathered}\end{equation}
with $\epsilon_v$ the valence eigenvalue $\hat{H} | \psi_v \rangle = \epsilon_v |\psi_v \rangle$. Similarly, let $\epsilon_c$ be the core eigenvalue. The TISE becomes
\begin{equation}\begin{gathered}
    \left[\hat{H} + \sum_c (\epsilon_v - \epsilon_c ) | \psi_c \rangle \langle \psi_c |\right]  | \tilde{\psi}_v \rangle =  \epsilon_v | \tilde{\psi}_v \rangle. 
    \label{eq:pseudo_tise}
\end{gathered}\end{equation} 
The bonus term, $\hat{V}_{rep} \equiv  \sum_c (\epsilon_v - \epsilon_c ) | \psi_c \rangle \langle \psi_c | $ is strictly repulsive since $\epsilon_v - \epsilon_c > 0$ by definition. This repulsive potential largely cancels the Coulomb potential between valence electrons and nuclei \cite{cohen1961cancellation}. Then the nuclear potential is shallow and we don't need short-ranged basis functions; more-over, this is why electrons in metals behave like free electrons.

The biggest issue with the OPW method so far is that the repulsive potential depends on the eigenvalues. Various methods exist to handle this, e.g. the "linearized, augmented plane wave" (LAPW) method where the energy dependence is removed by linearizing $\hat{V}_{rep}$ with respect to a guess for eigenvalue \cite{singh2006planewaves,martin2020electronic}. This is an all-electron method that explicitly includes core states. 

Different classes of methods, collectively called "pseudopotential" methods, replace the core electrons and nuclear potentials with a combined shallow potential that removes the core electrons from the Hilbert space. The projector-augmented-wave (PAW) method \cite{blochl1994projector} does this by defining the projection operator in $\hat{\mathcal{T}}$ as a set of tabulated functions and solves the TISE for only the pseudo wave functions, \cref{eq:pseudo_tise}. The eigenvalue dependence is eliminated by imposing a "norm-conserving" constraint that requires the norm of the pseudo electron wave function (and its derivative) to match the all-electron valence wave function norm (and its derivative) near the cores \cite{singh2006planewaves,thijssen2007computational}.
A more drastic (and very simple) method replaces the combined nuclear and repulsive interactions with a simple, effective potential: this is the famous norm-conserving pseudopotential (NCPP) method \cite{hamann1979norm} which is what people usually mean when they say "pseudopotentials" in the context of DFT. The hierarchy of accuracy of these methods is, broadly speaking, as follows: LAPW methods are the most accurate, with PAW next, and then NCPP. The computational cost is similarly LAPW methods are the most expensive, with PAW next, and then NCPP. 

It is satisfying to note that the PAW method is a well-defined approximation to the LAPW method \cite{holzwarth1997comparison} and that the different flavors of NCPP's are well-defined approximations to the PAW method \cite{blochl1994projector}. Modern pseudopotentials are very accurate and usually one of the only reasons to resort to more expensive methods is when the physics near the core is needed explicitly, e.g. to model core-spectroscopy, or when certain methods are easier to implement in one of the more expensive methods. E.g. DFT+U and spin-orbit coupling are much easier to model in the PAW method compared to the pseudopotential method. 

The technical details of these methods are outside the scope of this thesis. Nowadays, general and transferable pseudopotentials are tabulated for nearly every element in the periodic table and either come with publicly available computational packages or can be downloaded from public repositories such as the \texttt{pseudodojo} project. They are "plug-and-play" and usually work without user intervention. The same is true for PAW datasets. The details of the inner-workings aren't important for what follows, so we won't discuss them any further here.

\subsection{The self consistent field method}

We return to discussing the solution of \cref{eq:hartree_hamiltonian}. The difficulty is that the Hamiltonian depends explicitly on the density. Let's generalize from the Hartree approximation and study any Hamiltonian $\hat{H}[n(\bm{r})] = \hat{T} + \hat{V}[n(\bm{r})] $ in which the potential depends explicitly on the electronic density. This may or not use pseudopotentials: it won't matter in what follows. The density is calculated from the wave functions, so the solution, in a sense, depends on itself. We say that the correct solution has to be \emph{self consistent}. 

We can still express the TISE as an eigenvalue problem
\begin{equation}
    \hat{H}[n(\bm{r})] \psi_\alpha(\bm{r}) = \epsilon_\alpha \psi_\alpha(\bm{r}) 
\end{equation}
The Hamiltonian depends on the solutions $\psi_\alpha(\bm{r})$ (or more precisely on the density which depends on the solutions) so it's a self consistent problem. Since the potential depends on the solution, it is called a "self consistent potential". The method to solve it is called the "self consistent field" (SCF) method \cite{girvin2019modern,thijssen2007computational,martin2020electronic}.

In the Hartree approximation, the self consistent potential is the Hartree potential. But we don't need to be specific right now. All that matters is that the potential \emph{can} be calculated once the density is known. The electronic number-density is
\begin{equation}\begin{split}
    n(\bm{r}) = \sum_i \vert \psi_i(\bm{r}) \vert^2 f(\epsilon_i) 
    \label{eq:den_1}
\end{split}\end{equation}
The sum runs over all basis functions. The \emph{charge}-density is $\rho(\bm{r})=e n(\bm{r})$. The number $f(\epsilon)$ is the "occupation" of the single-particle state. In equilibrium, only the lowest energy states are occupied (of course each state can only hold one electron, or two if we include spin). At finite temperature, $f(\epsilon)$ is Fermi-Dirac distribution function: 
\begin{equation}\begin{split}
    f(\epsilon) = \left[ \exp\left( \frac{\epsilon-\epsilon_F}{k_B T} \right)+1 \right]^{-1}
\end{split}\end{equation}
$k_B$ is Boltzmann's constant and $T$ is the temperature, and $\epsilon_F$ is the \emph{Fermi-energy}. The Fermi-energy is determined from the requirement that\footnote{The technical implementation I have used in my own codes is a bisection algorithm: pick the highest and lowest eigenvalues and take the mid-point between them as the Fermi energy. Calculate the integral. If $N$ is too large, pick the mid-point between the lower bound and $\epsilon_F$. The new upper bound for the next iteration is $\epsilon_F$. If $N$ is too small, pick the mid-point between the upper bound and $\epsilon_{F}$. The new lower bound for the next iteration is $\epsilon_F$.0 Repeat until the calculated $N$ agrees with the real one within desired precision.}
\begin{equation}\begin{split}
    \int d\bm{r} n(\bm{r}) = \sum_i f(\epsilon_i) = N
\end{split}\end{equation}
with $N$ the number of electrons in the system. I.e. the total number of occupied single-particle states must equal the number of particles. $\epsilon_F$ is the energy of the highest occupied state that satisfies this requirement. 

Now for the SCF method. Let's just state it in words
\begin{enumerate}
    \item We make a guess for the initial orbitals, $\psi_\alpha(\bm{r})$. In a condensed matter setting, a reasonable guess is atomic orbitals. 
    \item We then calculate the density and the potential and solve the secular problem. The solutions $\psi_\alpha(\bm{r})$ determine the density. 
    \item For the solution to make sense, the input and output densities had better be identical. So we inspect the differences between the input and output densities. 
    \item If they agree within some prescribed value, then all is good and we consider the problem solved: the final $\{ \psi_\alpha(\bm{r}) \}$ and $\{ \epsilon_\alpha \}$ are the solutions. 
    \item If they don't agree, we make a new guess for the density and try again. The process is iterated until "self consistency" is reached.
\end{enumerate}
There are many ways to "guess again." Usually some of the old density is mixed with the new and the mixture is used as the input density for the next SCF step. Since each step involves diagonalizing a (usually large) matrix, it is essential to reach self consistency as quickly as possible. The method for "mixing" the densities is crucial. Some common examples are simple mixing, where we naively throw some fraction of the new density in with the old. More advanced examples are the "Pulay method". On a more technical note, this type of problem where we are solving an equation like $f(x)=x$ is called a "fixed-point" problem. There is a large body of literature in the applied math and computational physics communities on fixed point problems. The technical details are outside the scope of this thesis.

\section{Density functional theory}\label{sec:dft}

We are finally ready to define "density functional theory" (DFT) \cite{girvin2019modern,thijssen2007computational,martin2020electronic,giannozzi2020numerical}. It turns out that DFT is not an approximation: it is a reformulation of the many-body \emph{ground state} solution in terms of a set of self consistent single particle equations that depend on charge density\footnote{We mention that the Hartree-Fock method is an (approximate) self consistent method that depends on the wave functions rather than the charge density}. We have seen how to solve single-particle SCF problems in practice, so this poses no problem. Our starting point is the full interacting Born-Oppenheimer electronic Hamiltonian:
\begin{equation}\begin{split}
    \hat{H} = \sum^{N}_i \frac{\hat{p}^2_i}{2 m_e} +  \sum^{N}_i v_{ext}(\bm{r}_i) + \frac{1}{2} \sum^N_{i \neq j} u(\bm{r}_i,\bm{r}_j) \equiv \hat{T}+\hat{V}+\hat{U}
    \label{eq:e_ham'}
\end{split}\end{equation}
Usually in condensed matter, the kinetic-energy and two-particle operators, $\hat{T}$ and $\hat{U}$ respectively, are the same regardless of what external potential, $\hat{V}$, we might be exposing the electrons to. The form of the Hamiltonian in different systems differs only by the external potential. Our method of attack is to study the most general external potential possible and try to solve the problem once-and-for-all. As an added bonus, since it doesn't matter what the explicit form of the two particle operator is, DFT is applicable to many types of problems: e.g. coupled electron-phonon systems and the superconducting state \cite{oliveira1988density,requist2019exact}. 

The first step is to note that since the external potential is a one-particle operator that can be written as 
\begin{equation}
    \hat{V} = \sum_i v(\bm{r}_i) = \sum_i \int d\bm{r} \delta(\bm{r}-\bm{r}_i) v(\bm{r}) = \int d\bm{r} \hat{n}(\bm{r}) v(\bm{r}) 
\end{equation}
where we substitute the density operator $\hat{n}(\bm{r})$ we encountered earlier. We have dropped the subscript on $v_{ext}$ to ease notation. $v$ should be assumed external for now. The expectation value of the external potential
\begin{equation}
    \langle \Psi | \hat{V} | \Psi \rangle = \int d\bm{r} \langle \Psi | \hat{n}(\bm{r}) | \Psi \rangle v(\bm{r}) = \int d\bm{r} n(\bm{r}) v(\bm{r}) 
\end{equation}
depends only on the expectation value of the density, $n(\bm{r})$. We haven't made any assumptions about $| \Psi \rangle$ yet. It might as well be the exact ground state solution to the many-body Hamiltonian in \cref{eq:e_ham'}. So we have found that the potential $\langle \Psi | \hat{V} | \Psi \rangle\equiv V[n(\bm{r})]$ can be represented as a \emph{functional} of the density. Inspired by this, we want to reformulate the expectation value of the Hamiltonian as a functional of the density. Can this be done? It turns out that it can, as proved in two theorems by Hohenberg and Kohn in the 1960's. These theorems essentially prove that the exact many-problem can be expressed in terms of a unique density and that the density minimizes the ground state energy. Let us lazily prove the Hohenberg-Kohn theorems. Our discussion follows Girvin and Yang \cite{girvin2019modern}. 

\subsection{The Hohenberg-Kohn Theorems}\label{sec:hk_theorems}

We first postulate that the ground state energy is a functional of the density. In math
\begin{equation}\begin{split}
    E[n(\bm{r})] \equiv H[n(\bm{r})]=\langle \Psi_0 | \hat{H} | \Psi_0 \rangle
    \label{eq:df_gs}
\end{split}\end{equation}
where $|\Psi_0\rangle$ is the ground state solution. We assume the ground state is not degenerate. The density is $n(\bm{r})=\langle \Psi_0 | \hat{n}(\bm{r}) | \Psi_0 \rangle$. We are looking for a way to formulate the ground state energy as a functional of the density. If the energy functional and density are in one-to-one correspondence, we can (in principle) map the ground state energy onto a functional of the density. If we are given $|\Psi_0\rangle$, there is only one $n(\bm{r})$. What is \emph{not} obvious is that, given a $n(\bm{r})$ there is only one $|\Psi_0\rangle$. This is what we need to prove. If it's true, there is a one-to-one map between the ground state energy and the density and the postulate in \cref{eq:df_gs} is true.

\paragraph{The $1^{st}$ Hohenberg-Kohn theorem} proves that the density $n(\bm{r})$ and the ground state solution $|\Psi_0\rangle$ are in one-to-one correspondence. In other words, two different wave functions can't result in the same density. The proof is by \emph{reductio ad absurdum}. Suppose that $|\Psi_1\rangle$ and $|\Psi_2\rangle$ are ground states of Hamiltonians with two different external potentials $V_1[n(\bm{r})]$ and $V_2[n(\bm{r})]$, but that $\langle \Psi_1 | \hat{n}(\bm{r}) | \Psi_1 \rangle=\langle \Psi_2 | \hat{n}(\bm{r}) | \Psi_2 \rangle$. The rest of the Hamiltonian is identical between them. The energies are $E_1= \langle \Psi_1 | \hat{H}_1 | \Psi_1 \rangle $ and  $E_2 = \langle \Psi_2 | \hat{H}_2 | \Psi_2 \rangle $. Earlier we proved that variational solutions are always above the true ground state energy. Then $E_2 < \langle \Psi_1 | \hat{H}_2 | \Psi_1 \rangle$. 
\begin{equation}
    E_2  < \langle \Psi_1 | \hat{H}_1 | \Psi_1 \rangle + \langle \Psi_1 | \hat{H}_2 - \hat{H}_1 | \Psi_1 \rangle \\
     < E_1 + \langle \Psi_1 | \hat{V}_2 - \hat{V}_1 | \Psi_1 \rangle
\end{equation}
Identical reasoning results in
\begin{equation}
    E_1 < E_2 + \langle \Psi_2 | \hat{V}_1 - \hat{V}_2 | \Psi_2 \rangle
\end{equation}
Adding these two, together
\begin{equation}
    E_1 + E_2 < E_1 + E_2 + \langle \Psi_2 | \hat{V}_1 - \hat{V}_2 | \Psi_2 \rangle + \langle \Psi_1 | \hat{V}_2 - \hat{V}_1 | \Psi_1 \rangle
\end{equation}
The last two terms become
\begin{equation*}
\begin{gathered}
    \langle \Psi_2 | \hat{V}_1 - \hat{V}_2 | \Psi_2 \rangle + \langle \Psi_1 | \hat{V}_2 - \hat{V}_1 | \Psi_1 \rangle \\ = \int d\bm{r} n(\bm{r}) \left( v_1(\bm{r}) - v_1(\bm{r}) + v_2(\bm{r})  - v_2(\bm{r}) \right) = 0
\end{gathered}
\end{equation*}
and we find 
\begin{equation}
    E_1 + E_2 < E_1 + E_2 
\end{equation}
which is a logical contradiction. Thus, we have proved that $\langle \Psi_1 | \hat{n}(\bm{r}) | \Psi_1 \rangle \neq \langle \Psi_2 | \hat{n}(\bm{r}) | \Psi_2 \rangle$, where $| \Psi_1 \rangle$ and $| \Psi_2 \rangle$ are solutions to Schr\"{o}dinger's equations that differ only by their external potentials. Two different wave functions cannot generate the same density. There is only one $|\Psi_0\rangle$ that solves a particular Hamiltonian and it solves only that one. 

\paragraph{The $2^{nd}$ Hohenberg-Kohn theorem} proves that $n(\bm{r})$ calculated from the ground state solution $| \Psi_0 \rangle$ minimizes the energy functional 
\begin{equation}\begin{split}
    E[n(\bm{r})] = \langle \Psi_0 | \hat{H} | \Psi_0 \rangle = E_{U}[n(\bm{r})]+\int d\bm{r} n(\bm{r}) v(\bm{r}) .
    \label{eq:energy_functional}
\end{split}\end{equation}
The "universal functional" is completely independent of the external potential:
\begin{equation}
    E_{U}[n(\bm{r})] = \langle \Psi_0 | \hat{T}+\hat{U} | \Psi_0 \rangle .
\end{equation}
The 1$^{st}$ Hohenberg-Kohn theorem essentially proves that the universal functional exists. How to calculate it is unknown, but we can forget about that for now. Let's suppose that $n_1(\bm{r})$ and $|\Psi_1\rangle$ are the solution and density of some other ground state. By virtue of the variational approximation,
\begin{equation}
    E[n_1(\bm{r})] = \langle \Psi_1 | \hat{H} | \Psi_1 \rangle > E[n(\bm{r})] 
\end{equation}
If $|\Psi_1\rangle$ is not the ground state solution, then its density doesn't minimize the energy-functional. The smallest value of the energy functional is then for $n(\bm{r})$, which is in one-to-one correspondence with $|\Psi_0\rangle$, the ground state solution. It's clear that the simplification of the many-body problem provided by the Hohenberg-Kohn theorems is enormous. We went from solving for a $3N$ dimensional complex valued wave function to solving for a $3$ dimensional real scalar function. Still, there is no free lunch: the wave function $\Psi(\bm{x}_1,\cdots,\bm{x}_N)$ is still a $3N$ dimensional quantity and how to calculate it from the density is unknown. This theoretical framework is called density functional theory. Virtually all "ab-initio" calculations in materials science today are done in DFT and the success of this method is exemplified by the fact that Walter Kohn was awarded the Nobel prize in chemistry for his work on DFT.

\subsection{The Kohn-Sham Equations}

So far, the discussion was very general and we haven't really said how to solve a problem in practice. Moreover, we mentioned that we don't even know what the universal functional $E_U[n(\bm{r})]$ is. All we know is that it exists. How do we make progress? First of all, we will separate out all the parts of $E_U[n(\bm{r})]$ that we can calculate. Then we will approximate the rest of it. Hopefully the things we can calculate will be a large fraction of the total energy, so any errors in "the rest of it" will be small. Second, we want to do the same thing we did in the Hartree: recast the problem as an auxiliary set of effective single-particle equations. We do this because we know how to solve single-particle equations. In DFT, these equations are the \emph{Kohn-Sham equations} \cite{thijssen2007computational,girvin2019modern,giannozzi2020numerical,martin2020electronic}

We start by ignoring interactions. It turns out that even for non-interacting systems, we don't know how to write down $E_U[n(\bm{r})]$ directly \cite{thijssen2007computational}. But we do know how to calculate the ground state energy $E[n(\bm{r})]$ for non-interacting particles. The non-interacting ground state wave function is a Slater determinant of single-particle wave functions $\psi_{i,\sigma}(\bm{r})$. The density is given by \cref{eq:den_1}. Ignoring interactions, the Hamiltonian only has single-particle operators and the exact energy functional is 
\begin{equation}
    E[n(\bm{r})] = T_{s}[n(\bm{r})]+\int d\bm{r} n(\bm{r}) v(\bm{r}) = \sum_{i} f_{i} \langle \psi_{i} | -\frac{\hbar^2}{2 m_e}\nabla^2_i + v(\bm{r}_i) | \psi_{i} \rangle
\end{equation}
$T_s[n(\bm{r})]$ is the single-particle kinetic energy functional. The external potential terms on both sides are exactly the same and we find
\begin{equation}
   T_{s}[n(\bm{r})] = - \frac{\hbar^2}{2 m_e}\sum_{i} f_{i} \langle \psi_{i} |\nabla^2_i | \psi_{i} \rangle
    \label{eq:ke_functional}
\end{equation}
This is the exact \emph{non-interacting} kinetic energy functional for a particular density $n(\bm{r})$. Now back to the interacting problem. We want to write down all the interactions that we know how to calculate as our best guess to the true many-body energy functional. We can't write the ground state wave function $\Psi_0(\bm{x}_1,\cdots,\bm{x}_N)$ as a Slater determinant, but we are still free to expand the density in any basis we want: we use the same form as \cref{eq:den_1}, where the spin-orbitals used to expand the density are not necessarily solutions of the true Schr\"{o}dinger's equation. Rather, they are auxiliary functions called "Kohn-Sham orbitals" and are solutions of the Kohn-Sham equations. Now let's put together our guess. The Kohn-Sham energy functional is  (see \S7.1 in Martin \cite{martin2020electronic})
\begin{equation}
    E_{KS}[n(\bm{r})] = T_s[n(\bm{r})]+ E_{ext}[n(\bm{r})] + E_{H}[n(\bm{r})] + E_{xc}[n(\bm{r})]
    \label{eq:ks_functional}
\end{equation}
$T_s[n(\bm{r})]$ is the non-interacting kinetic energy functional which we already looked at. We include it as our "best-guess" for the interacting kinetic energy functional since it is exact in the non-interacting limit. Of course the interacting kinetic energy functional isn't identical and we don't know what physics we are missing. But let's not worry about this yet. Let's go over the other terms first. $E_{ext}$ is the external energy we have seen many times:
\begin{equation}
    E_{ext}[n(\bm{r})] = \int d\bm{r} n(\bm{r}) v(\bm{r})
\end{equation}
This is true in the interacting and non-interacting problems. We have also seen (in a different form) the $E_{H}$ term before. It is the Hartree energy:
\begin{equation}
    E_{H}[n(\bm{r})] = \frac{1}{2} \int d\bm{r} d\bm{r}' n(\bm{r}) u(\bm{r},\bm{r}') n(\bm{r}')
    \label{eq:hartree_energy}
\end{equation}
If e.g. $u(\bm{r},\bm{r}')$ is the Coulomb interaction, this is our best guess for the electrostatic Coulomb interaction. Note that it includes the same self-interaction errors we mentioned before. The final term is called the "exchange-correlation" energy. We have crammed everything we don't know into it: it includes all of the interacting kinetic energy we are neglecting, all of the dynamical correlation (e.g. screening), and all exchange interactions. It even includes the corrections that remove self-interaction from the Hartree energy. The Hohenberg-Kohn theorems in \cref{sec:hk_theorems} prove that $E_{xc}[n(\bm{r})]$ exists so this is, in principle, an \textbf{exact} reformulation of the many-body problem. Unfortunately, the Hohenberg-Kohn theorems don't tell us what $E_{xc}[n(\bm{r})]$ is.

Now let's apply the variational method to solve this energy functional. Minimization  with respect to the density is equivalent to minimization with respect to the Kohn-Sham (KS) orbitals \cite{girvin2019modern}. The result is the Kohn-Sham equations:
\begin{equation}\begin{split}
    \hat{H}_{KS} \psi_{i}(\bm{r}) = \left[ -\frac{\hbar^2}{2m_e}\nabla^2 + v_{KS} \right] \psi_{i}(\bm{r}) = \epsilon_{i} \psi_{i}(\bm{r})
    \label{eq:ks_equations}
\end{split}\end{equation}
where we introduced the Kohn-Sham potential 
\begin{equation}\begin{split}
    v_{KS}(\bm{r}) = v(\bm{r}) + v_H(\bm{r}) + v_{xc}(\bm{r}) 
    \label{eq:ks_pot}
\end{split}\end{equation}
$v_{xc}(\bm{r})$ is called the exchange-correlation potential and $v_H(\bm{r})$ is the Hartree potential (the same as in \cref{eq:hartree_potential}). The KS equations \cref{eq:ks_equations} form a set of auxiliary equations for the basis functions of the density. There is one KS equation for every basis function and each of them looks exactly like a TISE with the Hamiltonian $\hat{H}_{KS}$ given in \cref{eq:ks_equations}. Solving the set of KS equations is equivalent to solving for the density that minimizes the KS energy functional, \cref{eq:ks_functional}. Note that, just like in the Hartree approximation, this is a SCF problem since the Hartree and exchange-correlation potentials are functionals of the density. The KS equations can easily be formulated as a secular equation just like in \cref{sec:secular_eq} and solved by diagonalization.

The KS eigenvalues, $\epsilon_{i}$ are not really energies but they are usually called energies. The KS orbitals formally have limited physical meaning. They are auxiliary functions, but they are often interpreted as physical electron wave functions (see below). The KS ground state energy in terms of th KS eigenvalues is (\S5.1 in \cite{thijssen2007computational}): 
\begin{equation}
\begin{gathered}
    E_{KS}[n(\bm{r})] = \sum_{i} f_{i}\epsilon_{i} - E_{H}[n(\bm{r})] + E_{xc}[n(\bm{r})] -\int d\bm{r} n(\bm{r}) v_{xc}(\bm{r})
    \label{eq:ks_total_e}
\end{gathered}
\end{equation}
where $E_{nn}$ is the nuclear-nuclear interaction between ions. It turns out that there is some subtlety in including partial occupations $f_{i\sigma}$ in the KS scheme. e.g. if arbitrary variations in $f_{i\sigma}$ are allowed, the energy functional is not variational \cite{haynes1998linear}. Or if the Fermi-Dirac function is used and we wish to interpret the results at finite temperature, we have to minimize the \emph{free-energy} instead of the energy (see e.g. \S7.3 in Martin \cite{martin2020electronic} or ref. \cite{kratzer2019basics}). But let's not worry about these issues now. Just assume we are minimizing at fixed occupation numbers which are determined by the self consistency conditions on the density. 

We note that we have achieved something special here! All along, we have been trying to calculate the potential seen by the atoms. The total ground state energy of the atoms in the Born-Oppenheimer approximation is given by the Hamiltonian \cref{eq:e_n_ham}. We note that we can the rewrite it as
\begin{equation}
    E_{tot}(\bm{R}) = E_{KS}(\bm{R})+\frac{k}{2}\sum_{I\neq J} \frac{e^2 Z_I Z_J}{|\bm{R}_I-\bm{R}_J|}
    \label{eq:total_energy}
\end{equation}
where $E_{KS}(\bm{R})$ is the KS energy functional in \cref{eq:ks_total_e} with the dependence on atom positions made explicit. The Coulomb energy is rather straight forward to calculate with e.g. Ewald summations \cite{thijssen2007computational,allen2017computer,dove1993introduction,ziman1979principles,lee2009ewald} so we ignore it here. The electronic contribution to the atom-atom interaction isn't easy to calculate at all! However, as we have seen in this section, the KS scheme provides an efficient method of attack. $E_{KS}(\bm{R})$ in \cref{eq:total_energy} is the potential energy of the atoms that we have been looking for all along (cf. the introduction to this chapter). Later, we will see how to calculate its derivatives, i.e. the forces on the atoms.

\subsection{The stupidity energy}

So far we have made no approximations; we have only re-arranged the ground state energy functional into a different from. In principle, this is an exact method. In practice, we don't know what $E_{xc}[n(\bm{r})]$ is. Richard Feynman suposedly called it the "stupidity energy" because it contains everything we don't know and we don't know what it is. It usually just called the exchange-correlation (XC) energy. Still, a \emph{lot} of effort has been devoted to coming up with approximations to the XC energy. It is outside of the scope of this thesis to discuss these approximations in detail. We merely provide some comments. The curious reader is referred to literature \cite{martin2020electronic,thijssen2007computational,kohanoff2006electronic}

Typically the XC energy is approximated by fitting a parameterization to exact (numerical) results for the interacting homogeneous electron gas. There is a hierarchy of approximations of increasing complexity. Roughly speaking, the exchange-correlation potential (or energy) is Taylor expanded in moments of the density:
\begin{equation}
    v_{xc}[n(\bm{r})] = v_{xc}[n(\bm{r}_0),\bm{\nabla}n(\bm{r}_0),\nabla^2 n(\bm{r}_0),\cdots]    
\end{equation}
Keeping only the lowest order term is the local density approximation (LDA). The exchange-correlation energy in the LDA is written
\begin{equation}
    E_{xc}^{LDA}[n(\bm{r})] = \int d\bm{r} \epsilon_{xc}[n(\bm{r})] n(\bm{r})
\end{equation}
where $\epsilon_{xc}[n(\bm{r})]$ is the exchange-correlation energy per particle of the homogeneous electron gas evaluated at the density $n(\bm{r})$. Parameterizations that keep the next moment of the density, its gradient $\bm{\nabla}n(\bm{r})$, are called generalized gradient approximations (GGAs). Approximations keeping the next order term, $\nabla^2 n(\bm{r})$, are called meta-generalized gradient approximations (meta-GGAs). There are many different types of LDAs, GGAs, and even meta-GGAs. Most of these are natively available in most codes and the ones that aren't are usually available through the exchange-correlation library \textsc{libxc}. For the most part, these are "plug-and-play" in modern DFT: you calculate a density, hand it off to \textsc{libxc}, and it returns the exchange-correlation potential and energy. The user rarely needs to know the details of the exchange-correlation approximation being used. Rather, the user usually knows that e.g. GGA-flavor "$X$" performs best for material "$Y$" and then asks the code to use that one. Everything else is automatic.

\subsection{Electronic states vs. Kohn-Sham states}

Since the KS equations are really an auxiliary set of equations used to calculate the density, the formal interpretation of KS states as electronic states is murky \cite{thijssen2007computational,martin2020electronic}. The interpretation is given credence by extending the KS method, at least formally, using many-body Green's functions. The KS states are used as the variational states of the electrons in an unperturbed system (i.e. not as "auxiliary functions" as in the KS scheme) and the explicit electron-electron interaction (with the exchange-correlation contribution removed) is treated perturbatively or (formally) exactly in e.g. the Lehmann representation. Then formally correct statements can be made about excitation energies, quasiparticle wave functions, and so-on. See the next section. In any case, this problem is usually ignored and KS band-structures, gaps, and wave functions are regularly used as electronic states in post-processing calculations.

\subsection{Energies, self-interactions, and the band-gap problem}\label{sec:band_gap_problem}

The Hartree potential in the KS equations results in so called "self-interaction" errors. These errors have important consequence in many cases, the most famous being the so called "band-gap problem" \cite{godby1988self,sham1985density,perdew1982density}. In short, the problem is that DFT significantly under estimates band-gaps in insulators (sometimes so badly that they are wrongly predicted to be metallic, c.f. Ge \cite{filippi1994all}). A heuristic argument for this disagreement is as follows  \cite{360454}: the Hartree potential depends on the density, and each occupied orbital contributes to the density, so in the Hartree interaction, the occupied orbitals are, in a sense, interacting with themselves. The interaction is purely repulsive, so the occupied states all gain an energy penalty due to self-interaction. On the other-hand, valence states are unoccupied and have no such-penalty. The occupied states are shifted "up", closing the gap.  

Corrections to DFT can remove the self-interaction error to some extent. One such correction, which we will discuss more below, is the DFT+U method. The self-interaction error is removed by subtracting (roughly speaking) the Hartree interaction and replacing it with a local repulsive Hubbard interaction. 

Even more advanced methods, e.g. the GW method  \cite{hedin1965new,aulbur2000quasiparticle,martin2016interacting}, go beyond adiabatic DFT; they use many-body Green's function methods to calculate the quasi-particle energies using DFT wave functions as a basis for perturbation theory. They are usually in good agreement with experiment for band-gaps, but fall short for modelling optical excitations in insulators. This is because they don't include Coulomb interactions between electrons and holes; i.e. they don't include excitonic effects. The so called BSE method improves upon this \cite{leng2016gw,martin2016interacting}. Note, however, that both GW and BSE are ridiculously expensive compared to DFT and many implementations don't calculate forces, so can't be used for geometry optimization or lattice dynamics calculations. I'm sure they will all do this eventually, but the computational cost currently limits these methods to very simple systems, even for pure electronic structure calculations.

\section{DFT+U}\label{sec:dft+U}

The self-interaction error in DFT discussed above tends to over-delocalize charge, resulting in a poor description of localized states: typically the $d$ and $f$ states in transition metals. The so-called DFT+U method was introduced to (i) correct the self-interaction error and (ii) study the physics of systems where a large Hubbard interaction is expected, but not accurately described by plain DFT \cite{himmetoglu2014hubbard}. 

The essence of DFT+U is to add a Hubbard interaction to the KS Hamiltonian: the Hubbard interaction is a local repulsive two-particle electron-electron interaction
\begin{equation}\begin{gathered}
    \hat{H}_U \sim U \hat{n}_{\uparrow} \hat{n}_{\downarrow}
\end{gathered}\end{equation}
where $U>0$ is the repulsive interaction strength and $\hat{n}_{\uparrow}$ is a density operator for a spin-up electron in the localized orbital and similarly for spin-down. We reiterate, the electrons are assumed localized on the atom. Terms like $\sim\hat{n}_{\uparrow}\hat{n}_{\uparrow}$ automatically vanish by Pauli exclusion. 

There are numerous many-body methods for the Hubbard interaction, but in the context of DFT+U, the interaction is treated in the Hartree approximation: $\hat{n}_\uparrow \hat{n}_\downarrow \approx \hat{n}_\uparrow \langle \hat{n}_\downarrow \rangle + \hat{n}_\downarrow \langle \hat{n}_\uparrow \rangle - \langle \hat{n}_\uparrow \rangle \langle \hat{n}_\downarrow \rangle$. The KS Hamiltonian containing this approximate term can be solved using exactly the same methods (secular equation, self consistency) discussed above. The only real complications are (i) how to define "localized orbitals" and (ii) how to correct for the fact that the XC functional already, to some extent, contains the Hubbard interaction (the so-called "double-counting correction"). The double-counting correction is expected to mitigate, to some extent, the self-interaction error. There are many formulas for (i) and (ii), but in the study presented in \cref{chp:lco_lda_u}, we use the method proposed by Dudarev et al. \cite{dudarev1998electron}. 

An added complication occurs when there are multiple orbitals on the atom (which is the case for e.g. $d$ electrons with 5 $m$ states). In this case, we must also include an \emph{exchange} term. Some of the DFT+U flavors do this explicitly with two parameters: $U$ (Coulomb) and $J$ (exchange). The method by Dudarev et al. combines both into an effective $U_{eff}=U-J$ interaction. In any case, the DFT+U method is considered "semi-empirical" since $U$ (and $J$) is a tunable parameter. In many cases, it is tuned to reproduce an experimental quantity, e.g. the band-gap. This is the approach we take in \cref{chp:lco_lda_u}. 

\section{Crystals}
\label{sec:crystals} 

The goal of energy sciences is to design materials with properties that are useful to energy technology: i.e. "energy materials". We might as well start with the easiest materials to work on... it turns out that these are \emph{crystalline} solids. 

This section is based on refs. \cite{ziman1972principles,starke2016response} and Section 2.1.3 in Martin \cite{martin2004many}.

\subsection{Crystal lattices} 

Crystals possess discrete translation symmetry classified by the crystalline lattice. The lattice is a set of points generated by integral combinations of the lattice vectors $\bm{a}_1$, $\bm{a}_2$, and $\bm{a}_3$. A generic point in the lattice is given by
\begin{equation}
    \bm{R}_{\{n_1,n_2,n_3\}}=n_1\bm{a}_1+n_2\bm{a}_2+n_2\bm{a}_2 .
\end{equation}
Notes on notation: once the lattice vectors are given, it is perfectly reasonable to specify a point in the crystal simply by the set of numbers $\bm{n}\equiv\{n_1,n_2,n_3\}$. Often one goes further and simply writes $\bm{R}_{\bm{n}}$, $\bm{R}_n$, or even just $\bm{R}$ with the "$n$" implied when discussing arbitrary lattice points. 

\begin{figure}
    \centering
    \includegraphics[width=0.9\linewidth]{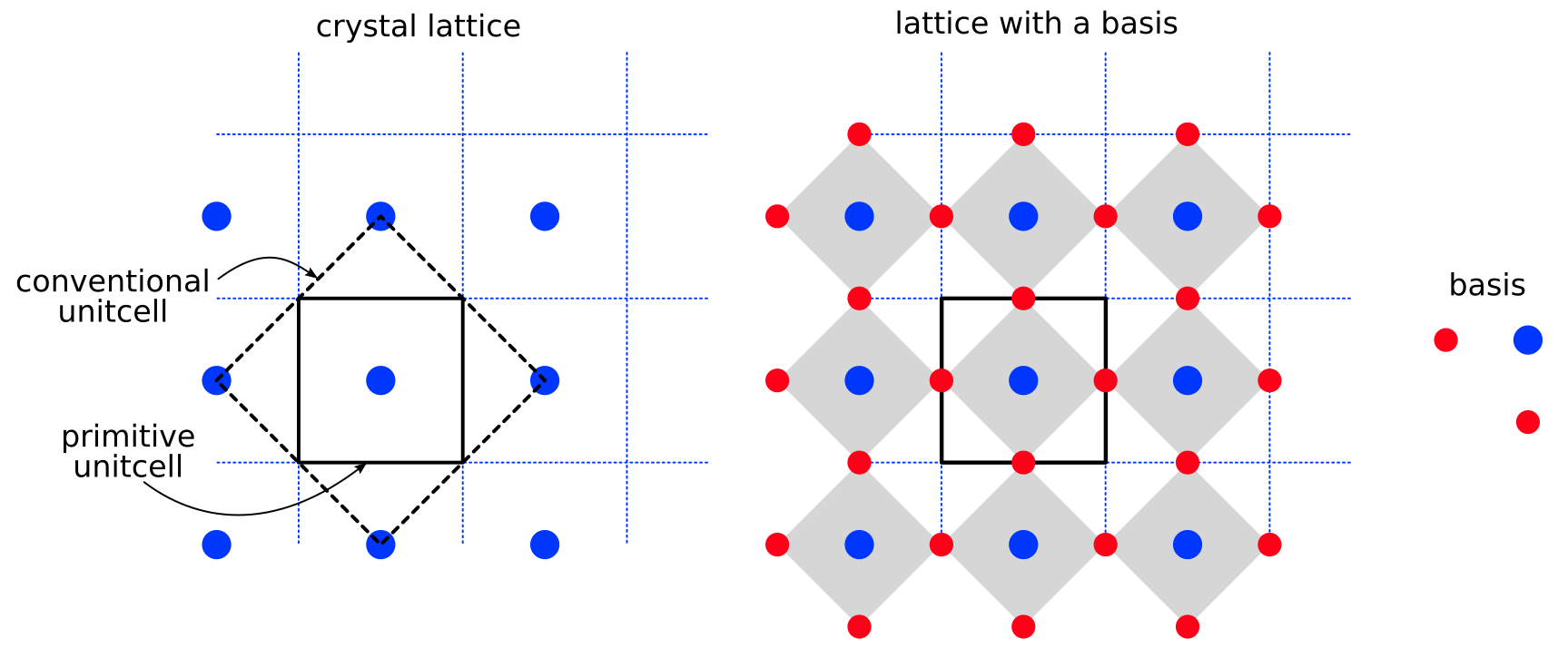}
    \caption{Diagram of a Bravais crystal lattice (left) and a lattice with a basis (right). The primitive (Wigner-Seitz) cell and a conventional cell are labeled in the figure.}
    \label{fig:lattice}
\end{figure}

It is also possible to associate a \emph{basis} with each point in the lattice. The basis is the fundamental "unit" of atoms that make up the crystal. The "cell" containing the basis is usually called the "unit cell". To be more precise, the unit cell is the volume of space contained between planes placed perpendicular to the lattice vectors and half-way between the lattice points. The boundaries of the unit cell are the lines formed by the intersections of the planes (see \cref{fig:lattice}). In many elemental crystals, like face-centered-cubic Ag, there is only one atom per unit cell. Such a lattice is called a "Bravais" lattice. This is very uncommon in nature, however. Most interesting energy materials nowadays have many atoms per unit-cell. e.g. Ba$_8$Ga$_{16}$Ge$_{30}$ has 54 atoms per unit cell (see \cref{chp:bgg}) and La$_2$CuO$_4$ has 28 (see \cref{chp:lco_lda_u}). The positions of the atoms are usually specified relative to the origin of the unit cell the atoms are in. For instance, suppose we want to know the absolute coordinate of the $i^{th}$ atom in unit cell $\bm{n}$. Call $\bm{\tau}_i$ the atom's position relative to the unit cell's origin. $\bm{\tau}$ is the same in every unit cell. The absolute position of the atom is then $\bm{r}_{ni}\equiv\bm{R}_{n}+\bm{\tau}_i$. If there are multiple atoms in the basis, then we have to specify the set of coordinates $\{\bm{\tau}\}$ (one for each atom in the basis).

The choice of lattice vectors isn't unique. There are many different ways to pick them that generate an equivalent crystal. A crystal is technically classified by its space group: this is the group theoretic name for the set of symmetry operations on a lattice and its basis that leave the crystal invariant. For a given space group, there exists a minimal volume \emph{Wigner-Seitz} cell called the primitive unit cell. But we don't have to pick the smallest one. We can pick a bigger unit cell that contains a basis (see \cref{fig:lattice}). For example, sometimes it is convenient to use a set of lattice vectors that are mutually orthogonal. This makes some parts of the physics simpler at the cost of a larger unit cell that contains more atoms. A unit cell that is not primitive is called a \emph{conventional} unit cell. We will see how this affects the physics later.

\subsection{Discrete translational symmetry}

\begin{figure}
    \centering
    \includegraphics[width=0.75\linewidth]{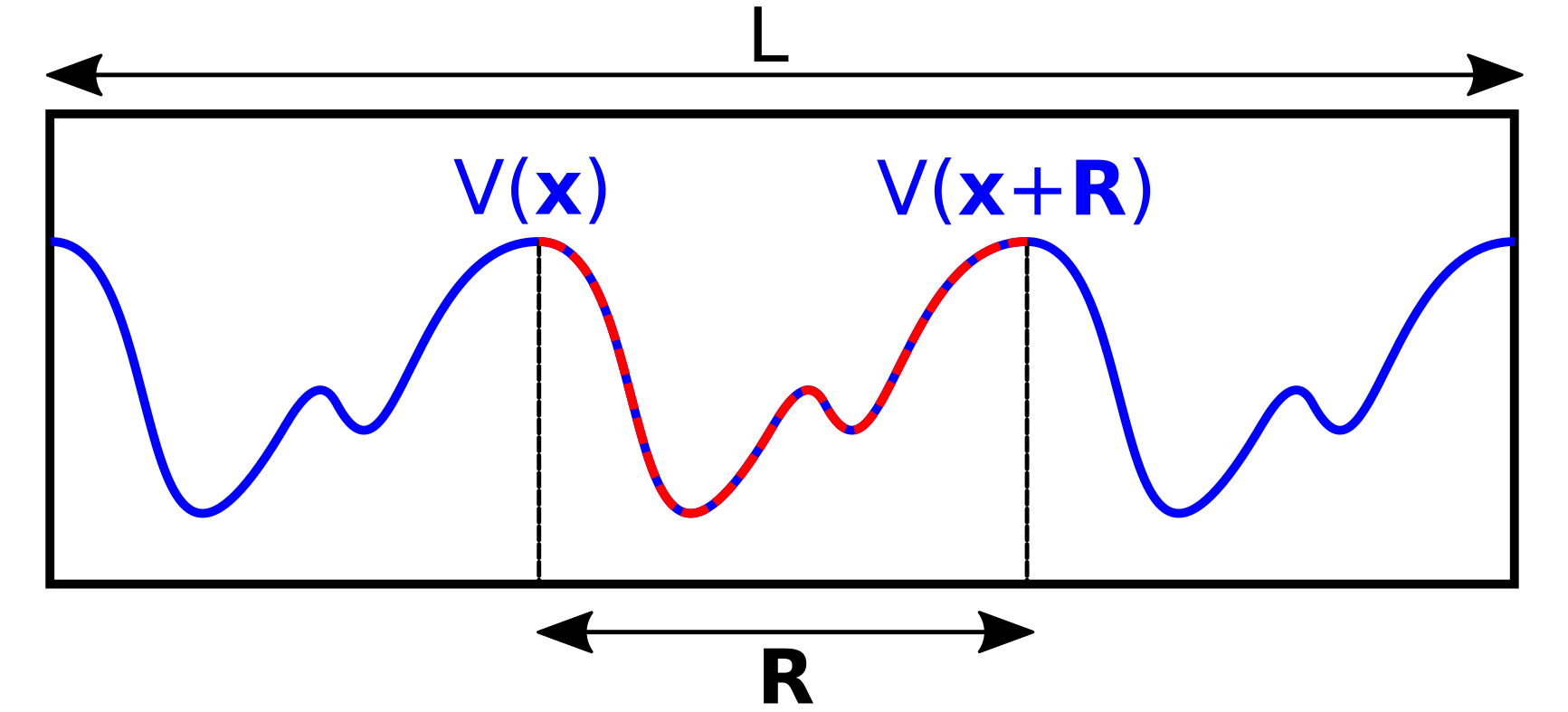}
    \caption{A periodic potential $V(\bm{r})$ with periodicity $\bm{R}$ in a box of length L. With Born-von Karman boundary conditions, the ends of the box are connected to each other.}
    \label{fig:periodic_pot}
\end{figure}

It turns out that doing physics in crystals offers some neat advantages. In physics we are used to using symmetries to make a problem easier. This is true in solid state physics too. For a crystal Hamiltonian of the form 
\begin{equation}
\begin{gathered}
    \hat{H}(\bm{r}) = -\frac{\hbar^2}{2m}\nabla^2 + V(\bm{r})
\end{gathered}
\end{equation}
with discrete translational symmetry, we expect it to be true that $V(\bm{r}+\bm{R})\equiv V(\bm{r})$ with $\bm{R}$ an arbitrary lattice vector. This is because each unit cell is identical (i.e. there is periodicity, see \cref{fig:periodic_pot}). Then it is true that $\hat{H}(\bm{r}+\bm{R})=\hat{H}(\bm{r})$. We say the Hamiltonian has "discrete translational symmetry". As we will now see, the discrete translational symmetry of the lattice allows us to invoke "Bloch's theorem", which greatly simplifies solving the Schr\"{o}dinger's equation in solids. \footnote{Group theory can also be applied to solid state physics problems in many much more technical ways; e.g. we can use our knowledge of the space group of a crystal to "block-diagonalize" its Hamiltonian. This reduces the problem of diagonalizing a large Hamiltonian into diagonalizing a set of smaller subspace Hamiltonians. This stuff is very cool, but very outside the scope of this thesis. See the amazing book by the amazing Mildred Dresselhaus for more details \cite{dresselhaus2007group}!}

\subsection{Bloch's Theorem}\label{sec:blochs_thm}

This section follows \S 4.3 in Sakurai \cite{sakurai1995modern}, and Chp. 1 in Ziman \cite{ziman1972principles}, but Bloch's theorem is studied in practically any book on solid state physics. Bloch's theorem is just a rather general statement about the form of eigenstates of a Hamiltonian in a periodic potential, e.g. in solids. Let's see what Bloch's theorem says. 

Assume we are dealing with a Hamiltonian of the form $\hat{H}(\bm{r})=\hat{\bm{p}}^2/2m+V(\bm{r})$ where the potential $V(\bm{r})=V(\bm{r}+\bm{R})$ is periodic under translation by a lattice vector $\bm{R}$. Define the \emph{translation} operator $\hat{T}_{\bm{R}}$ that formally changes the coordinates $\bm{r}\rightarrow \bm{r}+\bm{R}$. The adjoint $\hat{T}^\dagger_{\bm{R}} \equiv \hat{T}^{-1}_{\bm{R}}$ undoes the coordinate shift: $\bm{r}\rightarrow \bm{r}-\bm{R}$. It's action on the Hamiltonian is: $\hat{T}^\dagger_{\bm{R}} \hat{H}(\bm{r}) \hat{T}_{\bm{R}}=\hat{H}(\bm{r}) $ which follows from the periodicity of the potential. It's a unitary operator and its commutation relation with the Hamiltonian is 
\begin{equation}
    [\hat{T}_{\bm{R}},\hat{H}(\bm{r})] = 0 .
\end{equation}
The translation operator commutes with the Hamiltonian, so eigenstates of the translation operator are eigenstates of the Hamiltonian! Then, if we can find the eigenstates of the translation operator, we find the eigenstates of the Hamiltonian. 

Let $\psi(\bm{r})$ be an eigenstate of the translation operator. Since $\hat{T}_{\bm{R}}$ is unitary, it's eigenvalues can only be a phase. Specialize to translation by only a single basis vector:
\begin{equation}
    \hat{T}_{\bm{a}_1}\psi(\bm{r}) = \exp(i\theta_1)\psi(\bm{r}) .
\end{equation}
$\theta_1$ is a real number. Then translation by an arbitrary lattice vector $\bm{R}\equiv n_1\bm{a}_1+n_2\bm{a}_2+n_3\bm{a}_3$ has to give
\begin{equation}
    \hat{T}_{\bm{R}}\psi(\bm{r}) = \exp(i(n_1\theta_1+n_1\theta_2+n_3\theta_3))\psi(\bm{r}) \equiv \exp(i\bm{k}\cdot \bm{R}) \psi(\bm{r})
\end{equation}
where the sum was parameterized by the \emph{definition} $\bm{k}\cdot\bm{a}_i=n_i \theta_i$. So we can label solutions of the periodic Hamiltonian by a vector $\bm{k}$. Finally, define the following function that is periodic with respect to the crystal's unit cell:
\begin{equation}
    u_{\bm{k}}(\bm{r}) = \exp(-i\bm{k}\cdot \bm{r}) \psi_{\bm{k}}(\bm{r}) .
\end{equation}
That this is a unit cell-periodic function can be seen by
\begin{equation}\begin{split}
    u_{\bm{k}}(\bm{r}+\bm{R}) & = \exp(-i\bm{k}\cdot( \bm{r}+ \bm{R})) \psi_{\bm{k}}(\bm{r}+\bm{R})  \\
    & = \exp(-i\bm{k}\cdot \bm{r})\exp(i\bm{k}\cdot (\bm{R}-\bm{R}))\psi_{\bm{k}}(\bm{r}) \\
    & =  \exp(-i\bm{k}\cdot\bm{r})\psi_{\bm{k}}(\bm{r}) \\
    & = u_{\bm{k}}(\bm{r}) .
\end{split}\end{equation}
The statement of Bloch's theorem is that the solutions of the Schr\"{o}dinger's equation in a periodic potential can be represented by a plane wave times a unit cell-periodic function,
\begin{equation}
    \psi_{\bm{k}}(\bm{r})=\exp(i\bm{k}\cdot \bm{r})u_{\bm{k}}(\bm{r}),
    \label{eq:blochs_thm}
\end{equation}
as we have just verified.

\subsection{Plane waves} 

We introduced the vector $\bm{k}$ when deriving Bloch's theorem. What is this vector? In theoretical mechanics, Noether's theorem tells us that a system with continuous translational symmetry conserves momentum \cite{goldstein2002classical}. This is \emph{almost} true for crystals: instead, we have discrete translational symmetry, so \emph{crystal momentum} is conserved. The vector $\bm{k}$ is called crystal momentum\footnote{$\bm{k}$ actually has dimensions of a wave-number, $[\bm{k}]=1/L$. It is related to momentum by $\bm{p}=\hbar\bm{k}$. For convenience, the wave vector $\bm{k}$ is usually called crystal momentum anyway, though what is really meant is $\hbar \bm{k}$.}.

To make sense of this, let's recall some stuff about the solutions of a Hamiltonian with continuous symmetry: the simplest conceivable such Hamiltonian is the free-particle $\mathcal{\hat{H}}=\hat{\bm{p}}^2/2m=-\hbar^2 \nabla^2/2m$. Momentum is conserved so the solutions are momentum eigenstates $|\bm{k}\rangle$ with eigenvalue $E=\hbar^2\bm{k}^2 \equiv \bm{p}^2$. In position representation, $\langle \bm{r} | \bm{k} \rangle=c\exp(i\bm{k}\cdot\bm{r})$. The normalization factor, $c$, needs to be determined. There is an issue, however. If we put the system in infinite space and try to normalize in the usual way
\begin{equation}
    1=\langle \bm{k} | \bm{k} \rangle = \int d\bm{r} \langle \bm{k} | \bm{k} \rangle = |c|^2\int d\bm{r} = \infty
\end{equation}
we find that it is impossible. The way around this issue is to put the system into a finite "box" of volume $V\equiv L^3$. Then it is found that $c=V^{-1/2}\equiv L^{-3/2}$. Momentum eigenstates are often called "plane waves" and their quantum number is their momentum, $\bm{k}$. Explicitly, momentum eigenstates are
\begin{equation}
    \psi_{\bm{k}}(\bm{r}) = \frac{1}{V^{1/2}} \exp(i\bm{k} \cdot \bm{r}) .
    \label{eq:plane_wave}
\end{equation}
The wave functions in \cref{eq:plane_wave} are the solutions of the free-particle Hamiltonian. There is still an issue, however. Recall that the free-particle SE is a 2$^{nd}$ order differential equation, thus requiring two boundary conditions. One is taken care of by normalization. The other is given by the wave function's value on the box boundaries. For example, in the infinite well problem, we demand the wave function is 0 at the boundaries. In condensed matter physics, however, a more convenient choice is \emph{periodic boundary conditions}.

\subsection{Periodic boundary conditions; The Brillouin zone}
\label{sec:PBC}

The solutions of the SE with a periodic Hamiltonian satisfy Bloch's theorem (we call them "Bloch functions"). Bloch's theorem says the (unnormalized) solutions are
\begin{equation}
\begin{gathered}
    \psi_{\bm{k}}(\bm{r}) = \exp(i\bm{k}\cdot\bm{r}) u_{\bm{k}}(\bm{r}) \\
    u_{\bm{k}}(\bm{r}+\bm{R}) = u_{\bm{k}}(\bm{r})
\end{gathered}
\end{equation}
where $\bm{k}$ is a quantum number of the system. It is called \emph{crystal} momentum and is almost conserved like usual momentum: unlike momentum, crystal momentum is only conserved modulo a reciprocal lattice lattice vector. Let us see what this means. 

The system is placed into a "box" of volume $V$ and periodic boundary conditions (PBC) are imposed. There are $N=N_1 N_2 N_3$ unit cells with $N_i$ labeling the number of unit cells along the $i^{th}$ direction.  Then from the requirement $\psi_{\bm{k}}(\bm{R}_N)=\psi_{\bm{k}}(\bm{R}_0)$, with $\bm{R}_N$ a vector connecting a point on the opposite side of the box that must satisfy PBC with the point $\bm{R}_0$, we find
\begin{equation}
\begin{gathered}
    \exp(i\bm{k}\cdot\bm{R}_N) u_{\bm{k}}(\bm{R}_N) = \exp(i\bm{k}\cdot\bm{R}_0) u_{\bm{k}}(\bm{R}_0) = \\
    \exp(i\bm{k}\cdot \bm{R}_N) = \exp(i\bm{k}\cdot\bm{R}_0)
\end{gathered}
\end{equation}
where we used periodicity of the functions $u_{\bm{k}}(\bm{r})$. This equation must hold for all points on the edges of the box that are connected by a lattice vector, so let us just pick $\bm{R}_0\equiv 0$ to be the origin for convenience. We find that 
\begin{equation}
\begin{gathered}
    \exp(i\bm{k}\cdot\bm{R}_N) = 1
\end{gathered}
\end{equation}
which requires 
\begin{equation}
\begin{gathered}
    \bm{k}\cdot\bm{R}_N = 2 \pi (N_1 n_1 +  N_2 n_2 + N_3 n_3) = 2\pi n \\
    n \in  \mathbb{Z} 
\end{gathered}
\end{equation}
We chose to write $\bm{k}= n_1 \bm{b}_1 + n_2 \bm{b}_2 + n_3 \bm{b}_3$ in a basis such that $ \bm{a}_i\cdot \bm{b}_j \equiv 2\pi \delta_{ij}$. The $\bm{b}_i$ chosen this way are called \emph{primitive} reciprocal lattice vectors. They are \footnote{Another convenient (from a computational perspective) way to write the reciprocal lattice vectors is as follows: assemble the lattice and reciprocal lattice vectors into matrices, $\hat{a} = (\bm{a}_1,\bm{a}_2,\bm{a}_3)$ and $\hat{B} = (\bm{b}_1,\bm{b}_2,\bm{b}_3)$. The requirement $\bm{a}_i \cdot \bm{b}_j = 2\pi \delta_{ij}$ can be written as $\hat{A}^T \hat{B} = 2\pi \hat{I}$ with $\hat{I}$ the identity matrix. Then the reciprocal lattice vectors are $\hat{B}=2\pi[\hat{A}^T]^{-1}$.}
\begin{equation}
    \bm{b}_i = \epsilon_{ijk} \frac{2\pi}{\Omega} \bm{a}_j \times \bm{a}_k .
\end{equation}
$\Omega$ is the volume of the primitive unit cell. Then $n_i\equiv m_i/N_i$ and $k_i = 2\pi m_i / N_i a_i$ with $m_i \in \mathbb{Z}$. Evidently we don't need to consider wave vectors $k_i$ with $m_i \equiv (m'_i + N_i) > N_i $ since $\exp(i 2\pi (m'_i+N_i) /N_i ) = \exp(i2\pi m'_i/N_i)$ with $m'_i \leq N_i$ already included in the restricted range $m_i \leq N_i$. Note that $2\pi N_i/N_i a_i = 2\pi/a_i = b_i $ is a primitive reciprocal lattice vector. More generally, we see that adding any arbitrary reciprocal lattice vector $\bm{G} = N_1 \bm{b}_1 + N_2 \bm{b}_2 + N_3 \bm{b}_3$ to $\bm{k}$ has no affect\footnote{Actually, there is a small nuance. We note that $\bm{k}$ is equivalent to $\bm{k}+\bm{G}$, but the Bloch functions transform as $\psi_{\bm{k}+\bm{G}}(\bm{r}) =  \exp(i\bm{k}\cdot\bm{r})\exp(i\bm{G}\cdot\bm{r})u_{\bm{k+G}}(\bm{r})$. If we choose the \emph{periodic gauge} such that $\exp(i\bm{G}\cdot\bm{r})u_{\bm{k+G}}(\bm{r}) = u_{\bm{k}}(\bm{r})$, then  $\psi_{\bm{k}+\bm{G}}(\bm{r}) = \psi_{\bm{k}}(\bm{r})$ is periodic with respect to $\bm{G}$. More generally $\psi_{\bm{k}+\bm{G}}(\bm{r}) = \exp(i\bm{G}\cdot\bm{r})\psi_{\bm{k}}(\bm{r})$.}. This is what we mean when we say crystal momentum is only defined modulo a reciprocal lattice vector.

Restricting $m_i \leq N_i$ means that $k_i \leq b_i$. The region of reciprocal space spanned by $\bm{b}_i$ is called the \emph{first Brillouin zone} (1BZ). The 1BZ is the Wigner-Seitz cell of the reciprocal lattice. $\bm{k}$ are wave vectors that lie on a uniform grid in the 1BZ and there are $N=N_1 N_2 N_3$ of them. A $\bm{k}'$ that lies outside the 1BZ is equivalent to all points connected to it by a reciprocal lattice vector, including $\bm{k}$ in the 1BZ,  so that we only need to study wave vectors that lie in the 1BZ. This is the essence of crystal momentum. The importance of Bloch's theorem is that we only have to solve the SE for the finite number, $N$, of wave vectors $\bm{k}$ that lie in the 1BZ.

Usually instead of restricting $m_i \leq N_i$, we pick $-N_i/2 < m_i \leq N_i/2$. In "reciprocal lattice units," we specify the wave vector in units where $n_i= m_i/N_i$ so that we have $-1/2 < n_i \leq 1/2$ which is convenient notation because it is independent of the system size.

Let us also work out the normalization of the Bloch functions. Supposing that $\psi_{\bm{k}}(\bm{r})=c_{\bm{k}}\exp(i\bm{k}\cdot\bm{r})u_{\bm{k}}(\bm{r})$ we demand that 
\begin{equation}
\begin{gathered}
    \langle \psi_{\bm{k}} | \psi_{\bm{k}} \rangle  =|c_{\bm{k}}|^2 \int d\bm{r} |u_{\bm{k}}(\bm{r})|^2 \\ 
    = N |c_{\bm{k}}|^2 \int_\Omega d\bm{r} |u_{\bm{k}}(\bm{r})|^2 = 1
\end{gathered}
\end{equation}
The second line follows from periodicity. Integrating over "$\Omega$" means to only integrate over a single unit cell located at the origin. If we resolve to normalize the unit cell periodic part of the Bloch functions, $u_{\bm{k}}(\bm{r})$, such that $\langle u_{\bm{k}} | u_{\bm{k}} \rangle \equiv \int_\Omega d\bm{r} |u_{\bm{k}}(\bm{r})|^2 = 1$, then $c_{\bm{k}}=N^{-1/2}$.

\subsection{The infinite volume limit}

Another reason to impose PBCs is that we can systematically take the limit of an infinite volume box with all volume-dependent factors cancelling. In solid state physics, this means we can model an infinite crystal without ever having to care about what happens at its surface. With that in mind, we mention that is very common to encounter sums of the form
\begin{equation}
    V^{-1} \sum_{\bm{k}} 
\end{equation}
where the sum runs over all points in the 1BZ. This can be inconvenient to work with analytically; numerically, evaluating sums is straightforward... anyway, this is where taking the infinite volume limit comes in handy. Recall that $\bm{k}$ is a discrete set spaced $2\pi n /L $ apart. $2$ points spaced a single integer apart have spacing $\Delta n$. Then $\Delta k = 2 \pi \Delta n / L$. In the limit $L\rightarrow \infty$, $\Delta k$ becomes infinitesimal. Then $dn = L\cdot dk/(2\pi)$. Thus, we arrive at the "rule" for taking the infinite volume limit:
\begin{equation}
V^{-1}\sum_{\bm{k}}\rightarrow \int \frac{d\bm{k}}{(2\pi)^3} .
    \label{eq:inf_vol_limit}
\end{equation}
The rule in \cref{eq:inf_vol_limit} comes up very often in calculations.

\subsection{Fourier transforms}

As we have seen, we repeatedly try to generalize from the free-particle, which has continuous symmetry, to a crystal, which has discrete translational symmetry. What's the difference? The kinetic energy is manifestly translationally invariant, so the difference is in the potential. Ignoring interactions between electrons, the potential is the electrostatic interaction between an electron and the nuclei. But we already said that in a crystal, the nuclei form an ordered arrangement. This means that if we (or the crystal) move by an arbitrary number of lattice vectors, the potential is identical. i.e.   
\begin{equation}
    V(\bm{r})=V(\bm{r}+\bm{R})
\end{equation}
A diagram of $V(\bm{r})$ was shown in \cref{fig:periodic_pot}. Since the potential is periodic, it will be useful to express it (and many other quantities) in a Fourier representation. Let's do that now. 

Instead of specializing to a potential, we might as well study the Fourier transform of \emph{any} continuous periodic function with periodicity $\bm{R}$ in a box with size $V=L^3$. Call the function $f(\bm{r})$. It is expressible as a sum of plane waves, \cref{eq:plane_wave}
\begin{equation}
\begin{gathered}
    f(\bm{r}) = \sum_{\bm{G}} f_{\bm{G}} \exp(i\bm{G}\cdot\bm{r}) .
    \label{eq:fourier_transform_periodic}
\end{gathered}
\end{equation} 
Because $f(\bm{r})$ has the periodicity of the lattice, we only need to consider wave vectors that belong to the reciprocal lattice. Let $\bm{b}_i$ be the primitive reciprocal lattice vectors. Then $\bm{G} =m_1\bm{b}_1 +m_2 \bm{b}_2 + m_3 \bm{b}_3$ is a general reciprocal lattice vector. Let's check that these are the only components that contribute to $f(\bm{r})$. Consider a wave vector $\bm{Q}=\bm{q}+\bm{G}$ with $\bm{q}$ some wave vector other than a reciprocal lattice vector. Then translation by a lattice vector $\bm{R}$ results in $\exp(i \bm{Q}\cdot (\bm{r}+\bm{R})) = \exp(i \bm{Q}\cdot \bm{r}) \exp(i \bm{q}\cdot \bm{R}) \neq \exp(i\bm{Q}\cdot \bm{r})$ which is not periodic and thus it's Fourier coefficient must be identically 0. Note that in the special case $f(\bm{r})$ is real
\begin{equation}
\begin{gathered}
    f(\bm{r}) \equiv f^*(\bm{r}) = \sum_{\bm{G}} f^*_{\bm{G}} \exp(-i\bm{G}\cdot\bm{r}) = \sum_{\bm{G}} f_{\bm{G}} \exp(i\bm{G}\cdot\bm{r}) 
\end{gathered}
\end{equation} 
which implies that $f^*_{\bm{G}}=f_{-\bm{G}}$. This can be used to reduce the number of Fourier coefficients that have to be explicitly calculated.

The \emph{non-zero} Fourier coefficients in \cref{eq:fourier_transform_periodic} are determined using orthogonality. 
\begin{equation}
\begin{gathered}
    \int d\bm{r} f(\bm{r}) \exp(-i\bm{G}'\cdot\bm{r}) = \sum_{\bm{G}} \int d\bm{r} f_{\bm{G}} \exp(i(\bm{G}-\bm{G}')\cdot\bm{r}) \\
    = V \sum_{\bm{G}} f_{\bm{G}} \delta_{\bm{G},\bm{G}'}
\end{gathered}
\end{equation}
Then we have 
\begin{equation}
\begin{gathered}
    f_{\bm{G}} = V^{-1} \int d\bm{r} f(\bm{r}) \exp(-i\bm{G} \cdot\bm{r}) \\ = \Omega^{-1} \int_\Omega d\bm{r} f(\bm{r}) \exp(-i\bm{G}\cdot\bm{r})
    \label{eq:fourier_coeff}
\end{gathered}
\end{equation}
where, in the last line, periodicity of $f(\bm{r})$ was used to reduce the integral down to a single unit cell centered at the origin.

\section{Electronic structure of solids}

We now turn to apply the theory of crystals to the general discussion of electronic structure methods developed in this chapter. We saw that the goal of was to calculate the energy spectrum of the Born-Oppenheimer electronic Hamiltonian, \cref{eq:e_bo_hamiltonian}, and crystals are no exception. Our goal is solve the TISE. It turns out, however, that this takes an especially simple form in crystals. The periodicity means that crystal momentum, $\bm{k}$, is a good quantum number and solving the enormous Hamiltonian \cref{eq:e_bo_hamiltonian}, which is in principle an $N\times m$ matrix diagonalization with $N$ the number of unit cells in real space and $m$ the number of basis functions, can be reduced to solving $N$ number of $m$-dimensional \emph{Bloch} Hamiltonians (in other words, \cref{eq:e_bo_hamiltonian} is block diagonal). The energy spectrum (i.e. the eigenvalues) are functions of crystal momentum, $\bm{k}$. For each $\bm{k}$, there are $m$ eigenvalues. They are continuous functions of $\bm{k}$ (in the limit $N\rightarrow \infty$) and, as we saw in \cref{sec:PBC}, they are periodic with respect to the Brillouin zone. We call the eigenvalues "bands" because they form bands of allowed energy values in $\bm{k}$-space.

\subsection{The secular equation in crystals}

Let's explicitly specialize to planewaves for the basis functions: $w_{n}(\bm{r}) \equiv w_{\bm{G}}(\bm{r}) = \Omega^{-1/2} \exp(i\bm{G}\cdot\bm{r})$, where $\Omega^{-1/2}$ accounts for normalization. This choice is motivated by the fact these are the Fourier components of an arbitrary unit cell periodic function. The variational Bloch function is $\phi_{\bm{k}+\bm{G}}(\bm{r}) = N^{-1/2} \exp(i\bm{k}\cdot\bm{r}) w_{\bm{G}}(\bm{r}) = N^{-1/2} \exp(i(\bm{k}+\bm{G})\cdot\bm{r})$, justifying the notation $\bm{k}+\bm{G}$ in the subscript: $\bm{G}$ labels the basis functions and $\bm{k}$ the crystal momentum. This is a special case where the unit cell periodic part of the basis function doesn't depend on $\bm{k}$. For the eigenstate $\psi_{\bm{k}n}(\bm{r}) = N^{-1/2} \exp(i\bm{k}\cdot\bm{r}) u_{\bm{k}n}(\bm{r})$, $u_{\bm{k}n}(\bm{r})$ will depend on $\bm{k}$. The solutions to the secular equation are
\begin{equation}\begin{gathered}
    \psi_{\bm{k}n}(\bm{r})= \sum_{\bm{G}} c_{\bm{k}+\bm{G},n} \phi_{\bm{k}+\bm{G}}(\bm{r}) =  \frac{\exp(i\bm{k}\cdot\bm{r})}{\sqrt{N}}    \sum_{\bm{G}} \frac{c_{\bm{k}+\bm{G},n}}{\sqrt{\Omega}} \exp(i\bm{G}\cdot\bm{r}).
\end{gathered}\end{equation}
The unit cell periodic part is $u_{\bm{k}n}(\bm{r}) = \Omega^{-1/2} \sum_{\bm{G}} c_{\bm{k}+\bm{G},n} \exp(i\bm{G}\cdot\bm{r})$.

For convenience, define $\phi_{\bm{k}+\bm{G}}(\bm{r}) \equiv |\bm{k}+\bm{G}\rangle$. To derive the secular equation, we need to calculate matrix elements of the form 
\begin{equation}\begin{split}
    \langle \bm{k}'+\bm{G}' | \hat{H} | \bm{k}+\bm{G} \rangle &= \int d\bm{r} \phi^*_{\bm{k}'+\bm{G}'}(\bm{r}) \hat{H}(\bm{r}) \phi_{\bm{k}+\bm{G}}(\bm{r}) \\
    &= \left( \sum_{\bm{R}} \exp(i(\bm{k}-\bm{k}')\cdot \bm{R}) \right) \int_\Omega d\bm{r} \phi^*_{n\bm{k}'+\bm{G}'}(\bm{r}) \hat{H}(\bm{r}) \phi_{\bm{k}+\bm{G}'}(\bm{r}) \\
    &= N \delta_{\bm{k}\bm{k}'} \int_\Omega d\bm{r} \phi^*_{\bm{k}'+\bm{G}'}(\bm{r}) \hat{H} (\bm{r}) \phi_{\bm{k}'+\bm{G}'}(\bm{r})
    \label{eq:bloch_ham_1}
\end{split}\end{equation}
Some comments are in order at this point. Only $\bm{k}'=\bm{k}$ matrix elements are non-zero, which we could have guessed since $\bm{k}$ is a conserved quantity. The Hamiltonian, \cref{eq:e_bo_hamiltonian} has been reduced to $N$ Hamiltonians, one for each $\bm{k}$. We derive the secular equation for each one separately and perform $N$ diagonalization, calculating the eigenvalues $\epsilon_{\bm{k}n}$ and wave functions $\psi_{\bm{k}n}(\bm{r})$. $n$ labels the valence states that we explicitly consider in the calculation. The core states are replaced by a pseudopotential (see \cref{sec:pseudo}). In general, there are $m$ $\bm{G}$-basis functions and we want to calculate $p\leq m$ valence states; even when using pseudopotentials, we usually need $m\gg p$ for accurate calculations. In this case, explicit diagonalization is prohibitively expensive, scaling as $\mathcal{O}(m^3)$. Fortunately iterative diagonalization methods exist to calculate the $p$ lowest-lying states and the methods scale as $\mathcal{O}(m^2\ln m)$ which is a huge improvement (see \cref{sec:pseudo}). 

Back to business. The integral on the last line of \cref{eq:bloch_ham_1} further reduces to 
\begin{equation}\begin{split}
    \langle \bm{k}+\bm{G}' | \hat{H} | \bm{k}+\bm{G} \rangle &= N \int_\Omega d\bm{r} \phi^*_{\bm{k}+\bm{G}'}(\bm{r}) \hat{H} (\bm{r}) \phi_{\bm{k}'+\bm{G}'}(\bm{r}) \\
    & = \frac{1}{\Omega} \frac{\hbar^2(\bm{k}+\bm{G})^2}{2m} \int_\Omega d\bm{r} \exp(i(\bm{G}-\bm{G}')\cdot\bm{r}) \\
    &\qquad + \frac{1}{\Omega} \int_\Omega d\bm{r} \exp(-i(\bm{G}'-\bm{G})\cdot\bm{r}) V(\bm{r}) \\
    & = \frac{\hbar^2(\bm{k}+\bm{G})^2}{2m} \delta_{\bm{G}\bm{G}'} + V_{\bm{G}'-\bm{G}}
    \label{eq:bloch_ham_matrix_element}
\end{split}\end{equation}
In this thesis, we only consider \emph{local} potentials. A local potential only depends on a coordinate: e.g. $V(\bm{r})$. On the other hand, a \emph{non-local} potential depends on more than one coordinate or, in the case of NCPP and PAW methods \cite{thijssen2007computational,singh2006planewaves}, on an integral over space via the projection operator. The complications of a non-local potential aren't relevant to understanding the physics and are really just technical details, so we ignore them. Anyway, it doesn't matter what exactly the potential looks like: e.g. the difference between NCPP and PAW datasets is transparent to the user and simply providing one-or-the other is usually all that is needed from the user. All that matters is that the system is periodic and \cref{eq:bloch_ham_matrix_element} is applicable. Then $V_{\bm{G}'-\bm{G}}$ is just the Fourier transform of the periodic potential in a single unit cell, something that can be very easily calculated with fast Fourier transforms (FFTs). 

We also need the overlap matrix elements. They are simply
\begin{equation}\begin{gathered}
    \langle \bm{k}+\bm{G}' | \bm{k}+\bm{G} \rangle = \Omega^{-1} \int_\Omega d\bm{r} \exp(i(\bm{G}-\bm{G}')\cdot\bm{r}) = \delta_{\bm{G}\bm{G}'},
    \label{eq:bloch_ham_overlap_element}
\end{gathered}\end{equation}
i.e. the overlap matrix is the identity matrix and the secular equation reduces to an ordinary eigenvalue problem. In e.g. the PAW method, we still have to solve a generalized eigenvalue problem since $\hat{S}=\hat{\mathcal{T}}^\dagger\hat{\mathcal{T}}$ isn't diagonal (see \cref{sec:pseudo}), but this isn't a huge complication and codes deal with it automatically. 

If we care about magnetism, we have to allow spin-polarization. We put a spin-index $\sigma$ on all the wave functions. We use the same basis-states for spin-up and spin-down and put the spin-index on the coefficients: $\psi_{\bm{k}n\sigma}(\bm{r})=\sum_{\bm{G}} c_{\bm{k}+\bm{G},n\sigma} \phi_{\bm{k}+\bm{G}}(\bm{r})$. If we neglect spin-orbit coupling, which we do in this thesis, the spin-polarized Bloch Hamiltonians are block diagonal and we simply have to solve two Hamiltonians per $\bm{k}$-point: $\hat{H}_{\bm{k}\sigma} = \{ \hat{H}_{\bm{k}\uparrow}, \hat{H}_{\bm{k}\downarrow}\}$.

Finally, let us discuss the last nontrivial consideration with planewaves: self consistency. The density is calculated from $n(\bm{r}) = N^{-1} \sum_{\bm{k}n} f(\epsilon_{\bm{k}n}) |u_{\bm{k}n}(\bm{r})|^2$, where the factor $N^{-1}$ (from the normalization of Bloch functions) enforces that the density integrates to the number of electrons per unit cell, $N_e = \int_\Omega n(\bm{r})$. The potential depends on the density, $V(\bm{r})=V[n(\bm{r})]$, and the density is a sum over $\bm{k}$-points. So the general self consistent loop for planewaves is as follows: 
\begin{enumerate}
    \item make an initial guess for the input-density, $n(\bm{r})$.
    \item solve the secular equation for all $\bm{k}$-points. 
    \item calculate the output-density, $n(\bm{r})$, by summing over all $\bm{k}$-points. 
    \item if the output-density differs from the input, mix them and solve all $\bm{k}$-points again.
    \item iterate until the input and output agree within some tolerance.
\end{enumerate}
If the wave functions are spin-polarized, we also have to sum over spin at each $\bm{k}$-point. 

To summarize, in crystals, we can reduce \cref{eq:e_bo_hamiltonian} to $N$ $m$-dimensional ordinary eigenvalue problems that have very simple matrix elements given by \cref{eq:bloch_ham_matrix_element}. Furthermore, we can use iterative diagonalization methods to efficiently diagonalize the secular equations. The KS total energy in \cref{eq:ks_total_e} is easily calculated in terms of planewaves. All of this is done automatically by commonly used DFT codes and all the user has to do is tell the code the size of $N$ and $m$, as we will look at in more detail in \cref{sec:pw_dft_accuracy}.  

\begin{figure}
    \centering
    \includegraphics[width=0.8\linewidth]{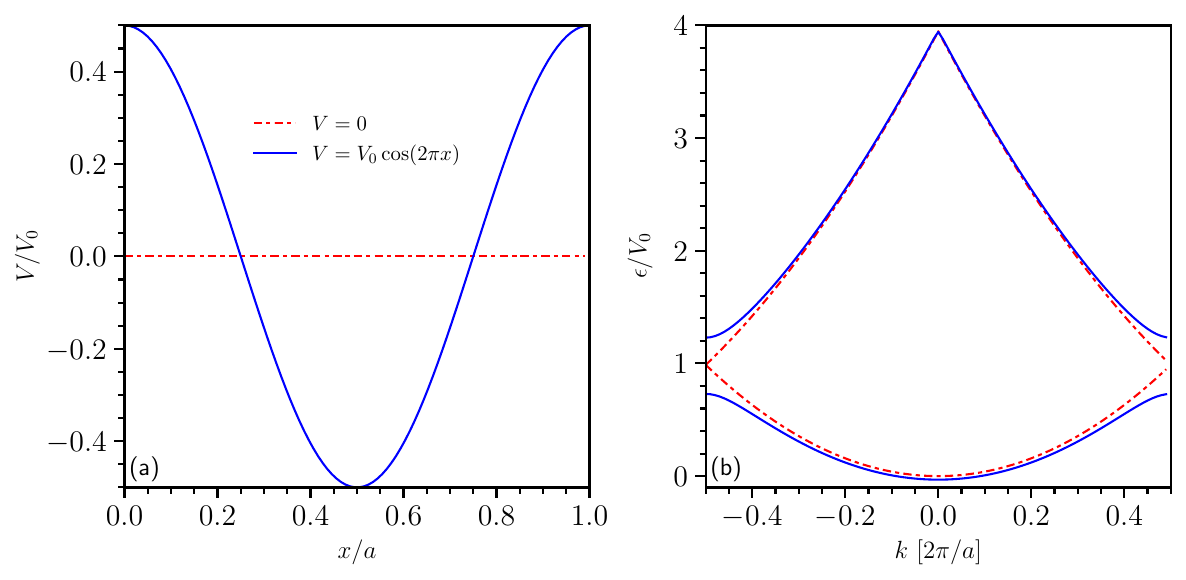}
    \caption{(a) Crystal potential for an empty lattice, i.e. free electrons (dashed line), and a periodic potential potential with wavelength $2\pi/a$ (solid line). (b) The energy bands for the free electron potential (dashed) and crystal potential (solid).}
    \label{fig:1d_ebands}
\end{figure}

\begin{figure}[t!]
    \centering
    \includegraphics[width=0.8\linewidth]{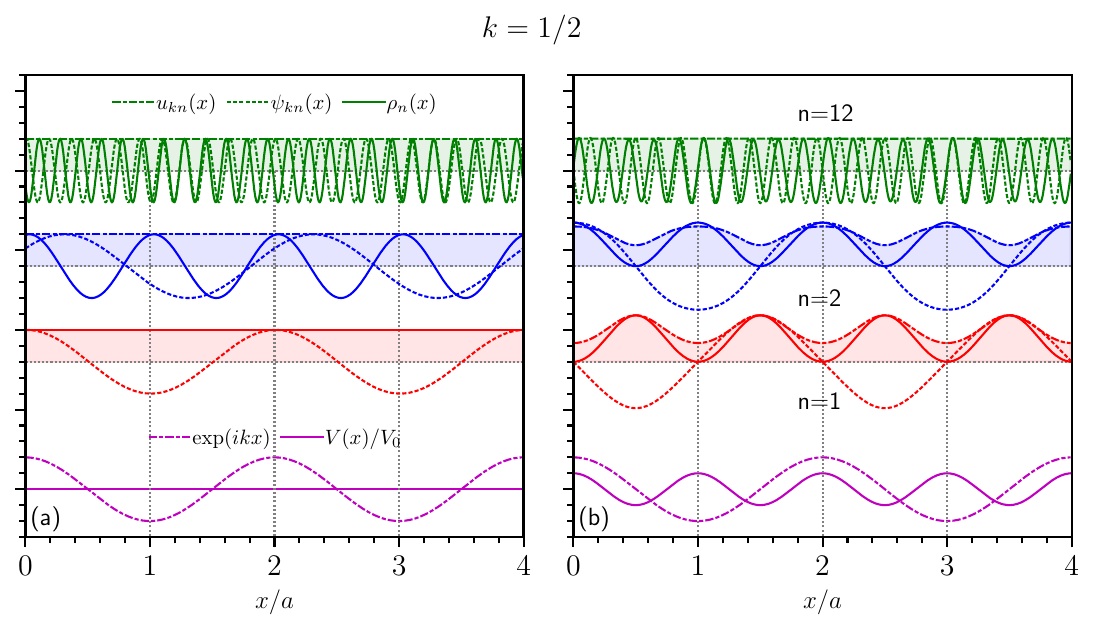}
    \caption{(a) wave functions of the free electrons potential and (b) the crystal potential in \cref{fig:1d_ebands}. The magenta lines at the bottom are the potential and the envelope functions $\exp(ikx)$ of the Bloch function $\psi_{kn}(x)\propto \exp(ikx)u_{kn}(x)$. In both panels, $k=1/2$. We plot each of $u_{kn}(x)$ (dash-dot), $\psi_{kn}(x)$ (dash), and $\rho_{n}(x)=N^{-1}\sum_k |u_{kn}(x)|^2$ (solid) lines above the potential. We plot $n=1$, $2$, and $12$.}
    \label{fig:1d_wave functions}
\end{figure}

\subsection{Band structure}\label{sec:band_structure}

Let us now discuss the band structure of solids. A simple example is sufficient to understand the effect of the crystal potential. As an example, specialize to a one-dimensional system with potential $V(x)=V_0 \cos(2\pi x/a)$. The potential is plotted in \cref{fig:1d_ebands}(a). This can be regarded as a toy model of an already converged DFT calculation with the Hartree, external, and XC potentials included in $V(x)$. We compare to free electrons in a periodic box of the same size as the primitive unit cell of crystal (the "empty lattice" potential). The band structure is plotted in \cref{fig:1d_ebands} (b). For the lowest band of the free electron potential, the (folded) band-structure is parabolic as expected. The crystal potential, on the other hand, has a "gap" at the zone boundary, $k\pm 1/2$. 

The gap is an effect of the crystal potential which can be understood in a simple way: the crystal potential breaks the continuous translational symmetry, so there must be loss of degeneracy, resulting in an anticrossing. We can use degenerate first order perturbation theory near the zone boundary to show this explicitly (see e.g. Chapter 9 in Ashcroft and Mermin \cite{ashcroftSolidStatePhysics2000}). The bands are flat at the zone boundary because $k=\pm 1/2$ are always degenerate and propagating in opposite directions: they form a standing wave, so their group velocity, $v_k = \partial \epsilon_{k} \partial k = 0$, vanishes.

The wave functions for the free electron and crystal potentials are plotted in \cref{fig:1d_wave functions}. For the empty lattice, the lowest laying state is constant, i.e. the electron is uniformly distributed as expected. Higher energy states have to be orthogonal to lower ones, so the wave functions have more nodes, but the density is still constant as we expect: there is no minimum in the potential, so the electron is equally likely to be everywhere. On the other hand, the crystal potential has a minimum at the middle of the unit cell ($x/a=1/2$). The lowest lying state has a maximum there. The $n=2$ state has to be orthogonal to the $n=1$ state, so has more nodes and its density is small near the minimum of the potential (and so on for the the other states). The lower lying states "screen" the potential, so the higher energy states look very much like free electrons (cf. the nearly flat density for $n=12$ in \cref{fig:1d_wave functions}(b)). This is consistent with the discussion of pseudopotentials earlier, c.f. \cref{sec:pseudo}. In all electron methods, valence states have to be orthogonal to all lower lying states, so are rapidly oscillating and require many basis-functions to represent accurately.

\subsection{Actual calculations: inputs and accuracy}\label{sec:pw_dft_accuracy}

We are finally ready to explicitly discuss DFT calculations of crystals. We focus on planewaves and implicitly assume NCPPs. The difference in using e.g. the PAW method is transparent to the user: simply providing a PAW dataset vs. a NCPP one is usually all that is needed. 

Actual DFT calculations accept many inputs from the user and many of them affect the accuracy of calculations. Most codes, e.g. \textsc{abinit}, allow the user to very precisely control many of the technical algorithms that run behind the scenes. These options aren't essential to understand in order to interpret the accuracy of a given calculation. Usually, the defaults are good and only two options that the user can (actually must) set make a significant difference. These options are (i) the energy cutoff and (ii) the $\bm{k}$-point grid, We discuss all of these topics in this section.  

We saw earlier that, for planewaves, the basis functions are labeled by the wave vector $\bm{G}$. For a smooth, shallow potential like NCPPs, the wave function is relatively smooth. Moreover, with pseudopotentials, we remove the core states so the valence states have zero or only a few nodes. Thus, we don't need to include fast-oscillating basis functions, i.e. those for which $|\bm{G}|$ is large. It is therefore natural to specify which basis functions to include by imposing a maximum for $G_{cut}\equiv |\bm{G}_{cut}|$, or, more conveniently, a cutoff energy $\epsilon_{cut}=\hbar^2 G_{cut}^2 /2m $. The maximum $G_{cut}$ is related to the energy cutoff by $G_{cut} = \sqrt{2m \epsilon_{cut}/\hbar^2 }$.  We are essentially defining a sphere with radius $\sqrt{\hbar^2 \epsilon_{cut}^2/2m}$ and we include all $\bm{G}$-points that lie within the sphere in our basis set. Since the accuracy of a variational calculation depends on the quality of the basis set, the energy cutoff $\epsilon_{cut}$ has the most direct affect on a calculation. Planewaves form a complete basis, so in principle, a calculation is exact in the limit of infinitely many planewaves. In practice, this is impossible and only a few hundred to a few thousand are needed. The typical way to check how many are needed is to pick a low value for $\epsilon_{cut}$ and calculate the groundstate energy of a material and then gradually increase $\epsilon_{cut}$ until the total energy is converged to within a desired value. For many planewaves, diagonalization is very expensive: iterative diagonalization methods are almost always used to make the problem tractable \cite{thijssen2007computational,5517}.

The other major factor for determining the accuracy of a DFT calculation of a crystal is the $\bm{k}$-point sampling. On one hand, we implicitly assume that a crystal is infinitely large. If there are $N$ unit cells in the crystal, there are $N$ $\bm{k}$-points (see \cref{sec:PBC}). In simulations of crystals, we use Bloch's theorem to re-write the TISE in $\bm{k}$-space and directly solve the equations on a set of $\bm{k}$-points. The simplest and most natural option is to use a uniform, discrete $\bm{k}$-point grid spanning the 1BZ. The number of $\bm{k}$-points along each reciprocal lattice vector determines the number of unit cells in the real-space crystal along that direction. To accurately model a realistic material, the $\bm{k}$-point grid should be as dense as possible so that the real-space crystal is as large as possible.  On the other hand, the density, which the KS equations self consistently depends on, is a sum over the $\bm{k}$-points. To accurately represent the density, we need to use sufficiently many $\bm{k}$-points. As already mentioned, a uniform grid is the most natural option. 

We note that there is a distinction between self consistent field (SCF) calculations and non-self consistent (NSCF) field calculations that is frequently made in the literature. An SCF calculation is first done on a uniform $\bm{k}$-point grid to converge the density and total energy. Then, an optional NSCF calculation can be done for an arbitrary $\bm{k}$-point set: e.g. a path thru the 1BZ to calculate band structure. In the NSCF calculation, the converged SCF density is read in from a file and is held fixed, i.e. there is no self consistency loop.

\section{Forces}\label{sec:dft_forces}

Finally, let us discuss one of the most useful results of DFT for this thesis: the Hellmann-Feynman forces. All along, we have been concerned with how to calculate the dynamics of the atoms and we've known that we need to calculate the forces on atoms. See e.g. \cref{eq:total_energy}. The KS scheme gives the electronic contribution to the forces on the ions as we now show. 

The derivative of the KS energy, which gives the electronic contribution to forces on the ions in the ground state, is 
\begin{align*}
    \bm{F}^{e}_i = -\frac{\partial E_{KS}}{\partial \bm{R}_i}.
\end{align*}
In Dirac notation,
\begin{equation}
\begin{gathered}
    -\frac{\partial}{\partial \bm{R}_i} E_{KS} = -\frac{\partial}{\partial \bm{R}_i} \langle \Psi | \hat{H}_{KS} | \Psi \rangle = -\langle \Psi | \frac{\partial \hat{H}_{KS} }{\partial \bm{R}_i}  | \Psi \rangle - E_{KS} \frac{ \partial \langle \Psi | \Psi \rangle } {\partial \bm{R}_i}  \\
    \equiv \bm{F}_{i,KS} + \bm{F}_{i,Pulay} .
\end{gathered}
\end{equation}
The first term is what we usually call the "force"; the second is called a "basis set correction" or "Pulay force". In a theorem due to Hellmann and Feynman, the Pulay force vanishes since, for a complete basis, However, in DFT calculations, we use a finite basis set and the Pulay force doesn't necessarily vanish. However, for certain finite basis sets, the Pulay term can be shown to vanish. This is the case for planewaves with NCPP's where the basis functions themselves are mutually orthogonal and don't depend explicitly on the coordinates. For atom-centered basis functions, i.e. where the basis-function depends explicitly on the coordinates like in the linear-combination of atomic-orbitals method (e.g. in the \texttt{cp2k} code), this term does \emph{not} vanish and it's contribution to the total forces is important. Note also that the PAW and LAPW methods have atom-centered contributions to the basis functions from the procetion operator (see \cref{sec:pseudo}) so the Pulay correction is non-vanishing for those methods too. In the planewave method with NCPP's, the Pulay correction is vanishing and the force is given by the first term only:
\begin{equation}
\begin{gathered}
    \bm{F}^{e}_i = \bm{F}_{i,KS} = -\langle \Psi | \frac{\partial \hat{H}_{KS} }{\partial \bm{R}_i}  | \Psi \rangle 
\end{gathered}
\end{equation}
Usually, the user of a code doesn't have to care that Pulay forces exists: the "force" written to the output file of a DFT calculation automatically includes the Pulay contribution. Still, it is worth knowing that this term exists: calculating the Pulay term adds computational cost that is large in some cases and might affect the choice of basis set (or pseudopotential method, e.g. NCPP vs PAW) for a particular calculation. 

Assuming that we are using planewaves and NCPP's, the only term in the KS total energy, \cref{eq:ks_total_e}, that depends explicitly on atomic positions is the external potential,
\begin{equation}
\begin{gathered}
    \bm{F}^{e}_i = -\frac{\partial E_{ext}[n(\bm{r})]}{\partial \bm{R}_i} = -\int d\bm{r} n(\bm{r}) \frac{v_{ext}(\bm{r})}{\partial \bm{R}_i}.
\end{gathered}
\end{equation}
The explicit form of the force depends on the explicit form of the pseudopotential which we don't specify here. Moreover, if the pseudopotential is non-local (\cref{sec:pseudo}), then only the first term is valid and the second term is replaced by something much more complicated. We don't focus on these details any further here.

\section{Summary}

To convince ourselves why all this was worth discussing and to get a feel for what we will use it all for later, let's make a few observations. The external potential in \cref{eq:ks_total_e} will almost always be the Coulomb field of atomic nuclei. Let's treat the nuclei as classical point-particles at zero temperature (so that the kinetic energy vanishes). The total energy of the crystal system 
\begin{equation}
    E_{tot}(\bm{R}) = E_{KS}(\bm{R})+\frac{k}{2}\sum_{I\neq J} \frac{e^2 Z_I Z_J}{|\bm{R}_I-\bm{R}_J|}
    \label{eq:total_energy_repeat}
\end{equation}
$E_{KS}(\bm{R})$ is the KS total energy in \cref{eq:ks_total_e}, here written explicitly as a function of the nuclear coordinates. The other term is the nuclear-nuclear Coulomb interaction. Minimizing the total energy with respect to $\bm{R}$ determines the equilibrium atomic structure of matter. Comparison to experimental data using x-ray or neutron scattering has shown that DFT is an extraordinarily accurate method. Derivatives of \cref{eq:total_energy_repeat} give us the forces on the nuclei. The derivative of $E_{KS}(\bm{R})$ is the electronic contribution to the forces on the ion (see \cref{sec:dft_forces}); the derivative of the ion-ion interaction is called the "Ewald" contribution to the force. It is straight forward to calculate with a trick called the Ewald summation, but the details are outside the scope of this thesis \cite{thijssen2007computational,allen2017computer,dove1993introduction,ziman1979principles,lee2009ewald}. 

We can use the forces to perform ab-initio molecular dynamics simulations and calculate many properties of matter. E.g. Fourier transforms of the atomic trajectories are related to the neutron scattering intensity. The details of this method are \cref{sec:classical_cross_section}. 

If the forces are taken around equilibrium, we can calculate second derivatives of the total energy, i.e. the force constants, that yield the phonon spectra. Comparing the phonon spectra to inelastic neutron scattering measurements has also confirmed the extraordinary success of DFT. Moreover, we can use the phonons to directly calculate the neutron scattering intensity in the harmonic approximation.

Beyond simple comparison to experiment, DFT also allows exploring new phases of matter. For instance, it is usually very difficult to study matter under extreme pressure in the laboratory. However, high pressure phases of matter can readily be studied in DFT by "relaxing" the atomic coordinates under external pressure. 

DFT also enables rapid materials discovery. Synthesizing new materials in the lab is usually an enormous effort; on the other hand, with DFT assessing the stability of a hypothetical new material is relatively easy and can even be done "high-throughput" to discover many materials rapidly (cf. the Materials Genome Initiative).

It is also important that \cref{eq:total_energy_repeat} doesn't depend on any assumptions about the structure of the matter we are studying. In normal condensed matter settings, DFT has been applied to study isolated atoms and molecules, gases and liquids, and ordered and disordered solids. Furthermore, we don't even have to assume anything about the interactions. They could be Coulomb interactions between charged particles, but could also be more exotic interactions e.g. the phonon-mediated attractive interaction between Cooper pairs. In fact there exists a density functional formulation of the superconducting state called "superconducting density functional theory" \cite{oliveira1988density,requist2019exact}.


\chapter{The Dynamics of the Atoms}\label{chp:phonons}

Many important properties of energy materials are directly related to the atomic structure: e.g. the electronic band structure and the optical properties depend on the equilibrium geometry of the lattice. Moreover, many properties depend on the \emph{dynamics} of the lattice (i.e. "lattice dynamics") too: e.g. resistivity, superconductivity, photoabsorption, etc. So, it's clear that we need to understand the dynamics of atoms in matter. As mentioned in \cref{chp:neutrons}, we can directly measure these dynamics with neutrons, but to make progress understanding the results of neutron scattering, we need a model for the dynamics. We learn how to model the dynamics of atoms in this chapter. 

This chapter loosely follows Dove \cite{dove1993introduction}, Ziman \cite{ziman1972principles}, and Ashcroft and Mermin \cite{ashcroftSolidStatePhysics2000}.

\section{Born-Oppenheimer forces}

We start off by assuming that the nuclei are classical; we write down the equations of motion and look for solutions. The most general expression (in the Born-Oppenheimer approximation) for the forces on the atoms can be derived from \cref{eq:total_energy}:
\begin{equation}
    U(\bm{r}) = E_{BO}(\bm{r})+\frac{k}{2}\sum_{i\neq j} \frac{e^2 Z_i Z_j}{|\bm{r}_i-\bm{r}_j|} .
    \label{eq:bo_crystal_energy}
\end{equation}
In this section, $\bm{r}_i$ is the position of the $i^{th}$ atom, etc. We won't need to write down the electron coordinates again for a while. The force on an atom is $\bm{f}_i=-\partial U/\partial \bm{r}_i$. The potential $E_{BO}(\bm{r})$ is the Born-Oppenheimer electronic energy "surface". It can be calculated in the Hartee-Fock approximation, in density functional theory (DFT) from the Kohn-Sham equations (\cref{eq:ks_total_e}), or other ways. The other term is just the mutual Coulomb repulsion between the nuclei. The electrons \emph{screen} the nuclei so that their interaction isn't purely electrostatic: the screening is contained in $E_{BO}(\bm{r})$ (though the screening is approximate since we assume the electrons respond instantaneously in the adiabatic approximation). In principle, this equation contains everything we need to calculate the classical trajectories of the nuclei in the Born-Oppenheimer approximation. 

Evaluating the Coulomb energy isn't a huge challenge: there are methods, e.g. Ewald summations, to handle this efficiently \cite{thijssen2007computational,allen2017computer,dove1993introduction,ziman1979principles,lee2009ewald}. The really difficult part is calculating $\partial E_{BO} / \partial \bm{r}_i$ since calculating $E_{BO}$ itself can be very expensive; e.g. calculating it from first principles implies solving the electronic structure problem. We saw in \cref{chp:electrons} that this is a non-trivial endeavour. For now, just assume that we already know how to calculate all contributions to $\partial U/\partial \bm{r}_i$ and just focus on solving the resulting equations of motion... but nobody knows how to solve Newton's equations analytically for more than two particles. What now? We have to resort to either approximations or numerical methods (or both): in this thesis, we employ the \emph{harmonic approximation} and \emph{molecular dynamics}.

\section{The harmonic approximation}

In the rest of this section, we focus on the harmonic approximation. We assume the structure of interest is crystalline. At low enough (but finite) temperature, the atoms only move a little bit and wiggle around their average positions. Then it is reasonable to describe the coordinates of the atoms as small displacements around the equilibrium geometry. Define $\bm{r}_{i\alpha}(t)=\bm{R}_i+\bm{\tau}_\alpha+\bm{u}_{i\alpha}(t)$ as the position of the $\alpha^{th}$ atom in the $i^{th}$ unit cell (cf. \cref{fig:lattice}). $\bm{R}_i$ is the origin of the $i^{th}$ unit cell and $\bm{\tau}_\alpha$ is the basis position of the $\alpha^{th}$ atom. $\bm{u}_{i\alpha}(t)$ is the instantaneous displacement which we assume is "small". Occasionally, we will call $\bm{r}^{(0)}_{i\alpha}=\bm{R}_i + \bm{\tau}_\alpha$. For now, we let the atoms be classical; then $\bm{u}_{i\alpha}$ is a classical coordinate. Later, we will consider the case of quantum particles: then $\hat{\bm{u}}_{i\alpha}$ is an operator. We will see that the quantum case is a generalization of the classical approximation with the same energies.

The potential energy of the system of atoms is $U(\bm{r})$. For many interesting systems, e.g. the Coulomb interaction, the potential energy is "pair-wise", i.e. only depends on interactions between pairs of atoms \footnote{In empirical methods, three-body interactions (i.e. the potential between atoms $i$ and $j$ depends on the coordinate of $k$) are also commonly used since they offer more freedom for fitting real interactions. This is the case in e.g. the Tersoff potential, but this won't matter in this thesis.}. This is the case in \cref{eq:bo_crystal_energy}. We specialize to pair-wise interactions from now on. Let the pair potential be $\phi_{\alpha\beta}(\bm{r}_{i\alpha}-\bm{r}_{j\beta})$. This could represent each term in \cref{eq:bo_crystal_energy} or could be an empirical approximation. No matter, the potential energy in general is written
\begin{equation}
\begin{gathered}
    U(\bm{r}(t)) = \frac{1}{2} \sum_{ij \alpha\beta} \phi_{\alpha\beta}(\bm{r}_{i\alpha}(t)-\bm{r}_{j\beta}(t)) = \frac{1}{2} \sum_{ij \alpha\beta} \phi_{\alpha\beta}\left([\bm{r}^{(0)}_{i\alpha}-\bm{r}^{(0)}_{j\beta}]+[\bm{u}_{i\alpha}(t)-\bm{u}_{j\beta}(t)] \right).
\end{gathered}
\end{equation}
The factor of $1/2$ compensates for "double counting" the pair interactions: once for $i\alpha \rightarrow j\beta$ and once for $j\beta \rightarrow i\alpha$. This is a reasonable model that accounts for most interactions in nature; however, for most $\phi_{\alpha\beta}(\bm{r}_{i\alpha}-\bm{r}_{j\beta})$, the equations of motion are still intractable. We make progress via \emph{the harmonic approximation}.

The harmonic approximation, which we now describe, depends on two assumptions: (i) displacements of the atoms, $\bm{u}_{i\alpha}$, from the \emph{average} positions, $\bm{R}_i+\bm{\tau}_\alpha$, are small and (ii) the \emph{average} positions are the equilibrium geometry. Then we Taylor expand the potential energy, $U(\bm{r})$, in the displacements and drop high order terms. We have 
\begin{equation}
\begin{gathered}
    U(\bm{r}) = U_0 + \sum_{i\alpha} \bm{u}^T_{i\alpha} \cdot \left. \left( \frac{\partial U(\bm{r})} {\partial \bm{u}_{i\alpha}} \right) \right|_{\bm{u}=0} + \frac{1}{2} \sum_{ij\alpha\beta} \bm{u}^T_{i\alpha} \cdot \left. \left( \frac{\partial^2 U(\bm{r})} {\partial \bm{u}_{i\alpha} \partial \bm{u}_{j\beta}} \right) \right|_{\bm{u}=0} \cdot \bm{u}_{j\beta}  + \cdots 
\end{gathered}
\end{equation}
The $0^{th}$ order term $U_0$ is just the electrostatic potential energy in the crystal's equilibrium geometry. It is a constant independent of the displacements so it won't affect the dynamics at all. We can ignore it in what follows. The next term contains $-\bm{f}^{(0)}_{i\alpha} = \partial U / \partial \bm{u}_{i\alpha} |_{\bm{u}=0}$ which is just the force on the $\alpha^{th}$ atom in the $i^{th}$ unit cell in equilibrium... by definition, the forces vanish in equilibrium! In the same sense, we know the energy is minimized in equilibrium: the condition for this to be true is that all first derivatives vanish. The lowest-order, non-trivial term is the $2^{nd}$ order one and we stop expanding there. Then 
\begin{equation}
\begin{gathered}
    U(\bm{r}) \approx U_{H}(\bm{r}) \equiv \frac{1}{2} \sum_{ij\alpha\beta} \bm{u}^T_{i\alpha} \hat{\Phi}_{ij,\alpha\beta} \bm{u}_{j\beta} \quad \textrm{with} \quad
    \hat{\Phi}_{ij,\alpha\beta} = \left. \left( \frac{\partial^2 U(\bm{r})} {\partial \bm{u}_{i\alpha} \partial \bm{u}_{j\beta}} \right) \right|_{\bm{u}=0}
\end{gathered}
\end{equation}
where $\hat{\Phi}_{ij,\alpha\beta}$ is called the "force constant matrix". For any general term in the series we have (in 3D)
\begin{equation}
\begin{gathered}
    \bm{u}^T_{i\alpha} \hat{\Phi}_{ij,\alpha\beta} \bm{u}_{j\beta} \equiv  \bm{u}^T_{i\alpha} \cdot \hat{\Phi}_{ij,\alpha\beta} \cdot \bm{u}_{j\beta} 
    = \sum_{\mu\nu} u^\mu_{i\alpha} \Phi_{ij,\alpha\beta}^{\mu\nu} u^\nu_{j\beta} = \\
    \begin{pmatrix}
        u^x_{i\alpha} & u^y_{i\alpha} & u^z_{i\alpha} 
    \end{pmatrix} 
    \begin{pmatrix}
        \dfrac{\partial^2 U}{\partial u^x_{i\alpha} \partial u^x_{j\beta}} &  \dfrac{\partial^2 U}{\partial u^x_{i\alpha} \partial u^y_{j\beta}} & 
        \dfrac{\partial^2 U}{\partial u^x_{i\alpha} \partial u^z_{j\beta}} \\
        \dfrac{\partial^2 U}{\partial u^y_{i\alpha} \partial u^x_{j\beta}} &  \dfrac{\partial^2 U}{\partial u^y_{i\alpha} \partial u^y_{j\beta}} & 
        \dfrac{\partial^2 U}{\partial u^y_{i\alpha} \partial u^z_{j\beta}} \\
        \dfrac{\partial^2 U}{\partial u^z_{i\alpha} \partial u^x_{j\beta}} &  \dfrac{\partial^2 U}{\partial u^z_{i\alpha} \partial u^y_{j\beta}} & 
        \dfrac{\partial^2 U}{\partial u^z_{i\alpha} \partial u^z_{j\beta}}
    \end{pmatrix} 
    \begin{pmatrix}
        u^x_{j\beta} \\ u^y_{j\beta} \\ u^z_{j\beta} 
    \end{pmatrix}
\end{gathered}
\end{equation}
which hopefully makes clear the meaning of calling the force constant matrix a "matrix". 

In the harmonic approximation, the force on atom $i\alpha$ is 
\begin{equation}
\begin{gathered}
    \bm{f}_{i\alpha} = - \frac{\partial U}{\partial \bm{u}_{i\alpha}} =  - \sum_{j\beta} \hat{\Phi}_{ij,\alpha\beta} \bm{u}_{j\beta} ,
\end{gathered}
\end{equation}
from which we can write down the equations of motion:
\begin{equation}
\begin{gathered}
    m_{\alpha} \ddot{\bm{u}}_{i\alpha} = - \sum_{j\beta} \hat{\Phi}_{ij,\alpha\beta} \bm{u}_{j\beta} .
    \label{eq:harmonic_equation_of_motion}
\end{gathered}
\end{equation}
The equations of motion are translationally invariant: shifting the unit cell labels by an integer $k$ results in exactly the same set of equations (due to PBC) with new labels. Then, we know from \cref{sec:blochs_thm}, that the solutions will be Bloch functions. The nuance here is that, rather than a continuous field $\psi_{\bm{k}}(\bm{r})$, we will have a discrete vector function: $\bm{u}_{\bm{q}}(\bm{r}_{i\alpha}) \propto \bm{\epsilon}_{\bm{q}\alpha}\exp(i\bm{q}\cdot\bm{r}_{i\alpha})$, which is only defined for each Cartesian direction at each lattice point $\bm{r}_{i\alpha}$. $\bm{\epsilon}_{\bm{q}\alpha}$ is the unit cell-periodic part: it depends on wave vector $\bm{q}$ and has a value for every atom in the unit cell, $\alpha$, but is the same in every unit cell. $\bm{\epsilon}_{\bm{q}\alpha}$ is a Cartesian vector. We will encounter this expression again later where we describe it in more detail.

It turns out that the we can solve the equations of motion in the harmonic approximation, \cref{eq:harmonic_equation_of_motion}, exactly. Before we do that in general, let us look at some simple examples that familiarize us with the basic physics.

\subsection{The linear harmonic chain}\label{sec:monatomic_chain}

\begin{figure}
    \centering
    \includegraphics[width=0.75\linewidth]{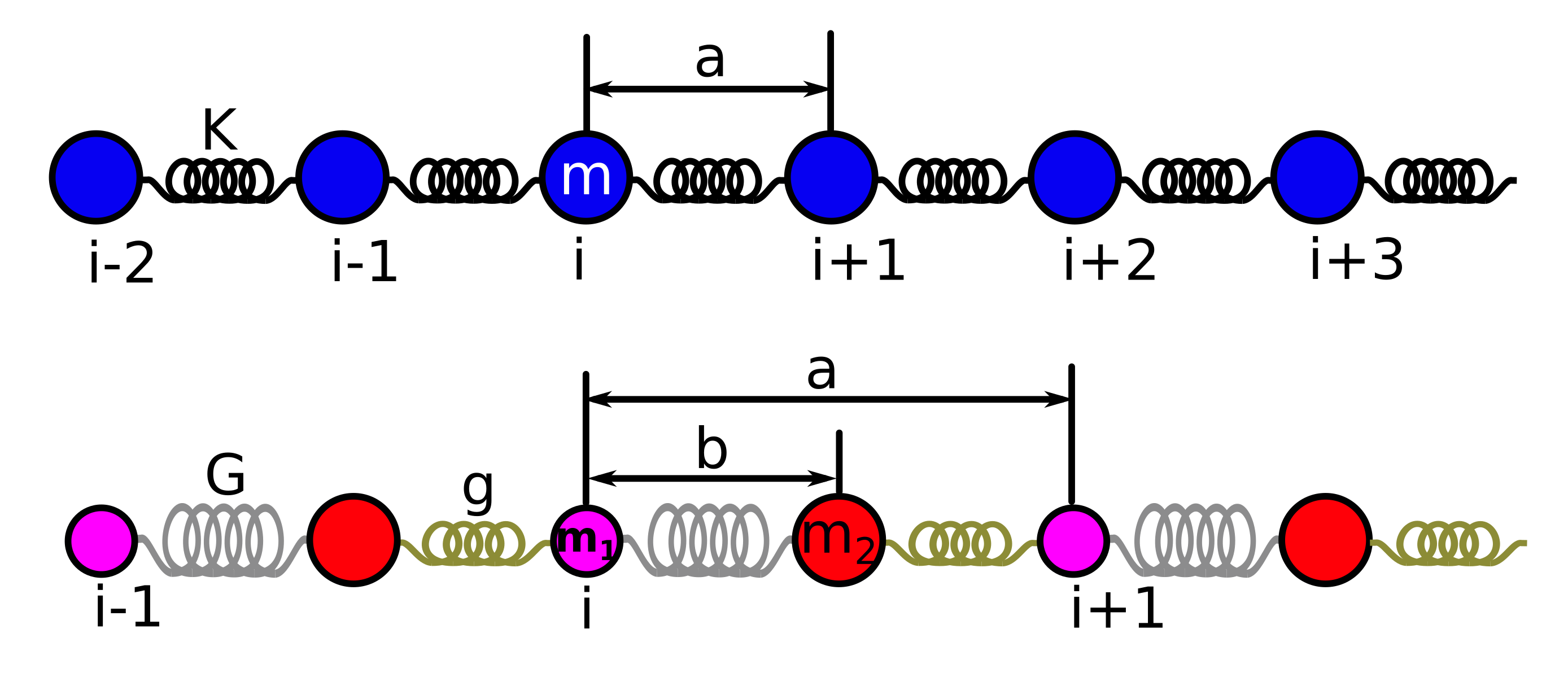}
    \caption{Diagram of a $1d$ harmonic Bravais chain (top) and a $1d$ harmonic diatomic chain (bottom). $a$ is the unit cell length and $b$ is the position of the second basis atom in the unit cell. In the Bravais chain, $m$ is the atomic mass and $K$ the spring constant. In the diatomic chain, $m_1$ and $m_2$ are the masses and $G$ and $g$ the spring constants.}
    \label{fig:harmonic_chain}
\end{figure}

\begin{figure}
    \centering
    \includegraphics[width=0.75\linewidth]{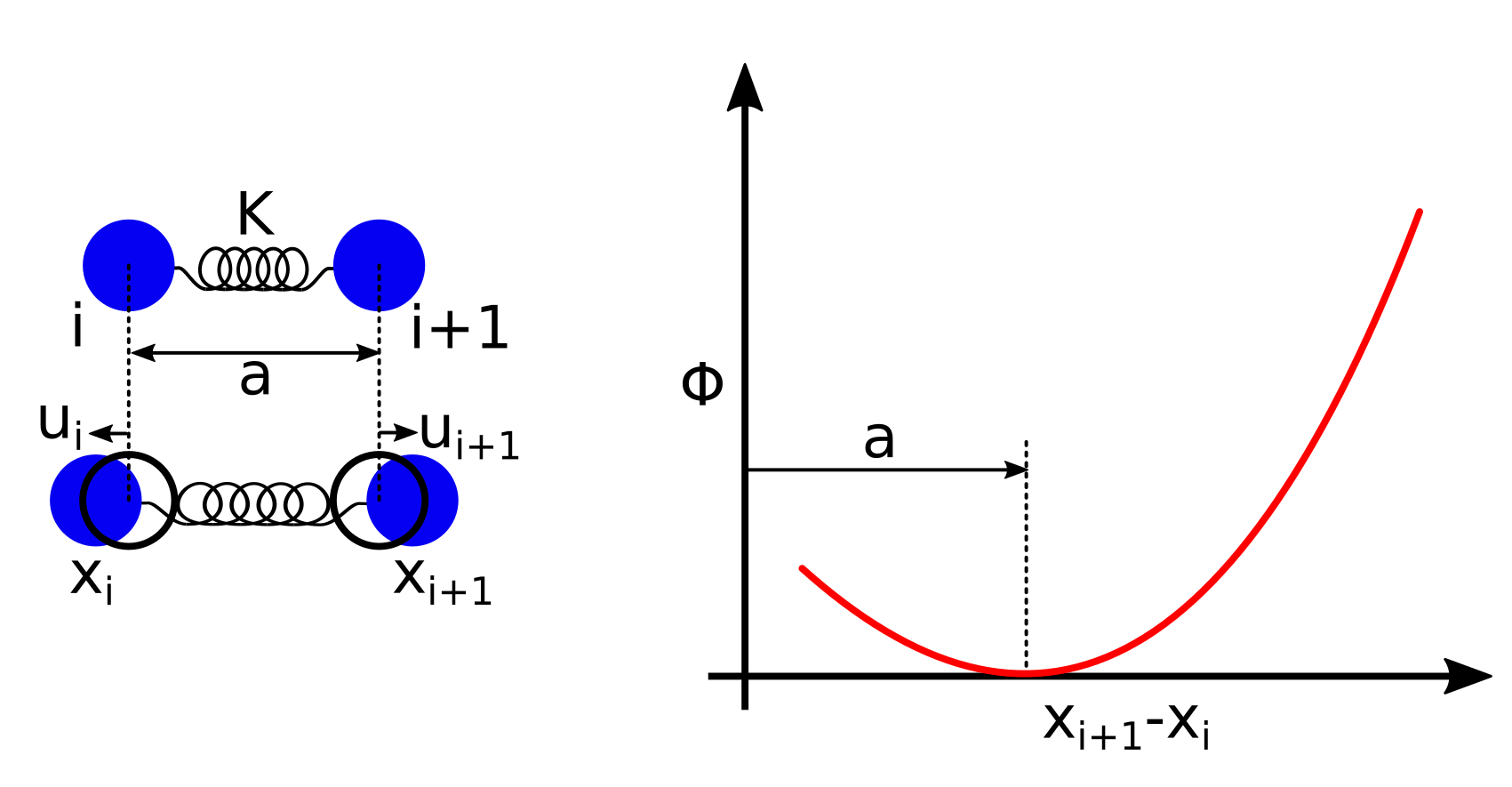}
    \caption{The bond-spring pair potential of the $1d$ harmonic chain in \cref{fig:harmonic_chain}: $\Phi = K(x_{i+1}-x_i-a)^2/2 = K(u_{i+1}-u_i)^2/2$. The potential is quadratic with minimum at $x_{i+1}-x_i = a$.}
    \label{fig:harmonic_potential}
\end{figure}

First, let us consider a linear, one-dimensional harmonic chain (\cref{fig:harmonic_chain}, top row). The lattice sites are labeled by numbers $i$, $i+1$, etc. For the pair potential, we assume a harmonic bond-spring potential: i.e. a potential quadratic in the distance between atoms, with an equilibrium bond length $a\neq 0$. See \cref{fig:harmonic_potential}. If we compress/stretch the bond between the two atoms, there is a linear restoring force. Specifically, for the potential energy between atom $i$ and atom $i+1$, we define $\phi(x_{i+1}-x_{i}) = K[x_{i+1} - x_{i}-a]^2/2$ with $a$ the lattice constant, $K$ the "spring constant", and $x_i = ai + u_i$ with position of atom $i$ with $u_i$ it's displacement. The mass of each atom is $m$. The total potential energy of the system is 
\begin{equation}
\begin{gathered}
    U = \frac{K}{2} \sum_i \left[ x_{i+1} - x_i - a\right]^2 = \frac{K}{2} \sum_i \left[ (ai+a+u_{i+1}) - (ai+u_i) - a\right]^2 \\
    = \frac{K}{2} \sum_i \left[ u_{i+1} - u_{i}\right]^2 .
\end{gathered}
\end{equation}
The only non-zero force constants are $\Phi_{i,i}$, $\Phi_{i,i-1}$, $\Phi_{i,i-1}$. They are 
\begin{equation}
\begin{gathered}
    \hat{\Phi}_{i,i} = 2K \qquad \hat{\Phi}_{i,i+1} = \frac{\partial^2 U}{\partial u_i \partial u_{i+1}} = -K \qquad \hat{\Phi}_{i,i-1} = \frac{\partial^2 U}{\partial u_i \partial u_{i-1}} = K 
\end{gathered}
\end{equation}
and the force on atom $i$ is 
\begin{equation}
\begin{gathered}
    f_i = -\Phi_{i,i}u_i -\Phi_{i,i+1}u_{i+1} -\Phi_{i,i-1}u_{i-1} = -K [2u_i - u_{i+1}-u_{i-1}] .
    \label{eq:1d_harmonic_force}
\end{gathered}
\end{equation}
We note that the reason the linear harmonic chain is special is that it is equivalent to the harmonic approximation, i.e. it is exactly solvable. If we calculate the force explicitly, it is
\begin{equation}
\begin{gathered}
    f_i = -\frac{K}{2}\frac{\partial}{\partial u_i} \left[ (u_i-u_{i+1})^2 + (u_{i-1}-u_i)^2   \right] = -K \left[ (u_i-u_{i+1}) - (u_{i-1}-u_i) \right] \\
    = - K [ 2u_i - u_{i+1} - u_{i-1}],
\end{gathered}
\end{equation}
which is identical to \cref{eq:1d_harmonic_force} above.

\begin{figure}
    \centering
    \includegraphics[width=0.75\linewidth]{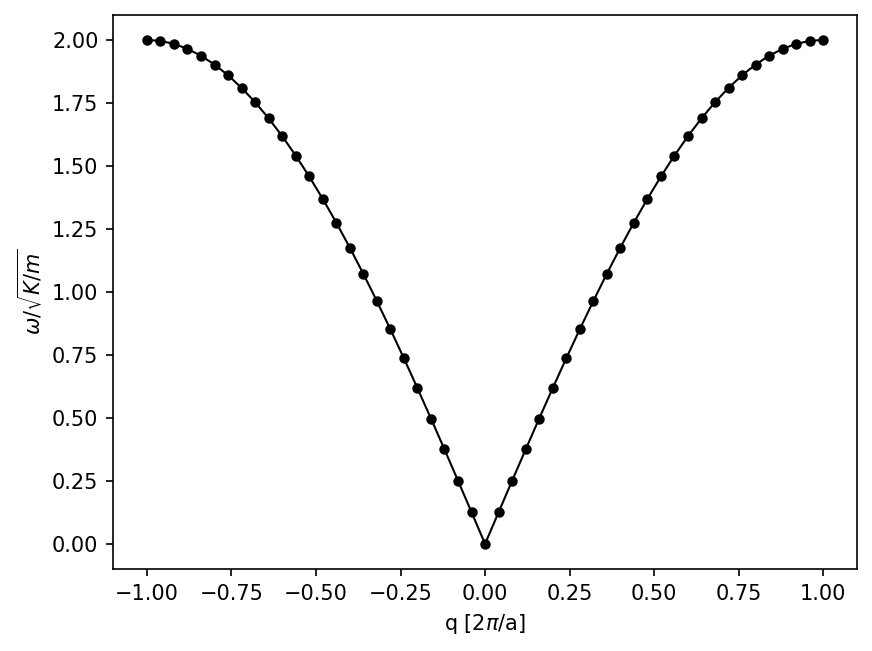}
    \caption{Dispersion, $\omega(q)$, of the $1D$ monatomic harmonic chain. There is only one branch: the acoustic branch.}
    \label{fig:1d_chain_dispersion}
\end{figure}

The equations of motion are $m a_i = F_i$ with $a_i$ the acceleration of the atom at site $i$.
\begin{equation}
\begin{gathered}
    m a_i = m \ddot{u}_i =  - K [ 2u_i - u_{i+1} - u_{i-1}].
    \label{eq:newtons_equation}
\end{gathered}
\end{equation}

The goal is to solve this equation for all $u_i(t)$. If we can do this, we know the trajectories of the atoms in the system and can, in principle, calculate their velocities, energies, etc. and all other physical observables of the system. So the goal is to solve these equations. However, the derivatives on the left-hand side make this hard to work with. Moreover, atoms at different sites are coupled which makes the problem harder. We can overcome both issues by Fourier transforming in both space and time. We assume that the linear chain forms a Born-Von Karman (BvK) ring with $N$ sites in it. The Fourier transform and its inverse are
\begin{equation}
\begin{gathered}
    u_q(\omega) = \frac{1}{\sqrt{2\pi N}} \sum_n \int dt u_n(t) \exp(-iqna+i\omega t) \\ 
    u_n(t) = \frac{1}{\sqrt{2\pi N}} \sum_q \int d\omega u_q(\omega) \exp(iqna-i\omega t).
\end{gathered}
\end{equation}
We want to transform Newton's equations to Fourier space where it will be easier to solve for $u_q(\omega)$. We can then invert the Fourier transform to determine $u_n(t)$, solving the equations of motion. Newton's \cref{eq:newtons_equation} become
\begin{equation}
\begin{gathered}
    m \omega^2 u_q(\omega) =  K [ 2 - \exp(iqa) - \exp(-iqa)] u_q(\omega) = 2K (1-\cos(qa)) u_q(\omega) \\ 
    = 4K \sin^2(qa/2) u_q(\omega) 
    \label{eq:dynamical_equations}
\end{gathered}
\end{equation}
Different wave vectors $q$ are decoupled and their equations of motion are identical so we have made a lot of progress. Note that for every $q$, there is a different $\omega\equiv \omega(q)$ that solves it; we call these solutions "phonons". Each one is a wave like modulation of the crystal that oscillates at a specific frequency, $\omega(q)$. What remains is to determine the wave vector dependence of $ \omega(q)$: we call this the \emph{dispersion}.
\begin{equation}
\begin{gathered}
    \omega(q) = 2\sqrt{\frac{K}{m}} \left| \sin \left( \frac{qa}{2} \right) \right|
    \label{eq:dispersion}
\end{gathered}
\end{equation}
We plot the dispersion in \cref{fig:1d_chain_dispersion}. The dispersion tells us the frequency of wave like oscillations in the crystal. Note that in the long-wavelength limit $qa \rightarrow 0$, $\omega(q) = |q| a\sqrt{K/m} $; i.e. the dispersion is linear with speed $\partial \omega / \partial q \equiv c=a\sqrt{K/m}$. In this limit, the length over which the atomic displacements vary is much larger than the lattice spacing and, from the perspective of the phonon, the lattice is continuous. In analogy to the theory of elasticity, this is called an \emph{acoustic} wave. In the opposite limit $ qa \rightarrow \pm \pi$, $\omega(q)\rightarrow 2\sqrt{K/M}$ with speed $\partial \omega/\partial q = 0$. 
\begin{figure}
    \centering
    \includegraphics[width=0.9\linewidth]{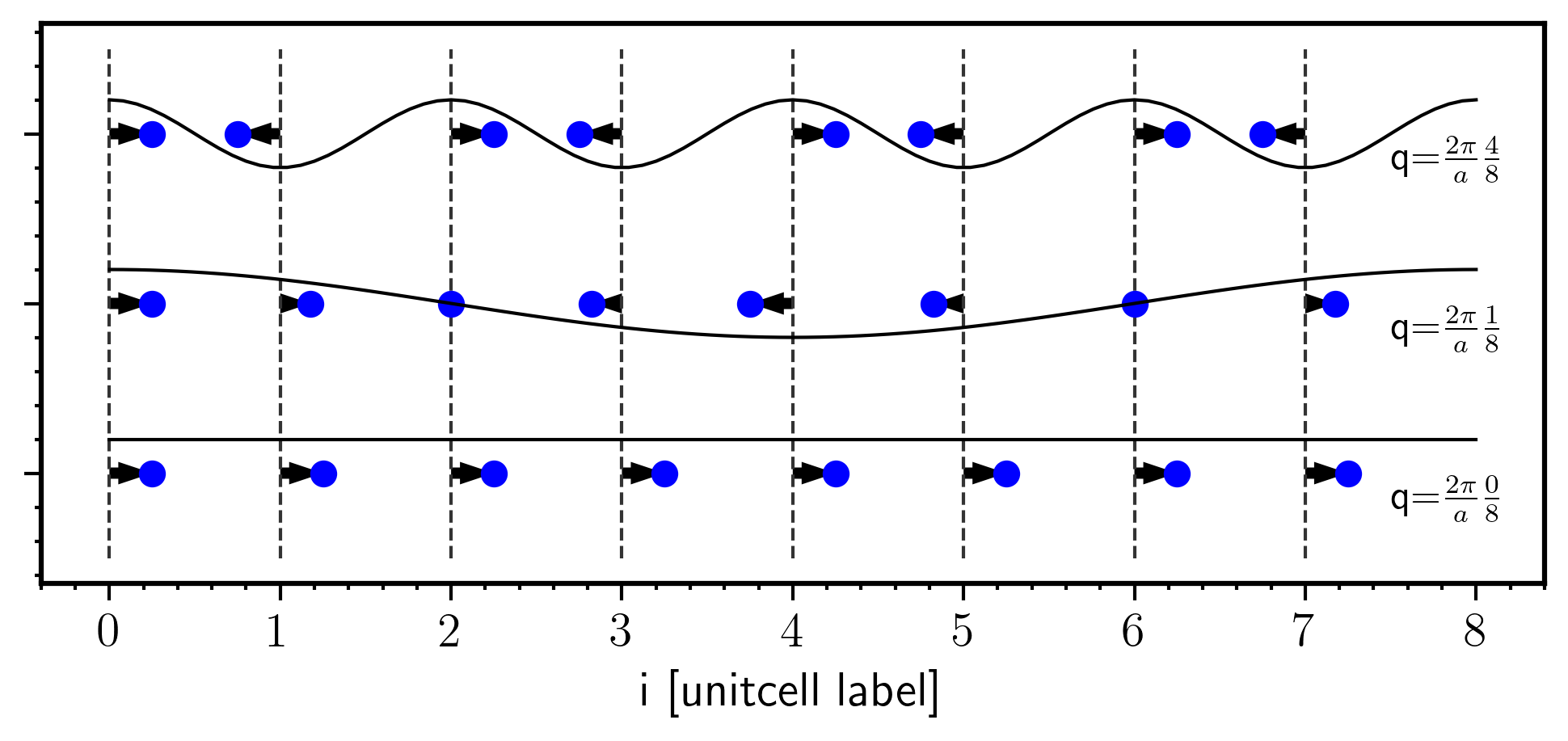}
    \caption{Phonon displacements of the $1D$ harmonic chain, $u^{(q)}_{n} = \Re[u_q\exp(iqna)]$ with $u_q = 0.25 a$. The chain has $N=8$ unit cells. The arrows label the displacements; the wave like modulation is $\Re[\exp(iqna)]=\cos(qna)$. We plot $q=0$, $q=2\pi/8a=1/8$ [rlu], and $q=\pi/a=1/2$ [rlu]. The atoms oscillate around their equilibrium positions $x_{n}=na$ with frequency $\omega(q)$.}
    \label{fig:1d_chain_disp}
\end{figure}
We have shown that the solutions are 
\begin{equation}
\begin{gathered}
    u_n(t) = \frac{1}{\sqrt{2\pi N}} \sum_q \int d\omega u_q(\omega) \exp(iqna-i\omega t)
\end{gathered}
\end{equation}
with $u^{(q)}_{n} = N^{-1/2} u_q \exp(iqna)$ the phonon Bloch functions: $u_q$ is the unit cell periodic part and $\exp(iqna)$ the wave like modulation. With only one solution per $q$, $u_q$ is arbitrary: its amplitude and phase will be fixed by boundary conditions. In \cref{fig:1d_chain_disp} we plot the phonon displacements $u^{(q)}_{n} = \Re [u_q \exp(iqna)]$ with $u_q = 0.25 a$. Note that, near $q=0$, the wave-length is long. At $q=0$, the whole crystal is rigidly displaced and there is no restoring force as discussed above: the frequency $\omega(q=0)=0$ vanishes. On the other hand, at the zone boundary, the displacement in each unit cell oscillates between $u_{n} = (-1)^{n} u_q$: the springs are maximally compressed and the restoring force on each atom is large. The zone boundary mode is the highest frequency $\omega(q)$ in the $1D$ monatomic harmonic chain.

\subsection{The linear diatomic chain}\label{sec:diatomic_chain}

If the unit cell atoms with different masses, different bond lengths, etc., then our lattice has a basis and the equations become more complicated. The bottom row in \cref{fig:harmonic_chain} shows a diatomic chain of different masses $m_1$ and $m_2$ coupled by different springs with spring constants $G$ and $g$. The potential energy is
\begin{equation}
\begin{gathered}
    U = \frac{1}{2} \sum_i \left[ G(u_{i1}-u_{i2})^2 + g (u_{i2}-u_{i+1,1})^2 \right].
\end{gathered}
\end{equation}
The force on atom $1$ in unit cell $j$ is 
\begin{equation}
\begin{gathered}
    F_{j,1} = -\frac{1}{2}\frac{\partial}{\partial u_{j,1}} \left[ G(u_{j1}-u_{j2})^2 + g(u_{j-1,2}-u_{j1})^2   \right] = - \left[ G(u_{j1}-u_{j2}) - g(u_{j-1,2}-u_{j1}) \right] \\
    = - [(G+g)u_{j1} -G u_{j2} - g u_{j-1,2} ] 
\end{gathered}
\end{equation}
and for atom $2$ in cell $j$
\begin{equation}
\begin{gathered}
    F_{j2} = -\frac{1}{2}\frac{\partial}{\partial u_{j2}} \left[ G(u_{j1}-u_{j2})^2 + g(u_{j2}-u_{j+1,1})^2   \right] = - \left[ -G(u_{j1}-u_{j2}) + g(u_{j2}-u_{j+1,1}) \right] \\
    = - [(G+g)u_{j2} -G u_{j1} - g u_{j+1,1} ] .
\end{gathered}
\end{equation}
The equations of motion are 
\begin{equation}
\begin{gathered}
    m_1 \ddot{u}_{i1} = - [(G+g)u_{j1} -G u_{j2} - g u_{j-1,2} ] \\
    m_2 \ddot{u}_{i2} = - [(G+g)u_{j2} -G u_{j1} - g u_{j+1,1} ]
    \label{eq:newtons_equations_diatomic}
\end{gathered}
\end{equation}
which in Fourier space become
\begin{equation}
\begin{gathered}
    m_1 \omega^2 u_{q1} = [(G+g)u_{q1} -G u_{q2} - g u_{q2} \exp(-iqa) ] \\
    m_2 \omega^2 u_{q2} = [(G+g)u_{q2} -G u_{q1} - g u_{q1} \exp(iqa) ]
    \label{eq:diatomic_chain_eom}
\end{gathered}
\end{equation}
We've decoupled the equations for neighboring unit cells into independent equations, one for each $q$. Still, atoms within the unit cell are coupled. The solutions of \cref{eq:diatomic_chain_eom} are not independent and will involve relative displacements of each atom in the unit cell. There are two equations above, so there will be two different sets of $u_{q1}$ and $u_{q2}$, one set that that solves each equation. The goal now is to find the sets of relative displacements that satisfy these equations. To start, we write the coupled set of equations in matrix form:
\begin{equation}
\begin{gathered}
\omega_q^2
\begin{pmatrix}
    ~~ m_1 ~~ & ~~ 0 ~~ \\
    ~~ 0 ~~ & ~~ m_2 ~~
\end{pmatrix} 
\begin{pmatrix}
    u_{q1} \\ u_{q2}  
\end{pmatrix} = 
\begin{pmatrix}
    (G+g) & -(G+g\exp(-iqa)) \\ -(G+g\exp(iqa)) & (G+g)
\end{pmatrix} 
\begin{pmatrix}
    u_{q1} \\ u_{q2}  
\end{pmatrix} 
    \label{eq:diatomic_chain_eom_matrix}
\end{gathered}
\end{equation}
Call the matrix on the right hand side $\hat{\Phi}_q$. Now define $u_{qi}(\omega)=\epsilon_{qi}(\omega)/\sqrt{m_i}$ with $\epsilon_{qi}(\omega)$ the (generally complex) frequency dependent amplitude of displacement of the $i^{th}$ atom in the unit cell. The appearance of $\sqrt{m_i}$ in the denominator will become clear soon. Plug this new equation into \cref{eq:diatomic_chain_eom_matrix} above:
\begin{equation}
\begin{gathered}
\omega_q^2
\begin{pmatrix}
    ~~ m^{1/2}_1 ~~ & ~~ 0 ~~ \\
    ~~ 0 ~~ & ~~ m^{1/2}_2 ~~
\end{pmatrix} 
\begin{pmatrix}
    \epsilon_{q1} \\ \epsilon_{q2}  
\end{pmatrix} = \hat{\Phi}_q
\begin{pmatrix}
    ~~ m^{-1/2}_1 ~~ & ~~ 0 ~~ \\
    ~~ 0 ~~ & ~~ m^{-1/2}_2 ~~
\end{pmatrix} 
\begin{pmatrix}
    \epsilon_{q1} \\ \epsilon_{q2}  
\end{pmatrix} 
\end{gathered}
\end{equation}
This is like an "eigenvalue problem" that we solve by diagonalization but differs by the matrix on the left hand side: we call that matrix $\hat{M}$. Also define the vector $\bm{\epsilon}_{q}=(\epsilon_{q1},\epsilon_{q2})^T$. Note that this is a generalized eigenvalue problem like that encountered in \cref{sec:variational_method}. We solve it by first inverting $\hat{M}$, multiplying the equation by $\hat{M}^{-1}$, and then diagonalizing the resulting eigenvalue problem. Since $\hat{M}$ is already diagonal , it's inverse is easy:
\begin{equation}
\begin{gathered}
\hat{M}^{-1} =
\begin{pmatrix}
    ~~ m^{-1/2}_1 ~~ & ~~ 0 ~~ \\
    ~~ 0 ~~ & ~~ m^{-1/2}_2 ~~
\end{pmatrix} 
\end{gathered}
\end{equation}
which we note is already present on the right hand side. Then 
\begin{equation}
\begin{gathered}
\hat{M}^{-1} \hat{\Phi}_q \hat{M}^{-1} =
\begin{pmatrix}
    \dfrac{G+g}{m_1} & -\dfrac{G+g\exp(-iqa)}{\sqrt{m_1 m_2}} \\ -\dfrac{G+g\exp(iqa)}{\sqrt{m_1 m_2}} & \dfrac{G+g}{m_2}
\end{pmatrix} \equiv \hat{D}_q
\label{eq:diatomic_chain_dynamical_matrix}
\end{gathered}
\end{equation}
The matrix $\hat{D}_q$ is called the "dynamical matrix". It has elements $D^{ij}_q=\Phi^{ij}_q/\sqrt{m_i m_j}$ where $\Phi^{ij}_q$ is defined by the matrix elements above. Note that $D^{21}_q = \bar{D}^{12}_q$ where the over-bar ($\bar{.}$) represents complex conjugation. It's clear that $\hat{D}^{ij}_{q}$ is Hermitian. It can be shown that the eigenvalues of a Hermitian matrix are real and that the eigenvectors are orthonormal\footnote{Since any scale factor for each row in \cref{eq:diatomic_chain_eom} will cancel, we are free to arbitrarily normalize the eigenvectors of $\hat{D}_{q}$. Doing so is convenient later.}. Then 
\begin{equation}
\begin{gathered}
    \hat{D}_q \bm{\epsilon}_q = \omega^2_q \bm{\epsilon}_q 
    \label{eq:diatomic_chain_secular_equation}
\end{gathered}
\end{equation}
is an ordinary eigenvalue problem. It is called the "dynamical equation". There are two distinct vectors $\bm{\epsilon}_{q\nu}$ and eigenvalues $\omega^2_{q\nu}$ that solve \cref{eq:diatomic_chain_secular_equation}. The subsript $\nu$ labels the solutions. More generally, if the dynamical matrix is $d$ dimensional, there are $d$ solutions. For $D$ dimensional space and $n$ atoms in the basis, the dynamical matrix is $d=nD$ dimensional. For $d>2$, solving the equation is hard by hand and instead it is usually done numerically on a computer. For $d=2$, we can diagonalize by hand and get an analytical solution. The eigenvalues $\omega^2_\pm$ are the roots of the equation
\begin{figure}[t!]
    \centering
    \includegraphics[width=0.9\linewidth]{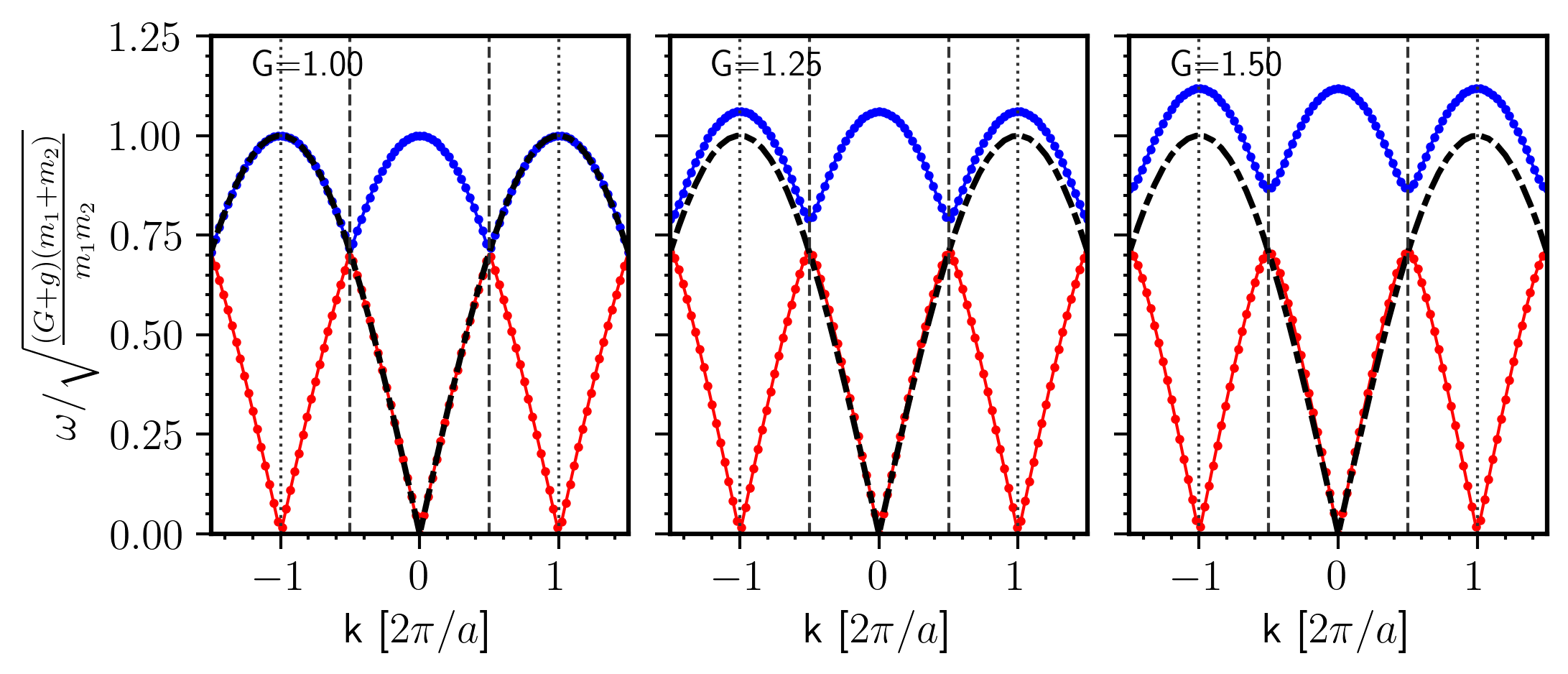}
    \caption{Dispersion, $\omega_\nu(q)$, of the $1D$ diatomic harmonic chain for different $G$. In all panels, $g=1$, $g=m_1=m_2=1$. There are two branches: the acoustic branch (lower) and optical branch (upper). The dashed vertical line is the 1BZ of the diatomic lattice; the dotted vertical line is the 1BZ of the monatomic lattice. The dash-dotted dispersion in each panel is the dispersion of the monatomic harmonic chain. }
    \label{fig:1d_diamotic_dispersion}
\end{figure}
\begin{equation}
\begin{gathered}
\det
\begin{pmatrix}
    D_q^{11}-\omega^2 & D_q^{12} \\ \bar{D}_q^{12} & D_q^{22}-\omega^2
\end{pmatrix} =(D_q^{11}-\omega^2)(D_q^{22}-\omega^2) - |D_q^{12}|^2 = 0.
\end{gathered}
\end{equation}
We find
\begin{equation}
\begin{gathered}
    \omega^2_{q\pm} = \frac{(G+g)(m_1+m_2)}{2m_1m_2} \pm \frac{\sqrt{(G+g)^2(m_1+m_2)^2-16 Ggm_1m_2 \sin^2(qa/2)}}{2m_1m_2}
    \label{eq:diatomic_chain_dispersion}
\end{gathered}
\end{equation}
where the two eigenvalues are labeled by the $\pm$. In the limit $qa\rightarrow 0$, $\sin^2(qa/2)\approx (qa/2)^2$ and
\begin{equation}
\begin{gathered}
    \omega^2_\pm \approx \frac{(G+g)(m_1+m_2)}{2m_1m_2} \pm \frac{ (G+g)(m_1+m_2)}{2m_1m_2} \left( 1- \frac{2 Ggm_1m_2 (qa)^2}{(G+g)^2(m_1+m_2)^2} \right)
\end{gathered}
\end{equation}
which has two solutions:
\begin{equation}
\begin{gathered}
    \omega_+ = \sqrt{\frac{(G+g)(m_1+m_2)}{m_1m_2}} \qquad
    \omega_- = |q| a \sqrt{ \frac{ G g }{(G+g)(m_1+m_2)} } .
\end{gathered}
\end{equation}
Just like for the monatomic chain above, one solution $\omega_- \rightarrow 0$ as $q \rightarrow 0$: this is the familiar acoustic mode. In fact, Goldstone's theorem requires that, in $D$ dimensions, there are $D$ modes that vanish as $q \rightarrow 0$ \cite{goldstein2002classical}. The other mode $\omega_+$ is constant as $q\rightarrow 0$. For a $d=nD$ dimensional dynamical matrix, there are $nD$ solutions: $D$ solutions are acoustic modes and $(n-1)D$ are \emph{optical} modes. See below. 

In \cref{fig:1d_diamotic_dispersion}, we plot the dispersion of the diatomic chain for several values of $G$. We fix $g=m_1=m_2=1$ in all panels. In the diatomic chain, the lattice constant is twice that of the monatomic chain: the 1BZ of the diatomic lattice is half as big. If we set $G=g$ and $m_1=m_2$, the resulting model is identical to the $1D$ monatomic harmonic chain (top row in \cref{fig:harmonic_chain}). We can "unfold" the smaller 1BZ of the diatomic chain onto the larger 1BZ of the monatomic chain, and the physics is identical. On the other hand, when $G\neq g$ or $m_1 \neq m_2$, a gap forms in the dispersion at the zone boundary: the dispersion is no longer identical to monatomic chain and cannot be unfolded (cf. \cref{sec:band_structure}). 

We can explicitly calculate the eigenvectors $\bm{\epsilon}_{q\pm}$ by substituting \cref{eq:diatomic_chain_dispersion} into 
\cref{eq:diatomic_chain_secular_equation}. 
\begin{equation}
\begin{gathered}
        \omega^2_{q\nu} \epsilon_{q\nu,1} = D_{11} \epsilon_{q\nu,1} + D_{12} \epsilon_{q\nu,2} \qquad \textrm{and} \qquad
        \omega^2_{q\nu} \epsilon_{q\nu,2} = \bar{D}_{12} \epsilon_{q\nu,1} + D_{22} \epsilon_{q\nu,2} 
\end{gathered}
\end{equation}
to find
\begin{equation}
\begin{gathered}
        \frac{\epsilon_{q\nu,1}}{\epsilon_{q\nu,2}} =  \frac{D_{12}}{\omega^2_{q\nu}-D_{11}}
\end{gathered}
\end{equation}
Hermiticity of the dynamical matrix means we only need to solve one of these equations. It's clear that we can't fix an overall scale in $\bm{\epsilon}'_{q\nu} = c(\epsilon_{q\nu,1},\epsilon_{q\nu,2})^T = (\epsilon'_{q\nu,1},\epsilon'_{q\nu,2})^T$ since $\epsilon'_{q\nu,1}/\epsilon'_{q\nu,2} = \epsilon_{q\nu,1}/\epsilon_{q\nu,2}$ where the overall scale has disappeared. To fix this ambiguity, we pick $c$ such that each eigenvector is normalized: $\bm{\epsilon}^\dagger_{q\nu}\cdot \bm{\epsilon}_{q\nu} = 1$. 
\begin{equation}
\begin{gathered}
    \bm{\epsilon}^\dagger_{q\nu}\cdot \bm{\epsilon}_{q\nu} = 1 = |c|^2 ( |\epsilon_{q\nu,1}|^2 + |\epsilon_{q\nu,2}|^2) \\
    c = \frac{1}{\sqrt{|\epsilon_{q\nu,1}|^2 + |\epsilon_{q\nu,2}|^2}}
\end{gathered}
\end{equation}
Moreover, the eigenvectors are orthogonal (which follows from Hermeticity of the dynamical matrix):
\begin{equation}
\begin{gathered}
    \bm{\epsilon}^\dagger_{q+}\cdot \bm{\epsilon}_{q-} = \bar{\epsilon}_{q+,1} \epsilon_{q-,1} + \bar{\epsilon}_{q+,2} \epsilon_{q-,2} 
    = \bar{\epsilon}_{q+,2} \epsilon_{q-,2} \left(\frac{|D_{12}|^2}{-|D_{12}|^2 } + 1 \right) = 0
\end{gathered}
\end{equation}
Finally, we can recover the atomic displacements: $\epsilon_{q,\alpha}=\sum_\nu \bm{\epsilon}_{q\nu,\alpha} \cdot \hat{e}_\alpha = \sum_\nu \epsilon_{q\nu,\alpha}$, where $\hat{e}_{\alpha}$ is a unit vector projecting onto the displacement of atom $\alpha$. Then 
\begin{equation}
\begin{gathered}
    u_{n\alpha}(t) = \frac{1}{\sqrt{ m_\alpha N}} \sum_{q\nu} A_{q\nu} \epsilon_{q\nu,\alpha} \exp(iqna - i\omega_{q\nu}t)
\end{gathered}
\end{equation}
where $A_{q\nu}$ is a complex amplitude that is fixed by the boundary conditions of the problem (e.g. initial atomic displacements). 

What we have found is that each atom participates in a wave like displacement of the crystal; the wave like displacements are the "phonons" or "normal modes" (normal meaning orthogonal). They are linearly independent and complete so we can express any arbitrary displacement (around equilibrium) of all of the atoms as a superposition over the phonon modes. There are $nD$ branches or "modes" ($n$ is the number of atoms in the primitive unit cell, $D$ the spatial dimension) at every $q$ point; there are $N$ number of $q$ points, where $N$ is the number of unit cells. I.e. we have as many degrees of freedom as in the original problem ($n\times D\times N$), but now each branch is linearly independent of all others at every $q$ point.   

\begin{figure}[t!]
    \centering
    \includegraphics[width=0.9\linewidth]{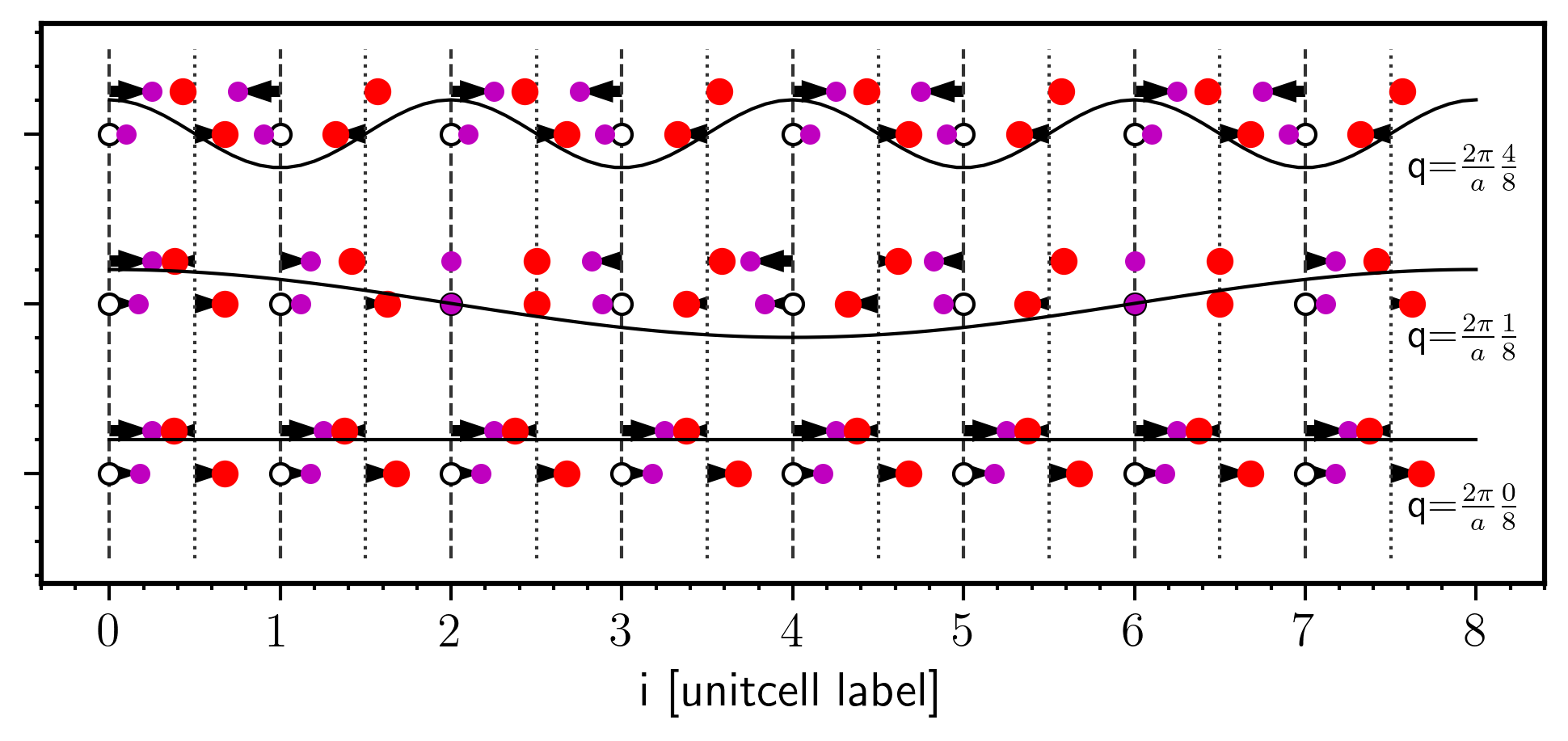}
    \caption{Phonon displacements of the $1D$ diatomic chain, $u^{(q\nu)}_{n\alpha} = A\Re[\epsilon_{q\nu,\alpha}/\sqrt{m_\alpha}\exp(iqna)]$ with $A = 0.25 a$. The chain has $N=8$ unit cells. The arrows label the displacements; the wave like modulation is $\Re[\exp(iqna)]=\cos(qna)$. The bottom/top row for each $q$ is the acoustic/optical branch. We plot $q=0$, $q=2\pi/8a=1/8$ [rlu], and $q=\pi/a=1/2$ [rlu]. The atoms oscillate around their equilibrium positions $x_{n}=na$ with frequency $\omega_\nu(q)$.}
    \label{fig:1d_diatomic_displacement}
\end{figure}

Basically we are saying that the atomic displacement can be expressed as Fourier series over the different modes. $\epsilon_{q\nu,\alpha}$ is the $\alpha^{th}$ component of a periodic wave like displacement of the atoms; i.e. it is the \emph{relative} amplitude of displacement of the $\alpha^{th}$ atom relative to all of the others. The coefficient $A_{q\nu}$ is the amplitude of the $\nu^{th}$ mode with wave vector $q$. The frequency of the oscillation is given by $\omega_{q\nu}$. Note that since $u_{n\alpha}(t)$ is generally real, it is required that $\bar{A}_{q\nu}(t)=A_{-q\nu}(t)$.

We can plot the displacements of the different phonons, $u^{(q\nu)}_{n\alpha}=\epsilon_{q\nu,\alpha}/\sqrt{m_\alpha}\exp(iqna)$, which are Bloch functions. The unit cell periodic part is $\epsilon_{q\nu,\alpha}/\sqrt{m_\alpha}$. In \cref{fig:1d_diatomic_displacement}, we plot both the acoustic and optical modes for several $q$. The difference between acoustic and optical modes is encoded in the unit cell-periodic part $\epsilon_{q\nu,\alpha}/\sqrt{m_\alpha}$. The acoustic phonons displace all atoms in the unit cell \emph{in the same direction}. The optical modes displace the atoms in the unit cell \emph{in opposite directions}. We mentioned before that the acoustic phonons are called "acoustic" in analogy to the theory of elasticity of continuous media. On the other hand, there are no optical modes in continuous media. Rather, they are called "optical" for historical reasons: in an ionic crystal, the optical modes can result in change in polarization of a crystal, which is detected with optical spectroscopy.

\section{General formulation of lattice dynamics}

We now return to the general case of an arbitrary, $3D$ crystal with a basis. The harmonic potential is
\begin{equation}
\begin{gathered}
    U_{harm}(\bm{r}) = \frac{1}{2} \sum_{ij\alpha\beta} \bm{u}^T_{i\alpha} \hat{\Phi}_{ij,\alpha\beta} \bm{u}_{j\beta} \\
    \hat{\Phi}_{ij,\alpha\beta} = \left. \left( \frac{\partial^2 U(\bm{r})} {\partial \bm{u}_{i\alpha} \partial \bm{u}_{j\beta}} \right) \right|_{\bm{u}=0} 
\end{gathered}
\end{equation}
where $\hat{\Phi}_{ij,\alpha\beta}$ is the force constant matrix or "force constants" for short. There are some interesting symmetries of the force constants that will be useful later. (i) First, the order of differentiation doesn't matter, so $\hat{\Phi}_{ij,\alpha\beta} \equiv \hat{\Phi}_{ji,\beta\alpha}$. (ii) Since translating the entire crystal by a vector $\bm{d}$ doesn't change the energy (i.e. it doesn't stretch any bonds), then $U(\bm{r}^{(0)})=U(\bm{r}^{(0)}+\bm{d})$ and
\begin{equation}
\begin{gathered}
    U(\bm{r}^{(0)})-U(\bm{r}^{(0)}+\bm{d}) = \frac{1}{2} \sum_{ij\alpha\beta} \bm{d}^T \hat{\Phi}_{ij,\alpha\beta} \bm{d}  = \frac{N}{2} \bm{d}^T \left( \sum_{\bm{R}\alpha\beta} \hat{\Phi}_{\bm{R},\alpha\beta} \right) \bm{d} = 0 
\end{gathered}
\end{equation}
where we used translational invariance to reduce to a single sum over lattice vectors $\bm{R}$; $\bm{r}^{(0)}$ are the equilibrium coordinates. Since the displacement $\bm{d}$ can be arbitrary, it must be true that
\begin{equation}
\begin{gathered}
    \sum_{\bm{R}\alpha\beta} \hat{\Phi}_{\bm{R},\alpha\beta} = 0 .
    \label{eq:asr_1}
\end{gathered}
\end{equation}
The (harmonic) force on the $\alpha^{th}$ atom in the $i^{th}$ cell is $\bm{f}_{i\alpha} = -\partial U / \partial \bm{u}_{i\alpha} = m_\alpha \ddot{\bm{u}}_{i\alpha}$. Explicitly 
\begin{equation}
\begin{gathered}
    m_\alpha \ddot{\bm{u}}_{i\alpha} = - \sum_{j\beta} \hat{\Phi}_{ij,\alpha\beta} \bm{u}_{j\beta} .
    \label{eq:lattice_dynamics_newtons_eq}
\end{gathered}
\end{equation}
Similar reasoning as for the force constants above shows that the force on any atom should vanish if the whole crystal is displaced by a vector $\bm{d}$
\begin{equation}
\begin{gathered}
    \bm{F}_{i\alpha} = 0 = 
    \hat{\Phi}_{\bm{R},\alpha\beta} = - \left(\sum_{j\beta} \hat{\Phi}_{ij,\alpha\beta}\right) \bm{d} .
\end{gathered}
\end{equation}
Since $\bm{d}$ is arbitrary, we find
\begin{equation}
\begin{gathered}
    \sum_{j\beta} \hat{\Phi}_{ij,\alpha\beta} = 0 .
\end{gathered}
\end{equation}
Some re-arranging (and shifting $\bm{R}_i$ to the origin) results in 
\begin{equation}
\begin{gathered}
    \hat{\Phi}_{\bm{0},\alpha\alpha} = -\sum'_{\bm{R} \beta} \hat{\Phi}_{\bm{R},\alpha\beta} 
    \label{eq:asr_2}
\end{gathered}
\end{equation}
where the $'$ on the sum means to sum over all terms but $\hat{\Phi}_{\bm{0},\alpha\alpha}$. \cref{eq:asr_2} (which is equivalent to \ref{eq:asr_1}) is called the \emph{acoustic sum rule}. It results from the fact that acoustic modes should have vanishing frequency in the infinite wavelength limit: an infinite wavelength acoustic phonon is just a rigid translation of the crystal and should have no restoring force.

Just like above for the harmonic chains, we make progress solving the equations of motion by Fourier transforming. Let\footnote{There is an ambiguity in the phase here. We could expand using either $\exp(-i\bm{q}\cdot\bm{R}_i)$ or $\exp(-i\bm{q}\cdot\bm{r}^{(0)}_{i\alpha}) = \exp(-i\bm{q}\cdot\bm{R}_i) \exp(-i\bm{q}\cdot\bm{\tau}_\alpha)$ since both are periodic functions. Since we sum over $\bm{R}_i$, we can pull the $\bm{\tau}_\alpha$ phase out. It is clear that the only difference is the phase of $\bm{\epsilon}_{\bm{q}\alpha}$, which we can remove by a unitary transformation. However, in some cases it is convenient to choose the second convention: e.g. when studying polarization. Moreover, in computational problems, I prefer the later version since it easier to work with the relative position vectors $\bm{r}^{(0)}_{j\beta} - \bm{r}^{(0)}_{i\alpha}$ rather than explicitly tracking the unit cell each atom is in via $\bm{R}_i$.}
\begin{equation}
\begin{gathered}
    \bm{u}_{i\alpha}(t) = \frac{1}{\sqrt{ m_\alpha N}} \sum_{\bm{q}} \int \frac{d\omega}{\sqrt{2\pi}} \bm{\epsilon}_{\bm{q}\alpha}(\omega) \exp(i\bm{q}\cdot\bm{r}^{(0)}_{i\alpha} - i\omega t ) \\
    \bm{\epsilon}_{\bm{q} \alpha}(\omega) = \sqrt{\frac{m_\alpha}{N}} \sum_{i} \int \frac{dt}{\sqrt{2\pi}} \bm{u}_{i\alpha}(t) \exp(-i\bm{q}\cdot\bm{r}^{(0)}_{i\alpha} + i\omega t ) 
\end{gathered}
\end{equation} 
and plug in to \cref{eq:lattice_dynamics_newtons_eq}
\begin{equation}
\begin{gathered}
    N \omega^2 \bm{\epsilon}_{\bm{q}\alpha} = \sum_{\bm{q}'\beta} \left[ \sum_{ij} \frac{\hat{\Phi}_{ij,\alpha\beta}}{\sqrt{m_\alpha m_\beta}} \exp(i\bm{q}'\cdot \bm{r}^{(0)}_{j\beta}) \exp(-i\bm{q} \cdot \bm{r}^{(0)}_{i\alpha}) \right] \bm{\epsilon}_{\bm{q}'\beta} \\
    = \sum_{\bm{q}'\beta} \exp(i(\bm{q}'\cdot\bm{\tau}_\beta-\bm{q}\cdot\bm{\tau}_\alpha)) \left[ \sum_{ij} \frac{\hat{\Phi}_{ij,\alpha\beta}}{\sqrt{m_\alpha m_\beta}} \exp(i\bm{q}'\cdot \bm{R}_{j}) \exp(-i\bm{q} \cdot \bm{R}_{i})  \right] \bm{\epsilon}_{\bm{q}'\beta}  .
\end{gathered}
\end{equation} 
We define $\hat{\Phi}_{ij,\alpha\beta}/\sqrt{m_\alpha m_\beta} = \hat{C}_{\alpha\beta}(\bm{R}_j-\bm{R}_i)$ and introduce the coordinate transformation $\bm{R}_j=\bm{R}+\bm{R}_i$, i.e. we shift $\bm{R}_j$ by a constant $\bm{R}_i$. Then instead of summing over $\bm{R}_i$, we sum over $\bm{R}=\bm{R}_j-\bm{R}_i$. Then
\begin{equation}
\begin{gathered}
    \sum_{ij} \hat{C}_{\alpha\beta}(\bm{R}) \exp(i\bm{q}'\cdot \bm{R}_j) \exp(-i\bm{q} \cdot \bm{R}_i) \\
    = \sum_{\bm{R}} \hat{C}_{\alpha\beta}(\bm{R}) \exp(i\bm{q}'\cdot \bm{R}) \sum_i \exp(-i(\bm{q} -\bm{q}') \cdot \bm{R}_i)  = N \delta_{\bm{q}\bm{q}'} \hat{D}_{\alpha\beta}(\bm{q}')
\end{gathered}
\end{equation}
and
\begin{equation}
\begin{gathered}
    \omega^2 \bm{\epsilon}_{\bm{q}\alpha} = \sum_{\beta} \hat{D}_{\alpha\beta}(\bm{q}) \bm{\epsilon}_{\bm{q}\beta} 
\end{gathered}
\end{equation}
with
\begin{equation}
\begin{gathered}
    \hat{D}_{\alpha\beta}(\bm{q}) =  \exp(i\bm{q}\cdot(\bm{\tau}_\beta-\bm{\tau}_\alpha)) \sum_{\bm{R}} \frac{\hat{\Phi}_{\bm{R},\alpha\beta}}{\sqrt{m_\alpha m_\beta}} \exp(i\bm{q}\cdot\bm{R}) 
    \label{eq:lattice_dynamics_dynamical_matrix}.
\end{gathered}
\end{equation}
Compactly, 
\begin{equation}
\begin{gathered}
    \omega^2 \begin{pmatrix}
        \bm{\epsilon}_{\bm{q}1} \\ \bm{\epsilon}_{\bm{q}2} \\ \vdots \\ \bm{\epsilon}_{\bm{q}N} 
    \end{pmatrix} =
    \begin{pmatrix}
        \hat{D}_{11}(\bm{q}) & \hat{D}_{12}(\bm{q}) & \cdots & \hat{D}_{1N}(\bm{q}) \\
        \hat{D}_{21}(\bm{q}) & \hat{D}_{22}(\bm{q}) & \cdots & \hat{D}_{2N}(\bm{q}) \\
        \vdots              &  \vdots               & \ddots & \vdots \\
        \hat{D}_{N1}(\bm{q}) & \hat{D}_{N2}(\bm{q}) & \cdots & \hat{D}_{NN}(\bm{q}) \\
    \end{pmatrix}
    \begin{pmatrix}
        \bm{\epsilon}_{\bm{q}1} \\ \bm{\epsilon}_{\bm{q}2} \\ \vdots \\ \bm{\epsilon}_{\bm{q}N} 
    \end{pmatrix} 
\end{gathered}
\end{equation}
represents the set of coupled equations for all atoms in the unit cell. The fact that we only have to look in a single unit cell is due to the lattice periodicity; the displacements in other unit cells is fixed by the phase $\exp(i\bm{q}\cdot\bm{r}^{(0)}_{i\alpha})$ in the Fourier transform of the displacements. Just like with the diatomic chain, we solve this by calculating the eigenvalues and eigenvectors. We assume we can easily diagonalize this equation on a computer (though we have to do it once for every $\bm{q}$). We call the eigenvectors that solve the equation $\bm{\epsilon}_{\bm{q}\nu}$ and label the different solutions by $\nu$. For an $n$ atom basis in $D$ dimensions, there are $nD$ solutions. Compactly, we write the secular equation as 
\begin{equation}
\begin{gathered}
     \hat{D}(\bm{q}) \bm{\epsilon}_{\bm{q}\nu} = \omega^2_{\bm{q}\nu} \bm{\epsilon}_{\bm{q}\nu} .
    \label{eq:lattice_dynamics_secular_equation}
\end{gathered}
\end{equation}
The eigenvectors are 
\begin{equation}
\begin{gathered}
    \bm{\epsilon}_{\bm{q}\nu} = (\epsilon^x_{\bm{q}\nu,1},\epsilon^y_{\bm{q}\nu,1},\epsilon^z_{\bm{q}\nu,1},\epsilon^x_{\bm{q}\nu,2},\epsilon^y_{\bm{q}\nu,2},\epsilon^z_{\bm{q}\nu,2},\cdots,\epsilon^x_{\bm{q}\nu,n},\epsilon^y_{\bm{q}\nu,n},\epsilon^z_{\bm{q}\nu,n})^T \\
    = (\bm{\epsilon}_{\bm{q}\nu,1},\bm{\epsilon}_{\bm{q}\nu,2},\cdots,\bm{\epsilon}_{\bm{q}\nu,n})^T .
\end{gathered}
\end{equation}
They are assumed orthonormalized such that $\bm{\epsilon}^\dagger_{\bm{q}\mu} \cdot \bm{\epsilon}_{\bm{q}\nu} = \delta_{\mu\nu}$; orthogonality is due to $\hat{D}(\bm{q})$ being Hermitian and normalization can be enforced by applying an arbitrary scale factor. The atomic displacements for each mode can be calculated from 
\begin{equation}
\begin{gathered}
    \epsilon^\mu_{\bm{q},\alpha} = \sum_\eta \bm{\epsilon}_{\bm{q}\eta} \cdot \hat{\bm{e}}^\mu_\alpha = \sum_\eta \epsilon^\mu_{\bm{q}\eta,\alpha}  
\end{gathered}
\end{equation}
with $\mu$ labeling the Cartesian direction, $\alpha$ labeling the basis atom, and $\eta$ labeling the phonon branch. Then the general displacement of any atom can be written as
\begin{equation}
\begin{gathered}
    \bm{u}_{i\alpha}(t) = \frac{1}{\sqrt{ m_\alpha N}} \sum_{\bm{q}\eta} A_{\bm{q}\eta} \bm{\epsilon}_{\bm{q}\eta,\alpha} \exp(i\bm{q}\cdot\bm{r}^{(0)}_{i\alpha} - i\omega_{\bm{q}\eta} t ) = \frac{1}{\sqrt{ m_\alpha N}} \sum_{\bm{q}\eta} Q_{\bm{q}\eta}(t) \bm{\epsilon}_{\bm{q}\eta,\alpha} \exp(i\bm{q}\cdot\bm{r}^{(0)}_{i\alpha})
    \label{eq:harmonic_approx_displacements}
\end{gathered}
\end{equation} 
where $ A_{\bm{q}\nu}$ is a (complex) scalar amplitude for each mode and $\bm{q}$ point that is fixed by the boundary conditions of the problem, e.g. the initial positions of the atoms. $\bm{\epsilon}_{\bm{q}\eta, \alpha} = \epsilon^x_{\bm{q}\eta,\alpha} \hat{\bm{e}}^x_\alpha + \epsilon^y_{\bm{q}\eta,\alpha} \hat{\bm{e}}^y_\alpha + \epsilon^z_{\bm{q}\eta,\alpha} \hat{\bm{e}}^z_\alpha$. $Q_{\bm{q}\eta}(t)$ is called a "normal coordinate" and is defined by the equation above. It contains all of the dependence on time and initial conditions. Since the displacements are real, $Q_{\bm{q}\eta}(t)=Q_{-\bm{q}\eta}(t)$. Inverting \cref{eq:harmonic_approx_displacements}, we find
\begin{equation}
\begin{gathered}
    Q_{\bm{q}\eta}(t) = \sum_{i\alpha}  \sqrt{\frac{m_\alpha}{N}}  \bm{\epsilon}^\dagger_{q\eta,\alpha} \cdot \bm{u}_{i\alpha}(t) \exp(-i\bm{q}\cdot\bm{r}^{(0)}_{i\alpha}) .
    \label{eq:inverted_normal_coordinate}
\end{gathered}
\end{equation} 
The normal coordinate are useful because they enable us to reformulate the coupled problem of $dND$ degrees of freedom into $dND$ independent harmonic oscillators. Rather than one coordinate per atom, per direction, per unit cell, we have one coordinate per phonon branch per $q$ point. Each normal coordinate corresponds to a particular displacement pattern of the entire crystal. The usefulness of this formalism is particularly apparent in the Hamiltonian formalism which we will do for quantum mechanical phonons in \cref{sec:quantum_phonons}; we defer further discussion until then. For details of the classical treatment, see Chapters 4 and 6 in Dove \cite{dove1993introduction}.

Just like in the $1D$ case, we are saying that the atomic displacement can be expressed as Fourier series over the different modes. The difference is that $\bm{\epsilon}_{\bm{q}\eta,\alpha}$ is now a $3D$ vector that gives the displacement of the $\alpha^{th}$ relative to the other atoms. The coefficient $A_{\bm{q}\eta}$ has the same meaning as before: it is the Fourier coefficient of the $\eta^{th}$ mode with wave vector $\bm{q}$. 

The combination $\bm{\epsilon}_{\bm{q}\eta,\alpha}\exp(i\bm{q}\cdot\bm{r}^{(0)}_{i\alpha})$ is a wave like modulation of the crystal lattice involving relative displacements of each atom in each unit cell; we call the combination a "normal mode". Its quantum mechanical analogue is a "phonon". Note that the normal mode is a Bloch function; the underlying periodic part is the modulation $\bm{\epsilon}_{\bm{q}\eta,\alpha}$. It is the same in every unit cell. The variation from unit cell-to-unit cell is from the complex exponential which gives a wave like modulation of the underlying periodic displacements: this is why the phonons are called "lattice waves". Each phonon solves \cref{eq:lattice_dynamics_secular_equation} and since they are linearly independent, each evolves in time independently of the others. The time evolution is a periodic oscillation of the displacement given by the phase $\exp(-i\omega_{\bm{q}\eta}t)$. Once we know the initial conditions for the atomic trajectories, we know the trajectories for all times from \cref{eq:harmonic_approx_displacements}. The field of study of the dynamics of lattice waves is called "lattice dynamics".

\section{Example: two dimensional perovskite}\label{sec:2d_perovskite}

\begin{figure}[t!]
    \centering
    \includegraphics[width=0.7\linewidth]{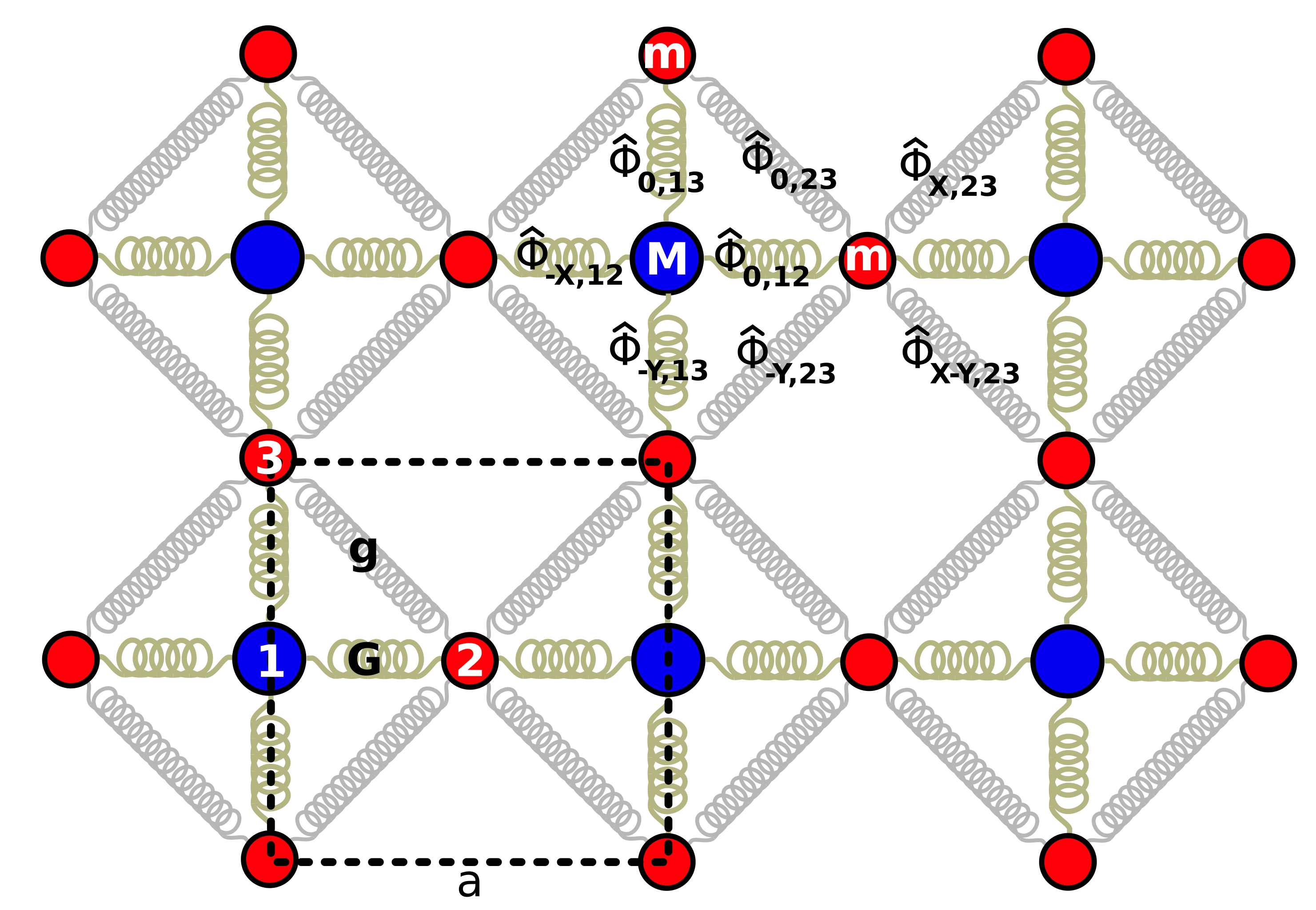}
    \caption{A simplified, $2D$ model of a perovskite. This particular model represents the Cu-O plane of a cuprate. The big, blue atoms (1) are Cu. The small, red atoms (2,3) are O. Each Cu atom is coupled to it's nearest-neighbor O atoms by massless springs, just as each O atom is coupled to its neighboring O atoms. $a$ is the lattice constant.}
    \label{fig:2d_square_lattice}
\end{figure}

We now wish to study an example of a lattice dynamical problem beyond $1D$. We will work in $2D$ since it is easier to visualize the displacements in $2D$; all of the complications of the general $3D$ calculation relative to the $1D$ chain are the same in $2D$. Let us work out the phonons for the particular model in \cref{fig:2d_square_lattice}. It is of scientific interest because it is supposed to model a Cu-O plane in a cuprate (see e.g. \cref{chp:lco_lda_u} and ref. \cite{sterling2021effect}): atom 1 with mass $M$ is supposed to be Cu and atoms 2 and 3 with mass $m$ are supposed to be oxygen. Each Cu atom is coupled to its 4 neighboring O atoms by massless springs with spring constant $G$; each O atom is coupled to its nearest-neighbor O atoms by massless springs with spring constant $g$. The energy of the Cu-O bond bond is 
\begin{equation}
\begin{gathered}
    U_{bond}=\frac{G}{2}(|\bm{r}_{1}(t)-\bm{r}_{2}(t)|-r_0)^2
\end{gathered}
\end{equation}
and similarly for the O-O bonds. This is the energy of a harmonic spring; displacing $\bm{r}_1$ or $\bm{r}_2$ from equilibrium stretches or compresses the spring, costing energy. This is just like the $1D$ harmonic chain above. $r_0$ is the equilibrium bond length: it is just half of a lattice constant $a/2$. \cref{fig:2d_square_lattice} is a schematic of this model.
Due to the periodic boundary conditions, atoms on the right edge are coupled to atoms on the left edge and vice versa. 

We need to calculate the force constants for these bonds. However, it will be convenient later to generalize a little and study a model that includes harmonic bonds to all neighbors. We define the pair potential as 
\begin{equation}
\begin{gathered}
    \phi^{\alpha\beta}_{ij}(\bm{r}_{i\alpha}-\bm{r}_{j\beta}) = \frac{1}{2} K^{\alpha\beta}_{ij} (|\bm{r}_{i\alpha}-\bm{r}_{j\beta}|-l^{\alpha\beta}_{ij})^2
\end{gathered}
\end{equation}
and the crystal potential energy as
\begin{equation}
\begin{gathered}
    U = \frac{1}{2} \sum_{ij,\alpha\beta} \phi^{\alpha\beta}_{ij}(\bm{r}_{i\alpha}-\bm{r}_{j\beta}) =  \frac{1}{4} \sum_{ij,\alpha\beta} K^{\alpha\beta}_{ij} (|\bm{r}_{i\alpha}-\bm{r}_{j\beta}|-l^{\alpha\beta}_{ij})^2 .
\end{gathered} 
\end{equation}
$K^{\alpha\beta}_{ij}$ is the spring constant between atoms $(i\alpha)$ and $(j\beta)$ ($\alpha^{th}$ atom in the $i^{th}$ cell etc.). $l^{\alpha\beta}_{ij}$ is the equilibrium bond length between these atoms. This type of potential can be regarding as a variational approximation to the true BO potential with $K_{ij}^{\alpha\beta}$ and $l_{ij}^{\alpha\beta}$ as variational parameters or can be considered as a toy model. In either case, if the distance between atoms is equal to the equilibrium length, the potential vanishes; if the atoms deviate from the equilibrium position, the potential is quadratic and this (approximate) model can be solved exactly.

The force constants are $\Phi^{\mu\nu}_{i\alpha,j\beta}= (\partial^2 U / \partial u^\mu_{i\alpha} \partial u^\nu_{j\beta})|_{\bm{u}=0}$. Note that $\partial/\partial u = ( \partial r / \partial u ) \partial /\partial r = \partial /\partial r$. We find
\begin{equation}
\begin{gathered}
    \frac{\partial^2 U}{\partial u^\mu_{i\alpha} \partial u^\nu_{j\beta}} = - K^{\alpha\beta}_{ij} \left( \delta_{\mu\nu} \left[1 - \frac{l^{\alpha\beta}_{ij}}{|\bm{r}_{i\alpha}-\bm{r}_{j\beta}|} \right] + l^{\alpha\beta}_{ij} \frac{(r^\mu_{i\alpha}-r^\mu_{j\beta})(r^\nu_{i\alpha}-r^\nu_{j\beta})}{|\bm{r}_{i\alpha}-\bm{r}_{j\beta}|^3} \right),
\end{gathered}
\end{equation}
which, for equilibrium positions ($|\bm{r}_{i\alpha}-\bm{r}_{j\beta}| = l^{\alpha\beta}_{ij}$), is
\begin{equation}
\begin{gathered}
    \Phi^{\mu\nu}_{i\alpha,j\beta} = \left( \frac{\partial^2 U}{\partial u^\mu_{i\alpha} \partial u^\nu_{j\beta}} \right)_0 = -K^{\alpha\beta}_{ij}  \frac{(r^\mu_{i\alpha}-r^\mu_{j\beta})(r^\nu_{i\alpha}-r^\nu_{j\beta})}{[l^{\alpha\beta}_{ij}]^2} 
    \label{eq:2d_perovskite_force_constants}.
\end{gathered}
\end{equation}

\begin{figure}[t!]
    \centering
    \includegraphics[width=0.9\linewidth]{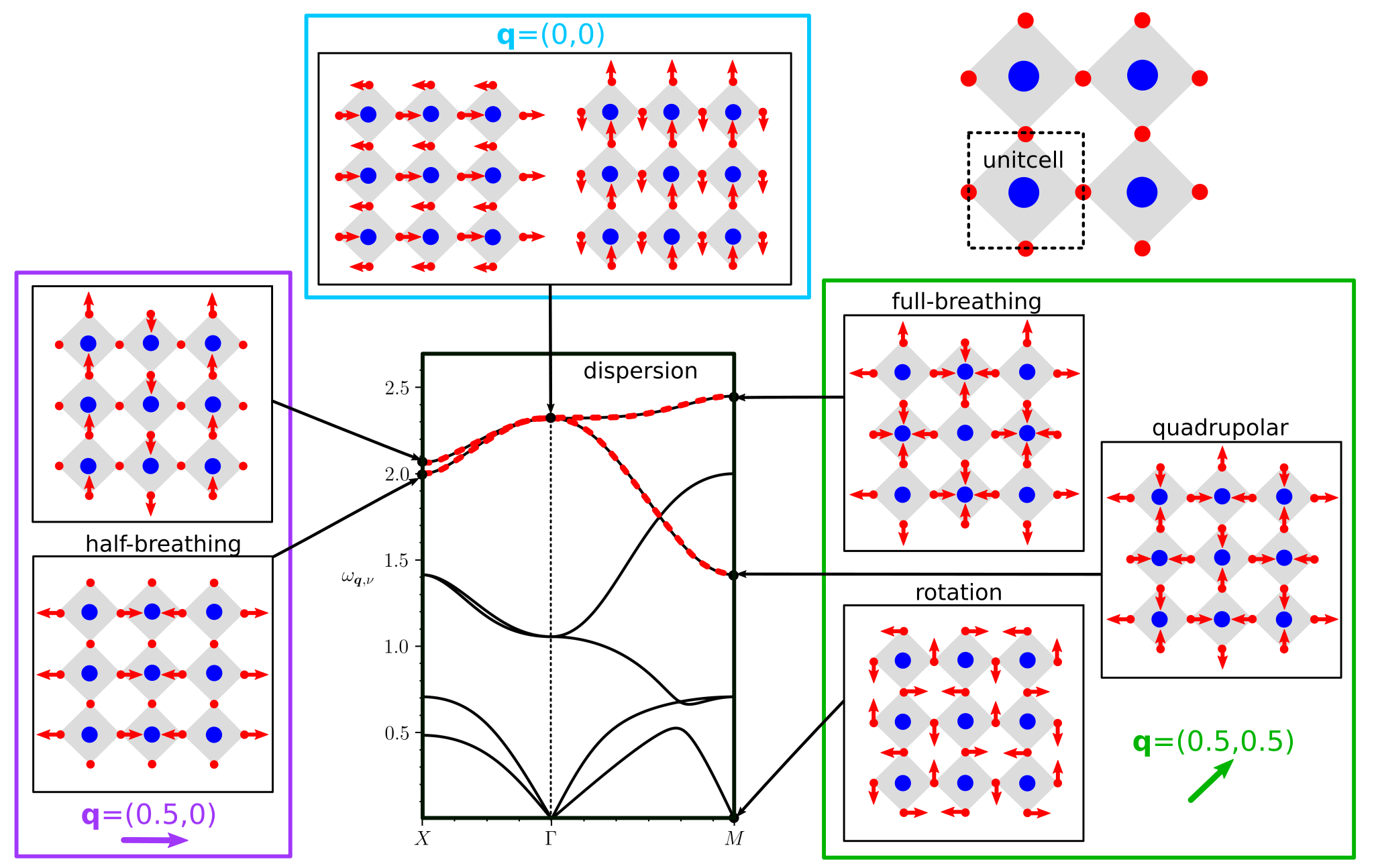}
    \caption{Dispersion and diagram of bond-stretching phonon eigenvectors of the model in \cref{fig:2d_square_lattice} at several wave vectors.}
    \label{fig:2d_perovskite_displacement}
\end{figure}

The phenomenological model introduced above doesn't provide a method to calculate $\hat{\Phi}_{ii,\alpha,\alpha}$ which is a valid force constant: it is called the \emph{self-term}. It arises from the interaction of the $\alpha^{th}$ atom in the $i^{th}$ unit cell and the rest of the crystal and we need it to satisfy the acoustic sum rule. The usual "trick" to calculate it is to use \cref{eq:asr_2}. Once we have calculated the force constants between all pairs, we can calculate the self-terms by calculating the sum in \cref{eq:asr_2}.

\begin{figure}[t!]
    \centering
    \includegraphics[width=0.9\linewidth]{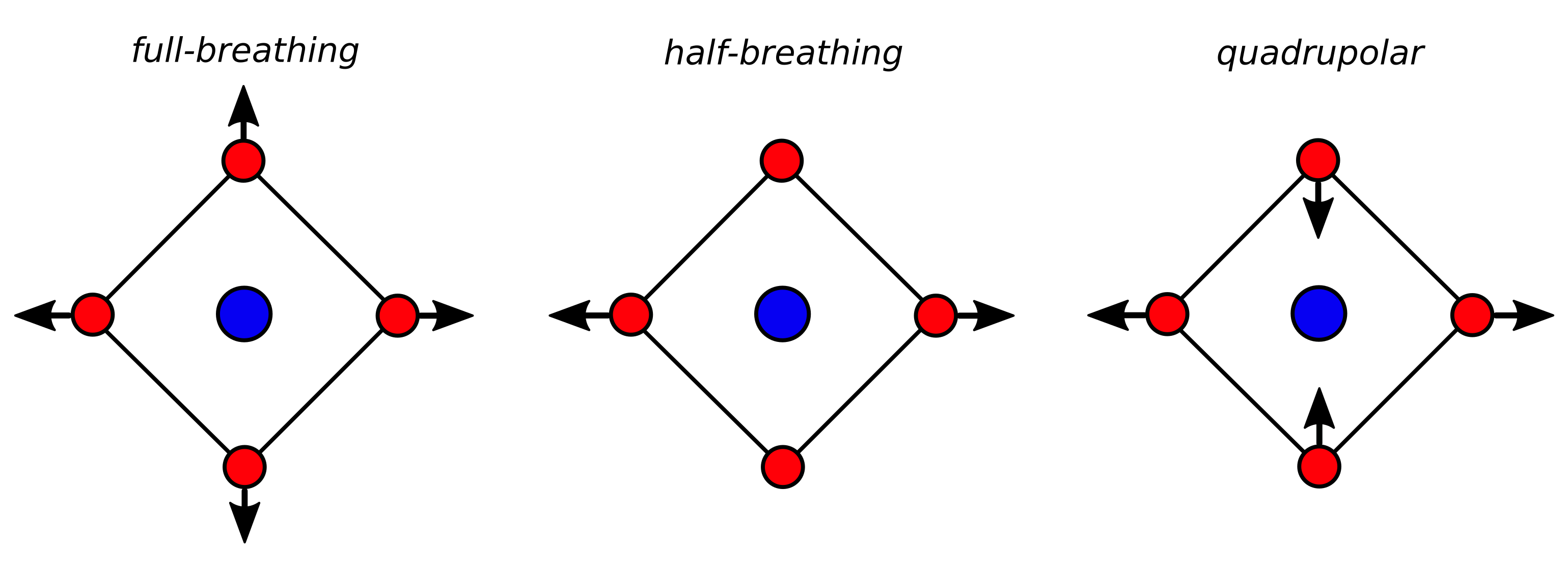}
    \caption{Bond-stretching optical phonon displacements in the 2d perovskite model with nearest-neighbor Cu-O and O-O coupling (cf. \cref{fig:2d_perovskite_displacement}).}
    \label{fig:2d_perovskite_optical}
\end{figure}

Except at very special points, it isn't possible to solve the model above analytically. However, it is easy to solve numerically. The dispersions and displacements are plotted in \cref{fig:2d_perovskite_displacement}. Of particular note are the bond-stretching phonons: the half-breathing, full-breathing, and quadrupolar modes. The zone boundary, unit cell-periodic part is also plotted in \cref{fig:2d_perovskite_optical}. These are of interest experimentally because both the half- and full-breathing modes exhibit strong electron-phonon coupling in many cuprates (e.g. parent compounds La$_2$CuO$_4$, HgBa$_2$CuO$_4$) as revealed by neutron scattering (see \cref{chp:lco_lda_u} or refs. \cite{park2014evidence,pintschovius2005electron,pintschovius2006oxygen,reznik2008photoemission,reznik2012phonon}). These branches also show significant doping and temperature dependence \cite{park2014evidence,pintschovius2005electron,pintschovius2006oxygen}. The quadrupolar mode is of interest as a contrasting example to the full-breathing mode: these modes are degenerate at the zone center, but the quadrupolar mode shows no signs of electron-phonon interactions. We argued in ref. \cite{sterling2021effect} that this coupling is due the way these modes modulate the "volume" around the Cu atoms. The electrons on the Cu atoms are strongly interacting and shrinking the volume "crams" more electrons onto the Cu, costing energy due to Coulomb repulsion. For details, see \cref{chp:lco_lda_u}. For now, let us focus on the volume modulation in general.

\subsection{Volume modulation and electron-phonon coupling}\label{sec:2d_volume_modulation}

\begin{figure}[t!]
    \centering
    \includegraphics[width=0.8\linewidth]{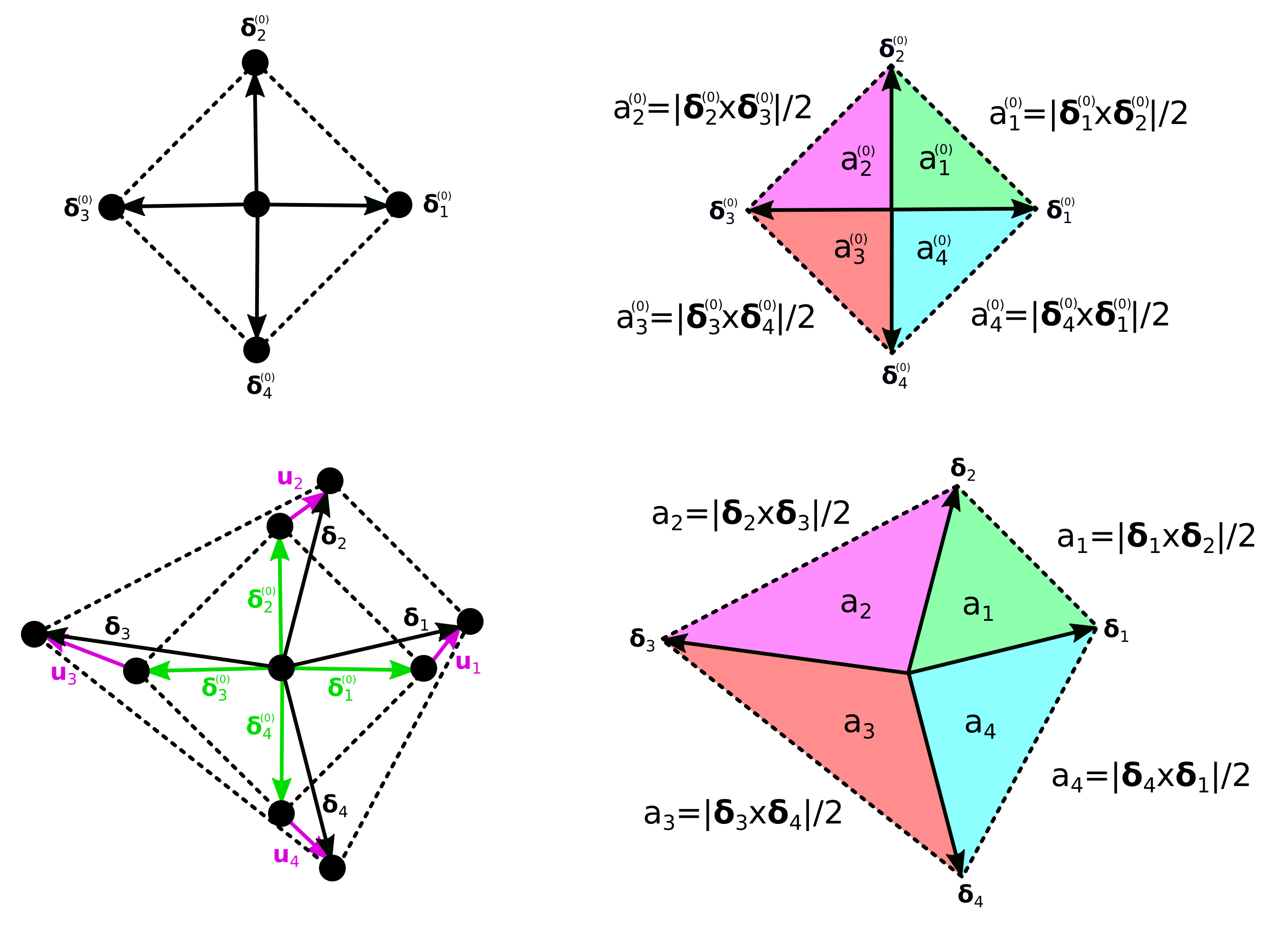}
    \caption{Schematic of how distortions of the $2D$ octahedra change the area. $\bm{\delta}^{(0)}$ are the nearest-neighbor vectors at the equilibrium geometry; $\bm{\delta}=\bm{\delta}^{(0)}+\bm{u}$ are the nearest-neighbor vectors in a distorted geometry.}
    \label{fig:2d_perovskite_delta_V}
\end{figure}

As an interesting exercise, we can quantify the volume modulation from the phonon displacements discussed above. However, instead of specializing the the breathing phonons (e.g. \cref{fig:2d_perovskite_optical}), let us work out the more general case of an arbitrary distortion (see \cref{fig:2d_perovskite_delta_V}). Note that in the case of a $2D$ model, we are calculating areas instead of volumes; however, the method here is conceptually equivalent to volume changes in $3D$ and I will use the words "volume" and "area" interchangeably in what follows. 

Assume the center of the octahedra contains a transition metal (TM) atom: e.g. Cu or Ni. We will place the center of the octahedra at the origin. There are 4 neighbors that make up the vertices; we label them $1-4$. The vectors from the TM atom to the vertices are labeled $\bm{\delta}_i$. The vectors in equilibrium (i.e. no distortion) are $\bm{\delta}^{(0)}_i$ and the displacements are $\bm{u}_{i}$. Then $\bm{\delta}_i=\bm{\delta}^{(0)}_i+\bm{u}_i$. 

The area of the octahedra is $A=\sum^4_{i=1} a_i = a_1 + a_2 + a_3 + a_4$ and similarly for $A^{(0)} = a^{(0)}_1 + a^{(0)}_2 + a^{(0)}_3 + a^{(0)}_4 = 4 a^{(0)}$. The \emph{relative} area modulation is $\xi \equiv \Delta A / A^{(0)} = (A-A^{(0)})/A^{(0)}$. Note that $\Delta A = A-4a^{(0)} = \sum_i (a_i - a^{(0)}) = \sum_i \Delta a_i$. We can calculate the area of any triangle by taking half of the magnitude of the cross product of the bond-vectors spanning it. This will give us half the area of the parallelogram spanned by the bond-vectors; i.e. the area the triangle. For example (careful to pick the $+$ sign)
\begin{equation}\begin{split}
    a_1 & = \frac{1}{2} ( \bm{\delta}_1 \times \bm{\delta}_2 ) \cdot \hat{\bm{z}} \\
    & = \frac{1}{2}( \delta^x_1 \delta^y_2  -  \delta^y_1 \delta^x_2 )\\
    a^{(0)}_1 & \equiv a^{(0)} = \frac{1}{2} ( \bm{\delta}^{(0)}_1 \times \bm{\delta}^{(0)}_2 ) \cdot \hat{\bm{z}} \\
    & = \frac{1}{2} ( \delta^{(0)x}_1 \delta^{(0)y}_2  -  \delta^{(0)y}_1 \delta^{(0)x}_2 )
\end{split}\end{equation}
In terms of the displacements, we write $\bm{\delta}_i = \bm{\delta}^{(0)}_i+\bm{u}_i$. The area of triangle $a_1 \equiv a_{12}$ is
\begin{equation}\begin{split}
    \Delta a_{12} & = \frac{1}{2} [ (\bm{\delta}^{(0)}_1 + \bm{u}_1 ) \times (\bm{\delta}^{(0)}_2 + \bm{u}_2) ] \cdot \hat{\bm{z}}  \\ 
    & = \frac{1}{2}(\delta^{(0)x}_1 + u^x_1)(\delta^{(0)y}_2 + u^y_2) -  (\delta^{(0)y}_1 + u^y_1)(\delta^{(0)x}_2 + u^x_2)  \\
    & = \frac{1}{2}( 2a^{(0)} + u^x_1 \delta^{(0)y}_2 + \delta^{(0)x}_1 u^y_2 - u^y_1 \delta^{(0)x}_2 - \delta^{(0)y}_1 u^x_2 ) + O(u^2).
\end{split}\end{equation}
Dropping $\mathcal{O}[u^2]$ terms (which is justified since we assume small displacements), we find
\begin{equation}
\begin{gathered}
    \Delta a_{12} = 
    \frac{1}{2}( \delta^{(0)x}_1 u^y_2  - \delta^{(0)y}_1 u^x_2 + u^x_1 \delta^{(0)y}_2 - u^y_1 \delta^{(0)x}_2  ) = \frac{1}{2}[ \bm{\delta}^{(0)}_1 \times \bm{u}_2 - \bm{\delta}^{(0)}_2 \times \bm{u}_1   ] \cdot \hat{\bm{z}} .
\end{gathered}
\end{equation}
For the change in area of any arbitrary $2D$ polygon polygon with $n$ sides (the 2d octahedra is the special case $n=4$):
\begin{equation}\begin{split}
    \Delta A & = \sum^n_{i=1} \Delta a_i \equiv [\Delta a_{12} + \Delta a_{23} + \cdots \Delta a_{n-1n} + \Delta a_{n1}] \\
    & = \frac{1}{2} [ \bm{\delta}^{(0)}_1 \times (\bm{u}_2 - \bm{u}_n)  +  \bm{\delta}^{(0)}_2 \times (\bm{u}_3 - \bm{u}_1) + \cdots \\ 
    & \qquad \bm{\delta}^{(0)}_{n-1} \times (\bm{u}_{n} - \bm{u}_{n-2}) + \bm{\delta}^{(0)}_n \times (\bm{u}_{1} - \bm{u}_{n-1})] \cdot \hat{\bm{z}} \\
    & = \frac{1}{2} \left( \sum^n_{i=1} \bm{\delta}^{(0)}_i \times ( \bm{u}_{i+1}-\bm{u}_{i-1} ) \right) \cdot \hat{\bm{z}} .
    \label{eq:2d_volume_modulation}
\end{split}\end{equation}
The index $i$ is cyclic; for $n$ neighbors, $i=n+1=1$. 

For the specific case of the $2D$ octahedra in \cref{fig:2d_perovskite_optical}, 
\begin{equation}
\begin{gathered}
    \Delta A = \frac{1}{2} [ \bm{\delta}^{(0)}_1 \times (\bm{u}_2 - \bm{u}_4)  +  \bm{\delta}^{(0)}_2 \times (\bm{u}_3 - \bm{u}_1) + \\
    \bm{\delta}^{(0)}_{3} \times (\bm{u}_{4} - \bm{u}_{2}) + \bm{\delta}^{(0)}_4 \times (\bm{u}_{1} - \bm{u}_{3})] \cdot \hat{\bm{z}} .
\end{gathered}
\end{equation}
There are 2 commonly studied optical modes that strongly modulate the volume; the half- and full-breathing modes. Another phonon close in energy and degenerate at the zone center is the quadrupolar mode; however, the quadrupolar phonon doesn't modulate the volume so doesn't couple to the electrons. Still, it is good to juxtapose the breathing phonons. Let us work out $\Delta A$ for each of these. Each arrow (i.e. component of the displacements $A_{\bm{q}\nu} \bm{\epsilon}_{\bm{q}\nu,\alpha}/\sqrt{m_\alpha}$) has the same amplitude: call it $d$. The rest of the displacements are 0. The bond-lengths are $l$. Then
\begin{equation}
\begin{gathered}
    \Delta A_{FB} = dl ( \hat{\bm{x}}\times\hat{\bm{y}} - \hat{\bm{y}}\times\hat{\bm{x}} + \hat{\bm{x}}\times\hat{\bm{y}} - \hat{\bm{y}}\times\hat{\bm{x}} ) \cdot \bm{z} = 4 dl ( \hat{\bm{x}}\times\hat{\bm{y}} ) \cdot \bm{z} = 4dl \\
    \Delta A_{HB} = dl ( - \hat{\bm{y}}\times\hat{\bm{x}} - \hat{\bm{y}}\times\hat{\bm{x}} ) \cdot \bm{z} = 2 dl ( \hat{\bm{x}}\times\hat{\bm{y}} ) \cdot \bm{z} = 2dl \\
    \Delta A_{QP} = dl ( -\hat{\bm{x}}\times\hat{\bm{y}} - \hat{\bm{y}}\times\hat{\bm{x}} - \hat{\bm{x}}\times\hat{\bm{y}} - \hat{\bm{y}}\times\hat{\bm{x}} ) \cdot \bm{z} = 0 .
\end{gathered}
\end{equation}
The full-breathing phonon couples most strongly to the charge on the TM atom; the half-breathing mode couples less strongly; the quadrupolar mode doesn't couple to the charge. These results are consistent with what we know about electron-phonon coupling in $2D$ perovskite like Mott-Hubbard systems: see \cref{chp:lco_lda_u}. 

The generalization of the "volume modulation" to $3D$ is straightforward, but the algebra is more tedious with uglier final expressions; I omit it here.

The idea that the volume modulation due to the phonons couples to localized electrons is characteristic of "Holstein phonons", which were introduced to represent a molecular crystal where a localized phonon really makes sense \cite{holstein1959studies}. The method of calculating volume change around a localized electron on a TM atom derived above allows us to generalize this idea to more ordinary crystals, where the phonons are not localized. The basic idea is that we can define a phenomenological electron-phonon coupling interaction as 
\begin{equation}
\begin{gathered}
    \hat{H}_{eph} \sim \xi(\hat{\bm{u}}) \hat{n}
\end{gathered}
\end{equation}
where $\xi(\hat{\bm{u}})$ is the relative volume modulation due to phonon displacements (it's an operator) and $\hat{n}$ is the electron density operator for the localized orbital on the TM atom. If we can partition a crystal into motifs (like in \cref{fig:2d_perovskite_optical}) such that it is sensible to discuss the volume change around a TM atom, then $\xi(\hat{\bm{u}})$ can be calculated by only summing over the neighboring atoms around the TM atom. We can reconnect this description to phonons by writing the displacements of the neighboring atoms in terms of quantized phonon operators (see \cref{sec:quantum_phonons} below); the relative volume modulation is given by \cref{eq:2d_volume_modulation}, where the sum is only over neighboring atoms and the displacements $\bm{u}$ are quantum mechanical operators. This is the subject of an ongoing project that I won't discuss in this thesis any further\footnote{I plan to publish something on this eventually and am happy to discuss it in more detail elsewhere.}.

\subsection{Soft modes; tetragonal instability}\label{sec:soft_modes}

At the $M$ point in \cref{fig:2d_perovskite_displacement}, there is a phonon that goes to $\omega=0$. In e.g. a diffraction experiment, this would show up as elastic intensity. Note, however, that $M$ is \emph{not} the zone center of the primitive cell. $\omega=0$ phonons with $\bm{q}\neq 0$ signal a structural instability: the displacement of the $\omega=0$ modes are static; i.e. the atomic displacements are "frozen in", resulting in a static structural distortion that generates \emph{elastic diffuse scattering} (see \cref{sec:diffuse_scattering}). 

In some materials, such modes are actually observable and signal a structural phase transitions. These modes are the subject of diffuse scattering experiments (cf. \cref{chp:mapi_diffuse}). However, from a computational perspective, such instabilities often signal an error in the model. This simple $2D$ model is supposed to resemble a cuprate, e.g. HgBa$_2$CuO$_4$, which has no such structural transition\footnote{Though La$_2$CuO$_4$ \emph{does} have structural transitions}. Then the issue must be with the model. It is easy to see why!

Examine the displacement of the unstable mode: it corresponds to rigidly rotating neighboring octahedra in opposite directions. With the model defined in \cref{fig:2d_square_lattice}, there are no bonds that stretch under this distortion; i.e. it costs no energy to rotate the octahedra and there is no restoring force. We can remove the instability in a few ways. E.g. we could add next-nearest neighbor coupling between the Cu and O atoms or we could put cations (e.g. Ba in HgBa$_2$CuO$_4$) into the model. Then rotating the CuO$_6$ octahedra stretches the bonds between the Ba and O atoms, costing energy. This will stabilize the rotational modes. 

In many cases, however, such modes \emph{do} show up experiments. Various mechanism lead to such modes, e.g. anharmonicity, electron-phonon interactions (charge-density waves), and entropy. We don't consider these mechanisms explicitly in this thesis, but we do look at the experimental detection of a different type of unstable mode  in \cref{chp:mapi_diffuse}.

\section{Quantum theory of lattice vibrations}\label{sec:quantum_phonons}

We now work out the quantum theory of a harmonic crystal. For generality, let us work on a lattice with a basis. We write down the Hamiltonian using the usual methods of first quantization: i.e. we replace position and momentum by their respective operators; the position or the $\alpha^{th}$ atom in the $i^{th}$ unit cell is $\hat{\bm{r}}_{i\alpha}=\bm{R}_i+\bm{\tau}_\alpha+\hat{\bm{u}}_{i\alpha}$. Since the unit cell and basis vectors are constant parameters, they are not operators; rather, the \emph{displacement} is the dynamical variable we care about and we promote it to an operator. The momentum conjugate to $\hat{u}^\mu_{i\alpha}$ is $\hat{p}^\mu_{i\alpha}$. $\mu$ labels the Cartesian direction. The commutation relation is $[\hat{u}^\mu_{i\alpha},\hat{p}^\nu_{j\beta}]=i\hbar\delta_{ij}\delta_{\alpha\beta}\delta_{\mu\nu}$. The Hamiltonian operator is\footnote{I am leaving the ``hat" off of $\Phi_{ij\alpha\beta}$ to avoid confusing the matrix with an operator; hopefully the context makes it clear} 
\begin{equation}
\begin{gathered}
    \hat{H} = \sum_{i\alpha} \frac{\hat{\bm{p}}_{i\alpha}\cdot\hat{\bm{p}}_{i\alpha}}{2 m_{\alpha}} + U(\hat{\bm{r}}) = \sum_{i\alpha} \frac{\hat{\bm{p}}_{i\alpha}\cdot\hat{\bm{p}}_{i\alpha}}{2 m_{\alpha}} + \frac{1}{2}\sum_{ij\alpha\beta} \hat{\bm{u}}^T_{i\alpha} \Phi_{ij,\alpha\beta} \hat{\bm{u}}_{j\beta} \\ 
    = \sum_{i\alpha,\mu} \frac{(\hat{p}^\mu_{i\alpha})^2}{2 m_{\alpha}} + \frac{1}{2}\sum_{ij\alpha\beta,\mu\nu} \Phi^{\mu\nu}_{ij,\alpha\beta} \hat{u}^\mu_{i\alpha} \hat{u}^\nu_{j\beta} 
    \label{eq:phonon_quantum_hamiltonian}
\end{gathered}
\end{equation}
We make progress by Fourier transforming. For the displacements, we have
\begin{equation}
\begin{gathered}
    \hat{u}^\mu_{i\alpha} =  N^{-1/2} \sum_{\bm{q}} \hat{u}^\mu_{\bm{q} \alpha} \exp(i\bm{q}\cdot\bm{r}^{(0)}_{i\alpha})  \\
    \hat{u}^\mu_{\bm{q}\alpha} = N^{-1/2}  \sum_{i} \hat{u}^\mu_{i \alpha} \exp(-i\bm{q}\cdot\bm{r}^{(0)}_{i\alpha}) 
\end{gathered}
\end{equation}
For the momentum operators, there is a minor subtlety: we \emph{define} the Fourier transform to have opposite the phase of the displacements so that the commutation relation of the Fourier transformed operators is preserved \cite{mahan2013many}. Explicitly
\begin{equation}
\begin{gathered}
    \hat{p}^\mu_{i\alpha} = N^{-1/2} \sum_{\bm{q}} \hat{p}^\mu_{\bm{q} \alpha} \exp(-i\bm{q}\cdot\bm{r}^{(0)}_{i\alpha})  \\
    \hat{p}^\mu_{\bm{q}\alpha} = N^{-1/2} \sum_{i} \hat{p}^\mu_{i \alpha} \exp(i\bm{q}\cdot\bm{r}^{(0)}_{i\alpha})
\end{gathered}
\end{equation}
and
\begin{equation}
\begin{gathered}
    [\hat{u}^\mu_{\bm{k}\alpha},\hat{p}^\nu_{\bm{q}\beta}] = N^{-1}  \sum_{ij} [\hat{u}^\mu_{i\alpha},\hat{p}^\nu_{j\beta}] \exp(-i\bm{k}\cdot\bm{r}^{(0)}_{i\alpha} ) \exp(i\bm{q}\cdot\bm{r}^{(0)}_{i\alpha} ) \\ 
    = N^{-1} \exp(-i\bm{k}\cdot\bm{\tau}_{\beta})\exp(i\bm{q}\cdot\bm{\tau}_{\alpha})  \sum_{ij} i\hbar\delta_{ij}\delta_{\alpha\beta}\delta_{\mu\nu} \exp(-i\bm{k}\cdot\bm{R}_i ) \exp(i\bm{q}\cdot\bm{R}_j ) \\ 
    = \frac{i \hbar\delta_{\alpha\beta}\delta_{\mu\nu}}{N} \exp(i(\bm{q}-\bm{k})\cdot\bm{\tau}_{\alpha}) \sum_{i}  \exp(i(\bm{q}-\bm{k})\cdot\bm{R}_i ) = i \delta_{\bm{q}\bm{k} }\delta_{\alpha\beta}\delta_{\mu\nu} 
\end{gathered}
\end{equation}
has the same structure as the real-space commutation relation. Alternatively, we could have picked that phase of the Fourier transform to be the same for momentum and displacement, in which case the commutator would have been $[\hat{u}^\mu_{\bm{k}\alpha},\hat{p}^{\dagger \nu}_{\bm{q}\beta}] = i \hbar \delta_{\bm{q}\bm{k} }\delta_{\alpha\beta}\delta_{\mu\nu} $. Whatever. 

Lets us work out the momentum space version of the kinetic and potential energy terms of the Hamiltonian separately:
\begin{equation}
\begin{gathered}
    \hat{T} = \sum_{i\alpha,\mu} \frac{(\hat{p}^\mu_{i\alpha})^2}{2 m_{\alpha}} = 
    N^{-1} \sum_{\bm{k}\bm{q}} \sum_{\alpha \mu} \frac{\hat{p}^\mu_{\bm{k}\alpha} \hat{p}^\mu_{\bm{q}\alpha}}{2 m_{\alpha}} \sum_i \exp(-i(\bm{k}+\bm{q})\cdot\bm{r}^{(0)}_{i\alpha}) = \sum_{\bm{q}\alpha,\mu} \frac{\hat{p}^\mu_{-\bm{q}\alpha} \hat{p}^\mu_{\bm{q}\alpha}}{2m_\alpha} .
\end{gathered}
\end{equation}
Since momentum is a Hermitian operator, we can also write $\hat{p}^\mu_{-\bm{q}\alpha}=(\hat{p}^\mu_{\bm{q}\alpha})^\dagger$ and similarly for the displacement operators. Now for the potential energy:
\begin{equation}
\begin{gathered}
    U= \frac{1}{2}\sum_{ij\alpha\beta,\mu\nu} \Phi^{\mu\nu}_{ij,\alpha\beta} \hat{u}^\mu_{i\alpha} \hat{u}^\nu_{j\beta} \\
    = \frac{1}{2N} \sum_{\bm{k}\bm{q}} \sum_{\alpha\beta,\mu\nu} \hat{u}^\mu_{\bm{k}\alpha} \hat{u}^\nu_{\bm{q}\beta} \exp(i(\bm{k}\cdot\bm{\tau}_\alpha+\bm{q}\cdot\bm{\tau}_\beta))  \sum_{ij} \Phi^{\mu\nu}_{ij,\alpha\beta} \exp(i\bm{k}\cdot\bm{R}_i)\exp(i\bm{q}\cdot\bm{R}_j) .
\end{gathered}
\end{equation}
Using translational invariance of the force constants and substituting $\bm{R}_i=\bm{R}_j-\bm{R}$
\begin{equation}
\begin{gathered}
    U = \frac{1}{2N} \sum_{\bm{k}\bm{q}} \sum_{\alpha\beta,\mu\nu} \hat{u}^\mu_{\bm{k}\alpha} \hat{u}^\nu_{\bm{q}\beta} \exp(i(\bm{k}\cdot\bm{\tau}_\alpha+\bm{q}\cdot\bm{\tau}_\beta)) \left( \sum_{\bm{R}}  \Phi^{\mu\nu}_{\bm{R},\alpha\beta} \exp(-i\bm{k}\cdot\bm{R}) \right) \sum_{j} \exp(i(\bm{k}+\bm{q})\cdot\bm{R}_i) \\ 
    = \frac{1}{2} \sum_{\bm{k}\bm{q}} \delta_{-\bm{k}\bm{q}} \sum_{\alpha\beta,\mu\nu} \hat{u}^\mu_{\bm{k}\alpha} \hat{u}^\nu_{\bm{q}\beta} \exp(i(\bm{k}\cdot\bm{\tau}_\alpha+\bm{q}\cdot\bm{\tau}_\beta))  \left( \sum_{\bm{R}}  \Phi^{\mu\nu}_{\bm{R},\alpha\beta} \exp(-i\bm{q}\cdot\bm{R}) \right) \\
    = \frac{1}{2} \sum_{\bm{q}} \sum_{\alpha\beta,\mu\nu} \hat{u}^\mu_{-\bm{q}\alpha} \hat{u}^\nu_{\bm{q}\beta} \left( \exp(i\bm{q}\cdot(\bm{\tau}_\beta-\bm{\tau}_\alpha)) \sum_{\bm{R}}  \Phi^{\mu\nu}_{\bm{R},\alpha\beta} \exp(i\bm{q}\cdot\bm{R}) \right) .
\end{gathered}
\end{equation}
So then finally
\begin{equation}
\begin{gathered}
    U = \frac{1}{2} \sum_{\bm{q}} \sum_{\alpha\beta,\mu\nu} \sqrt{ m_\alpha m_\beta} \hat{u}^\mu_{-\bm{q}\alpha}  \hat{u}^\nu_{\bm{q}\beta} D^{\mu\nu}_{\alpha\beta}(\bm{q})
\end{gathered}
\end{equation}
with $D^{\mu\nu}_{\alpha\beta}(\bm{q})$ the dynamic matrix elements defined earlier in \cref{eq:lattice_dynamics_dynamical_matrix}. The dynamical matrix in the quantum theory of lattice vibrations is identical to the classical theory. The Hamiltonian in momentum space is
\begin{equation}
\begin{gathered}
    \hat{H} = \frac{1}{2} \sum_{\bm{q}} \sum_{\alpha\beta,\mu\nu}  \left[  \frac{\hat{p}^\mu_{-\bm{q}\alpha} \hat{p}^\nu_{\bm{q}\beta}}{ \sqrt{ m_\alpha m_\beta} } \delta_{\mu\nu} \delta_{\alpha\beta} + \sqrt{ m_\alpha m_\beta} \hat{u}^\mu_{-\bm{q}\alpha} \hat{u}^\nu_{\bm{q}\beta} D^{\mu\nu}_{\alpha\beta}(\bm{q}) \right] .
\end{gathered}
\end{equation}
Let's define $\hat{P}^\mu_{\bm{q}\alpha} = \hat{p}^\mu_{\bm{q}\alpha}/\sqrt{m_\alpha}$ and $\hat{U}^\mu_{\bm{q}\alpha} = \sqrt{m_\alpha}\hat{u}^\mu_{\bm{q}\alpha} $. Then
\begin{equation}
\begin{gathered}
    \hat{H} = \frac{1}{2} \sum_{\bm{q}} \sum_{\alpha\beta,\mu\nu}  \left[  \hat{P}^\mu_{-\bm{q}\alpha} \hat{P}^\nu_{\bm{q}\beta} \delta_{\mu\nu} \delta_{\alpha\beta} + \hat{U}^\mu_{-\bm{q}\alpha} \hat{U}^\nu_{\bm{q}\beta} D^{\mu\nu}_{\alpha\beta}(\bm{q}) \right]  \\
    = \frac{1}{2} \sum_{\bm{q}} \sum_{\alpha\beta,\mu\nu}  \left[  \hat{P}^{\dagger\mu}_{\bm{q}\alpha} \hat{P}^\nu_{\bm{q}\beta} \delta_{\mu\nu} \delta_{\alpha\beta} + \hat{U}^{\dagger\mu}_{\bm{q}\alpha} \hat{U}^\nu_{\bm{q}\beta} D^{\mu\nu}_{\alpha\beta}(\bm{q}) \right] .
\end{gathered}
\end{equation}
The dynamical matrix couples position operators of different atoms in the unit cell. We can decouple them by changing basis; i.e. we diagonalize the dynamical matrix in the $(\alpha,\beta)$ indices. 

\subsection{Diagonalizing the dynamical matrix}


Define the ``vectors" 
\begin{equation}
\begin{gathered}
    \hat{\bm{U}}_{\bm{q}}=(\hat{U}^x_{\bm{q}1},\hat{U}^y_{\bm{q}1}\hat{U}^z_{\bm{q}1}, \cdots, \hat{U}^x_{\bm{q}n},\hat{U}^y_{\bm{q}n}\hat{U}^z_{\bm{q}n})^T \\
    \hat{\bm{U}}^\dagger_{\bm{q}}=(\hat{U}^{\dagger x}_{\bm{q}1},\hat{U}^{\dagger y}_{\bm{q}1}\hat{U}^{\dagger z}_{\bm{q}1}, \cdots, \hat{U}^{\dagger x}_{\bm{q}n},\hat{U}^{\dagger y}_{\bm{q}n}\hat{U}^{\dagger z}_{\bm{q}n})
\end{gathered}
\end{equation}
with $n$ the number of atoms in the unit cell. At each $\bm{q}$-point, we need to solve $ \hat{\bm{U}}^\dagger_{\bm{q}} D_{\bm{q}} \hat{\bm{U}}_{\bm{q}}$. Define $V_{\bm{q}}$ as the unitary matrix that diagonalizes $D_{\bm{q}}$ and $\Omega_{\bm{q}}$ the diagonal representation of the dynamical matrix; i.e. $V^\dagger_{\bm{q}} D_{\bm{q}} V_{\bm{q}}=\Omega_{\bm{q}}$. Note, $V_{\bm{q}}$ is the column-matrix of eigenvectors of $D_{\bm{q}}$. Then
\begin{equation}
\begin{gathered}
     \hat{\bm{U}}^\dagger_{\bm{q}} D_{\bm{q}} \hat{\bm{U}}_{\bm{q}} = \hat{\bm{U}}^\dagger_{\bm{q}} V_{\bm{q}} V^\dagger_{\bm{q}} D_{\bm{q}} V_{\bm{q}} V^\dagger_{\bm{q}} \hat{\bm{U}}_{\bm{q}} = \hat{\bm{Q}}^\dagger_{\bm{q}} \Omega_{\bm{q}} \hat{\bm{Q}}_{\bm{q}}
\end{gathered}
\end{equation}
with $\hat{\bm{Q}}^\dagger_{\bm{q}}=\hat{\bm{U}}^\dagger_{\bm{q}} V_{\bm{q}}$. Note that the diagonal elements of $\Omega_{\bm{q}}$ (which we call $\Omega_{\bm{q}\lambda}$) are exactly the same eigenvalues as in the classical lattice dynamical theory: i.e. $\Omega_{\bm{q}\lambda} \equiv \omega^2_{\bm{q}\lambda}$ with $\lambda$ labelling the ``branches" or ``modes". There are $nD$ modes; one for each atom and direction. We also define $\hat{\bm{\Pi}}^\dagger_{\bm{q}}=\hat{\bm{P}}^\dagger_{\bm{q}}V^\dagger_{\bm{q}}$ and $\hat{\bm{\Pi}}_{\bm{q}}=V_{\bm{q}} \hat{\bm{P}}_{\bm{q}}$. In this basis, the Hamiltonian can be written
\begin{equation}
\begin{gathered}
    \hat{H} = \frac{1}{2} \sum_{\bm{q}} \left[  \hat{\bm{P}}^{\dagger}_{\bm{q}} \cdot \hat{\bm{P}}_{\bm{q}} + \hat{\bm{U}}^\dagger_{\bm{q}} D_{\bm{q}} \hat{\bm{U}}_{\bm{q}} \right] = \frac{1}{2} \sum_{\bm{q} \lambda} \left[  \hat{\Pi}^{\dagger}_{\bm{q}\lambda} \hat{\Pi}_{\bm{q}\lambda} + \omega^2_{\bm{q}\lambda} \hat{Q}^\dagger_{\bm{q}\lambda} \hat{Q}_{\bm{q}\lambda} \right] .
\end{gathered}
\end{equation}
These new operators still satisfy the position-momentum commutation relation
\begin{equation}
\begin{gathered}
    [\hat{Q}_{\bm{k}\lambda},\hat{\Pi}_{\bm{q}\gamma}] = i \hbar \delta_{\bm{k}\bm{q}} \delta_{\lambda \gamma} .
\end{gathered}
\end{equation}
In index notation, ($\sigma,\delta$ are composite indices for basis atom index and direction)
\begin{equation}
\begin{gathered}
    \hat{Q}_{\bm{k}\lambda} = (V^\dagger_{\bm{k}} \hat{\bm{U}}_{\bm{k}})^\lambda = \sum_\sigma [V^\dagger_{\bm{k}}]^{\lambda \sigma } \hat{U}_{\bm{k}\sigma} \quad \textrm{and} \quad
    \hat{\Pi}_{\bm{q}\gamma} = (V_{\bm{q}} \hat{\bm{P}}_{\bm{q}})^\gamma = \sum_\delta V^{\gamma \delta}_{\bm{q}} \hat{P}_{\bm{q}\delta} .
\end{gathered}
\end{equation}
Then 
\begin{equation}
\begin{gathered}
    [\hat{Q}_{\bm{k}\lambda},\hat{\Pi}_{\bm{q}\gamma}] = [(V^\dagger_{\bm{k}}\hat{\bm{U}}_{\bm{k}})^{\lambda} ,( V_{\bm{q}} \hat{\bm{P}}_{\bm{q}})^{\gamma} ]  = \sum_{\sigma \delta} [V^\dagger_{\bm{k}}]^{\lambda \sigma } V^{\gamma \delta}_{\bm{q}} [\hat{U}_{\bm{k}\sigma},\hat{P}_{\bm{q}\delta}] = \\
    \sum_{\sigma \delta} [V^\dagger_{\bm{k}}]^{\lambda \sigma } V^{\gamma \delta}_{\bm{q}} \sqrt{\frac{m_\sigma}{m_\delta}} [\hat{u}_{\bm{k}\sigma},\hat{p}_{\bm{q}\delta}] = i\hbar\delta_{\bm{k}\bm{q}} \sum_{\sigma \delta} [V^\dagger_{\bm{k}}]^{\lambda \sigma } V^{\gamma \delta}_{\bm{q}} \sqrt{\frac{m_\sigma}{m_\delta}} \delta_{\sigma \delta} = i\hbar\delta_{\bm{k}\bm{q}} \sum_{\sigma} [V^\dagger_{\bm{k}}]^{\lambda \sigma } V^{\gamma \sigma}_{\bm{q}} .
\end{gathered}
\end{equation}
For $\bm{k}\neq\bm{q}$ the commutator vanishes as required. For $\bm{k}=\bm{q}$.
\begin{equation}
\begin{gathered}
    \sum_{\sigma} [V^\dagger_{\bm{k}}]^{\lambda \sigma } V^{\gamma \sigma}_{\bm{k}} = \sum_{\sigma} \bar{V}^{\sigma \lambda }_{\bm{k}} V^{\gamma \sigma}_{\bm{k}} = \delta_{\lambda \gamma}
\end{gathered}
\end{equation}
which follows from orthonormality of the unitary matrices (i.e. of the eigenvectors of the dynamical matrix). So then finally 
\begin{equation}
\begin{gathered}
    [\hat{Q}_{\bm{k}\lambda},\hat{\Pi}_{\bm{q}\gamma}] = i \hbar \delta_{\bm{k}\bm{q}} \delta_{\lambda \gamma} 
\end{gathered}
\end{equation}
which has the same structure as the usual displacement and momentum commutator, $[\hat{u},\hat{p}]=i\hbar$.

\subsection{Second quantization}

In our new basis, the Hamiltonian is \cite{dove1993introduction}
\begin{equation}
\begin{gathered}
    \hat{H} = \frac{1}{2} \sum_{\bm{q} \lambda} \left[  \hat{\Pi}^{\dagger}_{\bm{q}\lambda} \hat{\Pi}_{\bm{q}\lambda} + \omega^2_{\bm{q}\lambda} \hat{Q}^\dagger_{\bm{q}\lambda} \hat{Q}_{\bm{q}\lambda} \right] .
    \label{eq:normal_coord_hamiltonian}
\end{gathered}
\end{equation} 
We want to change basis one last time to ``second quantized" creation and annihilation operators:
\begin{equation}
\begin{gathered}
    \hat{a}^\dagger_{\bm{q}\lambda} = \sqrt{\frac{\omega_{\bm{q}\lambda}}{2\hbar}}\left( \hat{Q}^\dagger_{\bm{q}\lambda} - \frac{i}{\omega_{\bm{q}\lambda}} \hat{\Pi}_{\bm{q}\lambda} \right) \quad \textrm{and} \quad
    \hat{a}_{\bm{q}\lambda} = \sqrt{\frac{\omega_{\bm{q}\lambda}}{2\hbar}}\left( \hat{Q}_{\bm{q}\lambda} + \frac{i}{\omega_{\bm{q}\lambda}} \hat{\Pi}^\dagger_{\bm{q}\lambda} \right) .
\end{gathered}
\end{equation}
The commutation relation for these things is
\begin{equation}
\begin{gathered}
    [\hat{a}_{\bm{k}\gamma},\hat{a}^\dagger_{\bm{q}\lambda}] = -i \sqrt{\frac{\omega_{\bm{k}\gamma} \omega_{\bm{q}\lambda}}{4\hbar^2}} \left( \frac{1}{\omega_{\bm{q}\lambda}} [\hat{Q}_{\bm{k}\gamma}, \hat{\Pi}_{\bm{q}\lambda}] + \frac{1}{\omega_{\bm{k}\gamma}} [\hat{Q}^\dagger_{\bm{q}\lambda} , \hat{\Pi}^\dagger_{\bm{k}\gamma}] \right) \\
    = -i \sqrt{\frac{\omega_{\bm{k}\gamma} \omega_{\bm{q}\lambda}}{4\hbar^2}} \left( \frac{i \hbar \delta_{\bm{k}\bm{q}} \delta_{\lambda \gamma} }{\omega_{\bm{q}\lambda}} + \frac{i \hbar \delta_{-\bm{k},-\bm{q}} \delta_{\lambda \gamma} }{\omega_{\bm{k}\gamma}} \right) ,
\end{gathered}
\end{equation}
i.e. 
\begin{equation}
\begin{gathered}
    [\hat{a}_{\bm{k}\gamma},\hat{a}^\dagger_{\bm{q}\lambda}] = \delta_{\bm{k}\bm{q}} \delta_{\lambda \gamma} ,
\end{gathered}
\end{equation}
which is indeed the well-known commutation relation for second quantized phonon operators \cite{mahan2013many,ziman1969elements,kittel1963quantum}. Inverting the transformation (and recalling that $\hat{\Pi}^\dagger_{\bm{q}\lambda}=\hat{\Pi}_{-\bm{q}\lambda}$ and $\hat{Q}^\dagger_{\bm{q}\lambda}=\hat{Q}_{-\bm{q}\lambda}$):
\begin{equation}
\begin{gathered}
    \hat{Q}_{\bm{q}\lambda} = \sqrt{\frac{\hbar}{2\omega_{\bm{q}\lambda}}} (\hat{a}^\dagger_{-\bm{q}\lambda} + \hat{a}_{\bm{q}\lambda}) \qquad \qquad
    \hat{Q}^\dagger_{\bm{q}\lambda} = \sqrt{\frac{\hbar}{2\omega_{\bm{q}\lambda}}} ( \hat{a}^\dagger_{\bm{q}\lambda} + \hat{a}_{-\bm{q}\lambda} ) \\
    \hat{\Pi}_{\bm{q}\lambda} = i \sqrt{\frac{\hbar\omega_{\bm{q}\lambda}}{2}} (\hat{a}^\dagger_{\bm{q}\lambda} - \hat{a}_{-\bm{q}\lambda}) \qquad \qquad
    \hat{\Pi}^\dagger_{\bm{q}\lambda} = i \sqrt{\frac{\hbar\omega_{\bm{q}\lambda}}{2}} (\hat{a}^\dagger_{-\bm{q}\lambda} - \hat{a}_{\bm{q}\lambda} ) .
    \label{eq:second_quantized_normal_coord_1}
\end{gathered}
\end{equation}
Let's insert these into the Hamiltonian:
\begin{equation}
\begin{gathered}
    \hat{H} = \frac{1}{2} \sum_{\bm{q} \lambda} \left[  \hat{\Pi}^{\dagger}_{\bm{q}\lambda} \hat{\Pi}_{\bm{q}\lambda} + \omega^2_{\bm{q}\lambda} \hat{Q}^\dagger_{\bm{q}\lambda} \hat{Q}_{\bm{q}\lambda} \right] = \\
    \frac{1}{2} \sum_{\bm{q}\lambda} \frac{\hbar \omega_{\bm{q}\lambda}}{2}  \left( 2 \hat{a}^\dagger_{\bm{q}\lambda} \hat{a}_{\bm{q}\lambda} + 2 \hat{a}^\dagger_{-\bm{q}\lambda} \hat{a}_{-\bm{q}\lambda} + 2  \right) = \frac{1}{2} \sum_{\bm{q}\lambda} \hbar \omega_{\bm{q}\lambda} \left( 2 \hat{a}^\dagger_{\bm{q}\lambda} \hat{a}_{\bm{q}\lambda} + 1  \right) .
\end{gathered}
\end{equation} 
To summarize, the second quantized phonon Hamiltonian is 
\begin{equation}
\begin{gathered}
    \hat{H} = \sum_{\bm{q}\lambda} \hbar\omega_{\bm{q}\lambda} \left(  \hat{a}^\dagger_{\bm{q}\lambda} \hat{a}_{\bm{q}\lambda} + \frac{1}{2} \right) .
\end{gathered}
\end{equation}

For many purposes, e.g. working with the Hamiltonian in \cref{eq:phonon_quantum_hamiltonian} or studying electron-phonon coupling which is $\sim \hat{u}$, we need to write the displacements in terms of second quantized operators:
\begin{equation}
\begin{gathered}
    \hat{u}^\mu_{i\alpha} = \frac{1}{\sqrt{m_\alpha N}}  \sum_{\bm{q}\eta} \hat{Q}_{\bm{q}\eta} V^\mu_{\bm{q}\eta,\alpha}
    \exp(i\bm{q}\cdot\bm{r}^{(0)}_{i\alpha}) 
\end{gathered}
\end{equation} 
with $\eta$ labelling the $3n$ branches, $\alpha$ labelling the $n$ atoms, and $\mu$ labelling the Cartesian directions. In analogy with the classical theory above, we call $V^\mu_{\bm{q}\eta,\alpha} = \epsilon^\mu_{\bm{q}\eta,\alpha}$ the eigenvectors of the dynamical matrix. Using \cref{eq:second_quantized_normal_coord_1}, the displacements can be written:
\begin{equation}
\begin{gathered}
    \hat{\bm{u}}_{i\alpha} = \sum_{\bm{q}\eta} \sqrt{\frac{\hbar}{2 m_\alpha \omega_{\bm{q}\eta} N}} \bm{\epsilon}_{\bm{q}\eta,\alpha} (\hat{a}^\dagger_{-\bm{q}\eta}+\hat{a}_{\bm{q}\eta})
    \exp(i\bm{q}\cdot\bm{r}^{(0)}_{i\alpha}) 
    \label{eq:quantum_normal_mode_displacements}
\end{gathered}
\end{equation} 
with $\bm{\epsilon}_{\bm{q}\eta,\alpha}= \epsilon^x_{\bm{q}\eta,\alpha} \hat{\bm{e}}^x_\alpha + \epsilon^y_{\bm{q}\eta,\alpha} \hat{\bm{e}}^y_\alpha + \epsilon^z_{\bm{q}\eta,\alpha} \hat{\bm{e}}^z_\alpha $. Recalling the time dependence of the creation and annihilation operators is $\hat{a}^\dagger_{\bm{q}\eta}(t) = \hat{a}^\dagger_{\bm{q}\eta}\exp(i\omega_{\bm{q}\eta}t)$ and $\hat{a}_{\bm{q}\eta}(t) = \hat{a}_{\bm{q}\eta}\exp(-i\omega_{\bm{q}\eta}t)$, we can write
\begin{equation}
\begin{gathered}
    \hat{\bm{u}}_{i\alpha}(t) = \sum_{\bm{q}\eta} \sqrt{\frac{\hbar}{2 m_\alpha \omega_{\bm{q}\eta} N}} \bm{\epsilon}_{\bm{q}\eta,\alpha} [\hat{a}^\dagger_{-\bm{q}\eta}\exp(i\omega_{\bm{q}\eta}t)+\hat{a}_{\bm{q}\eta}\exp(-i\omega_{\bm{q}\eta}t)]
    \exp(i\bm{q}\cdot\bm{r}^{(0)}_{i\alpha}) \\
    = \frac{1}{\sqrt{ m_\alpha N}} \sum_{\bm{q}\eta} \hat{Q}_{\bm{q}\eta}(t) \bm{\epsilon}_{\bm{q}\eta,\alpha}
    \exp(i\bm{q}\cdot\bm{r}^{(0)}_{i\alpha}) 
\end{gathered}
\end{equation} 
with 
\begin{equation}
\begin{gathered}
    \hat{Q}_{\bm{q}\lambda}(t) = \sqrt{\frac{\hbar}{2\omega_{\bm{q}\lambda}}} [\hat{a}^\dagger_{-\bm{q}\eta} \exp(i\omega_{\bm{q}\eta}t)+\hat{a}_{\bm{q}\eta}\exp(-i\omega_{\bm{q}\eta}t)]
    \label{eq:second_quantized_normal_coord}
\end{gathered}
\end{equation}
the "normal coordinate" with explicit time dependence. 

Finally, referring back to \cref{eq:normal_coord_hamiltonian}, we see why normal coordinates are useful. \cref{eq:normal_coord_hamiltonian} is simply a sum over $nD$ independent harmonic oscillators; each oscillates at its own frequency independently of the others. The subtlety is that, rather than displacements of isolated oscillators, the displacements of the normal modes correspond to a distortion of the entire crystal in a particular pattern: see \cref{eq:inverted_normal_coordinate}. These are the "phonons".

\subsection{Remarks}

In this section, we saw that, for a crystal at low enough temperature, knowledge of the second derivatives of the potential energy, i.e. the force-constants, is enough to calculate the dynamics. Rather than explicitly calculating the trajectories, the goal of "lattice dynamics" is to calculate the phonon dispersions (i.e. the band-structure and eigenvalues). Given initial conditions, the eigenvectors and dispersions can be used to reconstruct the trajectories anyway. In \cref{sec:harmonic_cross_section}, we saw that we don't need trajectories to calculate the neutron scattering intensities in the harmonic approximation; the eigenvectors and dispersions are all that are needed. It doesn't even matter if the system is classical or quantum; the eigenvectors and dispersions are the same between the two. 

The force-constants can be calculated in a multitude of ways. The most accurate and direct method is to calculate the force-constants from DFT. We will look at this in more detail in the next chapter. A much cheaper method is to use an empirical potential like in \cref{sec:2d_perovskite}. If we have measured the neutron intensities, we can even fit the parameters of the empirical model to reproduce the measured intensities. This was the usual approach to lattice dynamics calculations before DFT calculations were common place.

\section{Molecular dynamics}
\label{sec:md}

In the last section, we saw how to efficiently study the dynamics of atoms in a crystalline solid. The result was "phonons". What if the material isn't perfectly ordered? Salient examples are e.g. liquids and gases, disordered materials, or even (as we will see later) hybrid crystals: materials with molecules on certain lattice sites, adding additional degrees of freedom (cf. \cref{chp:mapi_diffuse}). We now look at how to handle these materials.  

The approach is, conceptually, very straightforward: molecular dynamics is direct numerical integration of the atomic trajectories using e.g. the \emph{velocity-Verlet} method. Solving the trajectories is actually quite easy. The difficult part is calculating the forces on the atoms due to interactions. 

We mention that molecular dynamics solves the \emph{classical} dynamics of the atoms. At or above room temperature for most materials, this is a very good approximation (see the foornote in \cref{sec:classical_cross_section}). Regardless of whether we use quantum mechanics to calculate $U(\bm{r})$ (see below), the atomic dynamics.

The explicit form of $U(\bm{r})$ is arbitrary; as long as we can calculate the forces, we can do molecular dynamics. There are, however, specific names for the different ways we calculate the forces: if we calculate forces (i.e. $E_{BO}$) from DFT, we call it ab-initio molecular dynamics (AIMD). If we use a tight-binding approximation, we still considered it "quantum" and call the method tight-binding molecular dynamics (TBMD). If we go further and make an empirical approximation for the forces, we call it classical molecular dynamics (MD). Note, it common to approximate both the Coulomb energy and $E_{BO}$ as a single empirical term: e.g. the Tersoff, Morse, and Lennard-Jones potentials. 

No matter how we calculate the forces, solving the classical Newton equations once the forces are known is actually very cheap computationally. Let us do that now. 

\subsection{The velocity-Verlet algorithm}

This section follows Refs. \cite{tuckerman2023statistical,thijssen2007computational,allen2017computer}. Let $\bm{r}(t)$ represent the trajectory of atom. We make no assumptions about the structure of the matter (though the system is usually placed in a large box with periodic boundary conditions). The simplest method to study the time-dependence is Taylor-Expansion:
\begin{equation}\begin{split}
    \bm{r}(t+\delta t) = \bm{r}(t) + \delta t \dot{\bm{r}}(t) + \frac{1}{2}(\delta t)^2 \ddot{\bm{r}}(t) + \cdots .
\end{split}\end{equation}
The dot $\dot{\bm{r}}(t) = \partial \bm{r}(t)/\partial t \equiv \bm{v}(t)$ represents the time-derivative. We use Newton's equation to rewrite
\begin{equation}\begin{split}
    \bm{r}(t+\delta t) \approx \bm{r}(t) + \delta t \bm{v}(t) + \frac{(\delta t)^2}{2m} \bm{F}(t) .
    \label{eq:verlet_position}
\end{split}\end{equation}
There is a separate \cref{eq:verlet_position} for every atom in the simulation.
Similarly,
\begin{equation}\begin{split}
    \bm{r}(t) \approx \bm{r}(t+\delta t) - \delta t \bm{v}(t+\delta t) + \frac{(\delta t)^2}{2m} \bm{F}(t+\delta t)
\end{split}\end{equation}
evolves backwards in time from time $t+\delta t$. We can plug the former into the latter to find
\begin{equation}\begin{split}
    \bm{v}(t+\delta t) = \bm{v}(t) + \frac{\delta t}{2m} \left[ \bm{F}(t) +  \bm{F}(t+\delta t) \right] .
    \label{eq:verlet_velocity}
\end{split}\end{equation}
There are many ways to manipulate these equations resulting in different expressions, e.g. the conventional Verlet algorithm. The decomposition used here is called the "velocity-Verlet algorithm". 

We assume that we can calculate the forces, $\bm{F}(t)$, at any instant in time using DFT or an empirical potential as long as we know the positions at that time. The velocity-Verlet algorithm works as follows: 
\begin{enumerate}
    \item we make an initial guess for the positions, $\bm{r}(t_i)$, and velocities,  $\bm{v}(t_i)$. 
    \item We calculate the forces, $\bm{F}(t_i)$.
    \item We use \cref{eq:verlet_position} to update the positions to $t_i+\delta t = t_{i+1}$
    \item We calculate the forces at $\bm{F}(t_{i+1})$.
    \item We use \cref{eq:verlet_velocity} to update the velocities to $t_{i+1}$.
    \item We use the new positions, $\bm{r}(t_{i+1})$, and velocities, $\bm{v}(t_{i+1})$, to repeat the loop, and so on.
\end{enumerate}

The initial guess for the positions is made on physical grounds. Usually we guess the equilibrium configuration. The velocities are guessed according to Maxwell-Bolztmann distribution:
\begin{equation}\begin{split}
    f(v) = \sqrt{\frac{m}{2\pi k_B T}}\exp\left(-\frac{mv^2}{2k_B T}\right) .
\end{split}\end{equation}
I.e. the distribution of speeds is Gaussian. The width of the distribution is related to the temperature, $T$. $k_B$ is Boltzmann's constant. The real trouble is inverting the equation to generate a set of speeds, $\bm{v}_i$, that satisfy the distribution. Methods exist for this, e.g. the Box-Muller sampling method (see e.g. Section 3.8 in Tuckermann \cite{tuckerman2023statistical}). Usually in already available codes, you just tell it the temperature you want and it handles the rest. 

Of course, if we guess a speed distribution but the positions aren't consistent, the system will be \emph{out of equilibrium}. I.e. the potential energy will be zero (or something else that is wrong) while the kinetic energy is consistent with the guess for the speeds. If energy is conserved, some of the kinetic energy will dissipate into the bonds and the speeds will drop; the kinetic temperature of the system will be lower than what we requested. To fix this, we usually \emph{equilibrate} the system using a thermostat. 

\subsection{Finite temperature; thermostatting}

To control the temperature of the atoms in a molecular dynamics simulation, we have to couple them to a "bath". There are many formulations for this \cite{allen2017computer,tuckerman2023statistical,thijssen2007computational}. Most of these methods result in adding a friction term to the equations of motion that add or subtract energy to enforce that the kinetic energy reproduces the desired temperature. Here, we discuss the Nos\'e-Hoover method. In short, we add a friction term to the equations of motion:
\begin{equation}\begin{split}
    \frac{d\bm{r}}{dt} & = \frac{\bm{p}}{m} \\
    \frac{d\bm{p}}{dt} & = \bm{F} - \eta \bm{p} \\
    \frac{d \eta}{dt} & = \frac{1}{Q} \left[ \sum_i v^2_i - 3N \frac{k_B T}{2} \right] .
\end{split}\end{equation}
$\eta$ is a friction-like virtual degree of freedom that couples to the momenta of the real particles. Note that if the kinetic energy (divided by mass) of the system is equal to the desired kinetic energy (divided by mass), $3N k_B T /2$, then $\dot{\eta}=0$ and energy stops flowing between the bath ($\eta$) and the real system. $Q$ is the "mass" of the friction degree of freedom (it's called the Nos\'e mass). The Nos\'e mass must be chosen carefully since it controls the rate at which the bath and system exchange energy; if it is too large, the exchange is too fast and the system can be trapped in a region of phase space for a long time. If it is too small, the exchange is too slow and large oscillations between the bath and system can occur. 

Besides the normal interaction of the system trajectories, the virtual friction parameter also has to be updated. It turns out that verlocity-Verlet algorithm can be adapted to integrate the Nos\'e-Hoover equations of motion. Almost all codes do it automatically if you ask it to; all the user most do is specify the desired temperature and $Q$ (or a related parameter).

\subsection{Remarks}

In this section, we have seen that once we know how to calculate the forces on atoms (e.g. with DFT or an empirical potential) it is rather straightforward to solve numerically for their trajectories through time. I.e. we end up with a set of positions, $\{\bm{r}_1(t), \cdots, \bm{r}_N(t), \bm{r}_1(t+\delta t), \cdots, \bm{r}_N(t+\delta t), \bm{r}_1(t+n\delta t), \cdots, \bm{r}_N(t+n\delta t) \}$ for $n$ time-steps where $n$ determines the length of simulation in time. In \cref{sec:classical_cross_section}, we see how these trajectories can be used to calculate the intensity of neutrons scattered from a system of atoms. 

We now turn to apply the tools developed so far to study energy materials in the rest of this thesis.

\chapter{Bond-stretching phonons in an insulating cuprate from DFT+U}
\label{chp:lco_lda_u}

Typical density functional theory calculations that wrongly predict undoped cuprates to be metallic also predict Cu-O half and full-breathing phonon energies that are significantly softer than observed by neutron scattering, presumably because of weak on-site Coulomb repulsion on the Cu 3d orbitals (cf. \cref{sec:band_gap_problem}). I used DFT+U calculations with antiferromagnetic supercells of La$_\text{2}$CuO$_\text{4}$ to establish correlation between the on-site repulsion strength, tuned via adjusting the value of U (see \cref{sec:dft+U}), and phonon dispersions. I found that breathing and half-breathing phonons (see \cref{sec:2d_perovskite}) reach experimental values when U is tuned to obtain the correct optical gap and magnetic moments. I demonstrate that using distorted supercells within DFT+U is a promising framework to model phonons in undoped cuprates and other perovskite oxides with complex, interrelated structural and electronic degrees of freedom.

This section closely follows the corresponding paper, Sterling et. al \cite{sterling2021effect}, on which I was first author and Proj. Dmitry Reznik was PI. This project was purely computational, but made use of inelastic neutron scattering and inelastic x-ray scattering data from the literature.

\section{The Effect of Charge-Gap on Bond-Stretching Phonons in La$_2$CuO$_4$}

Phonons in the cuprates have been studied mostly in the doped phases in an effort to explain superconductivity and CDW physics \cite{McQueeney99,park2014evidence,reznik2008temperature,ahmadova2020phonon,giustino2008small,lebert20,le_tacon_inelastic_2014,pintschovius2005electron,reznik2012phonon}. However, it is becoming increasingly apparent that phonons in the \emph{undoped} cuprates merit further investigation. For example, La$_\text{2}$CuO$_\text{4}$ hosts interesting phonon physics even in the undoped phase. There are energy lowering structural distortions \cite{boni1988lattice,birgeneau1987soft,billinge1996probing,bozin2015reconciliation,sapkota2021reinvestigation}, spin-orbit induced magnetic behavior \cite{shekhtman1992moriya,thio1994weak}, and most recently a large thermal Hall effect has been observed and attributed to phonons \cite{grissonnanche2019giant,grissonnanche2020chiral}. Unfortunately, the overwhelming majority of previous phonon calculations for undoped cuprates used density functional theory (DFT) in either the local density approximation (LDA) or the generalized gradient approximation (GGA) \cite{cohen1990first,wang1999first,krakauer1993large,singh1996phonons,lebert2020doping,miao2018incommensurate,ahmadova2020phonon}.

The undoped cuprates are insulating and antiferromagnetic, but the LDA and GGA predict them to be metallic with either no magnetism or unrealistically small magnetic moments \cite{singh1991gradient,ambrosch1991local,giustino2008small,mattheiss1987electronic,yu1987electronically}. The disagreement is similar for the lattice dynamics. The LO bond-stretching phonons in undoped cuprates calculated from LDA or GGA do not agree with experiment but rather disperse steeply downward. Usually the bond-stretching dispersions calculated for undoped compounds fortuitously agree with experiments on overdoped compounds \cite{park2014evidence,pintschovius2006oxygen,pintschovius2005electron} and calculated dispersions are often presented along side experiments on overdoped materials \cite{ahmadova2020phonon,miao2018incommensurate,giustino2008small,lebert2020doping}. Most other branches are unaffected by doping and already match experiments in the LDA and GGA \cite{miao2018incommensurate,park2014evidence,pintschovius2006oxygen,pintschovius2005electron,ahmadova2020phonon,giustino2008small,lebert2020doping}. Calculating correct bond-stretching phonon dispersions of most undoped cuprates remains elusive.

The DFT+U method, which is an extension to DFT that includes an adjustable Hubbard-U like onsite repulsion on correlated orbitals (e.g. the Cu 3d orbitals in cuprates), is known to predict reasonable gaps and moments for the undoped cuprates and has been extensively applied to electronic structure calculations across a wide range of doping \cite{zhang2007electron,pesant2011dft+,czyzyk1994local,anisimov1992spin,wei1994electronic,svane1992electronic,anisimov2004computation,elfimov2008theory,puggioni2009fermi,oh2011fermi}. However, aside from an earlier DFT+U calculation that succeeded in calculating correct energies of a few zone boundary phonons in CaCuO$_\text{2}$ \cite{zhang2007electron}, the phonon spectrum of DFT+U calculations in cuprates is mostly unexplored. 

Inspired by the apparent success for the electronic structure, I use the DFT+U method to investigate the interplay between the Cu 3d on-site repulsion strength, tuned by varying U, and phonon dispersions. I already demonstrated spectacular agreement of the calculated acoustic phonons and nearby optic branches with experiment in another paper \cite{sapkota2021reinvestigation}. These low lying branches are nearly independent of U and agree well even in plain GGA. This is not surprising since the low energy phonons mainly involve motion of La and do not induce substantial charge redistribution around the Cu atoms. As such, dispersion of the low energy phonons are not a valuable metric for the accuracy of DFT+U. In this paper, I focus on the Cu-O bond stretching phonons which are expected to depend strongly on the on-site repulsion and are therefore most suitable to assess the accuracy of DFT+U. I demonstrate that tuning U to U=8 eV, which reproduces the experimental charge gap and magnetic moments, brings phonon dispersions in agreement with experiment when a realistic antiferromagnetic supercell of La$_\text{2}$CuO$_\text{4}$ is used. I also find that the charge fluctuations induced by the breathing phonons near the Cu atoms are reduced at U=8 eV, consistent with the hardening of the bond stretching phonons.

\section{Computational details}

I choose La$_\text{2}$CuO$_\text{4}$ for my study since the Cu-O bond stretching phonons have been already been measured by neutrons for all high symmetry wavevectors and ranges of doping \cite{park2014evidence,pintschovius2006oxygen,pintschovius2005electron,miao2018incommensurate}. La$_\text{2}$CuO$_\text{4}$, like many perovskite-oxides, contains energy-lowering distortions that are static in the insulating phase resulting in a symmetry-lowered supercell \cite{zhang2020competing,furness2018accurate,boni1988lattice,birgeneau1987soft,billinge1996probing,bozin2015reconciliation,sapkota2021reinvestigation,zhang2020symmetry,varignon2019origin}. The structural phases of La$_\text{2}$CuO$_\text{4}$ have been explored using the meta-GGA SCAN functional \cite{sun2015strongly} which also predicts gaps and moments in cuprates with accuracy comparable to DFT+U \cite{zhang2020competing,furness2018accurate}. It was shown that the insulating ground state predicted for undoped La$_\text{2}$CuO$_\text{4}$ is further stabilized by including static lattice distortions, consistent with the observed soft modes \cite{boni1988lattice,birgeneau1987soft,billinge1996probing,bozin2015reconciliation,sapkota2021reinvestigation}. The distortions occur either in the form of Cu-O octahedral rotations resulting in the low-temperature-orthorhombic phase (LTO) or octahedral tilts resulting in the low-temperature-tetragonal (LTT) phase \cite{varignon2019origin,varignon2019mott,trimarchi2018polymorphous,zhang2020symmetry,lane2018antiferromagnetic,furness2018accurate} (cf. \cref{sec:soft_modes}). Experiments show that the low-temperature structure of undoped La$_\text{2}$CuO$_\text{4}$ is LTO on average, but recent investigations showed that the local structure is likely fluctuating between the LTT and LTO phases \cite{lee2021hidden,bozin2015reconciliation,sapkota2021reinvestigation} and SCAN calculations showed that the energy difference between the LTO and LTT phases is comparable to computational error \cite{furness2018accurate}. Nevertheless, lattice dynamics experiments use wave vectors from the high-temperature-tetragonal (HTT) structure \cite{pintschovius2006oxygen,park2014evidence,miao2018incommensurate,sapkota2021reinvestigation} (\cref{fig:cells} a), so I used the LTT cell for convenience (\cref{fig:cells} b). 

\subsection{DFT and phonons}

For my DFT calculations, I used the projector augmented wave (PAW) method \cite{kresse1999ultrasoft,blochl1994projector} in the Vienna Ab-initio Simulation Package ($\textsc{vasp}$) \cite{kresse1996efficiency,kresse1996efficient,kresse1993ab}. I choose DFT+U over SCAN. Both predict reasonably accurate electronic structures, but DFT+U allows us to tune the on-site repulsion by changing the U value. The U correction was applied to the Cu 3d orbitals using the method proposed by Dudarev et al. where only a single parameter U$_{\text{eff}}\equiv$ U is used \cite{dudarev1998electron}. For exchange and correlation, I use the LDA \cite{perdew1981self} (i.e. LDA+U). To calculate the phonon dispersions, I used the finite-difference approach (see \cref{sec:finite_diffs}) in the code $\textsc{phonopy}$ \cite{phonopy} with 2x2x1 supercells for all calculations. To analyze the effect of U on the charge density, I also calculated the charge density redistribution, $\Delta n = n-n^{ph}$, where $n$ is the self-consistent charge density in the unmodulated (i.e. fully relaxed) structure and $n^{ph}$ is the self-consistent charge density calculated with a phonon eigenvector frozen into the unit cell with a small amplitude. 

\subsection{Phonon unfolding}

Since my calculations are based on the larger structurally distorted 28 atom cell of the low temperature AFM phases, the Brillouin zone is smaller with with 84 branches that are closely spaced in energy with numerous anticrossings (\cref{fig:cells} c). To relate the complicated dispersions in the LTT phase to the HTT cell, I calculated the inelastic neutron scattering structure factors S($\bf{Q}$,$\omega$) predicted by DFT+U in the reciprocal lattice units of the HTT cell (see \cref{sec:harmonic_cross_section}). The color-map in \cref{fig:cells} c)  shows that only a few branches contribute to the scattering intensity in agreement with experiments. The intensity around 85 meV in \cref{fig:cells} c) is from the half-breathing bond-stretching phonons. An alternative method, usually called "unfolding", has been used to calculate effective band structures from supercell electron \cite{ku2010unfolding,popescu2010effective,popescu2012extracting} and phonon \cite{allen2013recovering,ikeda2018temperature,ikeda2017mode,samolyuk2021role,mu2020unfolding,kormann2017phonon} dispersions in the past. My intuitive method benefits from direct comparability with experiment.

\section{Results}

\begin{figure}[t!]
\centering
\includegraphics[width=0.6\linewidth]{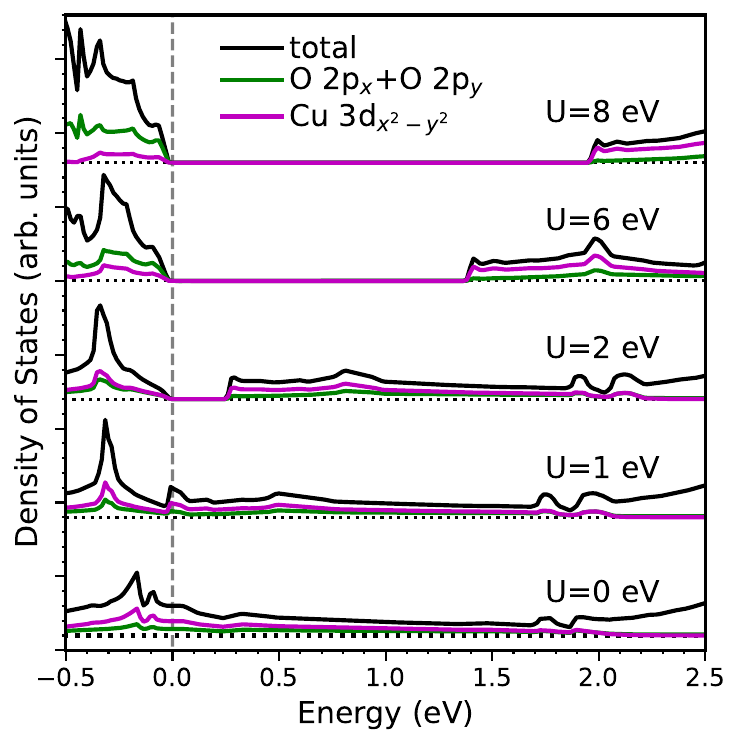}
        \caption{Electronic densities of states calculated using the U values indicated in the figure. The data are offset vertically for clarity. The colors indicate the orbitals the densities of states are projected onto. For U = 2, 6, and 8 eV the electronic structure is insulating with gaps $\approx$ 0.2, 1.4, and 2 eV respectively. The magnetic moments for U = 1, 2, 6, and 8 eV are $\pm$ 0.21, 0.33, 0.53, and 0.61 $\mu_{B}$ respectively. For U=0 and 1 eV, the ground state is metallic and non-magnetic}
\label{fig:dos}
\end{figure}

\subsection{Electronic structure}

U=0 eV (i.e. plain LDA), gives the expected non-magnetic, metallic ground state. There are four electronic regimes for nonzero U: (i) AFM but metallic (U=1 eV). (ii) AFM and insulating, but with unrealistically small gaps and moments (2$\geq$U$\geq$4 eV). (iii) AFM and insulating with reasonable gaps (5$\geq$U$\geq$8 eV). (iv) AFM and insulating with unrealistically large gaps and moments (U$>$8 eV). The electronic charge gap opens with U=2 eV ( \cref{fig:dos} ) increasing with increasing U. U=6 eV gives gap/magnetic moments of 1.4 eV / 0.53 $\mu_\text{B}$, and U=8 eV gives 2.0 eV / 0.61 $\mu_{B}$ respectively. These values agree with experiments \cite{yamada1987effect,tranquada1988antiferromagnetism,vaknin1987antiferromagnetism,mitsuda1987confirmation,thio1990determination,uchida1991optical,cooper1990optical,kastner1998magnetic,ono2007strong} and previous calculations using DFT+U \cite{zhang2007electron,pesant2011dft+,czyzyk1994local,anisimov1992spin,wei1994electronic,svane1992electronic}. Gaps and moments that come out of U$\approx$8 eV have the best agreement with experiment. Note that the ordered magnetic moment calculated in this way should be larger than measured, since quantum fluctuations that suppress the moment in real materials are not included in the calculation. I calculated the phonon dispersions using U = 0, 1, 2, 5, 6, and 8 eV.

\begin{figure}[t!]
\centering
\includegraphics[width=0.6\linewidth]{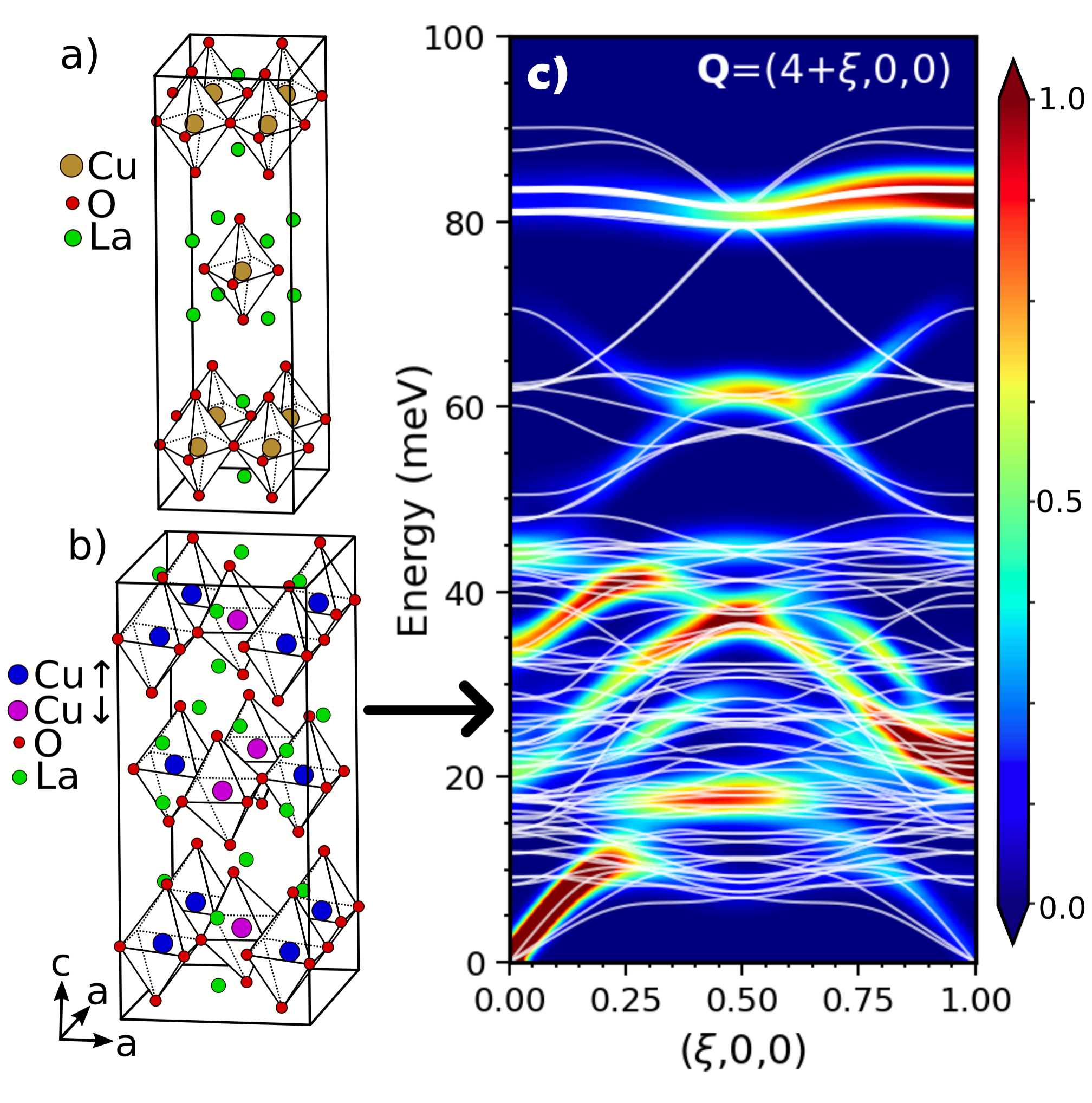}
        \caption{a) The HTT cell of La$_\text{2}$CuO$_\text{4}$. b) The LTT cell with correct AFM ordering and octahedral distortions. c) Phonon dispersions and dynamic structure factors, S($\bf{Q}$,$\omega$), calculated in the LTT phase with U=8 eV. The $\bf{Q}$=$(4+\xi,0,0)$ zone is where the bond-stretching branch was measured \cite{pintschovius2006oxygen}. The structure factors are broadened with a Gaussian with 3 meV width. The white lines are the phonon dispersions in the first Brillouin zone.}
\label{fig:cells}
\end{figure}

\subsection{Phonon vs. U}

Two Cu-O planes in the HTT and LTT cells give two bond-stretching branches. Both are degenerate in the HTT phase, but the octahedral tilts lift the degeneracy in the LTT phase. In the case of the half-breathing phonon (\cref{fig:cells} c), the lower energy branch stretches bonds that are bent, whereas the higher energy branch stretches bonds that are straight. Experiments do not observe the energy splitting which, due to weak coupling between the Cu-O layers, is too small (about 2 meV) to be resolved \cite{reznik2008temperature}. Lines in \cref{fig:summary} represent calculated dispersions of the bond-stretching phonons whose maximum intensity is in the BZ where experiments are performed. With U=8 eV, the dispersions are in striking agreement with experimental results. However the U=0 eV dispersion of the LO phonons (a,b) is significantly softer near the zone boundary than observed, consistent with previous DFT calculations \cite{cohen1990first,krakauer1993large,singh1996phonons,lebert2020doping,miao2018incommensurate,ahmadova2020phonon}. The calculated TO dispersion shown in (c) also agrees with experiment, but it is not affected by U. For U around 6 eV, the half-breathing phonon is weakly sensitive to small changes in U. The trend is nearly identical for the full-breathing branch. The improved agreement of the bond-stretching phonons in insulating La$_\text{2}$CuO$_\text{4}$ relative to the metallic U=0 eV ground state is consistent with a series of model calculations in La$_\text{2}$CuO$_\text{4}$ \cite{falter1997origin,falter1993effect,falter1995phonon,falter2000effect,falter2001nonlocal,falter2002influence,falter2006modeling,bauer2009impact}.

\begin{figure}[t!]
\centering
\includegraphics[width=0.5\linewidth]{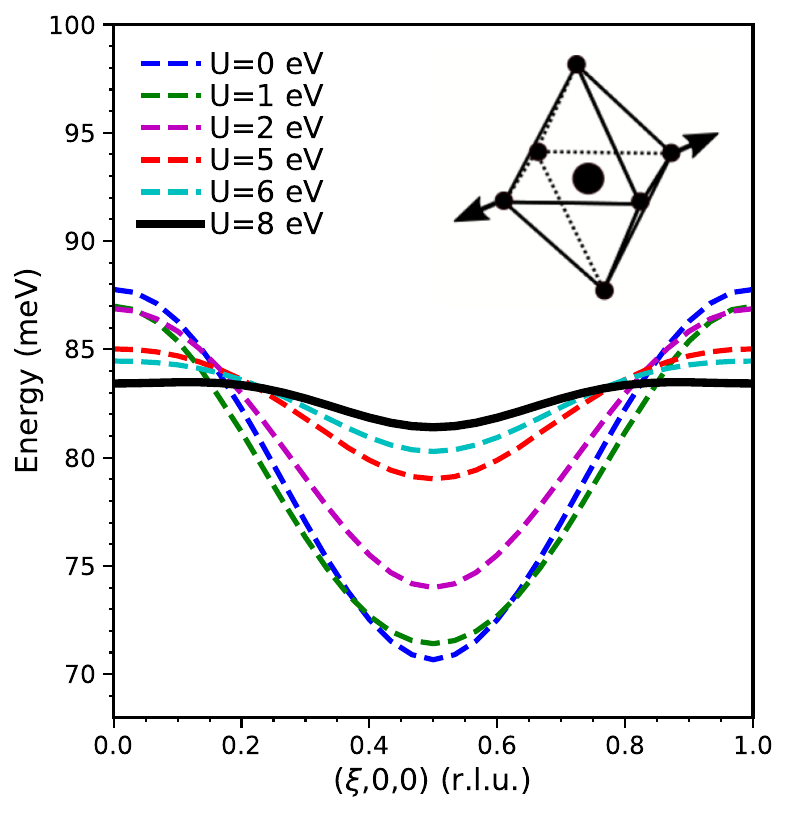}
        \caption{Half-breathing bond-stretching phonon calculated for the U values indicated in the figure. The zone boundary eigenvector is shown by the diagram in the plot. Only the branch that attained highest inelastic neutron scattering intensity in the $\bf{Q}$=$(4+\xi,0,0)$ zone is shown for each U.}
\label{fig:all_breathing}
\end{figure}

\section{Discussion}

In the bond stretching phonons, the motion of the in-plane O atoms modulates the charge density in the Cu-O bonds, pumping charge into/out-of the vicinity of the Cu atoms (see \cref{sec:2d_volume_modulation}). The amount of charge displaced by the bond stretching phonons at the zone boundary is presented in \cref{fig:chg_den_Cu}. The colormaps show the excess charge induced $\Delta n$ by bond-stretching atomic lattice displacements in a small region around the Cu atoms where the on-site potential is applied. Dark blue regions are positions in the unit cell where a (relatively) large amount of charge is depleted by the phonon; dark red regions are positions where a large amount of charge is added. I show the charge modulation calculated for the phase of the phonon displacement shown in the figure; if the phase were rotated by $\pi$, the O atoms moving away from the Cu sites would instead be moving towards them and the sign of the charge modulation would flip. However, the analysis below would be the same.

\begin{figure*}[t!]
\centering
\includegraphics[width=1\linewidth]{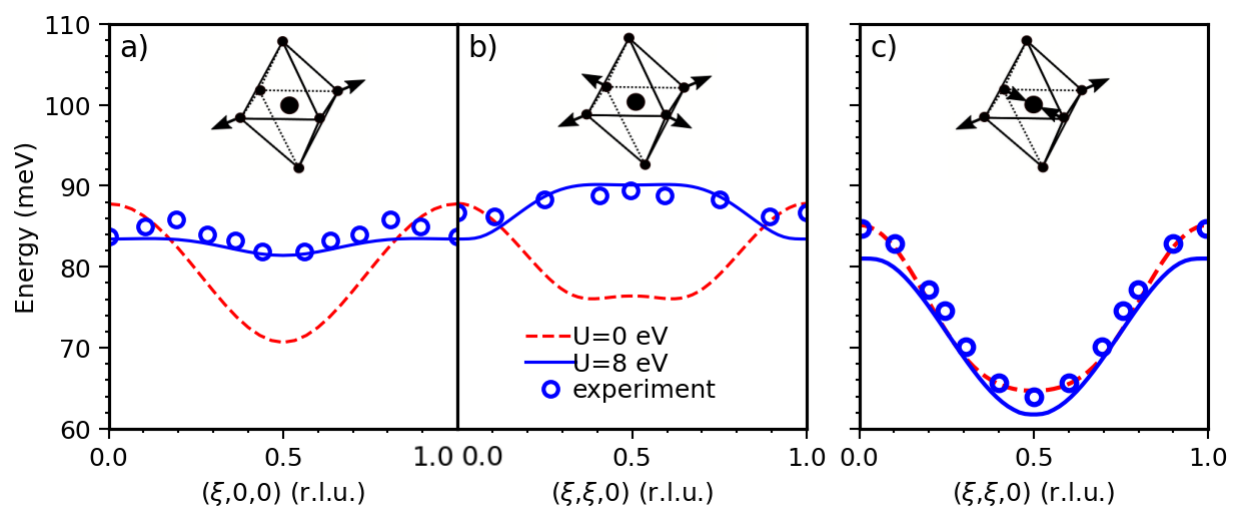}
        \caption{Bond-stretching phonon dispersions calculated using U=0 eV and U=8 eV. a) Dispersion of the half-breathing branch along the $(\xi,0,0)$ reciprocal lattice direction. b) Dispersion of the full-breathing branch along the $(\xi,\xi,0)$ reciprocal lattice direction. c) Dispersion of the quadrupolar branch along the $(\xi,\xi,0)$ reciprocal lattice direction. The dashed red lines are the dispersions calculated with U=0 eV and the solid blue lines are dispersions calculated with U=8 eV. The reciprocal lattice directions are consistent with the HTT cell used in the experiments. The blue circles are experimental results from La$_\text{2}$CuO$_\text{4}$. The experimental data along $(\xi,0,0)$ are from Park et al. \cite{park2014evidence} and along $(\xi,\xi,0)$ are from Pintschovius et al. \cite{pintschovius2006oxygen}. The zone boundary eigenvectors of these modes are indicated by the diagrams in each plot.}
\label{fig:summary}
\end{figure*}

For the LO bond stretching phonons, the amount of the charge that is pumped depends on U because it sets the energy cost to modulate the charge around the Cu atoms. If U is small, there is relatively little energy cost to pump charge into/out-of the vicinity of the Cu atoms. On the other hand, with U=8 eV, there is a substantial energy cost to change the amount of charge near the Cu atoms. The half and full-breathing phonons depend strongly on U since the motion of the in plane O atoms (top row in \cref{fig:chg_den_Cu}) changes the volume of the Cu octahedra, displacing the charge\footnote{In \cref{sec:2d_volume_modulation}, I derive explicit expressions for the volume change in a $2d$ model of the CuO plane. The derivation is conceptually useful for understanding this effect and can be used to write down a phenomenological electron-phonon interaction that is the subject of a different study.}. My calculations support this: there is a considerably larger amount charge displaced into/out-of the vicinity of Cu atoms by the LO phonons with U=0 eV (top row in \cref{fig:chg_den_Cu}) compared with U=8eV (bottom row in \cref{fig:chg_den_Cu}). On the other hand, the TO quadrupolar mode induces nearly the same charge modulation with U=0 eV and U=8 eV as discussed below. I also calculated the charge modulation using U=2 eV and the results are very similar to U=0 eV.

The phonon energy depends on the amount of screening, i.e. the energy is proportional to the amplitude of electronic charge fluctuations driven by atomic vibrations. Charges are free to redistribute between the Cu and O orbitals when U=0, but increasing U blocks these fluctuations for the half-breathing and the breathing modes, and the  amount of screening is reduced. As a result these modes harden with increasing U.

For the quadrupolar mode the motion of two in-plane O atoms outward is compensated by the inward motion of the other two in-plane O atoms, so that the eigenvector does not substantially modulate the volume of the octahedra. Note how charges depleted in two of the lobes due to this quadropolar displacement in \cref{fig:chg_den_Cu} are compensated by increased charge density in the other two, so the net occupation of the Cu site does not change. This is the reason that it is not sensitive to the value of U. A similar argument is true for the bond-stretching phonons near the zone center, whose energies also do not depend on U. I note that it was previously found that the symmetry of the zone boundary quadrupolar mode prohibits coupling to the Cu 3d orbitals \cite{falter2002influence}. On the other hand, the LO phonons are not prohibited from coupling to Cu 3d orbitals, consistent with their dependence on U \cite{giustino2008small}. 

\begin{figure}[t!]
\centering
\includegraphics[width=0.6\linewidth]{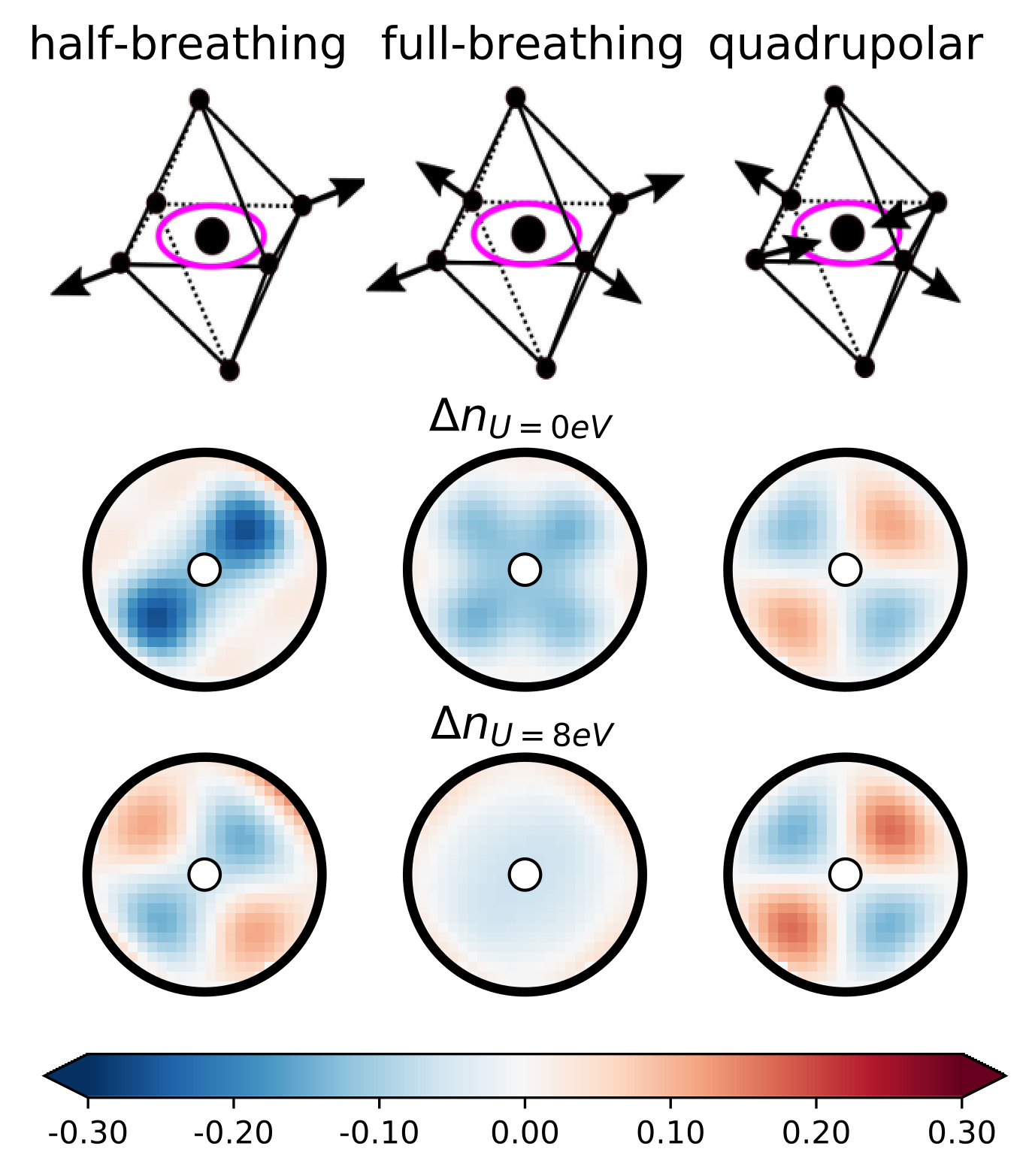}
	\caption{Comparison of the amount of charge pumped by the zone boundary phonons, $\Delta n$ (defined in the text), into the vicinity of the Cu atoms where the on-site potential is applied. Colorbar indicates the amount of charge pumped into/out-of a small volume around the Cu atoms, in units of $e/\text{\AA}^3$. The top and bottom rows of colormaps show the amount of charge displaced by the phonons calculated with U=0 eV (LDA) and U=8 eV, respectively. The displacement pattern of the Cu-O octahedra for each phonon is shown in the top row; the magenta circle shows the region around the Cu atom plotted in the colormaps. The subscript on the labels on the right indicates U used to calculate $\Delta n$ in that row.}
\label{fig:chg_den_Cu}
\end{figure}

In a nutshell, DFT+U points at the following mechanism: the hardening of the LO bond-stretching phonons with U can be understood by considering the charge density redistribution induced by the atomic displacements. With U=8 eV, modulating the charge on the Cu atom is unfavorable due to the imposed on-site repulsion. Rather than be compressed into the Cu-O octahedra, the charge in the Cu-O bonds tends to delocalize. In undoped La$_\text{2}$CuO$_\text{4}$, the delocalized charge wants to go into the Cu 3d orbitals on other sites, but transferring charge into those orbitals also costs an energy that scales with U. The charge around the Cu atoms becomes rigid (\cref{fig:chg_den_Cu}, bottom row) and the modulated charge has to be excited across the electronic gap (which also scales with U). The LO bond stretching phonons are coupled to the Cu 3d orbitals, so their energies increase. On the other hand, when holes are introduced into La$_\text{2-x}$Sr$_\text{x}$CuO$_\text{4}$ by doping, there are low-energy charge excitations that do not require changing the occupation of the Cu 3d orbitals \cite{uchida1991optical}. The charge fluctuations induced by the LO bond-stretching phonons are no longer frustrated by the on-site repulsion and the LO zone boundary phonons are soft \cite{park2014evidence,pintschovius2006oxygen}.  

For small U values, the agreement with the overdoped experimental results might lead one to form analogies between doping and varying U. In fact, it has been assumed that LDA or GGA calculations of undoped cuprates represent the overdoped compounds and the resulting bond-stretching dispersions do usually agree with experiments on overdoped cuprates \cite{krakauer1993large,miao2018incommensurate,giustino2008small,lebert2020doping}. I caution that agreement between the LDA or GGA dispersions and the overdoped compound is merely coincidental. Electron-phonon coupling quantities like line widths and spectral functions in hole doped La$_\text{2-x}$Sr$_\text{x}$CuO$_\text{4}$ are not reproduced by GGA \cite{giustino2008small,reznik2008photoemission}. A recent many-body perturbation theory calculation starting from DFT+U wave functions accurately reproduced the observed electronic spectral function and showed that the density of states at the Fermi level was different from GGA calculations \cite{li2021unmasking}. It has already been established that electronic structures from explicitly hole-doped DFT+U calculations are qualitatively different than undoped LDA or GGA Fermi surfaces \cite{anisimov2004computation,elfimov2008theory,puggioni2009fermi,oh2011fermi} and quantities like electron-phonon line widths, which result from integrals over the Fermi surface, will be qualitatively different too. In future work, I intend to validate this concept by calculating electron-phonon properties in cuprates using doped DFT+U ground states. 

\section{Summary}

I calculated the phonon spectrum of undoped La$_\text{2}$CuO$_\text{4}$ using the LTT supercell with the energy lowering distortions that are present in the real material and the DFT+U method with U=8 eV. My calculations reproduced the experimental band gap and antiferromagnetic moments. The calculated bond-stretching dispersions computed with U=8 eV agreed with experiment, demonstrating the sensitivity of the LO bond-stretching dispersions to the onsite repulsion, U, which frustrates modulating the charge around the Cu atoms. This is consistent with the experimental result that the LO bond-stretching branches soften with hole-doping, since doping permits low energy charge excitations that does not require pumping charge onto the Cu atoms. These results should be valid for transition metal oxides in general. I showed that the DFT+U method combined with the correct supercell is a robust framework for modeling phonons in the undoped cuprates and other perovskite oxides with complex, interrelated structural and electronic degrees of freedom.

\section{Acknowledgements}

This work benefited from discussion with J.M. Tranquada, A. Holder, and I.I. Mazin. The authors were supported by the DOE, Office of Basic Energy Sciences, Office of Science, under Contract No. DE-SC0006939. Portions of this research were performed using the Eagle computer operated by the Department of Energy's Office of Energy Efficiency and Renewable Energy and located at the National Renewable Energy Laboratory, and the RMACC Summit supercomputer which is supported by the National Science Foundation (awards ACI-1532235 and ACI-1532236), the University of Colorado Boulder, and Colorado State University. The Summit supercomputer is a joint effort of the University of Colorado Boulder and Colorado State University.

\chapter{Local atomic correlations in hybrid perovskites}
\label{chp:mapi_diffuse}

\begin{figure}[t!]
\centering
\includegraphics[width=0.75\linewidth]{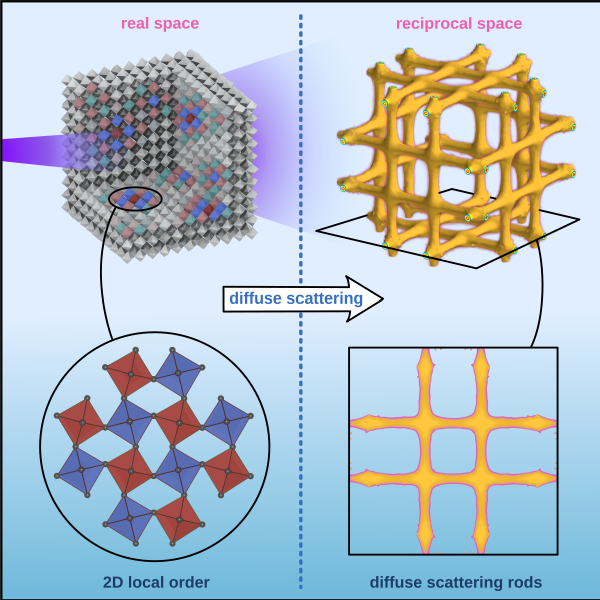}
    \caption{"Graphical summary" of the main results of this project. We use reciprocal space measurements, diffuse scattering, to understand the real-space correlations of CH$_3$NH$_3$PbI$_3$, which we calculate from molecular dynamics simulations.}
\label{fig:graphical_abstract}
\end{figure}

Hybrid lead halide perovskites (LHPs) are semiconductors with novel  energy-relevant properties that are distinctively governed by structural fluctuations. Diffraction experiments sensitive to long-range order reveal a cubic structure in the device-relevant, high-temperature phase ($\gtrsim 300 $ K). In this Chapter, we determine the true structure of two hybrid LHPs, CH$_3$NH$_3$PbI$_3$ and CH$_3$NH$_3$PbBr$_3$, using single-crystal diffuse scattering and inelastic spectroscopy combined with molecular dynamics simulations to calculate the scattering intensity (see \cref{sec:classical_cross_section}). The remarkable collective dynamics, not observed in previous studies, consist of a network of local two-dimensional, circular regions of dynamically tilting lead halide octahedra (lower symmetry) that induce longer range intermolecular CH$_3$NH$_3^+$ correlations. The dynamic local structure may introduce transient ferroelectric or antiferroelectric domains that increase charge carrier lifetimes, and strongly affect the halide migration, a poorly understood degradation mechanism. The main results of this section are summarized in \cref{fig:graphical_abstract}

This section is closely based on the corresponding article \cite{weadock2023nature} and its supplementary information on which I was co-first author with Nick Weadock. I contributed to the inelastic neutron scattering experiment and simulated the neutron and x-ray scattering from classical MD. The diffuse scattering experiments were done by Nick Weadock. The MD trajectories were calculated by Elif Ertekin's group at UIUC. This project was advised by Prof. Dmitry Reznik and Prof. Mike Toney. The research in this chapter uses the code \cite{sterling_pynamic} developed in the corresponding paper \cite{weadock2023nature}.

\section{The Nature of Dynamic Local Order in
CH$_3$NH$_3$PbI$_3$ and CH$_3$NH$_3$PbBr$_3$}

The crystal structure and associated symmetry of a material are key determinants of mechanical, electronic, optical, and thermal properties. One has to look no further than seminal condensed matter physics textbooks for derivations of these properties made possible by consideration of the long-range or average order determined by the crystal lattice and translational symmetry \cite{ashcroftSolidStatePhysics2000, kittelIntroductionSolidState2019}. However, in many materials including disordered rocksalts and intercalation compounds used for battery cathodes, relaxor ferroelectrics, thermoelectrics, and oxide and halide perovskites, the  important properties are not well described by the long-range structure. Instead it is short-range order that dominates aspects of the structure-function relationship \cite{rothSolvingDisorderedStructure2019, clementCationdisorderedRocksaltTransition2020, krogstadReciprocalSpaceImaging2020a, simonovHiddenDiversityVacancy2020a, xuThreedimensionalMappingDiffuse2004, krogstadRelationLocalOrder2018, rothSimpleModelVacancy2020a, lanigan-atkinsTwodimensionalOverdampedFluctuations2021}. In scattering experiments, short-range or local order manifests as diffuse scattering; a result of static and dynamic deviations from the average structure. Fixed chemical or local structural correlations result in static diffuse scattering, whereas thermal diffuse scattering arises from dynamic displacements due to lattice dynamics \cite{welberryOneHundredYears2016}. In an analogous fashion, short-range magnetic correlations generate magnetic diffuse scattering observable with neutron scattering \cite{rothModelfreeReconstructionMagnetic2018}. Resolving structural correlations in disordered materials has recently become more feasible with the development of high flux, single crystal X-ray and neutron diffuse scattering instruments and sophisticated modeling algorithms \cite{krogstadReciprocalSpaceImaging2020a,rosenkranzCorelliEfficientSingle2008,yeImplementationCrossCorrelation2018, proffenAdvancesTotalScattering2009, simonovYellComputerProgram2014, morganRmcdiscordReverseMonte2021}, opening up enormous opportunities to understand how local order impacts materials properties. Combining X-ray and neutron diffuse scattering reveals additional information about short-range order in materials with both light and heavy elements \cite{stocklerEpitaxialIntergrowthsLocal2022}.

\begin{figure}[t!]
\centering
\includegraphics[width=1\linewidth]{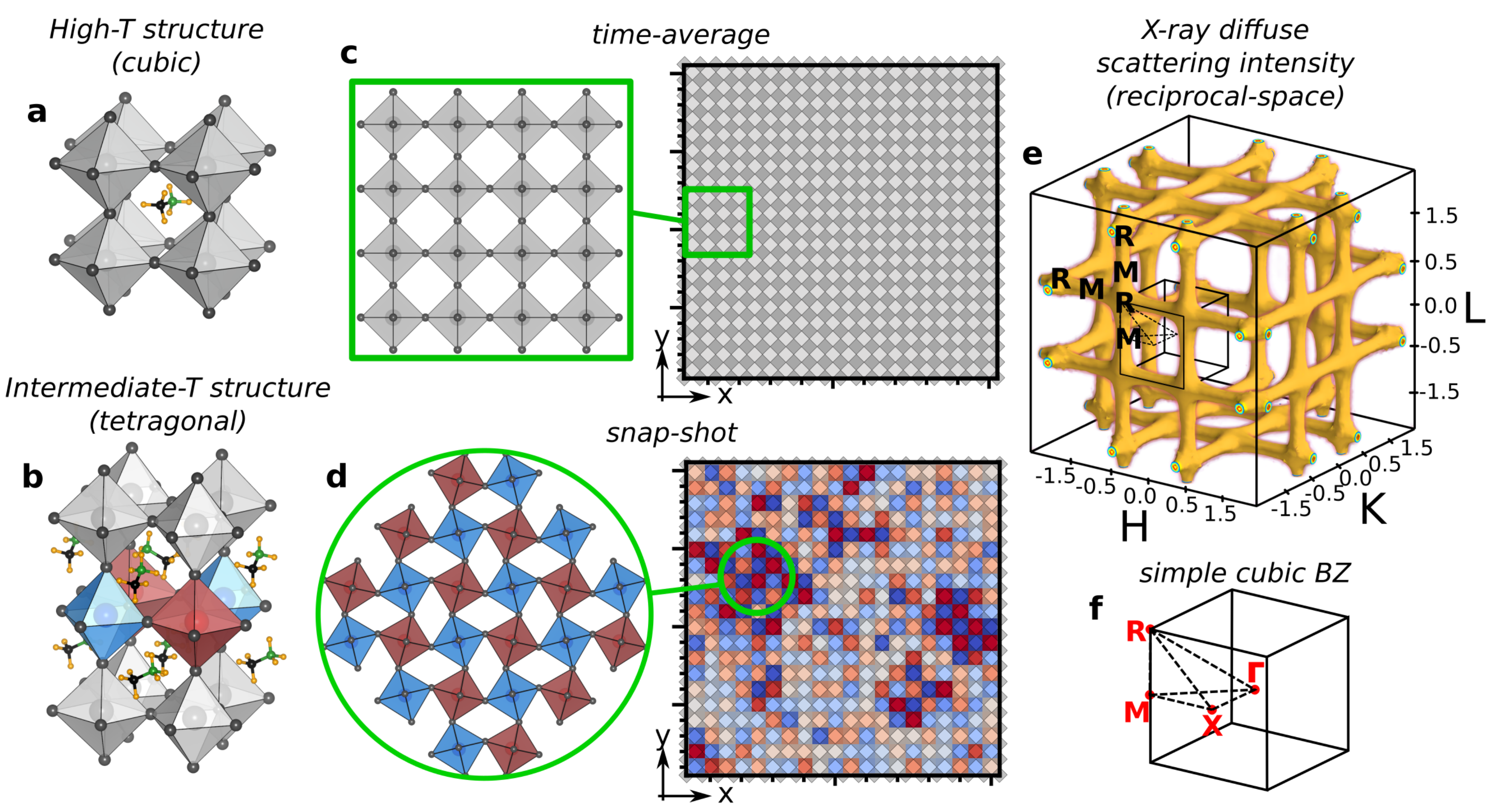}
    \caption{Visualizations of the high-temperature cubic structure (a, c) and the intermediate-temperature tetragonal structure (b) of MAPbI$_3$ and MAPbBr$_3$. MA$^+$ disorder has been omitted for clarity. The cubic-tetragonal transition in MAPbI$_3$ and MAPbBr$_3$ occurs at 328 and 236 K, respectively, by the freezing out of octahedral tilts in an out-of-phase pattern \cite{wellerCompleteStructureCation2015, guoInterplayOrganicCations2017}. Instantaneous snapshots (d) of the cubic structure from MD reveal dynamic two-dimensional correlations consisting of tilted PbX$_6$ octahedra that can orient to be perpendicular to any of the three principal axes. 3D visualization of the X-ray diffuse scattering intensity (e) generated by the structural correlations. The diffuse volume is indexed by Miller indices H,K,L, of which integer values enumerate each Brillouin Zone (BZ). The simple cubic BZ (f) with high symmetry points indicated in red. The diffuse rods in e run along the BZ edge, connecting M and R points.}
\label{fig:fig_1}
\end{figure}

Organic-inorganic metal halide perovskites are a recently re-invigorated class of semiconductors with remarkable optoelectronic performance that defies traditional intuition: Lead-based metal halide perovskites (LHPs) possess a mechanically soft, defect-tolerant crystal lattice with strong structural disorder and mobile ions at modest temperatures \cite{eggerWhatRemainsUnexplained2018}. Fluctuations in the orbital overlaps arising from large thermal displacements of iodide in methylammonium (CH$_3$NH$_3^+$, CD$_3$ND$_3^+$ = MA) lead iodide directly influence the temperature dependence of the electron (or hole) mobility and optical bandgap \cite{mayersHowLatticeCharge2018a}. X-ray and neutron diffraction measurements, which probe long-range order, report that the high-temperature phase is cubic with well-defined Bragg peaks \cite{wellerCompleteStructureCation2015, whitfieldStructuresPhaseTransitions2016, swainsonPhaseTransitionsPerovskite2003}. This cubic perovskite structure is shown in \cref{fig:fig_1}a and consists of corner-sharing PbX$_6$ (X = I$^-$, Br$^-$) octahedra surrounding a dynamically disordered MA$^+$ cation within the cuboctahedral interstice. Measurements probing short-range order in the high-temperature phase, however, suggest the local structure is of lower symmetry \cite{beecherDirectObservationDynamic2016,cominLatticeDynamicsNature2016,lauritaChemicalTuningDynamic2017, zhaoPolymorphousNatureCubic2020, weadockTestDynamicDomainCritical2020}. The lack of consensus regarding the exact symmetry and dynamics of this enigmatic local structure limits our understanding and control of optoelectronic properties and ion migration in LHPs. Short-range order arising from dynamic two-dimensional correlations of lead bromide octahedra has recently been identified in CsPbBr$_3$ \cite{lanigan-atkinsTwodimensionalOverdampedFluctuations2021}. These correlations are most prominent in the cubic phase above 433 K and may not play a significant role in CsPbBr$_3$-based device operation. The significance and prevalence of such correlations in hybrid LHPs, including contributions from the organic cations, is not resolved. 

Organic cation and halide anion migration in LHPs, especially under illumination, is detrimental to device performance and stability yet the origin is not well understood. Consequences of ion migration include formation of space charge potentials at interfaces, accelerated degradation and loss of constituent elements, and light-induced phase segregation in mixed-halide compositions \cite{yuanIonMigrationOrganometal2016}. A complete picture of the high-temperature structure and dynamics is essential to model ion migration pathways accurately \cite{holekevichandrappaCorrelatedOctahedralRotation2021}. It is also important to characterize the short-range order on the organic sublattice, as the rotating MA$^+$ molecular dipoles may screen band-edge charge carriers and extend carrier lifetimes \cite{chenOriginLongLifetime2017}.

We use single crystal X-ray and neutron diffuse scattering and neutron spectroscopy, combined with molecular dynamics (MD) simulations, to uncover the true structure and dynamics of the nominally simple cubic ($Pm\bar{3}m$) phases of MAPbI$_3$ ($>$ 327 K) and MAPbBr$_3$ ($>$ 237 K). We find that the cubic phase (\cref{fig:fig_1}a, c) is comprised of dynamic, two-dimensional roughly circular ``pancakes" of tilted lead halide octahedra which align along any of the three principal axes of the cubic structure and (\cref{fig:fig_1}b, d) induce additional structural correlations of the organic sublattice in the layers sandwiching the octahedra. Taken together, these regions of dynamic local order are several unit cells in diameter with lifetimes on the order of several picoseconds, resulting in a dynamic landscape for charge carriers and ion migration. This extended dynamic local order was neither observed nor predicted in previous studies of structural dynamics in hybrid LHPs \cite{ferreira_elastic_2018,gold-parkerAcousticPhononLifetimes2018,leguyDynamicsMethylammoniumIons2015,chenRotationalDynamicsOrganic2015}. Embedded within this dynamic structure we find additional static 3D droplets of the intermediate tetragonal phase, consistent with previous reports \cite{cominLatticeDynamicsNature2016,weadockTestDynamicDomainCritical2020}. The structural correlations on the MA$^+$ sublattice are a unique characteristic in these hybrid systems with implications discussed below. 

\subsection{Results and Discussion}\label{sec:mapi_results}

\begin{figure}[t!]
    \centering
    \includegraphics[width=1\linewidth]{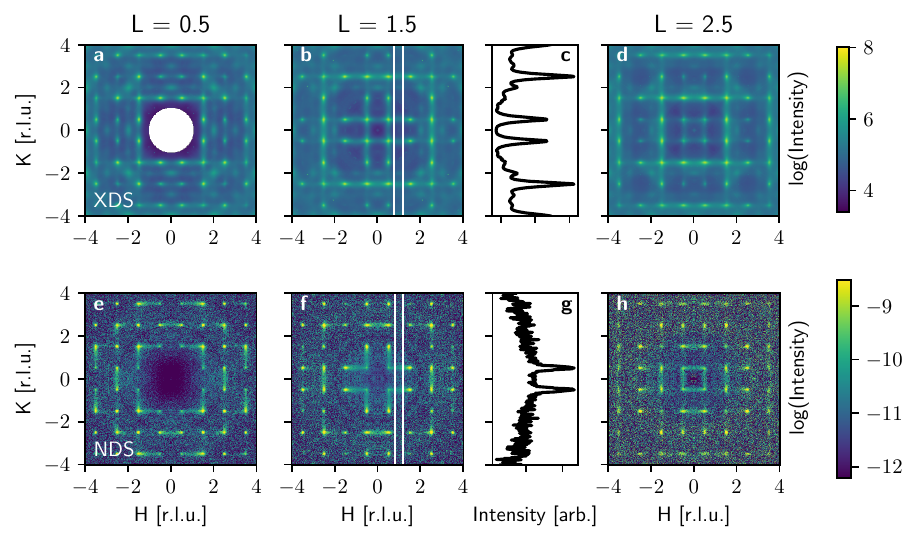}
    \caption{Experimental $S(\bm {Q})$ for protonated a)-d) and deuterated e)-h) MAPbBr$_3$ measured at 250 K (globally cubic phase) with x-ray and neutron diﬀuse scattering, respectively. The NDS data are limited to the elastic scattering channel of the CORELLI spectrometer (0.9 meV FWHM), whereas the XDS data are inherently energy integrated. Reciprocal space slices taken at L=0.5,1.5,2.5 contain only diﬀuse scattering intensity. Linecuts in c), g), are 
    obtained by integrating within the white boxes in b), f), respectively, and illustrate the diﬀerence in diﬀuse scattering intensities in XDS and NDS. The intensity in a), b), d), e), f), h) is plotted on a logarithmic scale}
    \label{fig:mapb_fig} 
\end{figure}

\begin{figure}[t!]
    \centering
    \includegraphics[width=0.9\linewidth]{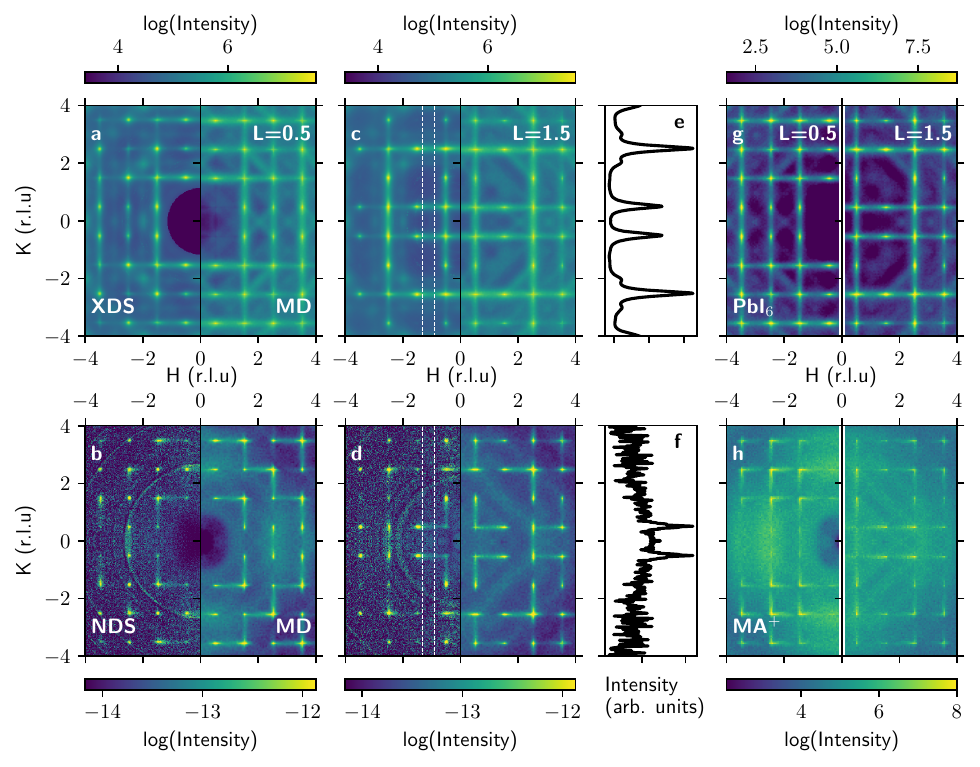}
    \caption{Experimental $S(\bm{Q})$ for MAPbI$_3$ measured with XDS (a,c left panels) and NDS (b,d left panels) at 345 K. The XDS data are inherently energy integrated whereas the NDS data are static with a 0.9 meV FWHM energy bandwidth. Panels (a, b) correspond to the L = 0.5 plane, and panels (c, d) the L = 1.5 plane. No Bragg intensity is expected in these half-order planes; therefore any intensity is diffuse scattering. The faint circular rings in the experimental NDS data are residual Debye-Scherrer diffraction rings from the aluminum sample holder. Theoretical $S(\bm{Q})$, calculated from MD simulations ($\pm$1 meV bandwith), are plotted in the right panels of (a-d). In (a, c), X-ray form factors were used whereas (b, d, g, h) were calculated using neutron scattering lengths. Panels (e, f) plot one-dimensional linecuts along K, integrating 0.7:H:1.3 and 1.4:L:1.6, which highlight the differences between XDS and NDS. Panels (g, h) plot the PbI$_6$ and MA$^+$ contributions to the calculated neutron $S(\bm{Q})$, respectively, with the left panels corresponding to L = 0.5 and the right panels L = 1.5. We note that the NDS plotted here corresponds to the elastic scattering channel of the CORELLI spectrometer ($\sim 0.9$ meV full width at half-maximum for $E_{\mathrm{i}} = 30$ meV\cite{yeImplementationCrossCorrelation2018}).}
    \label{fig:fig2} 
\end{figure}

In \cref{sec:exp_details_mapi}, we discuss the details of the experiments. The experimental scattering function $S(\bm{Q})$ from the cubic phase of MAPbI$_3$ at 345 K is shown in \cref{fig:fig2}. Specifically, we present the L = 0.5, 1.5 planes measured with both x-ray (XDS, \cref{fig:fig2}a,c, left panels) and neutron diffuse scattering (NDS, \cref{fig:fig2}b,d, left panels). $S(\bm{Q})$ calculated from atomic trajectories obtained with MD simulations of MAPbI$_3$ (see \cref{sec:classical_cross_section,MD}) is plotted in the right-hand panels of a-d and both panels of g,h for comparison. The experimental $S(\bm{Q})$ of the cubic phase of MAPbBr$_3$ at 250 K is plotted in \cref{fig:mapb_fig} and show diffuse scattering intensity profiles for XDS and NDS that are nearly identical to MAPbI$_3$.

Our approach to solving the true structure of the nominally simple cubic MAPbI$_3$ and MAPbBr$_3$ involves decomposing the observed diffuse scattering profile into four components and analyzing their energy and \textbf{Q}-dependence. These components include: (1) rods of constant (XDS) or varied (NDS) intensity spanning the BZ edge (M-R direction in \cref{fig:fig_1}, $q = [0.5, 0.5, L]$); (2) an additional contribution centered at the R-point [$q =(0.5, 0.5, 0.5)$]; (3) broad intensity centered at the X-point [$q =(0.5, 0, 0)$] observed previously and discussed in the SI; and (4) contributions from the MA$^+$ sublattice deduced from the alternating diffuse rod intensity profile observed in NDS but not XDS. This difference between XDS and NDS is highlighted with one-dimensional line cuts of XDS and NDS data shown in \cref{fig:fig2}e and f. Specifically, the peaks at K = $\pm2.5$ originating from the extended rod along H in XDS are not observed in the NDS data. Our analysis is complemented by MD simulations of MAPbI$_3$ which reproduce the experimental diffuse scattering. The experimental and calculated XDS (\cref{fig:fig2}) show remarkable agreement. Comparing the two results in a root-mean-square error of 3.6\% (see \cref{SI_fig:linecut_error} and related discussion) after adjusting only the background and scaling of the calculated $S(\bm{Q})$. Given the good agreement, we examine the real-space structure simulated with MD to determine the origin of the diffuse scattering (also see \cref{sec:md_corr}).

\begin{figure}[t!]
\centering
\includegraphics[width=0.75\linewidth]{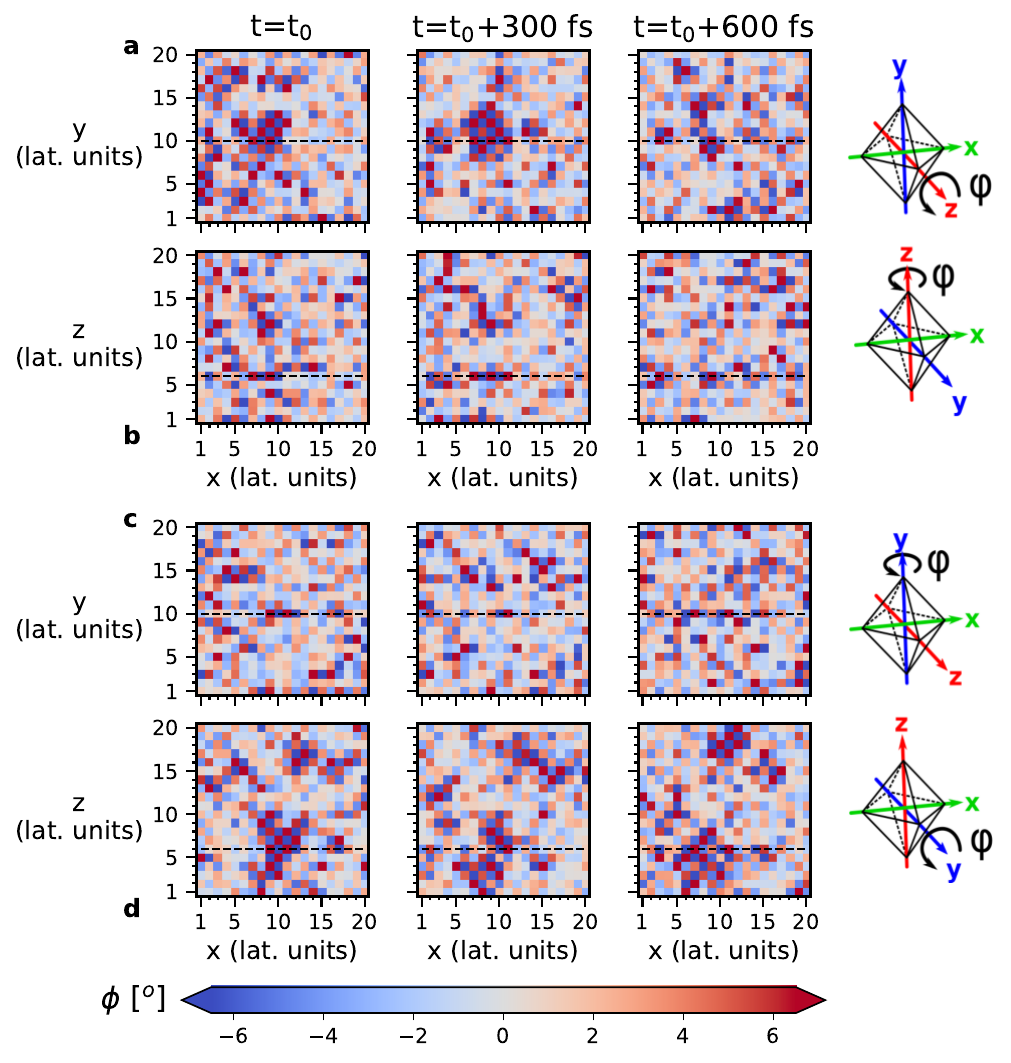}
    \caption{Spatiotemporal dynamics of PbI$_6$ octahedral rotations in cubic MAPbI$_3$. Each pixel encodes the octahedral rotation angle $\phi$ about the $z$-axis (a, b) and the $y$-axis (c,d). The columns are offset in time by $\Delta$t=200 fs to highlight the dynamic nature. The top rows (a, c) are in the $x$-$y$ plane and the bottom rows (b, d) are in the $x$-$z$ plane. In (a, d), the rotation is about the axis perpendicular to the plane as shown by the diagrams on the right. In (b, c), the rotation is about an axis parallel to the figure's vertical axis. The dashed lines indicate the intersection between the $x$-$y$ plane in (a, c) and the $x$-$z$ plane in (b, d). Note that the two-dimensional regions visible in (a, d) are one unit cell thick in the normal direction (b, c).}
\label{fig:snapshots}
\end{figure}

The XDS rods of diffuse scattering have constant intensity along the entire BZ edge (\cref{SI_fig:diffuse_MD}) and a width larger than the instrument resolution ($\approx 0.02 \textrm{\AA}^{-1}$). This lineshape arises from two-dimensional structural correlations in real-space. We show these correlations in \cref{fig:snapshots}, which presents a simplified visualization of the MD simulation. These plots track PbI$_6$ octahedral rotations, defined by azimuthal rotation angle $\phi$ about the [001] (\cref{fig:snapshots}a,b) or [010] (\cref{fig:snapshots}c,d) directions. Correlated regions of alternating large tilts are identified by neighboring red and blue pixels (representing octahedra) in \cref{fig:snapshots}a and d. Along the axis of rotation indicated in the corresponding schematic the tilts are uncorrelated between planes (\cref{fig:snapshots}b,c), revealing the existence of two-dimensional pancakes of tilted octahedra with a diameter on the order of 5 unit cells. These regions are reminiscent of tilts associated with the tetragonal structure in \cref{fig:fig_1}b. The tilts generate in-plane antiphase structural correlations with a periodicity of $2a$. The associated wave vector of the correlation is $q=2\pi / 2a =\pi/a$, which is the zone boundary for the simple cubic BZ. The tilted regions are confined to a single PbI$_6$ sheet and therefore the diffuse scattering intensity manifests as rods. The same structural correlations are present in the cubic phase of MAPbBr$_3$, since the XDS and NDS are qualitatively identical to that of MAPbI$_3$ (\cref{fig:fig2}).

Intermolecular structural correlations in the MA$^+$ sublattice are implied from the alternating intensity pattern of the diffuse rods present in NDS but not XDS. The MA$^+$ correlations are explored in the calculated $S(\bm{Q})$ in \cref{fig:fig2}g,h, and \cref{SI_fig:diffuse_MD,SI_fig:along_rods_neutrons,SI_fig:cross_term}. Specifically, we set the neutron scattering lengths of Pb and I to zero in the calculation to isolate the contribution from the MA$^+$ cation, and  C, N, and D to zero to isolate contributions from the inorganic octahedra. Diffuse scattering from the inorganic framework manifests as extended diffuse rods along the zone edge, resembling the experimental XDS as the X-ray atomic form factors for C, N, D are small compared to those for Pb and I and contribute little intensity. The calculated $S(\bm{Q})$ from CD$_3$ND$_3^+$ shows remarkable behavior (\cref{fig:fig2}h and \cref{SI_fig:diffuse_MD,SI_fig:along_rods_neutrons}), namely well-defined diffuse rods along the zone edges in addition to broad, isotropic intensity. This broad, isotropic component is a result of uncorrelated MA$^+$ dynamics \cite{leguyDynamicsMethylammoniumIons2015, chenRotationalDynamicsOrganic2015} commonly associated with the nominally cubic phases of LHPs. The presence of additional zone edge intensity, however, shows that previously unreported intermolecular structural correlations exist on the MA$^+$ sublattice. These local correlations are not purely 2D as the intensity varies along individual diffuse rods in the MA$^+$ $S(\bm{Q})$ (\cref{SI_fig:diffuse_MD,SI_fig:along_rods_neutrons}). Layers of PbX$_6$ octahedra are sandwiched between layers of MA$^+$ cations, therefore we expect the out-of-plane MA$^+$ structural correlations to extend at least two unit cells. Finally, the MA$^+$ and PbX$_6$ correlations are connected as evidenced by a nonzero interference term, shown in \cref{SI_fig:diffuse_MD,SI_fig:along_rods_neutrons}, which results in the alternating diffuse intensity profile observed in NDS.  

The lateral size of 2D pancakes, given by the correlation length, $\xi$, is obtained from the $\bm{Q}$-linewidth of the XDS and NDS diffuse rods and directly from MD real-space analysis (see \cref{sec:md_corr}). We fit the intensity of one-dimensional cuts in $\textrm{\AA}^{-1}$ across various diffuse rods to a Lorentzian lineshape in accordance with Ornstein-Zernike theory for exponentially decaying spatial correlations:
\begin{equation}
    S(\bm{Q}) \propto \frac{\Gamma}{\Gamma^2 + (q-q_0)^2}
\label{ozt}
\end{equation}

where $\Gamma = 1/\xi$ is the half-width at half-maximum (HWHM) of the fitted Lorentzian and $\xi$ is the correlation length \cite{xuThreedimensionalMappingDiffuse2004, valeCriticalFluctuationsSpin2019}. Correlation lengths of 4-6 unit cells ($\sim$3 nm) in diameter are obtained for MAPbBr$_3$ and MAPbI$_3$. Similar $\xi$ are obtained from NDS and XDS, indicating the in-plane $\xi$ of the 2D pancakes are nearly equivalent for the PbX$_6$ and MA$^+$ sublattices. We also calculate $\xi$ directly from the MD simulations (see \cref{sec:md_corr} and Figures therein) and find values consistent with experiment.

\begin{figure}[t!]
\centering
\includegraphics[width=1\linewidth]{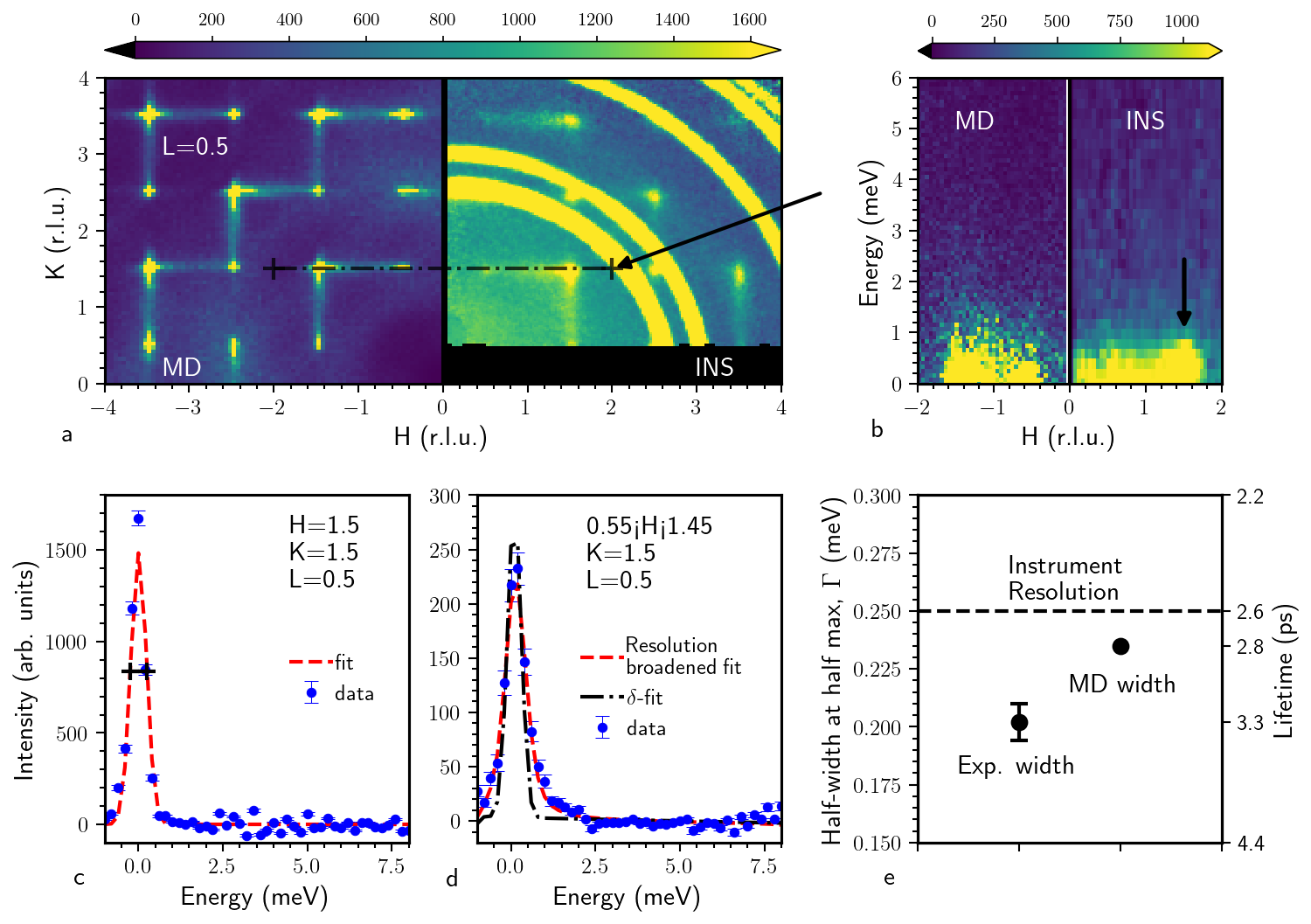}
    \caption{Neutron inelastic scattering data (right panels of (a,b); and (c,d) were measured at 340 K. In a, the $S(\bm{Q})$ from MD and experiment is integrated from $-1 \leq E \leq 1$meV. Powder rings from the aluminum sample holder are visible in a. Constant-\textbf{Q} scans at the R-point (c), integrated along 1.45:H:1.55, 1.45:K:1.55, 0.45:L:0.55) and along the diffuse rod (d), integrated along 0.55:H:1.45, 1.45:K:1.55, 0.45:L:0.55) are fit to Gaussian and resolution-broadened Lorentzian lineshapes with resolution linewidth of 0.5 meV FHWM, respectively (red dashed lines). The black line in (c) indicates the 0.5 meV resolution FWHM. In (d), we include the resolution lineshape (black dash-dot line), demonstrating a finite width beyond the resolution. Data in (c), (d) follow Poisson statistics where intensity is plotted as counts $n \pm \sqrt{n}$. In (e), the average HWHM (and corresponding lifetime) obtained from Lorentzian fits to constant-\textbf{Q} scans are plotted for experiments and MD. The dashed line represents the energy resolution of the MERLIN spectrometer. The experimental HWHM is the average $\pm$ standard deviation of fits for four separate diffuse rods.}
\label{fig:fig4}
\end{figure}

Molecular dynamics simulations show that the 2D structural correlations are transient and diffusive, with dynamics of the octahedral correlations tracked in \cref{fig:snapshots}. We investigate the dynamics of the PbI$_6$ and MA$^+$ correlations by evaluating the energy dependence obtained from the dynamical structure factor $S(\bm{Q}, E =\hbar\omega)$ measured with neutron inelastic spectroscopy (INS) on MAPbI$_3$ at 340 K. The $S(\bm{Q}, E)$ in the L = 0.5 plane, integrated from $-1\leq E\leq 1$ meV is shown in the right-hand panel of \cref{fig:fig4}a and matches that observed with NDS (\cref{fig:fig2}b) and that calculated from MD (left-hand panel of \cref{fig:fig4}b). A slice of $S(\bm{Q}, E)$ along the diffuse rod presented in \cref{fig:fig4}b shows that the intensity is centered at $E = 0$ meV with no inelastic component visible. Incoherent-background-subtracted energy scans (integrated along the length of the diffuse rod) show a finite, quasielastic component as exemplified in \cref{fig:fig4}d. Quasielastic scattering is a result of diffusive or relaxational motions involving small energy transfers, manifesting as a peak centered at $E = 0$ meV with a non-zero linewidth. The scattering is fit to a relaxational (Lorentzian) model, broadened by the resolution function, as shown in \cref{fig:fig4}d. The average energy HWHM, plotted in \cref{fig:fig4}e, is $0.20\pm0.01$ meV with a corresponding lifetime $\tau = \hbar/\mathrm{HWHM}$ of $3.25\pm0.13$ ps. Lifetimes are obtained in the same manner from the calculated $S(\bm{Q}, E)$ as shown in \cref{SI_fig:md_fit}, and the average lifetime of 2.8 ps is reported in \cref{fig:fig4}e. We note that these lifetimes are consistent with the mean residence time for rotations of the C-N bond in cubic MAPbI$_3$, suggesting that MA$^+$ reorientations are in part driven by the PbX$_6$ correlations \cite{chenRotationalDynamicsOrganic2015}. Overall, the structural correlations are dynamic (\emph{quasielasitc}) with finite lifetime of several picoseconds. 

R-point scattering ($\bm{Q} = [1.5,1.5,0.5]$, \cref{fig:fig4}c) noted above has been investigated previously and is attributed to static droplets of the intermediate-temperature tetragonal phase embedded within the high-temperature cubic phase, possibly nucleating about defects \cite{weadockTestDynamicDomainCritical2020}.  We observe no inelastic scattering at the R-points in both the INS or MD $S(\bm{Q} = (H, 1.5, 0.5) ,E)$ in \cref{fig:fig4}b. Incoherent-background-subtracted energy scans at a constant \textbf{Q} corresponding to R-points are best fit by a Gaussian lineshape with resolution limited linewidth (\cref{fig:fig4}c), hence this scattering is static with $\xi =2$ nm, consistent with previous reports \cite{weadockTestDynamicDomainCritical2020}. These droplets are distinct from the 2D pancakes but we are not able to resolve the spatial distribution of these two components. We expect that both structures influence MA$^+$ orientation.

The appearance of small regions of a lower temperature phase above the phase transition temperature is often a hallmark of critical scattering. For phase transitions with critical fluctuations, there is a marked jump in intensity of the associated low temperature phase as the system is cooled through the transition \cite{stirling_critical_1996}. The temperature dependence of the diffuse scattering through the cubic-tetragonal transition temperatures for MAPbI$_3$ and MAPbBr$_3$ is investigate in the supplementary information or the corresponding article, ref. \cite{weadock2023nature}. The intensity of the diffuse rods increases continuously with decreasing temperature through the transition, however no significant and abrupt jump in intensity is observed. Furthermore, the two-dimensional nature of the dynamic structural correlations is not typical of critical behavior at phase transitions. For the R-point scattering, however, there is a large jump in intensity as the cubic R-point transforms to the $\Gamma$-point of the tetragonal phase, giving Bragg scattering. The critical nature of the R-point scattering has been addressed previously \cite{weadockTestDynamicDomainCritical2020, cominLatticeDynamicsNature2016}.

We contrast our results to the 2D correlations in CsPbBr$_3$ \cite{lanigan-atkinsTwodimensionalOverdampedFluctuations2021}. These correlations, reportedly driven by soft, overdamped, anharmonic phonons at the BZ edge, occur within the high temperature cubic phase at 433 K. Anharmonic lattice dynamics are well documented in LHPs \cite{zhu2019mixed, gold-parkerAcousticPhononLifetimes2018, songvilayCommonAcousticPhonon2019, ferreira_elastic_2018}, therefore we expect that the correlations we observe in the hybrid LHPs MAPbI$_3$ and MAPbBr$_3$ arise from overdamped phonons as well. These phonons generate the non-dispersive, quasielastic intensity along the M-R BZ edge \cref{fig:fig4}b. Given the existence of 2D PbBr$_6$ correlations in CsPbBr$_3$, we propose that MA$^+$ orientation is driven by PbX$_6$ tilts; the MA$^+$ molecules orient to favor the lowest energy configuration defined by the distorted cuboctahedral geometry of the tilted regions and electrostatic interactions between lead halide octahedra and polar MA$^+$ molecules \cite{zhu2019mixed,lahnsteiner2016room}. The dynamic intermolecular correlations, forced by octahedral correlations, have been previously unreported and likely contribute to the excellent optoelectronic properties of hybrid LHPs. Furthermore, the correlations we observe in hybrid LHPs occur at temperatures relevant to device operation and have a longer lifetime than those in CsPbBr$_3$. Therefore, we expect this dynamic structure has a direct impact on device performance by influencing halide migration and charge carrier dynamics.

\subsection{Conclusion}

Our results show that the simple cubic phase of MAPbI$_3$ and MAPbBr$_3$ is in fact a composite structure containing dynamic 2D structural correlations and static tetragonal droplets with $\xi$ of a few nanometers. MA$^+$ correlations, induced by octahedral tilts, extend 2-3 unit cells in the normal direction. The simple cubic phase is only recovered when this composite is averaged over space and time (\cref{fig:fig_1} a-d). It is now critical to connect the structure observed here with the remarkable properties of LHP devices. The MYP potential used here has predicted optoelectronic properties and ionic conductivities that match well with experimental data \cite{mayersHowLatticeCharge2018a, barboniThermodynamicsKineticsIodine2018}; Mayers et al. conclude that the large amplitude dynamic iodide displacements are the main driver of the long charge carrier lifetime. In \cref{SI_fig:rappe}, we show that the $S(\bm{Q})$ calculated from the simulation of Mayers, et al., is in excellent agreement with our calculations and experimental data, demonstrating the same structure is present in their simulations. The large amplitude iodine displacements likely arise from the correlations we observe. Thus, the two-dimensional dynamic correlations are responsible for the unique charge carrier dynamics in LHPs. Based on the above assessment, we conclude that the ionic mobility reported in \cite{barboniThermodynamicsKineticsIodine2018} is similarly impacted by the observed dynamic correlations.

Finally, we hypothesize how the dynamic correlations and static droplets contribute to the optoelectronic properties and halide migration. We expect the structure  introduces spatial anisotropy to the electronic potential landscape with a longer lifetime than variations resulting from uncorrelated dynamic disorder. MA$^+$ has a large electric dipole moment of 2.3 Debye along the C-N axis \cite{chenOriginLongLifetime2017}. Uncorrelated dynamic disorder of the MA$^+$ sublattice, as previously reported, would generate a zero or negligible dipole moment. The MA$^+$ correlations observed here, however, may generate a net dipole moment, and this could lead to transient, local ferroelectric or antiferroelectric domains with lifetimes on the order of 3-6 ps. These domains influence electron-hole recombination and shift band alignment \cite{biEnhancedPhotovoltaicProperties2017, liuChemicalNatureFerroelastic2018}. The dynamic structural correlations also affect ion migration. Since the lifetime of the dynamic structural correlations (3-6 ps) exceeds the calculated time between attempted halide jumps (1 ps) \cite{frostWhatMovingHybrid2016}, a prefactor in calculating diffusivity, the correlations appear static and must be considered in future calculations. These correlations may impose a barrier to halide migration and reduce diffusivity \cite{holekevichandrappaCorrelatedOctahedralRotation2021}. The origin of the remarkable optoelectronic properties of LHPs continues to elude researchers; the local structure solved here is an important step in a complete understanding of these materials.

\section{Experimental details}\label{sec:exp_details_mapi}

\subsection{Diffuse X-ray scattering}\label{xds}

Reciprocal space maps, including Bragg and diffuse scattering contributions, were collected in transmission geometry at the Advanced Photon Source Sector 6-ID-D using monochromatic 86.9 keV X-rays. This X-ray scattering is inherently energy integrated, incorporating contributions from both static and thermal diffuse disorder. Single crystal samples of (CD$_3$ND$_3$)PbI$_3$, (CH$_3$NH$_3$)PbI$_3$, and (CH$_3$NH$_3$)PbBr$_3$, $\sim 500 \mu$m in size were mounted on the tip of Kapton capillaries. At each temperature, the sample was rotated $365^{\circ}$ at $1^{\circ} \mathrm{s}^{-1}$ with images collected every 0.1 s on a Pilatus 2M CdTe area detector. Sample temperatures were varied between 150 - 300 K with an Oxford N-HeliX Cryostream and 300 - 360 K with a hot nitrogen blower. The raw images are first processed with a peak finding algorithm to determine and refine an orientation matrix, and then rebinned into a reciprocal space volume $\pm$10 r.l.u. a side using the CCTW reduction workflow\cite{krogstadReciprocalSpaceImaging2020a,Cctw}. The dataset is symmetrized using cubic point group operations to remove missing data resulting from gaps between detector pixel banks. 

Sample damage is evident from changes in the scattering intensity with prolonged exposure. The damage thresholds were determined by identifying the onset of additional scattering intensity in subsequent measurements of the same sample at constant temperature. In MAPbI$_3$, damage is apparent after 90 minutes of measurement time (including time when the shutter is off in between each rotation for data collection), therefore all diffuse scattering analyzed here was collected on a fresh sample within 90 minutes.

\subsection{Diffuse neutron scattering}\label{nds}

Neutron diffuse scattering measurements were performed on the CORELLI spectrometer at the Spallation Neutron Source at Oak Ridge National Lab in Oak Ridge, Tennessee, USA\cite{rosenkranzCorelliEfficientSingle2008,yeImplementationCrossCorrelation2018}. Deuterated MAPbBr$_3$ (300, 350 mg, rectangular prism) and MAPbI$_3$ (220 mg, rhombic dodecahedron) crystals were used to reduce incoherent background contributions from hydrogen. The crystals were mounted in a CCR cryostat and oriented in the (HK0) scattering plane. The incident neutron energies range from 10 - 200 meV, and the 2$\theta$ coverage spans -30 to 145$^{\circ}$. MAPbBr$_3$ scattering data was collected from 150 - 300 K for 4-6 hours per temperature, and MAPbI$_3$ scattering data was collected from 300-400 K at 12 hours per temperature to account for smaller sample size and increased background from the thermal shielding. Diffuse scattering volumes $\pm$6 r.l.u. a side were obtained by reducing the raw data using defined workflows implemented in \texttt{MANTID}\cite{arnoldMantidDataAnalysis2014}. The contributions from inelastic scattering are removed by implementing the cross-correlation technique in the \texttt{MANTID} reduction workflow , \cref{sec:neutron_instruments}. The cross-correlation chopper selects elastically scattered neutrons within a resolution bandwidth of 0.9 meV full-width at half maximum \cite{yeImplementationCrossCorrelation2018}. Background artefacts including the aluminum sample enclosure are removed by subtracting the highest temperature datasets (300 K for MAPbBr$_3$, 400 K for MAPbI$_3$), normalized by the Bragg peaks, where little to no diffuse scattering is detected.

\subsection{Inelastic neutron scattering}\label{ins}

Neutron inelastic scattering measurements were performed on the Merlin direct geometry chopper spectrometer at the ISIS Neutron and Muon source at the Rutherford Appleton Laboratory in Didcot, UK\cite{bewleyMERLINNewHigh2006}. A deuterated single crystal of MAPbI$_3$ weighing 698 mg was oriented in the (HKK) scattering plane and mounted in a Brookhaven-style aluminum sample can. Temperature control was performed using a closed-cycle refrigerator with heaters attached directly to the sample can. Three incident energies of 11, 22, and 65 meV were selected by utilizing the repetition rate multiplication mode with the Fermi chopper set to 250 Hz. The $S(\bm{Q}, E)$ was collected at 340 K by rotating the sample 120$^{\circ}$ in 0.5$^{\circ}$ steps, and a radial collimator was used to reduce scattering from the aluminum sample can. Data reduction was performed using the \texttt{HORACE} data reduction suite and analyzed with Phonon Explorer and the National Institute of Standards and Technology (NIST) Center for Neutron Research (NCNR) \texttt{Data Analysis and Visualization Environment}  (DAVE) software packages \cite{ewingsHoraceSoftwareAnalysis2016a, reznikAutomatingAnalysisNeutron2020,phonon-explorer, azuahDAVEComprehensiveSoftware2009}.

Energy scans for linewidth analysis were sliced from the 11 meV incident energy $S(\bm{Q}, E)$ dataset. The resolution linewidth of this dataset was determined from constant-$\bm{Q}$ scans at $\bm{Q}$ values unique to the aluminum Debye-Scherrer rings. Several constant-$\bm{Q}$ scans obtained in this manner were fit to a Gaussian lineshape with constant background to obtain an average resolution linewidth of 0.5 meV FWHM. A representative constant-$\bm{Q}$ scan was used as the resolution function in subsequent analyses. The incoherent background was determined from constant-$\bm{Q}$ scans at $\bm{Q}$ values which contain only incoherent scattering with no contributions from diffuse rods, Bragg peaks, or aluminum Debye-Scherrer rings. Several background scans were averaged, then subtracted from the constant-$\bm{Q}$ scans along the diffuse rods and at the R-point to remove the incoherent background contribution and isolate the dynamics associated with these features.

Additional high-resolution neutron spectroscopy was performed on the cold neutron triple-axis spectrometer (SPINS) at the NCNR in Gaithersburg, MD, USA. The same deuterated single crystal of MAPbI$_3$ weighing 698 mg was oriented in the (HKK) scattering plane and mounted in a Brookhaven-style aluminum sample can. Constant-$\bm{Q}$ scans, in which we vary the energy transfer while maintaining constant momentum transfer $\bm{Q}$, were performed with a fixed final energy $E_\mathrm{F} = 5$ meV. We compare constant-$\bm{Q}$ scans taken along $\bm{Q} = (2, k ,0)$ and at $\bm{Q} = (1.5, 0, 0)$ at 300 and 140 K, to constant-$\bm{Q}$ cuts calculated from the MD S($\bm{Q},E$) as a way to validate the calculations.

\section{Molecular dynamics simulations}\label{MD}

MD trajectories for MAPbI$_3$ are generated using identical simulations to those in ref. \cite{zhu2019mixed}. We checked our calculations using both the mass of hydrogen and the mass of deuterium in the simulation and found no noticeable difference. We used a $20\times20\times20$ super-cell that is based on the 12-atom pseudocubic unit cell with cell length $a\approx6.3 ~\textrm{\AA}$. The simulations were done in the \textsc{LAMMPS} package \cite{plimpton1995fast}. The inter-atomic potential is from ref. \cite{mattoni2015methylammonium}. Trajectories were integrated using a 0.25 fs time step throughout. For the first 20 ps, the system was thermalized in the NPT ensemble at 345 K and 0 Pa, then it was equilibrated for another 20 ps in the NVE ensemble. After equilibration, the equations of motion were integrated for 20 ns and trajectories were written to a file every 25 fs. See \cref{sec:md} for more details about MD.

\section{S(Q,E) calculations}\label{calcsqe}

The dynamical structure factor, $S(\bm{Q},E)$, which is related to the cross section, can be calculated from MD. Given a set of suitably accurate trajectories, $\bm{r}_i(t)$, $S(\bm{Q},E)$ can be evaluated directly from \cref{eq:classical_cross_section},
\begin{equation}
    \frac{\partial^2 \sigma}{d\Omega dE'} = \frac{k'}{k} S(\bm{Q},E)
\end{equation}
with
\begin{equation}
    S(\bm{Q},E) = \Big \lvert \sum_i^N f_i(Q)\int \exp (i(\bm{Q}\cdot \bm{r}_i(t)-\frac{E}{\hbar}t))dt \Big \rvert^2.
    \label{eq:SQE_MD}
\end{equation}
See \cref{sec:classical_cross_section,sec:md} for more details. The theoretical $S(\bm{Q},E)$ calculations were done with a \texttt{python} code developed by us \cite{sterling_pynamic}. The calculated $S(\bm{Q},E)$ is weighted by X-ray atomic form factors or thermal neutron scattering lengths for direct comparison to experimental data \cite{hazemann2005high,brown2006intensity}. If the experiment uses neutrons, $f_i(Q)\equiv b_i$ is the $Q-$independent scattering length; if the experiment uses X-rays, $f_i(Q)$ is the $Q-$dependent atomic form factor where $Q$ is the magnitude of momentum transferred, $\bm{Q}$, from the incident beam of particles into the material. $E$ is the energy transferred into the material and the sum runs over all $N$ atoms. 

The reciprocal space resolution from MD is $\Delta Q \approx 0.05 ~ \text{\AA}^{-1}$. We do not integrate the calculation over reciprocal space unless otherwise specified, so the pixels in theoretical S($\bm{Q}$,$E$) plots have this spacing. In order to ensure we are far from the transient thermalization regime in MD, we only sample the end of the simulation. We split the last 0.25 nano-second of the simulation into 15 blocks that are $\approx$17 ps long; S($\bm{Q}$,$E$) is then calculated from 8 non-consecutive blocks and averaged over them.  For 17 ps blocks with a 50 fs sampling interval, the energy resolution is about $0.24$ meV and the maximum resolvable energy is 41.4 meV.

To compare to the experimental NDS data in \cref{fig:fig2}, theoretical S($\bm{Q}$,$E$) is integrated between $\pm0.5$ meV, which is comparable to the resolution in the experiment. To compare to the experimental XDS data in \cref{fig:fig2}, theoretical S($\bm{Q}$,$E$) is integrated between $\pm2$ meV.
To compare with INS data in \cref{fig:snapshots}, both the experimental and theoretical S($\bm{Q}$,$E$) are integrated $\pm1$ meV. Anywhere that purely theoretical data is compared, the integration is $\pm1$ meV. For excitations [e.g. \cref{fig:fig4}(b)], theoretical $S(\bm{Q},E)$ is not integrated over energy; the MD energy resolution is set by the length of the trajectory sampled in the calculation.

The workflow and analysis outlined above was performed on MD simulations of protonated MAPbI$_3$ and no significant differences were observed. The deuterated results are used in the main text for direct comparison to the experimental measurements.

\subsection{Calculated vs. experimental diffuse scattering}

\begin{figure}[t!]
\centering
\includegraphics[width=0.9\linewidth]{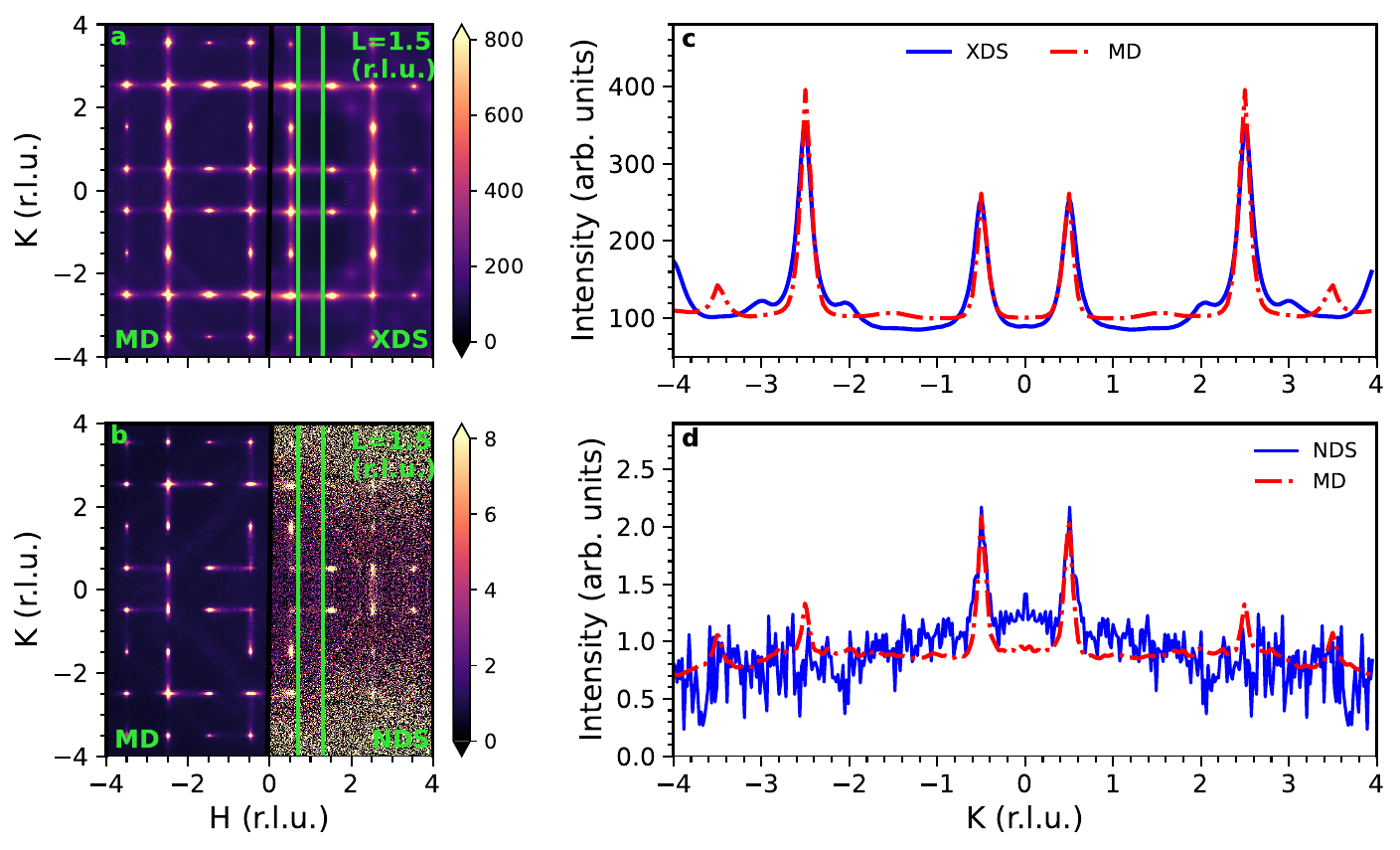}
    \caption{Linecut across the diffuse rods highlighting the agreement between MD and experiment. a) Diffuse scattering from MAPbI$_3$ in the $L=1.5$ plane from MD (left) and XDS at 345 K (right). c) A linecut through the intensity in a). Data in c) are integrated over the region between the vertical lines in a) [$0.7<H<1.3$, $K$, $1.4<L<1.6$ r.l.u.]. The height and scale of the MD data in c) were fit to the XDS data with an RMS error of 6.8$\%$. b) Diffuse scattering in the $L=1.5$ plane from MD (left) and MAPbI$_3$ NDS at 345 K (right). d) A linecut through the same region as in c) comparing calculated and MD NDS data. The height and scale of the MD data in d) were fit to the NDS data with an RMS error of 10.1$\%$. The MD data in a) and b) are scaled using the results of the fits in c) and d). In c), the peaks at integer values of K in the XDS data are likely phonons.}
\label{SI_fig:linecut_error}
\end{figure}

The relative scale of the theoretical and experimental $S(\bm{Q},E)$ is arbitrary since there are no incident X-ray or neutron fluxes in the MD simulation and this quantity was not explicitly measured in the experiments. Moreover, there are background contributions from the experimental setup that are not present in MD. In the comparisons between the calculated and experimental $S(\bm{Q},E)$ in the main text, the calculated data are multiplied by an arbitrary scaling factor to account for these differences and to allow a qualitative comparison. To obtain a more quantitative measure of the accuracy of our MD results, we calculated the root-mean-squared (RMS) error between experimental and computed $S(\bm{Q})$, adjusted for a constant background and scaled for incident flux. We compare MD data and experiment using linecuts across diffuse rods in \cref{SI_fig:linecut_error}. Impressively, we find that the diffuse rods agree very well between MD and experiment. 

\begin{figure}[t!]
\centering
\includegraphics[width=0.75\linewidth]{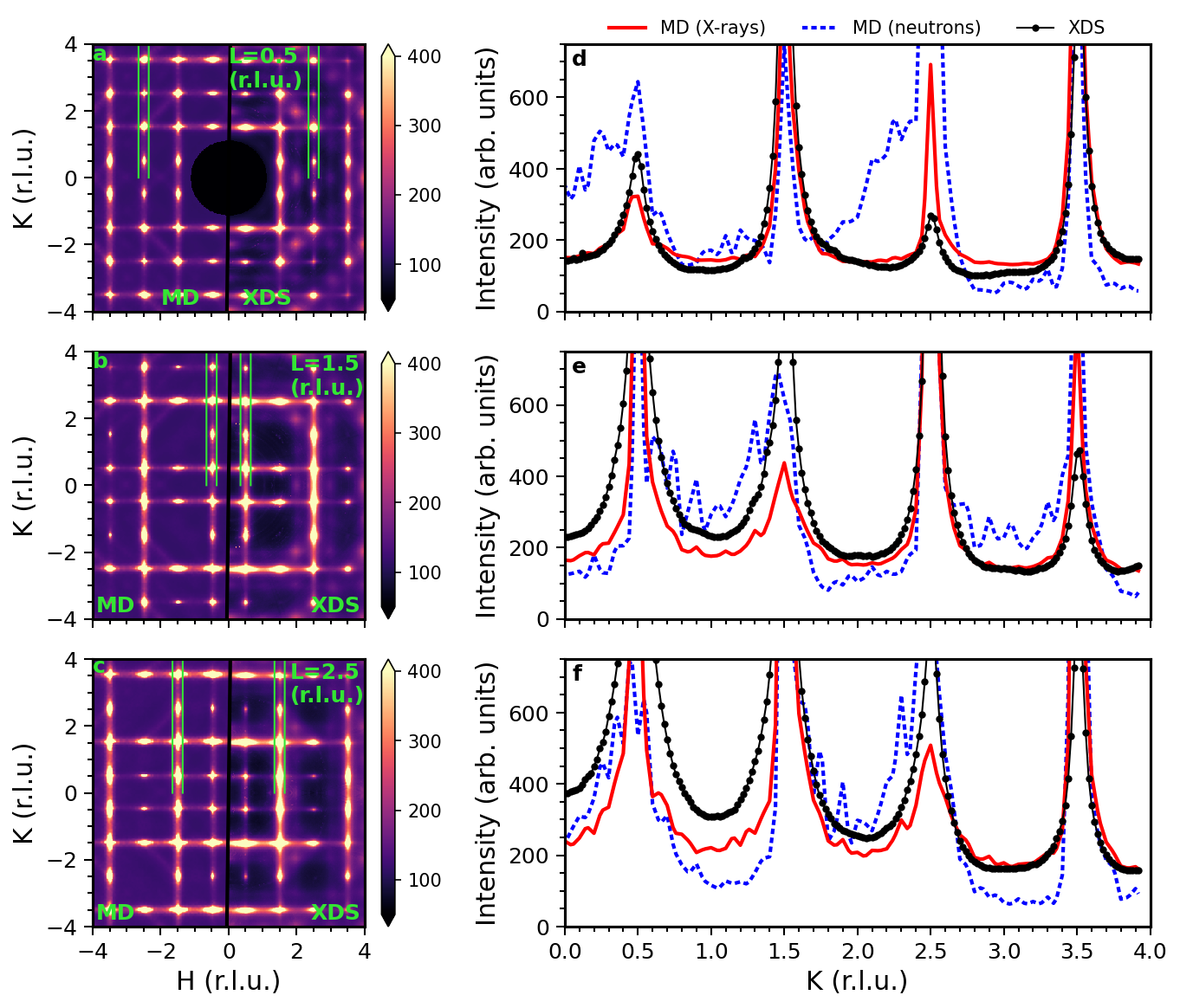}
    \caption{Comparison between the diffuse intensity along the rods in MAPbI$_3$ from MD and XDS at 345 K. The MD data (with X-ray form-factors) are on the left and XDS data on the right in a) - c). Linecuts integrated between the green lines in a) - c) are plotted in d) - f); MD (X-ray) data are red lines, MD (with neutron scattering lengths) are blue dotted lines, and the XDS data are black dots. MD (neutron) data are not plotted in a) - c) but provided for comparison. See \cref{SI_fig:diffuse_MD} for the corresponding data. The shift and scale applied to the MD data is the same as determined in \cref{SI_fig:linecut_error}. The XDS data in d) - f) are integrated $\pm0.5$ r.l.u. while the MD are not integrated over momentum.}
\label{SI_fig:along_rods_xrays}
\end{figure}

We noted in \cref{sec:mapi_results} that the XDS and NDS diffuse rods have qualitatively different intensity profiles: the magnitude of XDS intensity is relatively constant along several BZs, whereas for NDS the intensity is reduced for every other BZ as observed in \cref{fig:fig2}. Furthermore, along individual BZ edges the NDS diffuse intensity decays from the R-point. By contrast, the XDS intensity is constant except for the R-point contribution. \cref{SI_fig:along_rods_xrays} plots linecuts along the diffuse rods for the 345 K XDS data along $\bf{Q}$ = (2.5, K, 0.5), (0.5, K, 1.5) and (1.5, K, 2.5) in a,b,c, respectively. Linecuts along the rods extracted from the MD X-ray and neutron $S(\bm{Q})$ are also plotted for comparison. We see the experimental and MD XDS intensity is constant between the R-points, whereas the NDS intensity falls off asymmetrically between the R-points.

Since XDS shows no variation along the rod and NDS does, the variation in NDS likely arises from scattering from the MA$^+$ molecules which are nearly invisible to X-rays. We can verify this explicitly by calculating the different sublattice contributions to the diffuse scattering intensity as follows. Denoting $\rho(\bm{Q},E)=\sum_{i} f_i(Q)\int \exp (i(\bm{Q}\cdot \bm{r}_i(t)-tE/\hbar))dt$ with $i$ labeling the atom and the sum running over all atoms in a particular group, we separate the intensity into terms corresponding to the inorganic PbI$_6$ (cage) and organic (MA$^+$) sublattices:
\begin{equation}
\begin{split}
    S(\bm{Q},E) & = \lvert \rho_{\text{MA}}(\bm{Q},E) + \rho_{\text{cage}}(\bm{Q},E) \rvert^2 \\
    & = \lvert \rho_{\text{MA}}(\bm{Q},E) \rvert^2 + \lvert \rho_{\text{cage}}(\bm{Q},E) \rvert^2 \\
    &+ \rho_{\text{MA}}^*(\bm{Q},E)\rho_{\text{cage}}(\bm{Q},E) +\rho_{\text{cage}}^*(\bm{Q},E)\rho_{\text{MA}}(\bm{Q},E)
   \end{split}
\end{equation}
where $\rho_{\text{MA}}(\bm{Q},E)$ and $\rho_{\text{cage}}(\bm{Q},E)$ are only summed over the atoms in the groups denoted by the subscript: MA$^+$ (methylammonium) are C, D (deuterium), and N atoms while cage atoms are Pb and I. We call $\lvert \rho_{\text{MA}}(\bm{Q},E=0) \rvert^2\equiv S_\text{MA}(\bm{Q})$ and similarly for the cage atoms. The interference term is calculated from $S_\text{int.}(\bm{Q})=S(\bm{Q})-(S_\text{MA}(\bm{Q})+S_\text{cage}(\bm{Q}))$.

\begin{figure}[t!]
\centering
\includegraphics[width=1\linewidth]{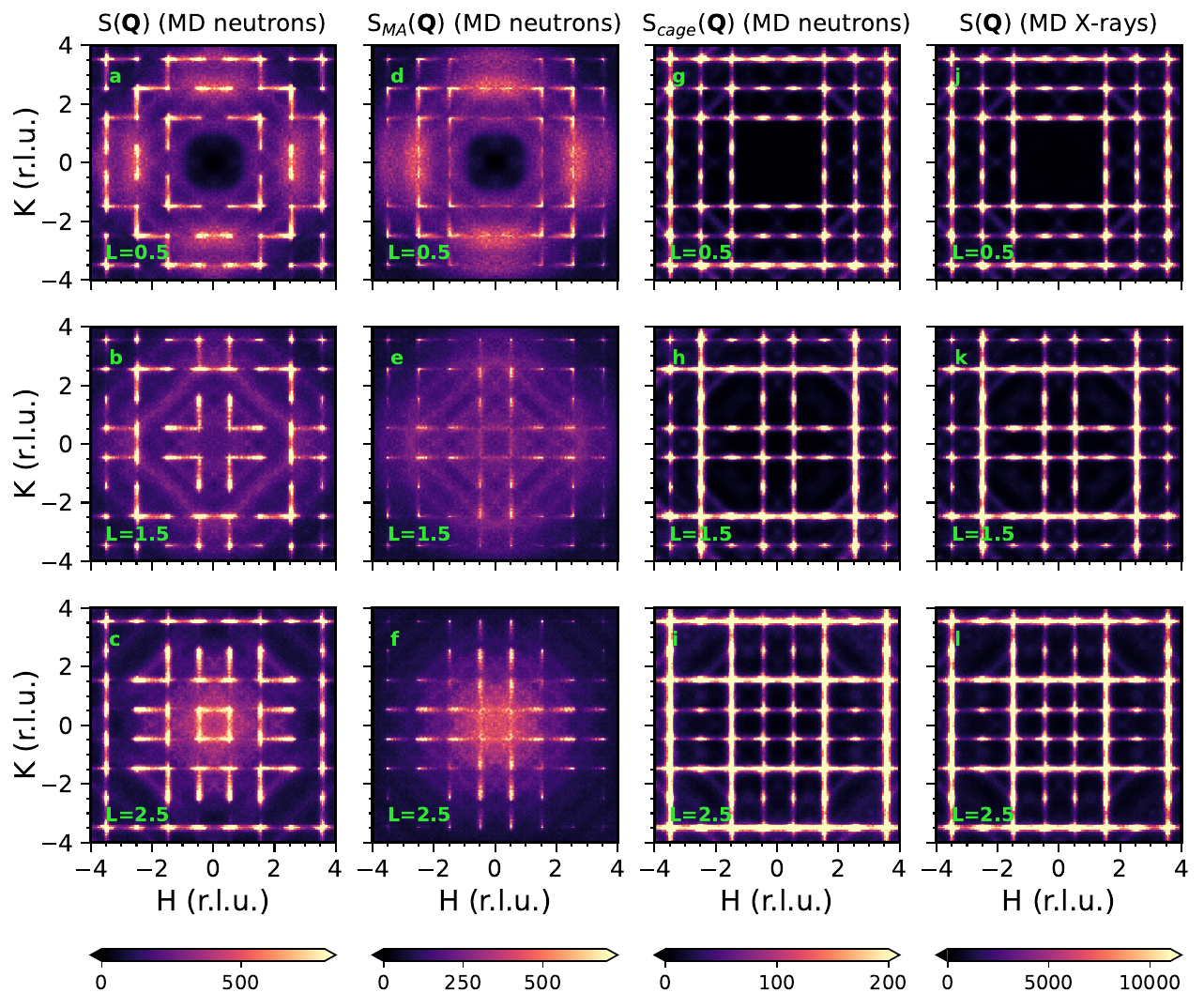}
    \caption{Different sublattice contributions to the total diffuse scattering intensity calculated from MD. The total NDS intensity is in a) - c) and the total XDS intensity is in j) - l). Scattering from only the MA and cage sublattices is in d) - f) and g) - i), respectively.}
\label{SI_fig:diffuse_MD}
\end{figure}

\begin{figure}[t!]
\centering
\includegraphics[width=0.9\linewidth]{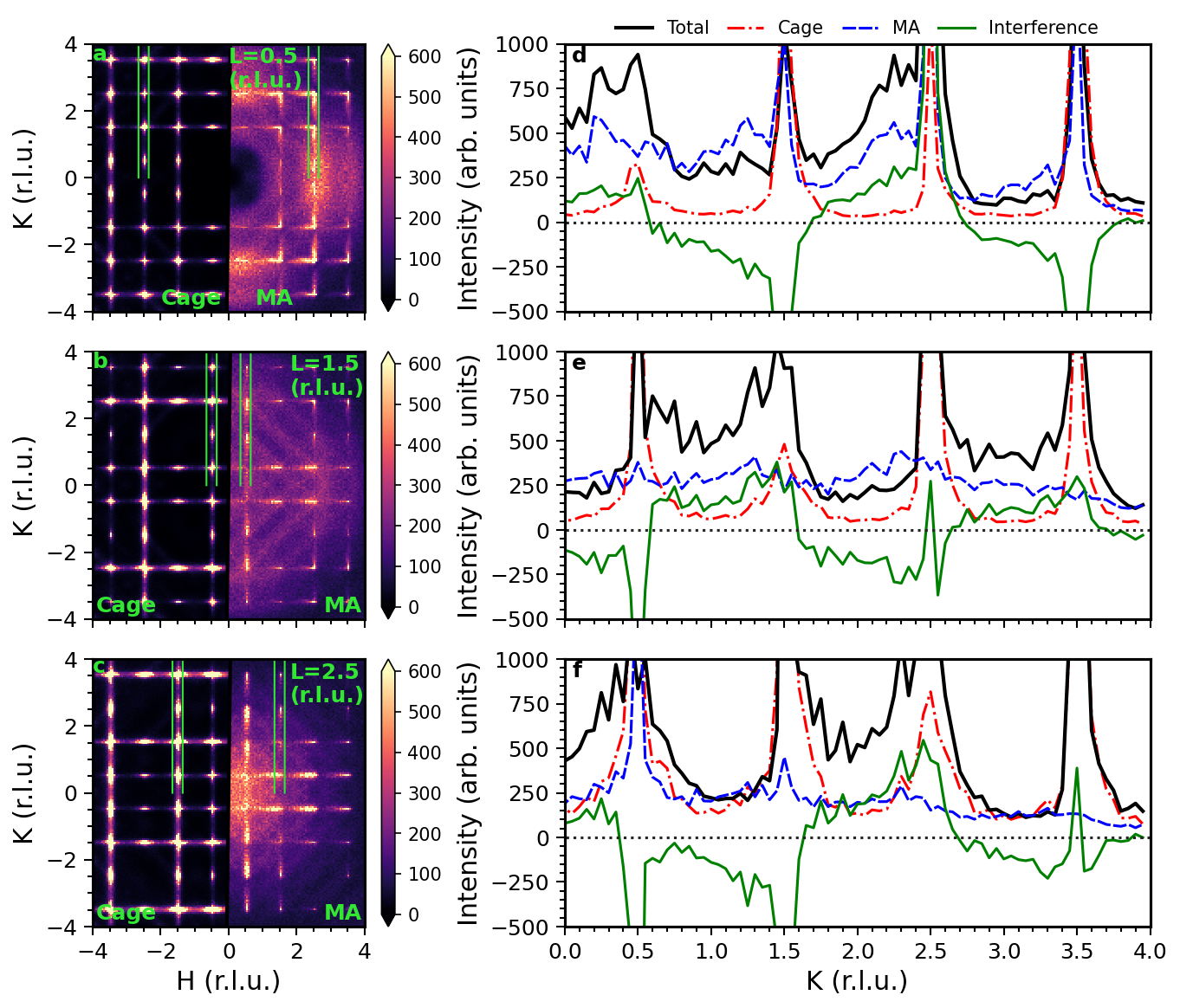}
    \caption{Different sublattice contributions to the intensity along the diffuse rods in MAPbI$_3$ from MD. The cage and MA$^+$ sublattice contributions to the theoretical neutron scattering intensity, $S_\text{cage}(\bm{Q})$ and $S_\text{MA}(\bm{Q})$, are on the left and right respectively in a) - c). In d) - f), the cage and MA$^+$ contributions are labeled in the legend above. Also shown are the total intensity and interference components from MD. The MD data are integrated $\pm0.5$ meV.} 
\label{SI_fig:along_rods_neutrons}
\end{figure}

The different contributions to $S(\bm{Q})$ are plotted in \cref{SI_fig:diffuse_MD}. We find that $S_\text{cage}(\bm{Q})$ is nearly indistinguishable from the calculated X-ray intensity (\cref{SI_fig:diffuse_MD} g-i).
i.e. the XDS measurement only probes the correlations of the PbI$_6$ octahedra. We also extract linecuts from each $S(\bm{Q})$ contribution along $Q$ = (2.5, K, 0.5), (0.5, K, 1.5) and (1.5, K, 2.5) and plot them in \cref{SI_fig:along_rods_neutrons}. The linecuts corresponding to $S_\text{cage}(\bm{Q})$ are in good agreement with the experimental and MD XDS in \cref{SI_fig:along_rods_xrays}. Moreover, because the XDS and cage-only intensities show these rods of constant intensity (\cref{SI_fig:along_rods_xrays,SI_fig:along_rods_neutrons}), it is clear that the PbI$_6$ octahedra structural correlations have a two-dimensional character. 

The plots of $S_\text{MA}(\bm{Q})$ in \cref{SI_fig:diffuse_MD,SI_fig:along_rods_neutrons} show two components of MA$^+$ diffuse scattering. First is the broad, spherical component corresponding to uncorrelated disorder arising from such molecular reorientations as rotations of methyl or amine groups about the C-N axis of MA$^+$. This component manifests as the nearly isotropic, circular intensity in \cref{SI_fig:diffuse_MD}d-f. Second are the strong, distinct diffuse rods with varying intensity along the rod direction. The rods orient parallel to one of the H,K,L directions and indicate the presence of intermolecular structural correlations similar to the pancakes of tilted PbI$_6$ octahedra. The varying intensity, however, implies that the intermolecular correlations extend beyond one unit cell in the out-of-plane direction. 

\begin{figure}[t!]
\centering
\includegraphics[width=1\linewidth]{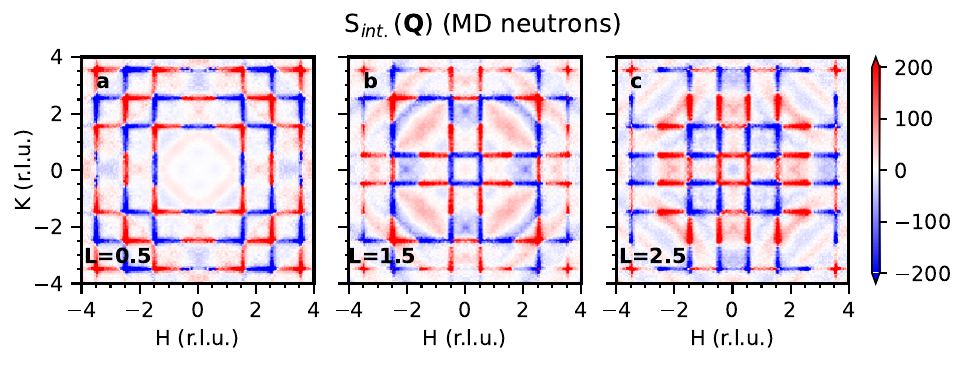}
    \caption{The interference part of S($\bm{Q}$) calculated from MD. The data in a) are in the $L=0.5$ r.l.u plane, in b) are in the $L=1.5$ r.l.u plane, and in c) are in the $L=2.5$ r.l.u plane.}
\label{SI_fig:cross_term}
\end{figure}

The interference term $S_\text{int.}(\bm{Q})$ is plotted in \cref{SI_fig:along_rods_neutrons,SI_fig:cross_term} and indicates that the separate sublattice correlations are correlated with one another. If they were entirely uncorrelated, $S_\text{int.}(\bm{Q})= 0$ everywhere and the total $S(\bm{Q}) = S_\text{cage}(\bm{Q}) + S_\text{MA}(\bm{Q})$ . Instead, there are regions of constructive and destructive interference, with the constructive component resembling the total NDS intensity (a-c in \cref{SI_fig:diffuse_MD}). We propose that the structural correlations between the two sublattices result from MA$^+$ molecules reorienting into the lowest energy configuration in the cubocathedral region between PbI$_6$ octahedra, as discussed in the \cref{sec:mapi_results}. Due to the nonzero $S_\text{int.}(\bm{Q})$ term, there is no clear way to isolate intermolecular MA$^+$ correlation lengths from the experimental or calculated diffuse scattering. Instead, we evaluate it explicitly below from the MD trajectories in \cref{sec:md_corr}.

\subsection{S(Q) from trajectories in Mayers et al.}

\begin{figure}[t!]
\centering
\includegraphics[width=0.9\linewidth]{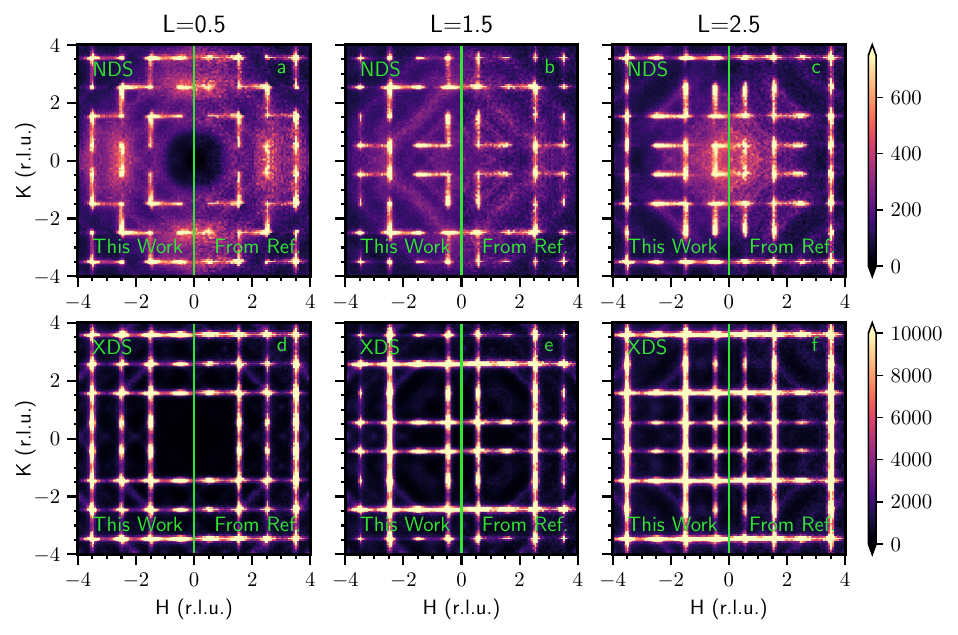}
    \caption{Simulated $S(\bm{Q})$ for both NDS a) - c) and XDS d) - f) calculated in the $L=0.5$ r.l.u. a), d), $L=1.5$ r.l.u. b), e), and $L=2.5$ r.l.u. c), f) planes. We compare $S(\bm{Q})$ calculated from our MD trajectories (left in each panel) to $S(\bm{Q})$ calculated from the trajectories in ref. \cite{mayersHowLatticeCharge2018a} (right in each panel). Besides the different momentum resolution due the different super cell sizes ($20\times20\times20$ in our calc. vs $16\times16\times16$ in ref. \cite{mayersHowLatticeCharge2018a}), the intensity is nearly indistinguishable between the two calculations.}
\label{SI_fig:rappe}
\end{figure}

Establishing a direct connection between the two-dimensional dynamic structural correlations and remarkable optoelectronic properties of LHP-based devices would be ideal, however there are several practical difficulties to fabricating or simulating LHP devices in which the correlations are not present. Instead, we can examine the MD simulations from other publications which calculate LHP properties to determine if the same structural correlations are present. In one example, Mayers, et al., accurately reproduced electronic transport properties in MAPI by combining tight binding calculations with the same classical MD potential that we have used \cite{mayersHowLatticeCharge2018a}. We calculated $S(\bm{Q})$ from their MD trajectories and find the same rods of diffuse intensity visible in our calculations and experiment. The two MD $S(\bm{Q})$ are compared in \cref{SI_fig:rappe}. The unmodulated rods arising from two-dimensional PbI$_6$ correlations are clearly visible in the XDS calculations ($d-f$). In the NDS calculations ($a-c$), the modulation of the rods due to interference from scattering from the MA$^+$ sub-lattice is well produced too. Moreover, direct real space visualization clearly shows the two-dimensional nature of the correlations. 

Considering the exceptional agreement between our results and those in Mayers et. al, we conclude that the same correlations present in our studies are present in theirs and the electronic properties they predict include the effects of the structural correlations. Specifically, they find that the large amplitude dynamic displacements of the iodine atoms modulate the electronic potential in such a way to scatter charge carriers. The spatial correlations of the iodide displacements, namely whether they are stochastic or due to short-range order, was not investigated. It is likely that the relevant lattice fluctuations that generate the disordered potential in ref. \cite{mayersHowLatticeCharge2018a} are the two-dimensional structural correlations or pancakes which are characterized by correlated large amplitude displacement of the iodine sub-lattice through the rotation of the PbI$_6$ octahedra. Their results combined with our analysis demonstrates that the correlations induced by anharmonic lattice dynamics have a large effect on the electronic and optical properties, showing the relevance of the correlations to device performance.

\section{Correlation lengths from MD}\label{sec:md_corr}

Correlation lengths $\xi$, corresponding to the widths of the two-dimensional structural correlations, are evaluated as stated in the Main Text and reported in \cref{tab:cl} below. For XDS experiments the $\xi$ correspond to PbX$_6$ correlations only, because the X-ray scattering intensity from light elements like C, N, D is significantly weaker.

\begin{table}[t!]
\begin{center}
\caption[]{Correlation lengths obtained from $\bm{Q}$ linewidths of the diffuse rods and R-points. We report average values and standard deviation determined from several diffuse rods or R-points. For all R-point linewidths fitted here, contributions from the diffuse rod intensity have been subtracted prior to fitting. All reported lengths are in $\textrm{\AA}$, and correspond to the radius of the structural correlation.}
\begin{tabular}{|c|c|c|c|c|c|c|c|}
\toprule
 & \multicolumn{2}{c|}{MAPbBr$_3$} & \multicolumn{5}{c|}{MAPbI$_3$} \\
\midrule
 & \multicolumn{2}{c|}{250 K} & 330 K & 335 K & \multicolumn{2}{c|}{345 K} & 360 K\\
 & XDS{\footnote{Experiments performed on protonated MAPbBr$_3$.}} & NDS & XDS & NDS & XDS & NDS & XDS\\
\midrule
Diffuse rods& $10(2)$ & $10(2)$ & $13(4)$ & $15(4)$ & $12(3)$ & $15(8)$ & $11(3)$\\
R-points& $19(2)$ & $17(7)$ & $23(4)$ & $25(7)$ & $19(4)$ & $21(6)$ & $17(2)$\\
\bottomrule
\end{tabular}
\label{tab:cl}
\end{center}
\end{table}

In addition to determining the PbI$_6$ correlation lengths from the diffuse scattering data, we evaluate the PbI$_6$ and MA$^+$ $\xi$ directly from the MD trajectories. To do this, we define a general correlation function between two dynamical variables, $\alpha(\bm{r},t)$ and $\beta(\bm{r},t)$ as \cite{allen2017computer}
\begin{equation}
\begin{gathered}
    G_{\alpha \beta}(\bm{r},t) = \langle \alpha(\bm{r},t) \beta(0,0) \rangle = \int \alpha(\bm{r}+\bm{r}',t+t') \beta(\bm{r}',t') d\bm{r}' dt'. 
    \label{eq:corr_func}
\end{gathered}
\end{equation}
The space-integral at equal time extends over the entire volume of the simulation box and the time integral over a long enough period to sample the correlations adequately. For a system with long-range order, $G_{\alpha \beta}(\bm{r},t)$ will be finite at infinite distance, whereas a system with short-range order exhibits a peak which decays towards zero at large $\bm{r}$ and $t$.

\begin{figure}[t!]
\centering
\includegraphics[width=0.4\linewidth]{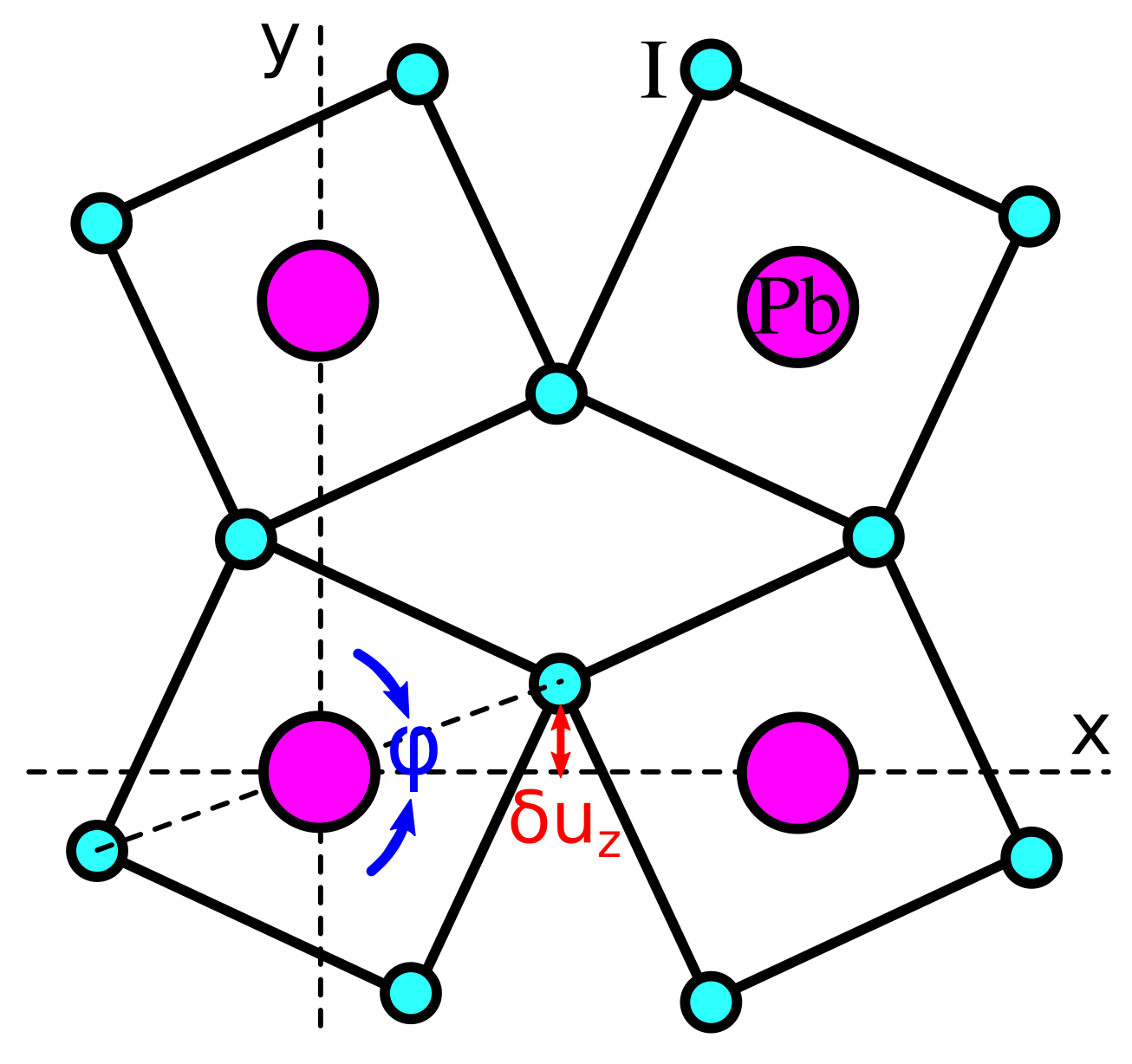}
    \caption{Diagram depicting the dynamical variable $\delta u_z(\bm{r},t)$ for explicit calculation of the PbI$_6$ correlations. $\delta u_z$ are the displacements of the iodine atoms due to rotation of the octahedra about the $z$-axis by angles $\phi$.} 
\label{SI_fig:iodine_displacement}
\end{figure}

For the PbI$_6$ correlations, we use the ``in-plane" iodine atom displacements, $\delta u_z(\bm{r},t)$, as dynamical variables (see \cref{SI_fig:iodine_displacement}). The in-plane displacement is defined as that of the iodine atom perpendicular to the Pb-I bond arising from rotations of the octahedra about the axis perpendicular to the plane. We limit our analysis to displacements in the $x-y$ plane; i.e. rotations about the $z-$axis, but note that these correlations are present within the $y-z$ and $z-x$ planes as well, as discussed in the Main Text. We define $\delta u_z(\bm{r},t) = \sum_i \delta(\bm{r}-\bm{r}_i) \sin(\phi_i(t))$ with the bond-length set to unity and the octahedra fixed at the center of the unit cell. Thus, the PbI$_6$ correlation function only incorporates azimuthal rotations of the octahedra. 

In the case of the MA$^+$ cations, the variables $\alpha(\bm{r},t)$ and $\beta(\bm{r},t)$ are the Cartesian components of the MA$^+$ molecular orientation: $\bm{n}(\bm{r},t) =  n_x(\bm{r},t) \bm{\hat{x}} + n_y(\bm{r},t) \bm{\hat{y}} + n_z(\bm{r},t) \bm{\hat{z}}$. The vectors $\bm{n}_i(t)$ are parallel to the relative position vector between the C and N atoms belonging to the same MA$^+$ cation. The orientation field is $\bm{n}(\bm{r},t)\equiv \sum_i \bm{n}_i(t) \delta(\bm{r}-\bm{r}_i)$ with $\bm{r}_i$ the center-of-mass coordinate of the molecule. The sum runs over all MA$^+$ cations in the simulation cell (using periodic boundary conditions). For simplicity, we also assume $\bm{n}_i(t)$ are dimensionless unit vectors and that their centers-of-mass are fixed at the unit cell center. This isolates the MA$^+$ correlation function to measuring the MA$^+$ orientations only.

\cref{eq:corr_func} can be evaluated directly, but the computational cost scales poorly with system size and time interval. It is more efficient to evaluate in reciprocal space where, by the convolution theorem, convolution (e.g. \cref{eq:corr_func}) becomes simple multiplication. The reciprocal space analogue of \cref{eq:corr_func} is 
\begin{equation}
\begin{gathered}
    G_{\alpha \beta} (\bm{q},\omega) = \alpha^*(\bm{q},\omega)  \beta(\bm{q},\omega) .
    \label{eq:corr_func_Qw}
\end{gathered}
\end{equation}
$\alpha(\bm{q},\omega)$ is the space and time Fourier transform of the $\alpha(\bm{r},t)$ and similarly for $\beta(\bm{q},\omega)$. $^*$ denotes complex conjugation. The real space version is recovered by inverse Fourier transforming (assuming the volume and time interval are infinite for notational convenience):
\begin{equation}
\begin{gathered}
    G_{\alpha \beta}(\bm{r},t) = \int G_{\alpha \beta} (\bm{q},\omega)  \exp(-i\bm{q}\cdot\bm{r}+i\omega t) \frac{d\omega }{2\pi} \frac{d\bm{q}}{(2\pi)^3} .
    \label{eq:corr_func_fft}
\end{gathered}
\end{equation}
\cref{eq:corr_func_fft} is efficiently evaluated using numerical fast Fourier transforms. 

\begin{figure}[t!]
    \centering
    \includegraphics[width = \columnwidth]{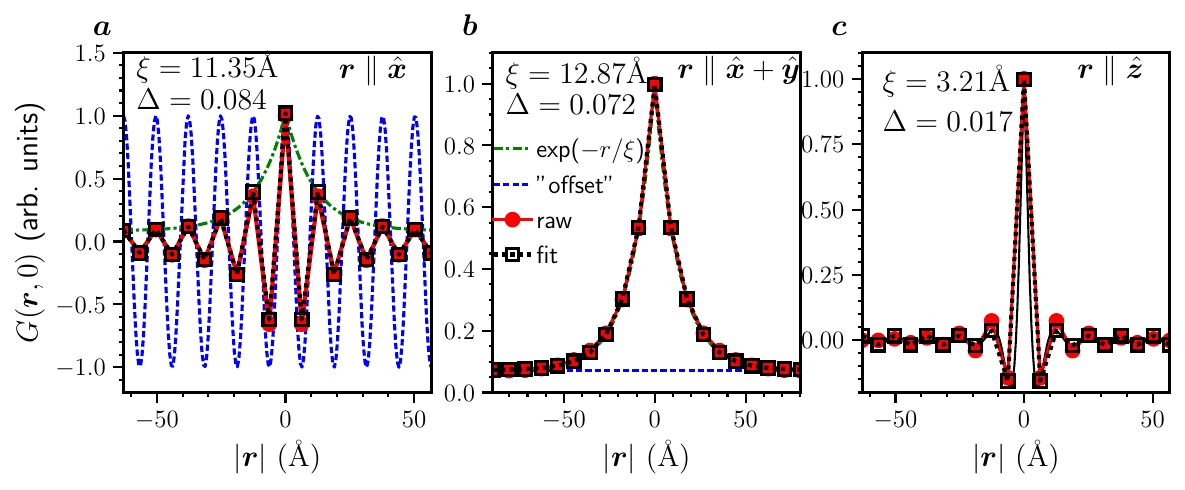}
    \caption{Equal-time correlations of the PbI$_6$ rotations along an in-plane lattice vector a), along the in-plane diagonal b), and along the axis perpendicular to the plane c). Components of the fit are plotted as dashed (offset) and dot-dash ($\exp(-\vert r\vert /\xi)$ lines as detailed in b). The correlation lengths, $\xi$, are plotted with each panel.}
    \label{SI_fig:MDcorr_PbI6}
\end{figure}

The PbI$_6$ correlations in \cref{SI_fig:MDcorr_PbI6} exhibit a localized peak superposed with an oscillatory or constant offset. The offset corresponds to partial long-range order arising from a population of PbI$_6$ octahedra with non-zero tilts. We note that these tilt angles are less than those observed within the 2D pancakes, as reflected in \cref{fig:snapshots}, and therefore the correlations in \cref{SI_fig:MDcorr_PbI6} are smaller at longer length scales. Along the in-plane lattice vector (\cref{SI_fig:MDcorr_PbI6}a), neighboring octahedra rotate in opposite directions as required by the PbI$_6$ bond coupling them. As a result, the correlations alternate between positive and negative for each successive unit cell. Similarly, along the in-plane diagonal, neighboring octahedra rotate in the same direction (\cref{SI_fig:MDcorr_PbI6}b), so there are no negative correlations. The $\xi$ are determined by fitting an exponential plus an oscillatory or constant offset depending on the direction. The in-plane $\xi$ are both $\sim 12~\textrm{\AA}$, close to the $12(3)~\textrm{\AA}$ determined by fitting the diffuse rods in the XDS data. 

Along the out-of-plane direction, \cref{SI_fig:MDcorr_PbI6}c, the localized peak is extremely narrow: less than a unit cell. Since the dynamical variables for the PbI$_6$ correlations are the in-plane iodine displacements, larger correlations correspond to displacements of similar magnitude. Thus, the largest peaks in \cref{SI_fig:MDcorr_PbI6} correspond to localized regions of octahedra with similar tilts. These extend several unit cells in the in-plane direction but only a single unit cell in the out-of-plane direction. Moreover, the in-plane $\xi$ in \cref{SI_fig:MDcorr_PbI6}a, b, are nearly identical, suggesting the correlated regions are nearly circular. The correlations in these PbI$_6$ calculations correspond to the 2D pancakes with large octahedral tilts as discussed in the Main Text.

\begin{figure}[t!]
    \centering
    \includegraphics[width=1\linewidth]{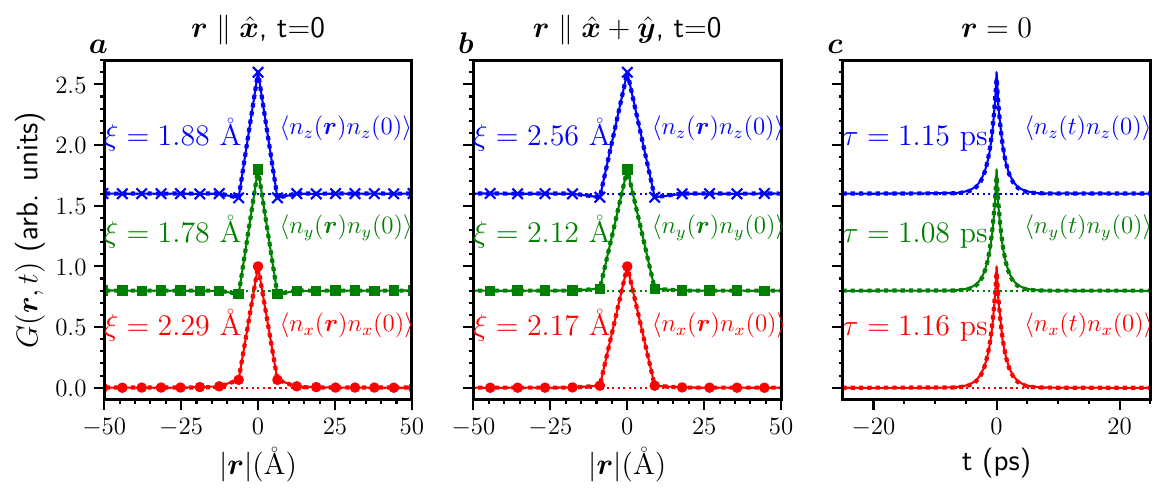}
    \caption{Equal-time correlations of the MA$^+$ orientations along a lattice vector a), along the diagonal spanned by two lattice vectors b), and at equal-position but separated in time c). Correlations between Cartesian components of the MA$^+$ orientation vector are labeled $\langle n_i(\bm{r}) n_i(0) \rangle$ with $i$ labeling the component. The correlation length, $\xi$, and correlation times, $\tau$, are plotted in the panels.}
    \label{SI_fig:MDcorr_MA_1}
\end{figure}

\begin{figure}[t!]
    \centering
    \includegraphics[width=1\linewidth]{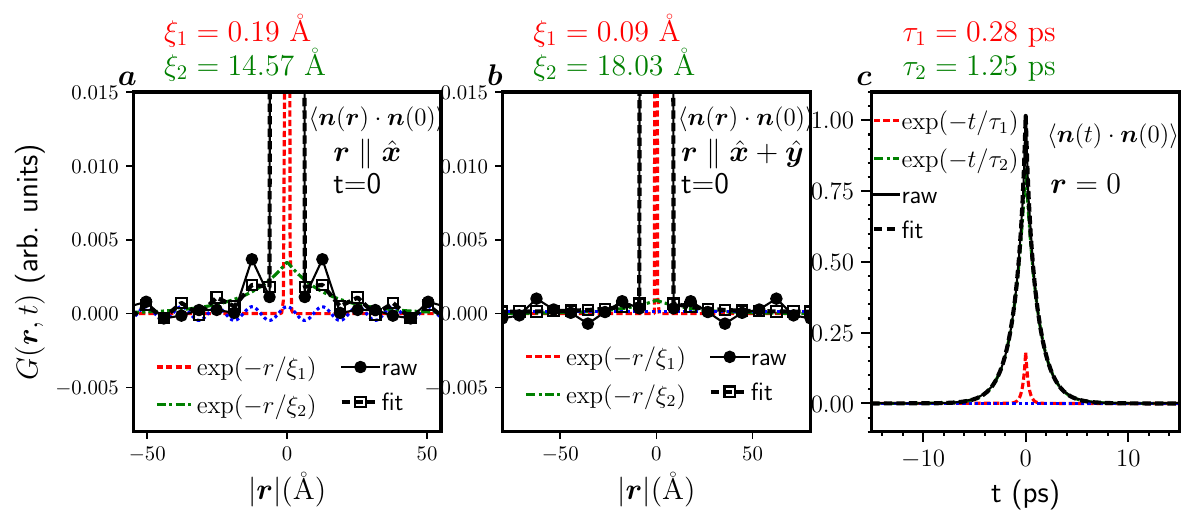}
    \caption{Equal-time correlations of the MA$^+$ orientations along a lattice vector a), along the diagonal spanned by two lattice vectors b), and at equal-position but separated in time c). Only the dot products $\langle \bm{n}(\bm{r}) \cdot \bm{n}(0) \rangle = \langle \sum^3_{i=0} n_i(\bm{r}) n_i(0) \rangle $ are shown. The $y-$axis scale is set to highlight the long-ranged length scale component of the correlations. The two correlation lengths, $\xi_1$ and $\xi_2$, and the correlation times, $\tau_1$ and $\tau_2$, are plotted in the panels.}
    \label{SI_fig:MDcorr_MA_2}
\end{figure}

At a glance, the equal time MA$^+$ correlations in \cref{SI_fig:MDcorr_MA_1} look like narrow, sharp peaks indicating primarily uncorrelated orientations. The peaks are fit as $\exp(-\vert r\vert /\xi)$ with correlation lengths of $1/2$ a unit cell. However, we expect MA$^+$ correlations to have $\xi$ similar to those measured experimentally ($\sim 15 \textrm{\AA}$, see \cref{tab:cl}). Correlation lengths of $\sim 1/2$ a unit cell are inconsistent with this. However, there is a small, additional component to the MA$^+$ correlations with a second length scale, highlighted in \cref{SI_fig:MDcorr_MA_2} which expands the y-axis. Here, we fit two exponential decays and, where appropriate, an oscillatory or constant background to the correlations. The fit is improved and RMS error reduced. The longer length scale is $\sim 15 \textrm{\AA}$, corresponding to in-plane MA$^+$ correlations consistent with both the PbI$_6$ 2D pancakes discussed above and the experimental data. 

In ref. \cite{mattoni2015methylammonium}, MAPbI$_3$ was simulated using the same MD potential and the \emph{time} correlations of the MA$^+$ orientations are fit to a model including \emph{three} time scales: (i) diffusive dynamics of coupled molecules, (ii) free rotation of the molecules, and (iii) coupling of the molecules to phonons. The local 2D nature of the PbI$_6$ dynamics were unknown at the time, so it is unclear if one of the three time scales corresponds to the pancakes. Still, direct comparison of the time correlation curves in \cref{SI_fig:MDcorr_MA_1}c, \ref{SI_fig:MDcorr_MA_2}c to those in reference \cite{mattoni2015methylammonium} reveals very good agreement. The short ranged component could be related to free rotation of the molecules or to inter-molecular interactions.

\section{Inelastic scattering from MD}

\begin{figure}[t!]
\centering
\includegraphics[width=0.8\linewidth]{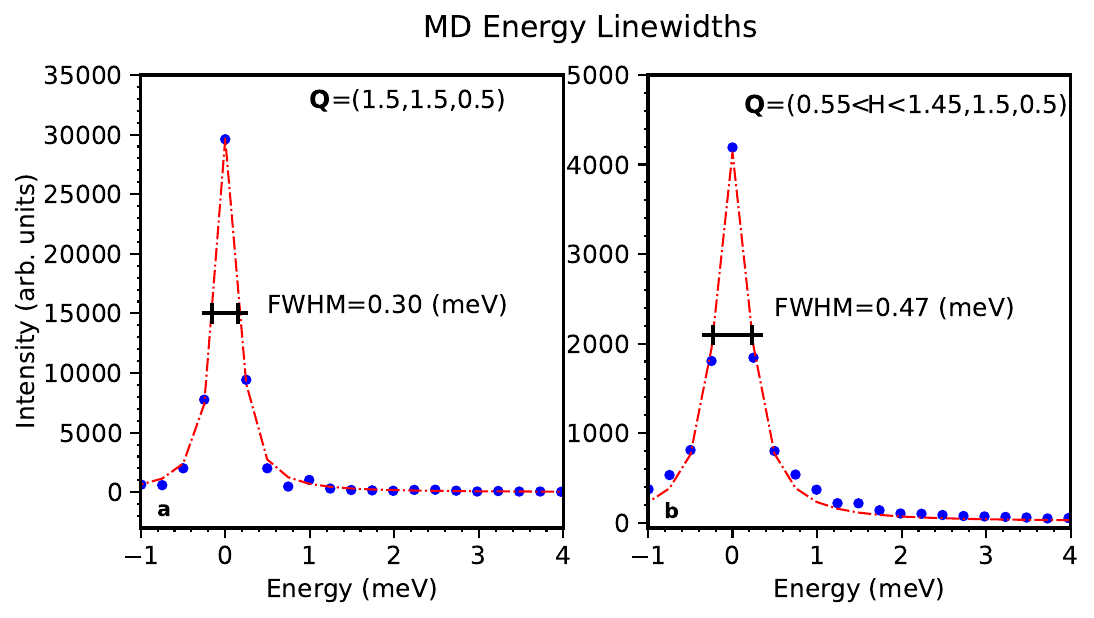}
    \caption{Energy linewidth fits to simulated INS intensity calculated from MD at the R-point, a), and to calculated intensity integrated along a diffuse rod, b). The data in a) are integrated between (1.45:H:1.55, 1.45:K:1.55, 0.45:L:0.55) and in b) between (0.55:H:1.45, 1.45:K:1.55, 0.45:L:0.55). The lineshapes are assumed to be Lorentzian. The FWHM's are shown by the markers and labels in the plots. }
\label{SI_fig:md_fit}
\end{figure}

In \cref{SI_fig:md_fit} we show the results of fitting Lorentzian functions to $S(\bm{Q},E)$ calculated from MD to determine theoretical energy linewidths at the R-point and along a diffuse rod. The MD lifetimes are compared to experiment in \cref{fig:fig4} in \cref{sec:mapi_results}. MD predicts lifetimes consisted with those observed in the experiment. 

\subsection{Phonon dispersions}

\begin{figure}[t!]
\centering
\includegraphics[width=0.75\linewidth]{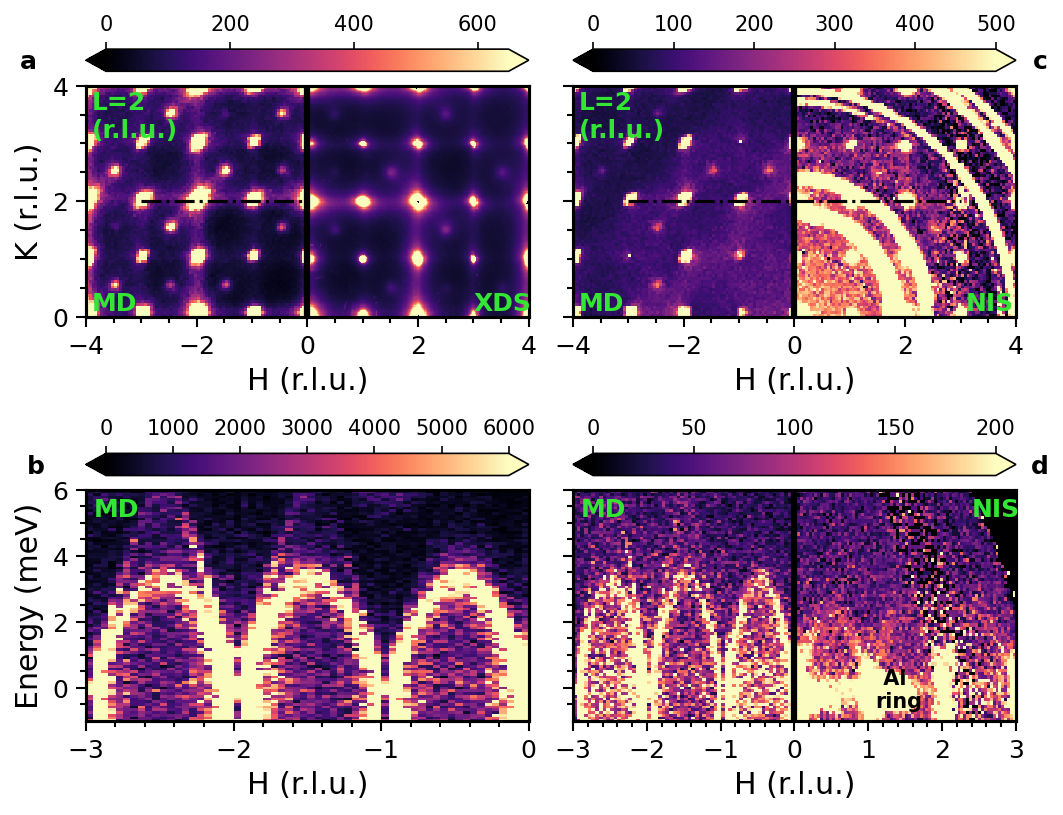}
    \caption{a),c) Diffuse scattering intensity in the L=2 plane from MD and experiment. a) - b) are x-rays and c) - d) are from neutrons. b) and d) show inelastic scattering from MD and experiment. Note, we do not have experimental inelastic x-ray data so the inelastic x-ray plot only contains theoretical results.}
\label{SI_fig:x_phonons}
\end{figure}

The acoustic phonon dispersion along the $\Gamma$-X direction has been well studied in MAPbI$_3$ and MAPbBr$_3$ \cite{gold-parkerAcousticPhononLifetimes2018,ferreira_elastic_2018} and is reproduced in our MD $S(\bm{Q},E)$ calculations in \cref{SI_fig:x_phonons} with both X-ray form-factors and neutron scattering lengths. These dispersions are also present in the neutron inelastic scattering (INS) $S(\bm{Q},E)$ in \cref{SI_fig:x_phonons}d. Given the low energy of these phonons, we expect them to contribute thermal diffuse intensity in both the XDS and NDS/INS experiments. In the L = 2 scattering plane, we indeed see rods of diffuse intensity along the $\Gamma$-X-$\Gamma$ high symmetry direction. A careful inspection of the $S(\bm{Q},E)$ in \cref{SI_fig:x_phonons}b and d show additional quasielastic intensity at $E = 0$. This intensity varies between BZs in the calculated neutron $S(\bm{Q},E)$ in \cref{SI_fig:x_phonons}d, suggesting additional diffuse scattering contributions from the MA$^+$ cations.

Scattering is also observed at the X-point in the L = 0.5, 1.5, 2.5 planes in XDS and the energy-integrated MD $S(\bm{Q})$, but not NDS. This indicates an inelastic process with at least 0.5 meV energy transfer. X-ray and neutron spectroscopy have identified transverse and longitudinal acoustic phonons at the X-point with energies ranging from 2 - 5 meV in several hybrid LHPs \cite{gold-parkerAcousticPhononLifetimes2018, songvilayCommonAcousticPhonon2019, fujii_neutron-scattering_1974, beecherDirectObservationDynamic2016, ferreira_elastic_2018}. Therefore, it is likely that the broad diffuse intensity observed at the X-points originates from these phonons.

\subsection{Acknowledgements}

This research benefited from helpful discussions with Xixi Qin, Volker Blum, and Alex Zunger. We thank Maximilian Schilcher and David Egger (both TU Munich) for providing their MD simulations for comparison. This work (X-ray and neutron scattering, interpretation) was supported by the Center for Hybrid Organic Inorganic Semiconductors for Energy, and Energy Frontier Research Center funded by the Office of Basic Energy Sciences, an office of science within the US Department of Energy (DOE). J. A. V. acknowledges fellowship support from the Stanford University Office of the Vice Provost of Graduate Education and the National Science Foundation Graduate Research Fellowship Program under Grant No. DGE – 1656518 (sample preparation, NDS and XDS data collection). H. I. K. acknowledges funding through the DOE Office of Basic Energy Sciences, Division of Materials Science and Engineering, under Contract No. DE-AC02-76SF0051 (sample preparation). B.A. and E.E. acknowledge funding from the National Science Foundation, award OAC 2118201 (MD simulations). T.C.S. and D.R. acknowledge funding by the DOE Office of Basic Energy Sciences, Office of Science, under Contract No. DE-SC0006939 (INS data collection, S(Q) calculations, interpretation). A.M.R. acknowledges support from the National Science Foundation, Science and Technology Centers Program, under grant number DMR-2019444 (supplemental MD simulations). A portion of this research used resources at the Spallation Neutron Source, a DOE Office of Science User Facility operated by the Oak Ridge National Laboratory. Use of the Advanced Photon Source at Argonne National Laboratory was supported by the U.S. Department of Energy, Office of Science, Office of Basic Energy Sciences, under Contract No. DE-AC02-06CH11357. Experiments at the ISIS Pulsed Neutron and Muon Source were supported by a beamtime allocation from the Science and Technology Facilities Council. Any mention of commercial products here is for information only; it does not imply recommendation or endorsement by the National Institute of Standards and Technology.

\chapter{Disorder and inhomogeneity broadening in thermoelectric clathrates}\label{chp:bgg}

In the search for high-performance thermoelectrics, materials such as clathrates have drawn attention due to having both glass-like low phonon thermal conductivity and crystal-like high electrical conductivity. Ba$_{8}$Ga$_{16}$Ge$_{30}$ (BGG) has a loosely bound guest Ba atom trapped inside rigid Ga/Ge cage structures. Avoided crossings between acoustic phonons and the flat guest atom branches have been proposed to be the source of the low lattice thermal conductivity of BGG. Ga/Ge site disorder with Ga and Ge exchanging places in different unit cells has also been reported. We used time-of-flight neutron scattering to measure the complete phonon spectrum in a large single crystal of BGG and compared these results with predictions of density functional theory to elucidate the effect of the disorder on heat-carrying phonons. Experimental results agreed much better with the calculation assuming the disorder than with the calculation assuming the ordered configuration. Although atomic masses of Ga and Ge are nearly identical, I found that the small disordered perturbation lifts the degeneracy of otherwise highly-degenerate, flat optical modes, leading to effective broadening of phonon peaks and the introduction of many anticrossings in that  strongly reduces acoustic phonon group velocities, significantly reducing thermal conductivity. Our work points at a new path towards optimizing thermoelectrics 

This section is closely based on the corresponding article \cite{roy2023occupational}, on which I was co-author and contributed all of the computational and theoretical work. Susimata Roy and Prof. Dmitry Reznik performed the experimental analysis. Dmitry Reznik was PI.

\section{Occupational Disorder as the Origin of Flattening of the Acoustic Phonon Branches in the Clathrate Ba$_{8}$Ga$_{16}$Ge$_{30}$}

Thermoelectric materials enable environmentally-friendly waste-heat to electricity conversion \cite{snyder2008complex, zhang2015thermoelectric}. The efficiency of a thermoelectric is determined by a dimensionless quantity called the figure of merit: $ZT=\sigma S^{2} T/ \kappa$, in which $\sigma$ is the electrical conductivity, $S$ is the Seebeck coefficient, $T$ is the temperature, and $\kappa$ is the thermal conductivity \cite{ziman1972principles}. The thermal conductivity $\kappa=\kappa_L+\kappa_e$ is the sum of lattice thermal conductivity ($\kappa_{L}$) and electronic thermal conductivity ($\kappa_{e}$). It is classically known that the lattice component is the relevant one for thermoelectric performance \cite{chasmar1959thermoelectric}. According to the kinetic theory of gases, $\kappa_{L}$ is proportional to the lattice heat capacity $C$, average phonon velocity $v$, and the average relaxation time $\tau$ of phonons. 
In most cases low, glass-like $\kappa_{L}$ is attributed to reduced $v$ and/or $\tau$.

\begin{figure}[t!]
\centering
\includegraphics[width=1\linewidth]{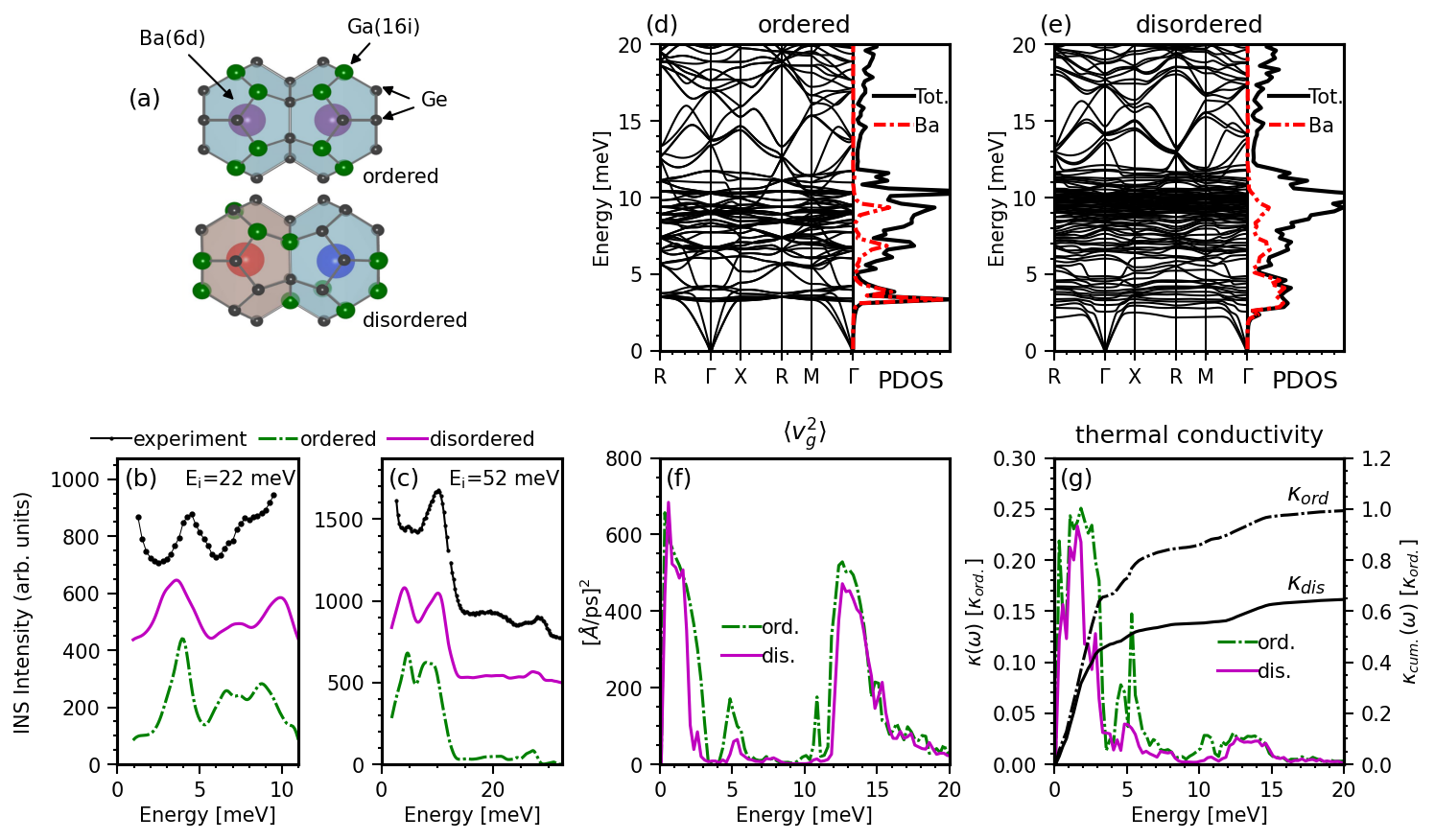}
    \caption{Summary of the lattice dynamical phenomena considered in this work. (a) Ba-6d containing (Ga, Ge)$_{24}$ cage portion of the 3d unit cells of perfectly ordered (top) and disordered, with scrambled Ge/Ga atom occupations (bottom) Ba$_8$Ga$_{16}$Ge$_{30}$.(b,c) Black lines show neutron-weighted densities of states (DoS) (inelastic neutron scattering intensities integrated over the 3$<$H$<$6, 3$<$K$<$7, 4.5$<$L$<$6.5 r.l.u. region of reciprocal space) measured with incident energy E$_i$=22 meV (b) and E$_i$=52 meV (c). The green/pink lines are the same quantity calculated from DFT using the ordered/disordered unit cells and broadened with an approximate resolution function fit to the experimental intensity. The disordered calculation and the experimental curves for E$_i$=22 meV are offset vertically by 350 and 400 counts respectively. The experimental curve has an additional large ($\sim$250 counts) background that is not present in the calculations. For E$_i$=52 meV, the disordered and experimental curves are offset by 500 and 700 counts respectively. Here, the experimental background is about $\sim$ 600 counts. (d,e) The ordered (d) and disordered (e) phonon dispersions and densities of states (PDOS). The dash-dot red curve is the projection onto the Ba atoms; the black curve is the total DOS. (f) The average group velocities squared, $\langle v^2_g\rangle$, calculated from the dispersions in (d) and (e) by averaging over all modes and $\bm{q}$-points. (g) Black curves represent the spectral, $\kappa(\omega)$, and colored curves represent the cumulative, $\kappa_{cum.}(\omega)$, thermal conductivities at 300 K calculated from the group velocities and densities of states in (f) and (d),(e). Only Umklapp processes are considered; $\tau(\omega)=\tau_0 \omega^{-2}$ with $\tau_0$ the same for the ordered and disordered calculations. The data in (g) are in units of the total thermal conductivity of the ordered crystal.}
    \label{fig:summary_fig}
\end{figure}

Glass-like low lattice thermal conductivity and crystal-like high electrical conductivity coexist in so-called phonon-glass electron-crystal (PGEC) materials \cite{slack1995crc}. One way to realize this concept is by designing materials such that loosely bound "guest'' atoms sit inside empty spaces of a rigid atomic lattice with good electronic conductivity \cite{nolas1998semiconducting,keppens1998localized,hsu2004cubic,poudeu2006high}. In this case interactions between the low-energy rattling motion of the guest atoms and acoustic phonons reduce $v$ and/or $\tau$ depending on specific materials. In particular, the reduction of $v$ is achieved through avoided crossings (anticrossings) between low-lying optic branches with acoustic branches that result from coupling between the phonons in branches that would cross in the absence of such a coupling. 
In this case the dispersion curves never cross instead "avoiding" each other. As a result the branches become more flat and $v$ decreases. 

The X$_8$Ga$_{16}$Ge$_{30}$ (X=Ba,Eu,Sr) clathrates have low lattice thermal conductivity ($\sim$ 1 W/mK at room temperature), which makes them promising candidates for efficient thermoelectrics \cite{nolas1998semiconducting,bentien2004thermal,cohn1999glasslike}. Their structure is characterized by tetrakaidecahedral (Ga, Ge)$_{24}$ and dodecahedral cages (Ga, Ge)$_{20}$ with loosely bonded Ba, Eu, or Sr guest atoms inside, which makes them a classic type of PGEC materials \cite{kovnir2004semiconducting,toberer2010zintl}. Evidence of strong occupational disorder of the Ga/Ge sites has also emerged recently, i.e. Ga and Ge are distributed nearly randomly on the cage vertices [\cref{fig:summary_fig} (a)]  \cite{chakoumakos2001structural,christensen2006crystal,bentien2000experimental,bentien2005crystal,bentien2002maximum}. Since Ga and Ge have similar atomic masses, the effect of their occupational disorder on phonon dispersions was implicitly assumed to be minimal. 

Here we combined comprehensive time-of-flight (TOF) inelastic neutron scattering measurements of Ba$_{8}$Ga$_{16}$Ge$_{30}$ with calculations based on the density functional theory (DFT) assuming both the ordered and the disordered structures and found that the disorder splits the degenerate rattler branches into multiple nearly flat branches. The new branches produce many more avoided crossings with the acoustic modes, which significantly lowers their average group velocities and, as a consequence, reduces thermal conductivity. Our result demonstrates that occupational disorder control represents a new direction in the design of thermoelectric materials based on clathrates.

\section{Experimental Details}

\begin{figure}[t!]
\centering
\includegraphics[width=0.65\linewidth]{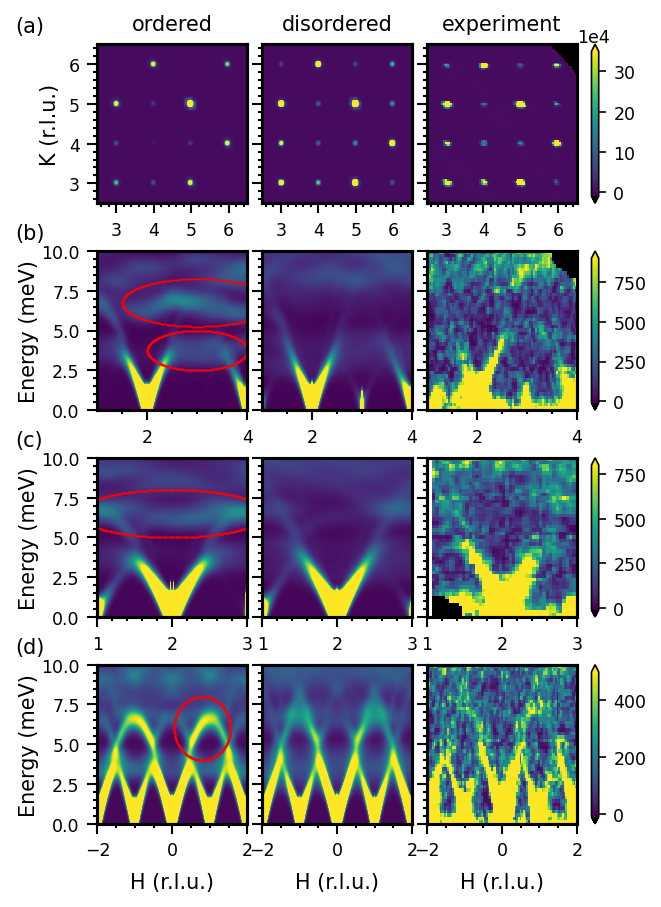}
    \caption{Ordered (left column) and disordered (middle column) neutron spectra compared to experiment (right column). (a) Bragg peaks (-1$<$E$<$1 meV) in the (H,K,L=6) plane. In the ordered cell, coherent scattering from the ordered arrangement of atoms results in certain Bragg peaks (e.g. $\bm{Q}$=(4,4,6)) having no intensity. On the other hand, scattering from the disordered arrangement of atoms results in some remaining intensity at these Bragg peaks, in excellent agreement with the experimental Bragg pattern. The remaining rows show inelastic scattering spectra S($\bm{Q},\omega$) along the $\bm{Q}$=(H,6,6) (b), $\bm{Q}$=(H,2,8) (c), and $\bm{Q}$=(H,5,3) (d) reciprocal lattice directions. The disordered calculation is averaged over all otherwise equivalent directions as explained in the text. Red ovals in the ordered calculation indicate intensity from excitations that is not visible in the disordered calculation and experiment as discussed below.}
    \label{fig:bragg_sqw}
\end{figure}

\begin{figure}[t!]
\centering
\includegraphics[width=0.6\linewidth]{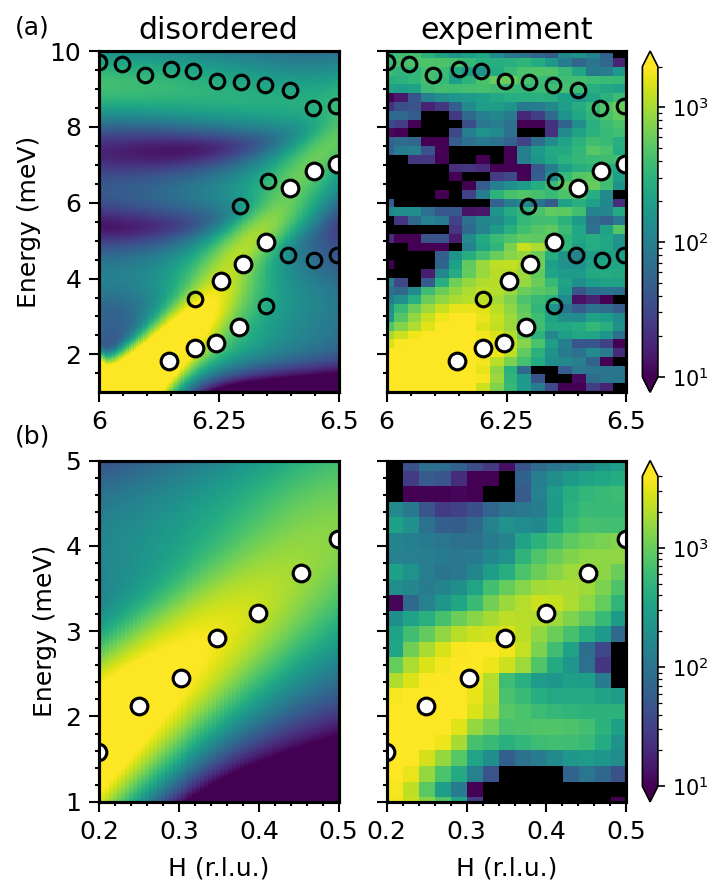}
    \caption{(a) LA and (b) TA phonon scattering intensities (color maps) at $\bm{Q}$=(5.5+h,0,0) (a) and $\bm{Q}$=(6,h,0) (b) and experimental phonon energies (circles) obtained from the multizone fit of experimental inelastic neutron scattering intensity S(Q,$\omega$) (see fig. $5$ in supplementary information). The big circles indicate acoustic phonons and the small circles are optic modes. We do not include the small intensity near 2.5 meV in our phonon fits as it does not appear in any other zones and may be an artifact. }
    \label{fig:dispersions}
\end{figure}

Neutron scattering measurements were performed on the MERLIN direct geometry chopper spectrometer at the ISIS Neutron and Muon source at the Rutherford Appleton Laboratory in Didcot, UK \cite{bewleyMERLINNewHigh2006}. The Ba$_{8}$Ga$_{16}$Ge$_{30}$ crystal used for the inelastic neutron scattering experiment is the same sample that was used in ref. \cite{christensen2008avoided}. MERLIN has high flux and a large detector area, collecting the four dimensional inelastic neutron scattering data set $S(\bm{Q},\omega)$ across many Brillouin zones (BZ). For data analysis, we used the Phonon Explorer software \cite{phonon-explorer,reznik2020automating}, which enables efficient search for wavevectors where a particular effect is observed most clearly, and performs multizone fitting to efficiently separate phonon branches that are much more closely spaced than the experimental resolution \cite{parshall2014phonon}. I used \textsc{phonopy} to solve the lattice dynamical equations based on the force-constants from Ref. \cite{ikeda2019kondo}. The inelastic neutron scattering structure factors, $S(\bm{Q},\omega)$ were simulated using the  \textsc{snaxs} \cite{parshall} and \textsc{euphonic} \cite{euphonic} softwares. To represent the finite line widths and experimental resolution broadening, the structure factors in \cref{fig:bragg_sqw,fig:dispersions} were convolved with a Gaussian function assuming the full-width at half-maximum (FWHM) is 0.75 meV. The intensity scales in arbitrary units for all $S(\bm{Q},\omega)$ calculations are the same.

\section{Computational details}

The ordered phase of Ba$_{8}$Ga$_{16}$Ge$_{30}$ typically used in DFT calculations (\cref{fig:summary_fig} (a), top) has cubic symmetry and intensities along reciprocal lattice directions with permuted axes are identical. On the other hand, the real structure of Ba$_{8}$Ga$_{16}$Ge$_{30}$ is disordered with Ga and Ge atoms randomly distributed on the cage vertices (\cref{fig:summary_fig} (a), bottom) \cite{chakoumakos2001structural,christensen2006crystal,bentien2000experimental,ikeda2019kondo,bentien2005crystal,bentien2002maximum}. The disorder breaks the cubic symmetry (Pm-3n$\rightarrow$P1) and intensities along directions that are equivalent in the ordered crystal are no longer identical. Still, the experimental $S(\bm{Q},\omega)$ is averaged over the large ($\sim 10^{23}$) number of different disordered unit cell configurations of the macroscopic crystal, which results in an apparent cubic symmetry.

In DFT the force constants of Ba$_{8}$Ga$_{16}$Ge$_{30}$ were calculated from a single unit cell in both the ordered and disordered phases \cite{ikeda2019kondo} because Ba$_{8}$Ga$_{16}$Ge$_{30}$ has a large enough unit cell that force-constants fall off sufficiently at the cell boundary (see \cref{sec:finite_diffs}). Still, disorder breaks translational symmetry, so the computational unit cell should be sufficiently large that atoms on opposite sides of the cell are uncorrelated. Moreover, one should calculate phonons using an ensemble of disordered configurations and average over all ensembles. However, this is too computationally expensive in our case. 

I calculated $S(\bm{Q},\omega)$ by evaluating \cref{eq:one_phonon_creation} from a single configuration (from ref. \cite{ikeda2019kondo}) chosen using the "special quasi-random structure" (SQS) method \cite{zunger1990special}. The goal of SQS is to pick a small computational supercell that best matches the disorder in a very large (ideally infinite) supercell. This was achieved by fitting correlation functions calculated from a single unit cell with scrambled Ge/Ga occupations to true "random" correlation functions; e.g. from experiment or calculated from a very large supercell. The "disorder" (in the case of Ba$_8$Ga$_{16}$Ge$_{30}$, the Ga/Ge site occupancies) is varied to minimize the difference between the correlation function(s) of the computational cell and the true random correlation function(s). The structure with the best match was chosen for the calculations. To approximate the implicit directional averaging, the theoretical $S(\bm{Q},\omega)$ calculated from this disordered cell were averaged over all directions that are equivalent assuming cubic symmetry.

\section{Results}

Calculation based on the ordered structure predict that the optical branches bunch into several narrow energy intervals in the (generalized) phonon density of states. Projecting the phonon densities of states onto the individual atoms shows primarily Ba - 6d, character, thus these are the rattler modes of Ba in the 6d site. The introduction of the Ba atom into an otherwise pristine crystal generates anticrossings in the acoustic modes \cite{christensen2008avoided}. Disorder lifts the degeneracy of the other-wise highly degenerate flat optical branches and spreads them out in energy (\cref{fig:summary_fig}d,e). In particular, the ordered calculation gives avoided crossings between acoustic branches and the nearly flat rattler branches near $\sim$4 meV, which tend to reduce the phonon group velocity and, as a consequence, the thermal conductivity. Due to disorder, these avoided crossings become distributed throughout a broad energy interval between 2 and 5 meV, which further suppresses thermal conductivity by a large amount. The main effect of the splitting of the rattler modes is to create additional avoided crossings with the acoustic phonons (see \cref{sec:bgg_kappa} below for more details).

\begin{figure}[t!]
    \centering
    \includegraphics[width=0.6\linewidth]{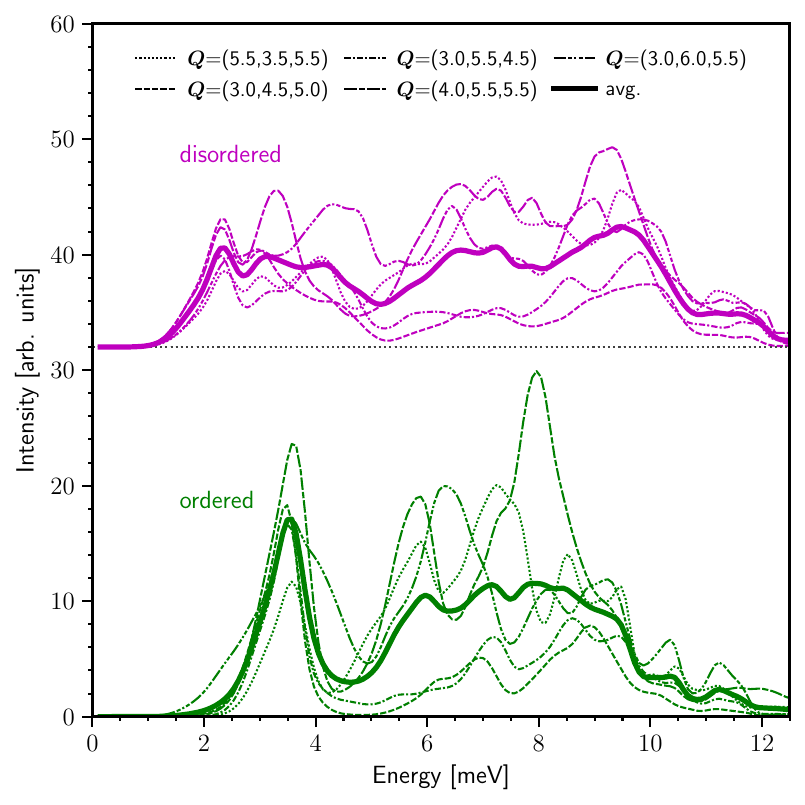}
    \caption{Selected (theoretical) constant-$\bm{Q}$ cuts from the range $3<H<6$, $3<K<7$, $4.5<L<6.5$ r.l.u., i.e. the range integrated for the generalized densities of states in Fig. 1 in the main text. The $\bm{Q}$-points are labelled in the figure. To approximate the experimental $\bm{Q}$-resolution, each point is integrated on an $11\times 11\times 11$ grid within $\bm{Q}\pm 0.25$ r.l.u. The energy resolution is the same as used in Fig. 1 in the main text. Notably, the degenerate peaks at about 3.5 meV in the ordered calculation split into numerous peaks in the range 1.5-5 meV in the disordered calculation.}
    \label{fig:const_Q_subset}
\end{figure}

\cref{fig:summary_fig}b-c shows that the calculated phonon density of states (DOS) corrected for the neutron cross section agrees much better with experiment when the disordered structure is used. In both calculations the energy of the lowest DOS peak is about 3meV vs 4.5 meV in experiment, consistent with the known tendency of the DFT to underestimate the force constants overall. However, the width of this peak, which comes from the gaps at the avoided crossings of acoustic phonons and rattler modes, is increased in the disordered calculation to closely match experiment. Moreover, the calculated Bragg peak intensities agree much better with experiment if disorder is included in the calculation (\cref{fig:bragg_sqw}). The same applies to select phonon spectra (\cref{fig:bragg_sqw}b-d). Color maps of the simulated and background-subtracted experimental spectra of acoustic phonons shown in Figure \ref{fig:dispersions} are similar, highlighting the accuracy of the calculation. Peak positions obtained from the multizone fit agree very well with the color maps of both the simulation and the experiment. In particular, signatures of avoided crossings of the LA branch with optic modes are clearly visible in the simulation and the data.

Comparison of my calculations to the Raman results of Ref. \cite{takasu2006dynamical} shows that the disorder-induced inhomogeneity broadening of the Raman phonon peaks far beyond the experimental resolution is a natural consequence of disorder-induced splitting of the modes (see supplementary info of the corresponding ref. \cite{roy2023occupational}). Moreover, detailed analysis of the $\bm{Q}$-integrated neutron spectra in \cref{fig:summary_fig}b) and c) confirm the apparent broadening in the disordered calculation, consistent with the experimental data, is due to inhomogeneity. In \cref{fig:const_Q_subset}, I calculate the INS spectra for several $\bm{Q}$-points and compare their average between the ordered and disordered calculations. Notably, the sharply peaked mode near $3.5$ meV in the ordered cell becomes a broad, flat peak in the disordered cell.

\subsection{Thermal conductivity calculations}\label{sec:bgg_kappa}

\begin{figure}[t!]
    \centering
    \includegraphics[width=\linewidth]{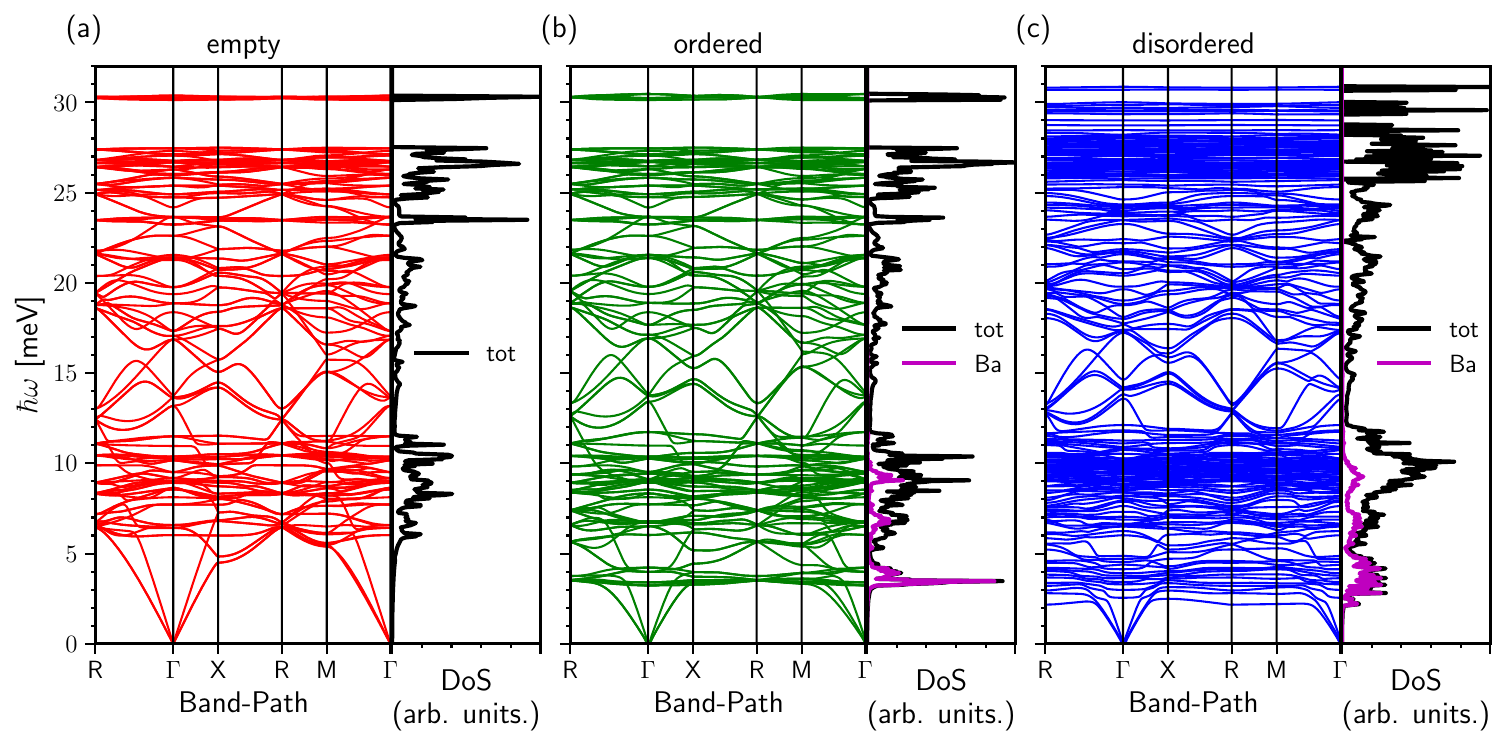}
    \caption{Phonon dispersions and densities of states in (a) the empty clathrate Ga$_{16}$Ge$_{30}$, (b) ordered Ba$_8$Ga$_{16}$Ge$_{30}$, and (c) disordered Ba$_8$Ga$_{16}$Ge$_{30}$. In the empty clathrate there are no Ba atoms, i.e there are no rattler modes intersecting the acoustic phonons. In the ordered compound, the (degenerate) ratter modes intersect the acoustic phonons at $\sim3.5$ meV, forming avoided crossings. In the disordered compound, the rattler modes split, forming many more avoided crossings across a larger energy range.}
    \label{fig:bands}
\end{figure}

My calculation of thermal conductivity (\cref{fig:summary_fig}f-g) considering only Umklapp (\cref{fig:summary_fig}f-g) illustrates a profound effect of the disorder, which suppresses phonon thermal conductivity mostly around 3-5meV where the effect of the rattler branches is the strongest and the cumulative thermal conductivity the largest. The acoustic phonons are the only branches in BGG with significant group velocities; thus, the acoustic phonons contribute significantly to the thermal conductivity. In the ordered case, and ignoring the rattler mode, the acoustic branches disperse nearly linearly up to "Debye frequency" $\sim6$ meV (see \cref{fig:bands}a). With the rattler atom present, the rattler modes intersect the acoustic branches at about $\sim4$ meV (\cref{fig:bands}b), flattening the acoustic modes over a region only $\sim1$ meV wide ( $\sim17\%$ of the acoustic region). The rest of the acoustic region maintains steep dispersion. This effect has already been attributed to lowering the thermal conductivity \cite{christensen2008avoided,tadano2015impact}. 

In the disordered cell, the rattler modes split and suppress group velocities in a much larger region that spans $\sim2$ to $\sim6$ meV ( $\sim 66\%$ of the acoustic region!) (\cref{fig:bands}c). If it is to be believed that the rattler modes in the ordered crystal have a large effect, then the disorder-induced splitting of the rattler modes ought to have an equally (if not more) significant effect.

To prove these claims I estimate the thermal conductivity in the ordered, disordered, and \emph{empty} ordered (i.e. Ga$_{16}$Ge$_{30}$) cells using a simple model. I include the empty ordered cell to determine whether or not the rattler atoms actually have an impact in the first place (spoiler: they do, see \cite{tadano2015impact,christensen2008avoided}). Note that the empty clathrate isn't actually stable. See Tadano et. al \cite{tadano2015impact}. They modeled the empty clathrate using the same data as the Ba$_8$Ga$_{16}$Ge$_{30}$ calculation, but they fit the force constants with the coupling to the Ba atoms constrained to be 0. I used exactly the same procedure here and my empty clathrate dispersions (\cref{fig:bands}a) agree perfectly with theirs.

For the thermal conductivity, I can use (eq. (2) in \cite{zhou2020thermal} or (3) in \cite{zhao2021phonon})
\begin{equation}
    \kappa(T) \propto \int^{\omega_{max}}_0 d\omega c(\omega,T) \bar{v}^2_g(\omega) \tau(\omega,T).
\end{equation}
The integral is over all phonon energies $\hbar \omega$. $\omega_{max}$ is the largest frequency of any phonon. $\bar{v}_g(\omega)$ is the group velocity averaged over all modes and $\bm{q}$ points. $\tau(\omega,T)$ is the average lifetime of all modes with energy $\hbar \omega$; in general, the lifetimes depend on temperature, $T$. $c(\omega,T)$ is the heat capacity. Now let 
\begin{equation}
\begin{gathered}
    u(\omega,T) = \hbar \omega\left[ \frac{1}{2} + g(\omega) n(\omega,T)  \right]
\end{gathered}
\end{equation}
be the energy density of the phonon system. $g(\omega)$ is the density of states and $n(\omega,T)$ the Bose-Einstein distribution:
\begin{equation}
\begin{gathered}
    n(\omega,T) = \frac{1}{\exp \left( \frac{\hbar\omega}{k_B T} \right)-1}.
\end{gathered}
\end{equation}
$g(\omega) n(\omega,T)$ is the "number of phonons" with energy $\hbar \omega$ at temperature $T$. The specific heat is 
\begin{equation}
\begin{gathered}
    c(\omega,T) = \frac{\partial u(\omega,T) }{\partial T} = \hbar \omega g(\omega) \frac{\partial n(\omega,T) }{\partial T} = k_B  \left( \frac{\hbar \omega}{k_B T} \right)^2 \exp\left( \frac{\hbar \omega}{k_B T} \right) n^2(\omega,T) g(\omega). 
\end{gathered}
\end{equation}
Then the thermal conductivity is
\begin{equation}
    \kappa(T) = k_B \int^{\omega_{max}}_0 d\omega  \left( \frac{\hbar \omega}{k_B T} \right)^2  \exp\left( \frac{\hbar \omega}{k_B T} \right) n^2(\omega,T)  g(\omega) \bar{v}^2_g(\omega) \tau(\omega,T).
\end{equation}
If I assume that the only modes in the system are linearly dispersing acoustic phonons with speed of sound, $c$, and use the Debye model for the density of states, $g(\omega)=\omega^2 /(2 \pi^2 c^3)$, I arrive at the Callaway model used in e.g. Ikeda et. al \cite{ikeda2019kondo}. Rather, since our claims depend precisely on the form of the group velocities, I will stick to using the densities of states and average group velocities calculated from the ab-initio force constants. So then the only remaining ingredients I need are the lifetimes, $\tau(\omega,T)$. 

Ikeda et. al \cite{ikeda2019kondo} give a good discussion of $\tau(\omega,T)$ and which terms are important. I \emph{could} use their estimates for the lifetimes in our model. However, I choose a simpler approach: I assume Umklapp scattering, $\tau(\omega,T)\equiv \tau_0 / \omega^{2}$, is the only important contribution to reducing lifetimes. This is usually true at high temperature in semiconductors. I also assume $\tau_0$ is the same for the empty, ordered, and disordered cells. This way, I truly isolate the dependence of the thermal conductivity on the disorder-suppressed group velocities. In fact, this will \emph{under estimate} the reduction in the thermal conductivity of the disordered cell relative to the ordered cell and likewise for the ordered vs empty cell \cite{tadano2015impact}. That this is true for the ordered vs. disordered cell has been shown by Ikeda et. al \cite{ikeda2019kondo} who explicitly calculated the Gr\"{u}neisen parameters and showed that anharmonicity is enhanced in the disordered case: i.e. the lifetimes $\tau(\omega)$ are \emph{smaller} in the disordered cell relative to the ordered one. Similarly, another study on (ordered) Ba$_8$Ga$_{40}$Au$_6$ and Ba$_8$Ga$_{41}$Au$_5$ where one Ge atom was randomly substituted for an Au atom in the ordered cell found a 3-fold reduction in the thermal conductivity of Ba$_8$Ga$_{41}$Au$_5$ relative to the ordered cell \cite{lory2017direct}. They attribute this to an increase in the phase-space for  multi-phonon scattering.

It should be noted that, while Tadano et. al \cite{tadano2015impact} conclude that the anharmonicity is more important than the suppression of group velocities going from the empty cell to ordered Ba$_8$Ga$_{16}$Ge$_{30}$, in Ikeda et. al \cite{ikeda2019kondo} it was shown that anharmonicity only has a minor impact on reducing the thermal conductivity; thus, the suppression of group velocities is critical in correctly describing the low thermal conductivity in disordered compound Ba$_8$Ga$_{16}$Ge$_{30}$.

So I let $\tau(\omega,T) \equiv \tau_0 / \omega^2$. Then I have
\begin{equation}
\begin{gathered}
    \kappa(T) \sim \int^{\omega_{max}}_0 d\omega \exp\left( \frac{\hbar \omega}{k_B T} \right)   n^2(\omega,T) g(\omega) \bar{v}^2_g(\omega) T^{-2}  \\
    \kappa(\omega,T) \equiv \exp\left( \frac{\hbar \omega}{k_B T} \right) n^2(\omega,T)  g(\omega) \bar{v}^2_g(\omega) T^{-2} 
    \label{eq:kappa_simple}
\end{gathered}
\end{equation}
where I ignore the constant pre-factors for simplicity. $\kappa(\omega,T)$ is the so-called "spectral thermal conductivity". I will call $\kappa_{E}(T)$, $\kappa_{O}(T)$, and $\kappa_{D}(T)$ the thermal conductivities of the empty, ordered, and disordered compounds respectively. Another useful quantity to calculate is the "cumulative thermal conductivity" 
\begin{equation}
\begin{gathered}
    \kappa_{cum.}(T,\omega) =  \int_0^{\omega} d\omega' \kappa(T,\omega')
    \label{eq:kappa_cum}
\end{gathered}
\end{equation}
where the integral extends up to some finite $\omega \leq \omega_{max}$. $\kappa_{cum.}(\omega,T)$ characterizes how much the phonons below $\omega$ contribute to the total thermal conductivity. For $\omega=\omega_{max}$, $\kappa_{cum.}(T)=\kappa(T)$. 

Obviously the prescription above will give a bad quantitative estimate of the total thermal conductivity since I let $\tau_0$ be arbitrary. Moreover, $\tau(\omega,T)$ generally depends on temperature, with lifetimes decreasing with temperature. Then I will also predict completely wrong temperature dependence (in fact Ikeda et. al \cite{ikeda2019kondo} show that none of the conventional scattering processes produce the correct temperature dependence). Still, \cref{eq:kappa_simple} is a useful way to compare the relative thermal conductivities of the different configurations. For these reasons, it will be convenient to measure the thermal conductivities relative to the maximum, converged value of the empty clathrate. I call it $\lim_{T\rightarrow \infty} \kappa_E (T) \equiv \kappa^\infty_E$.  

\begin{figure}[t!]
    \centering
    \includegraphics[width=\linewidth]{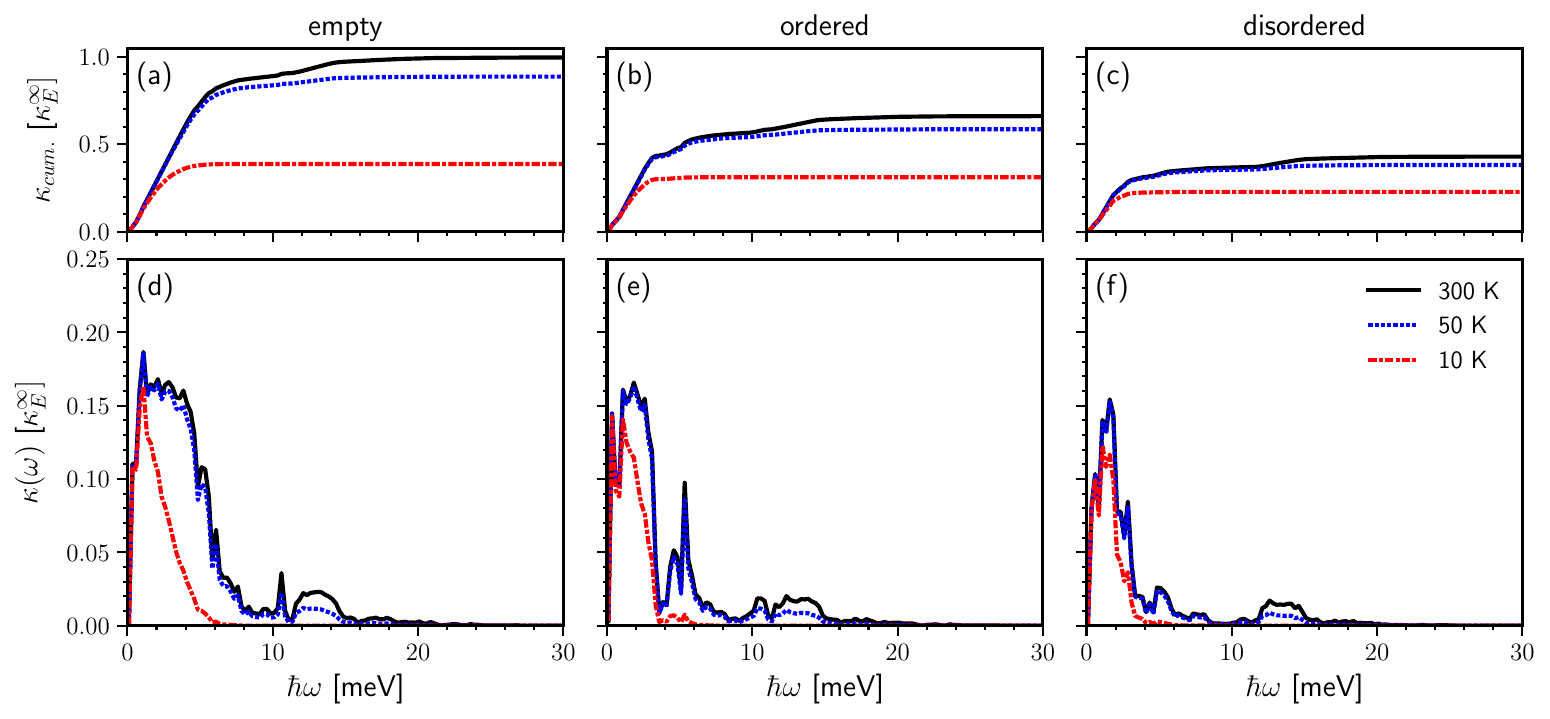}
    \caption{(a)-(c) Cumulative and (d)-(f) spectral thermal conductivities in empty Ga$_{16}$Ge$_{30}$, ordered Ba$_8$Ga$_{16}$Ge$_{30}$ , and disordered Ba$_8$Ga$_{16}$Ge$_{30}$. The data are plotted at 10 K (red dash-dotted lines), 50 K (blue dotted lines), and at 300 K (black solid lines).}
    \label{fig:kappa_freq}
\end{figure}

\cref{fig:kappa_freq} shows the temperature dependence of the thermal conductivity in each configuration. At 50 K, the thermal conductivity is already near its maximum value in each crystal. This is because the occupation of the low-energy, dispersive phonons have saturated (see \cref{fig:kappa_freq}d-f). Higher energy modes are flat, so contribute little to the thermal conductivity. In reality, the thermal conductivity in each compound should reach their maxima at some intermediate temperature, then begin decreasing with temperature as scattering rates increase. This effect is completely neglected here. 

The solid black lines in \cref{fig:kappa_freq}(a)-(c) show the cumulative thermal conductivity at room temperature. The most important feature is that the primary divergence of the thermal conductivities in the different compounds happens below about 10 meV. The physical interpretation of this is as follow: (i) when going from the empty to ordered cells, introducing the rattler modes flattens part of the acoustic branches, suppressing group velocities of the important carriers. See \cref{fig:kappa_freq}(d)-(e), in particular look at the region below 10 meV. (ii) When going from the ordered to disordered cell, the disordered lifts the degeneracy of the rattler modes, flattening a much larger region of the acoustic phonons. See \cref{fig:kappa_freq}(e)-(f).

I have shown that the flattening of the low-energy acoustic branches is significant. For reference, in units of $\kappa^\infty_E$, the converged thermal conductivities of the ordered and disordered clathrates are $\kappa^\infty_O=0.66$ and $\kappa^\infty_D=0.43$ respectively. There is a $34\%$ reduction in the thermal conductivity going from the empty to the ordered cell and $35\%$ reduction from the ordered to disordered cell. I reiterate that this reduction is \emph{solely} due to the disorder-induced suppression of the group-velocities; any disorder-induced suppression of the lifetimes will only reduce the thermal conductivities further \cite{ikeda2019kondo,tadano2015impact,lory2017direct}. I conclude that disorder is an essential ingredient for the low thermal conductivity in Ba$_8$Ga$_{16}$Ge$_{30}$; it is at least as important as the reduction induced by the rattler atoms already discovered by others \cite{christensen2008avoided,tadano2015impact}. 

\section{Discussion}

In Ba$_8$Ga$_{16}$Ge$_{30}$ and similar semiconducting clathrates, most DFT calculations investigating the avoided crossing regions assume that the Ga/Ge site occupancies are fully ordered with Ga only in the 16i site and Ge in the remaining cage vertex positions \cite{tadano2015impact,euchner2019predicting,gonzalez2017estimating,tadano2018quartic,blake1999clathrates,dong2000chemical}. However, there is extremely strong evidence that the Ga/Ge occupancies on the cages vertices are disordered \cite{chakoumakos2001structural,christensen2006crystal,bentien2000experimental,ikeda2019kondo,bentien2005crystal,bentien2002maximum}. 
Unfortunately modeling disordered materials from first principles greatly increases the computational workload, which limited the number of studies of the lattice dynamics in clathrates \cite{he2014nanostructured,gao2017giant}.

Ikeda et al. \cite{ikeda2019kondo} investigated the \emph{temperature-dependence} of the thermal conductivity and specific heat of Ba$_8$Ga$_{16}$Ge$_{30}$ with and without disorder. Only when accounting for correlation in a Kondo-like phonon effect could the experimental temperature dependence be explained \cite{ikeda2019kondo}. They noted that whereas the temperature dependence is robust against disorder, the absolute value of the thermal conductivity is lowered in the disordered calculation. The decrease was attributed to shorter lifetimes due to increased anharmonicity \cite{ikeda2019kondo}, similar to Ba$_8$Ga$_{41}$Au$_{5}$ where disorder increases the available phase-space for 3-phonon scattering,  \cite{lory2017direct}. Here I showed that another important effect of disorder is to lower the phonon group velocities.

Disorder-induced dispersion flattening was theoretically predicted in K$_8$Al$_8$Si$_{38}$, with Al/Si occupational disorder \cite{he2014nanostructured}. Another study found that isotope disorder in the guest atom sites in Si$_x$ and Ba$_8$Si$_x$ ($x\in\{46,230,644\}$) played a major role in reducing thermal conductivity \cite{gao2017giant} due to both lifetimes and group velocity reduction.  However, these effects have not been confirmed in experiment, which we did here for Ba$_8$Ga$_{16}$Ge$_{30}$.

Our ordered cell calculations show that the flat branches in \cref{fig:bragg_sqw,fig:dispersions} near $\sim$4 and $\sim$6 meV come mainly from pure Ba guest atom modes (See \cref{fig:summary_fig} and the atom projected INS intensities in the supplementary information of ref. \cite{roy2023occupational}). These multiply degenerate branches all contribute some intensity to the spectrum. Their intensities appear as a weak but visible flat branch across the BZ (red ovals in \cref{fig:bragg_sqw}). It also shows up as a narrow, pronounced peak in the ordered cell neutron-weighted DoS in \cref{fig:summary_fig}. However, absence of this intensity in the experiment above the background and the significantly improved agreement of the Bragg patterns of the disordered cell over the ordered cell, necessitates the disordered calculation.

In the disordered cell calculations, broken symmetry splits the Ba rattler atom branches. Since the wave vectors $\bm{Q}$ with permuted axes are no longer equivalent, $S(\bm{Q},\omega)$ is averaged over all equivalent cubic directions in the simulation suppressing and broadening the simulated intensity of the flat modes. As a result, the only substantial intensity is near the acoustic branches, consistent with experiment. The sharp, pronounced peaks near $\sim$4 and $\sim$6 meV in the neutron-weighted DoS (\cref{fig:summary_fig}) become broad and flat when calculated from the disordered cell. Notably, the neutron-weighted DoS curve calculated from the disordered cell agrees with the experimental data much better than the simulation without disorder.

In Ba$_{8}$Ga$_{16}$Ge$_{30}$, an avoided crossing between an acoustic phonon mode and the flat guest modes was theoretically predicted and it was claimed that the Ba guest atoms behave as local resonating scattering centers for the acoustic phonon modes in the region of the avoided crossing \cite{dong2001theoretical}. Subsequent triple axis neutron scattering measurements along the [h h 0] direction \cite{christensen2008avoided} found that the Ba rattler atoms flatten the phonon dispersions, reducing the group velocity $v$, rather than acting as a local resonant scattering center and reducing $\tau$.
In our data focusing on the [h 0 0] direction large-gap avoided crossings with different well separated rattler atom branches are apparent in the LA phonon dispersion (\cref{fig:dispersions}a).
Meanwhile, the intensity in \cref{fig:dispersions}b depicts a TA mode that is linear and continuously dispersing in both the experiment and in the simulation based on disordered calculation. However, my DFT calculations predict many 'small-gap' avoided crossings near $\sim$4 and $\sim$6 meV, that are washed out by the instrument resolution both in the experiment and in the simulation for both the TA branch and, away from the large-gap avoided crossings, the LA branch.

Ga and Ge differ in nuclear-charge and mass by only $\sim3.5\%$, so the perturbation to the force-constants induced by mass or nuclear-charge disorder is insignificant. However, the bonds formed with Ga and Ge have different valence configurations. The electronic structure, which affects the Born-Oppenheimer electronic-energy part of the interatomic force constants, is thus qualitatively different between the ordered and disordered configurations.  The impact of disorder can be seen in the substantially lowered average group velocities due to small-gap avoided crossings, between $\sim 2$ and $\sim 4$ meV in \cref{fig:summary_fig}). See the next section for a heuristic explanation of inhomogeneity broadening. Regardless of any reductions in $\tau$ (which are probably also present \cite{ikeda2019kondo}), the velocity $v$ in the kinetic theory model is substantially reduced, which might explain the anomalously low thermal conductivity in Ba$_{8}$Ga$_{16}$Ge$_{30}$. 

\subsection{Inhomogeneity broadening}\label{sec:}

The large-gap avoided crossings apparent in the experimental data already flatten the acoustic modes, which reduces the average group velocity. These avoided crossings are due to the presence of the Ba rattler atom and are not from disorder \cite{christensen2008avoided}. This flattening would already contribute to lowering the lattice thermal conductivity in Ba$_{8}$Ga$_{16}$Ge$_{30}$. Moreover, my calculations suggest that there may be more avoided crossings present in the material due to the splitting of the rattler modes than are apparent in the experimental INS spectra. There are no splittings visible in the either the experimental or (with reasonably assumed broadening) theoretical INS spectra. Sufficiently "small-gap'' avoided crossings can not be resolved by the experiment.  

\begin{figure}[t!]
\centering
\includegraphics[width=0.5\linewidth]{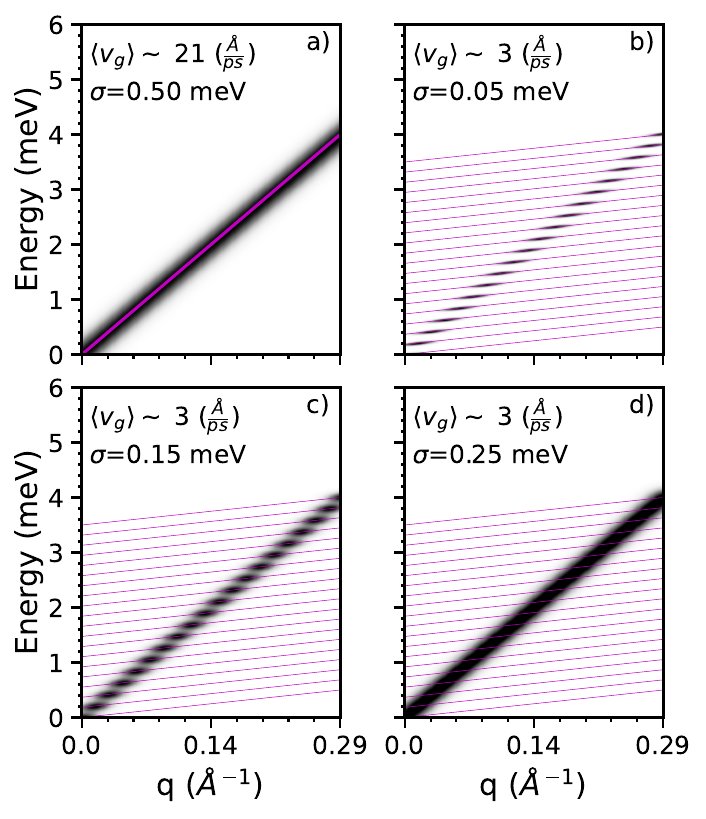}
    \caption{Heuristic figure indicating how an acoustic phonon with many avoided crossings can appear to be continuously dispersing. a) shows a model acoustic phonon with group velocity $\sim$4 ($\textrm{\AA} /s$), comparable to an acoustic phonon in the main text. b)-d) show closely spaced model optical phonons with intensity only in the region of the acoustic phonon in a). These branches are comparable to the split rattler atom modes in the. The colormaps are structure factors broadened with the FWHMs (indicated in the figures as $\sigma$). With relatively small 0.25 meV broadening in d), the small splittings in c) are already concealed and the intensity is indistinguishable from a continuously dispersing acoustic phonon. However, the average group velocity in d) is 7 times smaller than in a).}
    \label{fig:heuristic}
\end{figure}

These additional avoided crossings would significantly reduce the average group velocity of the acoustic phonons beyond what is apparent by naively fitting a straight line (\cref{fig:heuristic}). For sufficiently small gaps and moderate line widths, acoustic branch like intensity composed of numerous avoided crossings between split optical phonons would appear identical to a continuous, linear dispersing acoustic branch (\cref{fig:heuristic} a) and d)). The intensity in \cref{fig:heuristic} is from a model acoustic phonon with group velocity $\sim$21 ($\textrm{\AA}$/ps). The intensity in \cref{fig:heuristic} d) is from numerous model optical branches with large structure factors only in the vicinity of the acoustic phonon in a). Importantly, the model INS intensity in d) is indistinguishable from a), while the average group velocity in d) ($\sim3~\textrm{\AA}$/ps) is 7 times smaller.

Since we cannot observe the splittings in the experimental spectra, it is not clear that the additional small-gap splittings predicted by DFT in the disordered phase are present in the experimental spectrum. However, the significantly improved agreement of the disordered phase calculation with experiment over the ordered phase calculation is a strong indication that the splittings are there. Moreover, as explained above in \cref{fig:const_Q_subset}, averaging over disordered configurations suppresses and broadens the rattler atom intensity, consistent with the disordered calculation and experiment. My DFT calculations show that additional splittings due to disorder are present in both the LA and TA phonons along both the [h 0 0] directions in this work and the [h h 0] directions in ref. \cite{christensen2008avoided}. If the splittings are present, the severely reduced average group velocity of acoustic bands away from the already flattened large-gap avoided crossing regions could explain the anomalously low thermal conductivity in Ba$_{8}$Ga$_{16}$Ge$_{30}$.

\section{Conclusion}

To conclude, my calculations using the disordered unit cell validated by experiments show that Ga/Ge occupational disorder has a large effect on phonon dispersions through strong influence on electronic structure that underlies interatomic force constants. The disorder introduces many closely spaced optic branches with many more avoid crossings with acoustic modes. These small gaps are essential for understanding the observed low thermal conductivity of Ba$_{8}$Ga$_{16}$Ge$_{30}$ and other similar materials. 

\section{Acknowledgements}

All work at the University of Colorado was supported by U.S. Department of Energy, Office of Basic Energy Sciences, Office of Science, under Contract No. DE-SC0006939. EST acknowledges the support from NSF DMR 1555340. We thank the ISIS Facility for beam time RB1410509. We would like the thank Holger Euchner and Silke Paschen for providing us with the DFT force-constants from ref. \cite{ikeda2019kondo}.

\chapter{Summary and outlook}
\label{chp:outro}

In this thesis, I have endeavoured to show how neutron scattering can be used to investigate the physics of energy materials. Neutrons scatter from the atoms in a crystal and the distribution of scattered neutrons depends on the structure and dynamics of the atoms. The dynamics of the atoms is relevant to practically all properties that make energy materials useful. Understanding the scattering distribution enables us to understand the dynamics of atoms. Progress understanding the scattering is made with computational methods: we compared calculations of the scattering distribution to experiment to analyze the corresponding real space behavior of the atoms.

I applied these methods to several interesting energy materials: cuprate La$_2$CuO$_4$ (LCO) \cite{sterling2021effect}, solar (hybrid) perovskite CH$_3$NH$_3$PbI$_3$ (MAPI) \cite{weadock2023nature,sterling_pynamic}, and thermoelectric clathrate Ba$_{8}$Ga$_{16}$Ge$_{30}$ (BGG) \cite{roy2023occupational}. In all cases, anomalies in the neutron scattering data were explained using computational methods, advancing the state of knowledge about each material. Hopefully the reader agrees that neutron scattering combined with computation is a powerful method to study energy materials. Further support is provided by several other successful projects, either related to these \cite{sapkota2021reinvestigation,ahmadova2020phonon} or not \cite{sterling2024structural,sukhanov2020lattice}, that these tools have been applied to but have been omitted from this thesis. I am also still working on ongoing projects that have been spurred by the methods or research presented here. 

A secondary goal of this thesis was to serve as a "manual" for using the methods presented here. In learning these methods, I developed a substantial set of notes that are presented in \cref{chp:neutrons,chp:electrons,chp:phonons}. I hope to have my own PhD students one day and believe that these tools, notes, and the resulting expertise will be invaluable for me to help my students become successful, just as those I have learned from have helped me.


\begin{singlespacing}
\printbibliography[
  title={References},
  heading=bibintoc,
]
\end{singlespacing}


\end{document}